\documentclass[12pt,vi,twoside]{mitthesis}
\usepackage[T1]{fontenc}
\usepackage{tocloft}
\usepackage{pifont}
\usepackage{marvosym}
\usepackage{lgrind}
\usepackage{color}
\usepackage{amsbsy}
\usepackage{amstext}
\usepackage{amssymb}
\usepackage{mathrsfs}
\usepackage{graphicx}
\usepackage{longtable}
\usepackage{acronym}
\usepackage{etoolbox}
\preto\chapter\acresetall
\usepackage[toc,page]{appendix}
\usepackage[skip=5pt]{caption}
\usepackage{dcolumn}
\usepackage{bm}
\usepackage{mcite}
\usepackage[textfont=scriptsize]{subfig}
\usepackage{placeins}
\usepackage{threeparttable}
\usepackage{lastpage}
 \usepackage{epigraph}
 \usepackage{multicol}
 \usepackage{blindtext}
\DeclareGraphicsExtensions{.jpg, .pdf, .png}
  \usepackage[Bjornstrup]{fncychap}
  \definecolor{royalpurple}{RGB}{102,51,153}
   \ChNumVar{\fontsize{76}{80}\usefont{OT1}{pzc}{m}{n}\selectfont\color{royalpurple}}
   \ChTitleVar{\raggedleft\Large\sffamily\bfseries\selectfont}
\usepackage{fancyhdr}

\usepackage[super,comma,numbers,compress]{natbib}
\usepackage[unicode=true,
 bookmarks=true,bookmarksnumbered=false,bookmarksopen=false,
 breaklinks=false,pdfborder={0 0 0},backref=false,colorlinks=true]
 {hyperref}
\hypersetup{pdftitle={\emph{ab initio} Calculations of Optical Properties of Clusters: A Ph.D thesis submitted to IIT Mumbai},
 pdfauthor={Ravindra Shinde},
 pdfsubject={Computational Condensed Matter},
 unicode=false,
 pdftoolbar=true,
 pdfmenubar=true,
 pdffitwindow=true,
 pdfstartview={FitH},
 pdfnewwindow=true,
 pdfcreator={RevTeX},
 linkcolor=[rgb]{0.0,0.2,0.0},
 citecolor=blue,
 urlcolor=[rgb]{0.18,0.03,0.33}}

\usepackage{wallpaper}
\usepackage{natmove}
\usepackage{array}

\makeatletter
\makeatletter
\newcommand*{\citenst}[2][]{%
  \begingroup
  \let\NAT@mbox=\mbox
  \let\@cite\NAT@citenum
  \let\NAT@space\NAT@spacechar
  \let\NAT@super@kern\relax
  \renewcommand\NAT@open{[}%
  \renewcommand\NAT@close{]}%
  \citep[#1]{#2}%
  \endgroup
}
\makeatother

\begin{document}
\pagenumbering{roman}
%
%
%
%
%
%
%
%
%
%
%

\title{\emph{Ab initio} Calculations of Optical Properties of Clusters}

\author{\href{http://home.iitb.ac.in/~ravindra.shinde}{Ravindra Shinde}}
\department{Department of Physics}

\degree{Doctor of Philosophy \and Master of Science}

\degreemonth{February}
\degreeyear{2014}
\thesisdate{\today}

\supervisor{Alok Shukla}{Professor}


\maketitle



\pagestyle{empty}
\cleardoublepage
 \pagestyle{empty}








\begin{center}
 {\bf {\Huge Abstract}}
\end{center}
Atomic and molecular clusters have been a topic of great interest for last few decades, mainly because of their 
unusual properties, tunability and vast technological applications. In this thesis, we have explored the optical properties of 
few clusters using first principles calculations. \par

 We have performed systematic large-scale all-electron correlated calculations
on boron B$_{n}$, aluminum Al$_{n}$ and magnesium Mg$_{n}$ clusters (n=2--5), to study their linear optical
absorption spectra. Several possible isomers of each cluster were considered, and their geometries were optimized at the coupled-cluster
singles doubles (CCSD) level of theory. Using the optimized ground-state geometries, excited states of different clusters were computed using
the multi-reference singles-doubles configuration-interaction (MRSDCI) approach, which includes electron correlation effects at a sophisticated
level. These CI wavefunctions were used to compute the transition dipole matrix elements connecting the ground and various excited states
of different clusters, eventually leading to their linear absorption spectra. The convergence of our results with respect to the basis
sets, and the size of the CI expansion was carefully examined. \par

Isomers of a given cluster show a distinct signature spectrum, indicating a strong-structure
property relationship. This fact can be used in experiments to distinguish between different isomers of a cluster. Owing to the sophistication of
our calculations, our results can be used for benchmarking of the absorption spectra and be used to design superior time-dependent density
functional theoretical (TDDFT) approaches. The contribution of configurations to many-body wavefunction of various excited states 
suggests that in most cases optical excitations involved are collective, and plasmonic in nature. \par

In addition, we have calculated the optical absorption in various isomers of neutral boron B$_{6}$ and cationic boron B$_{6}^{+}$ clusters using 
computationally less expensive configuration interaction singles (CIS) approach, and benchmarked these results against more sophisticated 
equation-of-motion (EOM) CCSD based approach. In all closed shell systems, a complete agreement on the nature of configurations 
involved is observed in both methods. On the other hand, for open-shell systems, minor contribution from double excitations are observed, 
which are not captured in the CIS method. \par

Optical absorption in planar boron clusters in wheel shape, B$_{7}$, B$_{8}$ and B$_{9}$ computed using EOM-CCSD approach, have been 
compared to the results obtained from TDDFT approach with a number of functionals. This benchmarking reveals that range-separated 
functionals such as $\omega$B97xD and CAM-B3LYP give qualitatively as well as quantitatively the same results as that of EOM-CCSD.
\cleardoublepage

\pagestyle{plain}
\tableofcontents
\newpage

\chapter*{List of Acronyms}
\vspace{-8ex}
\addcontentsline{toc}{chapter}{List of Acronyms}
\begin{acronym}[EOM-CCSD   ]
\setlength{\parskip}{0ex}
 \setlength{\itemsep}{0.5ex}
 \acro{HOMO}{Highest Occupied Molecular Orbital}
 \acro{LUMO}{Lowest Unoccupied Molecular Orbital}
 \acro{SOMO}{Singly Occupied Molecular Orbital}
 \acro{SCF}{Self-Consistent Field}
 \acro{HF}{Hartree-Fock}
 \acro{RHF}{Restricted Hartree-Fock}
 \acro{ROHF}{Restricted Open Shell Hartree-Fock}
 \acro{UHF}{Unrestricted Hartree-Fock}
 \acro{CI}{Configuration Interaction}
 \acro{FCI}{Full Configuration Interaction}
 \acro{CIS}{Configuration Interaction Singles}
 \acro{CISD}{Configuration Interaction Singles Doubles}
 \acro{MRSDCI}{Multi-Reference Singles Doubles Configuration Interaction}
 \acro{DFT}{Density Functional Theory}
 \acro{LDA}{Local Density Approximation}
 \acro{GGA}{Generalized Gradient Approximation}  
 \acro{TDDFT}{Time-Dependent Density Functional Theory}
 \acro{ALDA}{Adiabatic Local Density Approximation}
 \acro{CCSD}{Coupled Cluster Singles Doubles}
 \acro{EOM-CCSD}{Equation-of-Motion Coupled Cluster Singles Doubles}
 \acro{NTO}{Natural Transition Orbitals}
 \acro{MP4}{M\o{}ller - Plesset Perturbation Theory 4$^{th}$ order}
\acro{ARPES}{Angle-Resolved Photo-Emission Spectra}
\end{acronym}

 	\pagestyle{fancy}
 	\renewcommand{\sectionmark}[1]{\markright{\thesection\ #1}}
 	\fancyhf{}
 	\lhead{\nouppercase{\leftmark}}
  	\rhead{\href{home.iitb.ac.in/~ravindra.shinde}{ Ravindra Shinde}}
 	\cfoot{Page \thepage\ of \pageref{LastPage}}
\pagenumbering{arabic}
\hypersetup{linkcolor=black}
  \chapter{\label{chap:main-intro}Introduction}
\vspace{-4em}
\begin{figure}[h!]
 \centering
 \includegraphics[width=15cm]{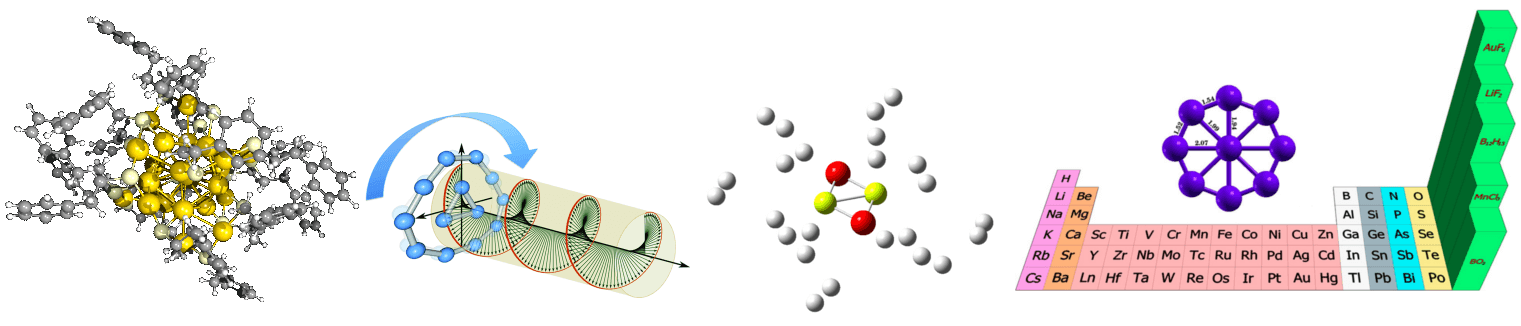}
 \end{figure}

Clusters are nothing but a collection of atoms. Even in the medieval age, people had used the so-called clusters to make colored glasses, 
without any scientific knowledge. The number of atoms can vary from the lowest possible value of two to tens or hundreds 
of thousand atoms. These species bridge the gap between atoms and their respective bulk systems. There has been tremendous progress in the scientific 
exploration of properties of these clusters, especially in the recent few decades. \cite{julius_book, clustnano_book, alonso_book, johnston_book, 
nanoclusters_bridge_jena_book, deheer_rmp} Interestingly, the properties exhibited by clusters are often 
different from that of their bulk counterparts. Also, clusters offer a great tunability or tailoring the properties of materials, which is 
otherwise not possible in simple molecules. Owing to tremendous tunability of properties, clusters are favored in technological applications. 
A plethora of synthetic molecules can be explored to investigate science, which is otherwise difficult with normal elements. 
Many clusters have ability to store hydrogen molecules, thereby suggesting the possibility of solid-stage hydrogen energy storage devices. \cite{chemphys-li2o} 
Clusters are also promising candidates as catalysts. \cite{jena-pnas-castleman-clusters} Gold-coated silica nanoparticles have been found out 
to be useful in bioscience, as they absorb infrared light enough to locally destroy the cancer cells. \cite{application_cancer_pnas,application_cancer_prl} 
For certain types of cancer, boron neutron capture therapy is used, which involves capture of thermal neutrons by boron nuclei $^{10}$B. Instead 
of administering boron to the tumor via conventional boron compounds, boron clusters are used, as they offer higher cell selectivity. \cite{kiran_wang}
Some clusters mimic the properties of elements in the periodic table. Hence, in the future they can be used instead of real elements, whose supply is 
ending. \cite{superatom_perspective_jpcl} But what makes the clusters different from very well-known molecules? 
In principle, all molecules are clusters, but the reverse is not always true. In spite both being a 
collection of atoms, clusters are generally metastable as compared to molecules at ambient conditions. Molecules have a well-defined 
stoichiometry, whereas clusters' composition depends on production conditions. Clusters tend to coalesce when brought in close vicinity 
of each other, and they often react with ambient gases.

\section{Atomic and Molecular Clusters}
Since clusters are constituted of atoms, a natural question arises: when will a cluster behave as bulk material of parent atoms? Owing to large
number of electrons, bulk systems form bands of energy, whereas in atoms, energy levels are discrete. In case of clusters, these energy levels are 
neither too discrete nor do they form bands. The size at which the properties of clusters will approach their bulk counterparts, may depend 
upon which property is being investigated. The unusual electronic structure of clusters is due to the quantum confinement of electrons belonging to
molecular orbitals. The energy gap between \ac{HOMO} and \ac{LUMO} determines the various properties of clusters and their stability. The 
magnitude of this gap varies with size and composition of the cluster, and how the molecular orbitals are occupied by electrons. 
\subsection{Physical Properties}
The evolution of electronic structure of various clusters was studied rigorously.\cite{structure_evolution_TiN_jcp, structure_evolution_WO_jcp,
structure_evolution_metallicity_arpc, na_dehaar_prl, rao_jena_ele_struct_al, evolution_magnesium_prl,evolution_magnesium_jellinek_prl,
evolution_nickel_jpca, evolution_be_clusters_jcp} For example, how metallic are the small sodium clusters or, at which size does 
non-jellium to jellium transition occur in aluminum clusters, are few topics to mention. However, there is no single answer to these questions 
in general because different materials display different evolution pattern. Also, the evolution trend too is different for different properties 
under consideration. Most of the covalently bonded carbon or silicon clusters form icosahedral structures. This five-fold symmetric nature is never seen in the bulk 
systems. Fullerene, for example, has buckyball structure, but its bulk system graphite or diamond have completely different structures. 
On the other hand, in many ionically bonded systems such as alkali halides and metal carbides, nitrides and oxides show 
symmetry similar to that of their bulk crystalline structures.\cite{structure_evolution_TiN_jcp, structure_evolution_WO_jcp} Evolution of electronic structure
is relatively easier to define in metal clusters. Several groups have proposed different criteria to address this issue. For example, 
von Issendorff and Cheshnovsky suggested that the clusters can be considered metallic when the gap between occupied 
and unoccupied states at Fermi energy is consistently smaller than or equal to the Kubo band gap. \cite{structure_evolution_metallicity_arpc}
On the other hand, Walt de Heer studied the ability of sodium clusters to screen electric fields as a criterion for evolution of metallicity. \cite{na_dehaar_prl}
Rao and Jena studied evolution of various properties such as binding energy, relative stability, fragmentation channels, ionization potential and 
vertical and adiabatic electron affinities of neutral and cationic clusters of aluminum as a function of size. \cite{rao_jena_ele_struct_al} 
The \emph{s--p} band gaps were observed in anionic magnesium cluster at size n = 18, which suggests metallic behavior. 
\cite{evolution_magnesium_prl,evolution_magnesium_jellinek_prl} In case of nickel clusters, binding energy per atom increases
 monotonically, but the clusters does not mimic the bulk structure. \cite{evolution_nickel_jpca} The onset of bulk behavior is observed 
at different sizes of beryllium clusters. The interatomic distance rapidly approaches the bulk value, but binding energies and ionization 
potential show a slow evolution towards cohesive energy and work function respectively. \cite{evolution_be_clusters_jcp}

An experimental mass spectra of Na clusters revealed that there are pronounced peaks for particular number of sodium atoms in a cluster. 
\cite{shell_structure_sodium_prl} The numbers for which the cluster was relatively stable resembled to that of 
nuclear shell fillings of 2, 8, 20, 40, \textit{etc}. On similar lines, a jellium model was proposed in which electronic charges 
are taken as a uniform quantity spread evenly in space as do the positive background of atomic nuclei. 
The same model was applied to sodium clusters by considering a sphere of uniform positive background charge density and valence 
electrons fill the energy levels.  It successfully showed that clusters containing 2, 8, 20, 40 electrons are very stable as they complete the shell.
As Na atom has one valence electron, it was predicted that cationic clusters with 1, 3, 9, 21, \textit{etc}. will have pronounced stability. This fact was later 
confirmed and also established that the stability of clusters can be altered by changing the number of valence electrons. 
\cite{exp_shell_structure_na_prb} Also, a bigger stable or magic cluster will fragment in such a way that the fragments will again 
be smaller magic clusters. \cite{fragmentation_jena_prl, fragmentation_exp_prl}

Even though many elements in the periodic table have partially filled valence orbitals, not all of them show magnetism. It can be well understood by
knowing how net spin magnetic moments couple each other. Clusters offer a great flexibility of studying this phenomenon as it allows to change 
the its size as well as geometry. Rao and Jena predicted that geometry will play a role in determining the magnetism in lithium clusters. 
\cite{magnetism_topology_prb1,magnetism_topology_prb2} For up to five atom cluster of lithium, planar geometry is the most 
stable one. However, they are relatively less magnetic as compared to three-dimensional clusters as governed by Hund's rule. 
Clusters of another set of non-magnetic transition elements such as V, Rh and Pd can also show magnetism. \cite{magnetism_vanadium_prb, 
magnetism_4d_clusters_prl, magnetism_exp_rho_prl, magnetism_palladium_prl} A highly symmetric rhodium cluster showed giant magnetic moments mainly due to 
enhanced electronic degeneracy caused by symmetry.\cite{magnetism_exp_rho_prl}
 The antiferromagnetic bulk manganese shows ferromagnetism in small clusters. \cite{magnetism_exp_mn_prl} 
Clusters of magnetic elements not only exhibit superparamagnetism, but their magnetic moments are also larger than that 
of bulk values. \cite{magnetism_exp_co_prl,magnetism_exp_ni_prl,magnetism_ferro_science,magnetism_iron_prl}

Since clusters usually have high surface-to-volume ratio, their melting points should be lower as compared to bulk. This can be understood from
the fact that surface atoms will have less coordination number, which melts much earlier than the core part. This phenomenon was widely 
studied and was confirmed in many cases. \cite{nature_melting,nature_size_selected} However, small clusters of gallium behave differently.
These small Ga clusters have melting points much higher than their bulk counterparts. \cite{melting_hot_solid_prl, melting_kanhere_ga_prl, 
melting_kanhere_magic_prl} The purely covalent nature of bonding between gallium atoms in the cluster as compared to
 covalent-metallic bonding in bulk, is responsible for such an anomaly.

A phenomenon analogous to thermionic emission in bulk systems can also be seen in clusters. Under certain circumstances, the ionization of 
clusters is delayed. This study also can help in understanding the evolution of cluster properties towards bulk. Two conditions must be met in order to 
observe the delayed ionization; first, the ionization potential of the cluster must be less than its dissociation energy, and cluster should be able 
to access vibrational and rotational states to store the energy in excess of the cluster's ionization potential. The former condition favors 
ionization over dissociation, and is met in many systems such as, C$_{60}$ fullerene and transition metal carbides and oxides.
\cite{delayed_ionization_TM_jcp, delayed_ionization_zirconium_jcp, delayed_ionization_Ti_V_cpl}

\subsection{Chemical Properties}
One of the many interesting properties of clusters is, organic molecules can also bind to various sites of inorganic or metal clusters. It can have 
metal atoms, metal clusters and metal surfaces binding to various organic molecules. A wealth of information is now available in the field of 
organometallics. \cite{organometallics_benzene_organo, organometallics_acetylene_jpc, organometallics_ethylene_jpc, organometallics_benzene_jpc, 
organometallics_benzene_cpl, organometallics_benzene_jacs, organometallics_benzene_jcp, organometallics_thiol_hakkinen_pnas}
In most of the cases, transition metal clusters are passivated by such organic molecules to achieve exceptional stability. 
Favorite docking positions of metal atoms on a given organic molecule, or changes in the structure of clusters 
as multiple organic molecules attach to metal clusters are few examples that have been studied rigorously. For example, structures of 
various 3d transition metal atoms such as Sc, Ti, V, Cr, Mn, Fe, Co and Ni attached to benzene ring or 
coronene (a benzene ring surrounded by another six benzenes) were studied.\cite{organometallics_benzene_organo, organometallics_coronene_jpca} 
The mass spectra of $M_{n}(Bz)_{m}^{+}$ revealed that only those structures are favored with m= n + 1 for $M=$ Sc, Ti, V. For Cr and Mn, a
single largest peak corresponding to (n=1, m=2) is observed. This indicates that transition metal is sandwiched between stacking of benzene rings.
Also, the number of metal atoms can exceed the number of benzene rings, but the maximum number of benzene rings in a stable 
cluster seldom exceed four. Because of such intercalation, the reactivity of transition metal decreases. Magnetism in such organometallic complexes
have also been found out to be unusual. Magnetic dipole moments of free atoms of Sc, V, Ti, Cr, Mn, Fe, Co and Ni are 1$\mu_{b}$, 2$\mu_{b}$,
3$\mu_{b}$, 6$\mu_{b}$, 5$\mu_{b}$, 4$\mu_{b}$, 3$\mu_{b}$ and 2$\mu_{b}$ respectively. When these atoms are supported on benzenes,
the magnetic moments change dramatically. \cite{organometallics_benzene_cpl, organometallics_benzene_jacs, organometallics_benzene_ni_jcp}
Magnetic elements (Fe, Co and Ni) exhibit reduced magnetic moments whereas Sc, V, Ti show enhanced moments. Magnetism in Cr stays unchanged.
This peculiar behavior suggests that magnesium in organometallic systems is greatly influenced by supporting molecules.

Certain clusters have such an electronic structure that they can be considered an artificial element, which mimics the physics and chemistry of 
a particular element in the periodic table. This is possible because many properties of cluster depend upon their size, shape, composition and charge.
There have been a lot of theoretical predictions backed by experimental evidence that, clusters behave as atoms. Such clusters are called 
superatoms which serve as building blocks of new three-dimensional periodic table. \cite{superatom_perspective_jpcl} 
Castleman and co-workers observed that Al$_{13}^{-}$ has 
very less reactivity than its neighboring clusters. \cite{superatom_castleman_jcp} Since Al atom has three valence electrons, Al$_{13}$ will have 
39 electrons, which is one short of magic abundance number 40, making electron affinity of Al$_{13}$ very large. It was proposed that this 
cluster can form salt by combining it with alkali metal, just as a normal salt. \cite{superatom_al13_khanna_cpl} This was experimentally 
confirmed by Wang\cite{wang_sp_hybrid_prl} and Bowen\cite{superatom_al13_bowen_jcp} and their co-workers. Hence, Al$_{13}$ became 
the first superatom, or rather superhalogen to mimic an element in the Mendeleev's periodic table. This led to an enormous exploration of 
possibilities of various giant atoms. Li$_{3}$O cluster has ionization potential (3.54 eV), lower than that of any alkali metal, and H$_{12}$F$_{13}$ 
recorded highest electron affinity (13.87 eV), higher than any halogen. \cite{superatom_lowest_ionization_jcp, superatom_largest_affinity_cpl}
Some boron clusters mimic the properties hydrocarbons\cite{kiran_wang} while thiol protected gold cluster 
[Au$_{25}$(SR)$_{18}$]$^{-}$ behaves as noble gas. \cite{superatom_thiol_nobel_jacs} Clusters consisting of all-inorganic elements can be used
as ligands. \cite{coord-chem-review,boron-8-ligand} 

Although jellium model is successful in describing the magical stability of alkali metal clusters, it cannot be applied to study the stability 
of covalently bonded systems, such as fullerene or planar boron clusters. However, for these systems a simple electron counting rule can give a
great insight into the stability. The H\"{u}ckel rule says that, if the system has delocalized $\pi$ electrons, and if they are 
equal to 4n + 2 ( n = 0, 1, 2 ... ), then the system is said to be aromatic and will be extra stable. If it is equal to 4n, then it is called antiaromatic 
and will destabilize the system. A famous example of this is benzene, which has 6 $\pi$ electrons, and is aromatic. A planarity is also implied 
by the H\"{u}ckel rule for aromaticity. This rule is successful applied to a large number of carbon- and boron-based clusters, and are found to be
aromatic. \cite{kiran_wang, wheel_wang, ring_wang, tube_wang, aromatic_toroid,ishida_jacs,borozene_sahu,coord-chem-review} 
Boron clusters B$_{n}$ (n $\leq$ 20) prefer to be planar, and are governed by aromatic nature. A three-dimensional structure of B$_{12}$ also
shows enhanced stability mainly because of largest \ac{HOMO}--\ac{LUMO} gap, and the most stable isomer of B$_{12}$ -a planar structure- 
having 6 $\pi$ electrons resembles to that of benzene. Based on the H\"{u}ckel rule, several metallic clusters are also found to exhibit 
aromaticity. Al$_{4}^{2-}$, for instance, is aromatic and square-planar owing to two $\pi$ electrons, whereas Al$_{4}^{4-}$ with four $\pi$ electrons is
antiaromatic. \cite{aromaticity_all_metal_science, aromaticity_al_li_science} An aromatic, planar boron cluster having wheel shape rotates
when shined by a circularly polarized light. \cite{wankel-motor-heine, alina-b13-wankel} This can be termed as the smallest aromatic nano-motor.

\subsection{Optical Properties}
Optical response of clusters can provide an insight into their electronic structure. An analogy is seen between photonuclear processes and optical 
responses of metal clusters. With valence electrons moving collectively against the jellium background, the photoabsorption in metal clusters 
exhibits excitation in dipole plasma mode. This is analogous to giant dipole resonance occurring in nuclei. \cite{optical_yanno_broglia_pra,
optical_yanno_jena_prb, optical_yanno_pacheco_prb} Mie theory of charge oscillations of classical metal spheres suggests that 
the photoabsorption spectra of alkali metal clusters will have single dominant peak. However, quantum chemical methods and other 
many-body techniques have shown that optical absorption in clusters like neutral Na$_{20}$ and Na$_{40}$ exhibit multi-peaks. \cite{optical_na_deheer}
Hence model calculations are not enough to study the photoabsorption spectra of all metal clusters and, of course, of that covalently bonded clusters.
In past, there have been various experimental as well theoretical studies of photoabsorption in atomic and molecular clusters. 
\cite{na_dehaar_prl,optical_na_deheer,na_opt_kappes_jcp,na_opt_bethe_salp,li_boustani_prb,li_na_tddft} 

Conventional mass spectrometry can distinguish between different clusters only according to their mass, but not according to 
their geometry. One has to rely on other theoretical or experimental data to be able to differentiate one isomer
from another. For example, using first principles calculations of vibrionic fine structure in C$_{20}^{-}$, and comparing it with experimentally
available data, Saito and Y. Miyamoto\cite{saito_fullerene} identified the cage and bowl structures. Optical absorption spectroscopy, coupled
with extensive theoretical calculations of the optical absorption spectra, can be used to distinguish between distinct isomers of clusters
produced experimentally because normally optical absorption spectra are sensitive to the geometries of the clusters.

An accurate description of both ground and excited states can lead to better understanding of photoabsorption processes. Accounting for electron
correlation in the calculation of ground and excited states energies is of paramount importance. Also, the method of calculation should be 
independent of the nature of the system. Such criteria lead to adopting computationally extensive methods --to be presented here-- for precise
 and accurate results. The results can then be used to benchmark other less accurate methods in order to study other larger systems.

To understand the nature of excitation in various photoabsorption spectra of metal as well as covalent clusters, we present a set of studies of 
optical absorption in clusters using state-of-the-art computational techniques. We have mainly used \ac{CI}, and its multi-reference version
\ac{MRSDCI} to compute ground and excited state energies. Unlike \ac{DFT}, this method gives us access to the many-body wavefunction
 of the system, thereby allowing us to compute transition probabilities, and various other expectation values. A large number of reference 
states were employed to incorporate
electron correlation effects. Photoabsorption spectra of small boron, aluminum and magnesium clusters computed using this sophisticated 
method not only revealed the nature of excitations, also it can be used to guide future experiments and theoretical methods. We have also 
computed photoabsorption spectra using a popular \ac{TDDFT} approach and compared it with a single-reference wavefunction-based 
\ac{EOM-CCSD} method. A benchmarking of various \ac{DFT} functionals against \ac{EOM-CCSD} is also carried out. 

\section{Summary} 
We carried out a relatively underexplored yet interesting and useful photoabsorption study of various types of clusters. 
A first principles calculations of optical absorption spectra
can help in identifying different isomers of a cluster. The peculiar behavior of clusters --properties anomalous to that of bulk-- is also observed
in context of photoabsorption. Owing to a better description of electron-correlation effects, these calculations could be treated as 
benchmarks, and be used to design better \ac{TDDFT} approaches. 

We briefly summarize our main results presented in this thesis.
\begin{itemize}
 \item 
Large-scale all-electron correlated calculations of photoabsorption spectra of small boron clusters indicate a strong structure-property relationship. 
The analysis of wavefunctions involved in photoabsorption spectra suggests plasmonic nature of photoexcited states. [Ravindra Shinde and 
Alok Shukla, Nano LIFE, \textbf{02}, 1240004 (2012)]

\item
Several new isomers of neutral and cationic B$_{6}$ clusters were found using coalesce and kick geometry optimization technique. 
\ac{NTO} as well as wavefunction analysis of photoabsorption spectra computed using \ac{CIS} revealed that collective excitation 
take place in open-shell clusters. [Ravindra Shinde and Alok Shukla, Eur. Phys. J. D, \textbf{67}, 98 (2013)]

\item
Photoabsorption spectra of planar boron clusters in wheel shape are computed. Most of the absorption takes place in high-energy range, thereby 
opening a possibility of using these clusters as ultra-violet absorbers. [Ravindra Shinde, Sridhar Sahu and Alok Shukla, \textit{to be submitted}]

\item
A benchmarking of various \ac{DFT} functionals against \ac{EOM-CCSD} for calculating photoabsorption spectra of molecular boron wheel 
clusters is presented. Some functionals exhibit surprisingly similar spectra as that of \ac{EOM-CCSD} with remarkably less computation.
[Ravindra Shinde, \textit{to be submitted}]

\item
Large-scale configuration interaction calculations of photoabsorption spectra of small aluminum clusters indicate collective excitation in 
some isomers of the clusters. These results could serve as theoretical tool to identify various isomers experimentally produced in the 
mass spectra. [Ravindra Shinde and Alok Shukla, \textit{submitted} (arxiv: 1303.2511)]

\item
Large-scale configuration interaction calculations of photoabsorption spectra of magnesium clusters exhibit collective excitation in 
some isomers of the clusters. These results also could serve as theoretical tool to identify various isomers experimentally produced in the 
mass spectra. [Ravindra Shinde and Alok Shukla, \textit{to be submitted}]
\end{itemize}

\section{Outline}
The rest of the thesis is organized as follows. In chapter \ref{chap:main_theory}
, we present theoretical approaches relevant to the calculations of 
ground and excited state energies as well as computation of photoabsorption spectra. We particularly focus on \ac{CI} methods and its variants,
as they do provide us with the main methodology for first principles photoabsorption calculations. Chapter \ref{chap:main_smallboron}
 discusses the results 
of large-scale all electron correlated calculations of photoabsorption spectra of small boron clusters. Convergence of calculations with respect to
basis sets, number of active orbitals and frozen-core approximation is also presented. In chapter \ref{chap:main_boron6}
, \ac{CIS} 
photoabsorption spectra of various isomers of neutral and cationic B$_{6}$ clusters are presented. Method of geometry optimization, \ac{NTO}
analysis, spin contamination is also discussed. The \ac{CIS} optical absorption spectra of few isomers is compared with that of 
obtained from \ac{EOM-CCSD}. Chapter \ref{chap:main_smallal} also presents an \textit{ab initio} all-electron account of photoabsorption spectra of various isomers of 
aluminum clusters obtained using \ac{MRSDCI} method. In chapter \ref{chap:main_benchmarking}
 , we discuss the benchmarking of various \ac{DFT} exchange-correlation 
functionals against \ac{EOM-CCSD} in light of photoabsorption spectra of planar boron clusters B$_{7}$, B$_{8}$ and B$_{9}$ in wheel shape. 
 In chapter \ref{chap:main_magnesium}, we present effect of basis sets and
number of active orbitals on photoabsorption spectra of magnesium dimer, computed using computationally demanding \ac{FCI} method. 
Optical absorption spectra of various isomers of bigger magnesium clusters calculated using \ac{MRSDCI} method, are also presented. 
Finally, in Chapter \ref{chap:main_conclusion}, we summarize our conclusions and discuss future directions. A detailed information about wavefunctions of excited
states contributing to various photoabsorption peaks of the cluster, are presented in Appendix.

  \chapter{\label{chap:main_theory}Theory and Computational Methods}
In this chapter, a brief introduction to the underlying theory and various computational methods is given. We discuss general
electronic structure methods employed here, followed by, various methods used to compute optical properties.

\section{Methods of Electronic Structure Calculations}
 A wealth of information about various properties of atomic and molecular systems can be obtained by solving the Schr\"{o}dinger equation for 
that system. However, this is a daunting task for many of the real systems, as they involve many electrons. First principles electronic 
structure calculations deal with solving such equations without depending on external parameters and using very basic information. Here in this
section, we briefly describe the methods of electronic structure calculations used for studying atomic clusters.

    \subsection{Hartree-Fock Approach}
Atomic clusters are classic examples of a many electron systems, where
we can benchmark the results of the available theoretical methods.
Since our ultimate aim is to find the approximate solutions of the non-relativistic
Schr\"{o}dinger's equation,
\begin{equation} \label{eq:many-body-equation}
\mathscr{H}|\Phi\rangle=\mathscr{E}|\Phi\rangle
\end{equation}
 where \ensuremath{\mathscr{H}} is Hamiltonian operator for the system
of electrons and nuclei and $\Phi$ is the combined many-body wavefunction
of the system. The Born -- Oppenheimer approximation makes this Hamiltonian
separable into two parts - nucleus and electronic. This approximation
rests on the fact that kinetic energy of nuclei is much smaller than
that of the electrons. Hence, effectively, we will be dealing with
following kind of Hamiltonian (in atomic units), 
\begin{equation} \label{eq:born-oppenheimer}
\text{\ensuremath{\mathscr{H}}}_{elec}=-\frac{1}{2}\sum_{i=1}^{N}\nabla_{i}^{2}-\sum_{i=1}^{N}\sum_{A=1}^{M}\frac{Z_{A}}{r_{iA}}+\sum_{i=1}^{N}\sum_{j>i}^{N}\frac{1}{r_{ij}}
\end{equation}
The electronic wavefunction now explicitly depends on electron coordinates
and depends parametrically on nuclear coordinates. 

Let $\psi(r)$ be spatial one-electron wavefunction and ${\alpha(\omega)}$ or ${\beta(\omega)}$ be the spin part of electron wavefunction. 
The combined spatial and spin wavefunction is called spin orbital and is given by ${\chi(\textbf x)}$. 
\begin{equation}
\chi(\mathbf{x})=\psi(\mathbf{r})\alpha(\omega)\label{eq:spin_orbital}
\end{equation}
The ground state of N-electron system can be approximated as a single antisymmetrized product known as Slater determinant. In general, the solution 
will be a linear combination of Slater determinants. It uses single electron wavefunctions (spin orbitals) i.e. $\chi_i(\textbf x_1)$.
\begin{equation} \label{eq:slater_determinant}
\Phi_0{(\textbf x_1,\textbf x_2, ..., \textbf x_N)} ={\frac{1}{\sqrt{N!}}} \left| \begin{array}{cccc}
\chi_i(\textbf x_1) & \chi_j(\textbf x_1) & \ldots & \chi_k(\textbf x_1) \\
\chi_i(\textbf x_2) & \chi_j(\textbf x_2) & \ldots & \chi_k(\textbf x_2) \\
\vdots      & \vdots      &        & \vdots      \\  
\chi_i(\textbf x_N) & \chi_j(\textbf x_N) & \ldots & \chi_k(\textbf x_N) \end{array} \right|
\end{equation}
The rows of the determinant are labeled by electrons, and columns by spin orbitals. Interchanging the coordinates of two electrons corresponds 
to interchanging two rows of the determinant, which changes sign of the determinant. Hence the wavefunction is antisymmetric. 
Since, a determinant with two identical columns is zero, it naturally obeys Pauli exclusion principle.
The variational method is used to choose best set of spin orbitals so as to minimize the ground state energy of many-body system. \cite{szabo_book}

The energy functional is given by,
\begin{equation}
E[\chi] = \sum_i^N{ \langle{i|h|i}\rangle } + \frac{1}{2} \sum_i \sum_j {\left ( {\langle{ij|ij}\rangle } -  {\langle{ij|ji}\rangle } \right ) }
\end{equation}
The ${\langle{i|h|i}\rangle }$ denotes the matrix element of one-electron operator between the spin orbitals and so on.
 This functional is varied till we get lowest energy subjected to the constraint that the spin orbitals remain orthogonal. $\int \chi_i^*(1) \chi_i(1)\,dx_1 = 1$ 
Then we get the Hartree Fock equations,\cite{szabo_book}
 \begin{displaymath}
 h(1) \chi_i(1) +{ \left [ \sum_{i\neq j}^{N} \int dx_2 \chi_j^*(2) \frac{1}{r_{12}} \chi_j(2) \right ] } \chi_i(1)   \quad \quad \quad \quad \quad \quad 
\end{displaymath}
 \begin{equation}
 - \left [ \sum_{i \neq j}^{N} \int dx_2 \chi_j^*(2) \frac{1}{r_{12}} \chi_i(2) \right ]  \chi_j(1)   =  \sum_{j=1}^{N} \varepsilon_{ji} \chi_j(1)  
\end{equation}
 \begin{equation}
 													      f|\chi_{i} \rangle	  = \sum_{j=1}^{N} \varepsilon_{ji} | \chi_j \rangle 
 \end{equation}
The energy functional of a single determinant gives equation in a non standard (non-canonical) form, which can be transformed into a canonical one by
unitary transformations on spin orbitals. These unitary transformations do not alter the total energy Fock operator $f$. The sum of Coulomb integrals and sum of exchange integrals also 
remain invariant. So, after unitary transformation, the Hartree-Fock equation takes the following form.
\begin{equation} \label{eq:canonical-hf}
 f|\chi_{i}^{'} \rangle = \varepsilon_{i}^{'} | \chi_{i}^{'} \rangle 
\end{equation}
The description of electronic wavefunction in terms of a single Slater determinant is equivalent
to saying that the electrons move independently of each other except
the Coulomb repulsion due to average effect of all the electrons (also
exchange interaction due to antisymmetrization). Exchange interaction
prevents the electrons with same spin from occupying the same point
in space. The Density functional theory also has a similar formulation.
So we can say that electrons are correlated to each other in some
way. But are we computing all possible correlation effects? 

The correlation energy is defined as,
\begin{equation}
 E_{corr} = \mathscr{E}_{0} - E_{HF}
\end{equation}
where, $E_{HF}$ is energy in the Hartree-Fock limit and $\mathscr{E}_{0}$ is exact non-relativistic energy of the system. 
Since $E_{HF}$ is always an upper bound to $\mathscr{E}_{0}$, the correlation energy is always negative. Since Hartree-Fock treats 
inter-electron repulsion in an averaged manner, one would expect the magnitude of correlation energy going down, when atoms in a molecule 
are pulled apart. However, this is not always true. In the case of water molecule, the electron correlation energies (for DZ basis set) 
increases when H -- O bonds are stretched.\cite{harrison95}. So, the nature of correlation energy not only lies in the electrons ``avoiding'' each
other, but also has some intrinsic nature. The former case accounts for the so-called dynamical electron correlation, and latter is termed as
``static'' or non-dynamical one. Hartree-Fock formulation takes just one Slater determinant in account, it
implicitly assumes that this single reference configuration is the only dominant term in the expansion of the wavefunction, and it fails
when it is not. For example, restricted Hartree Fock (closed shell) cannot describe the dissociation of a molecule in two open shell fragments.
An open shell HF does it, incorrectly. Methods which are based on single reference description, such as single-reference perturbation theory,
DFT, coupled cluster may fail at such points. It cannot also describe rearrangements of electrons in partially filled shells. The failure to account 
for the static electron correlation effects by these single reference electronic structure methods demands multi-reference description of the 
molecular state.

 \subsection{Configuration Interaction Approach}

In general, instead of representing the N-electron wavefunction by
a single determinant, we can expand the wavefunction in all possible
determinants formed from complete set of spin orbitals. And if the
set of spin orbitals is complete then we can get the exact ground state of the
system. Each such Slater determinant is called a configuration.
Since a combination of such determinants are taken, the method
is called Configuration Interaction. 

Let N-electron basis functions be denoted by $|\Phi_{i}\rangle$,
the eigenvectors of \ensuremath{\mathscr{H}} can be expressed as, 
\begin{equation}
|\Psi_{j}\rangle=\sum_{i}^{L}c_{ij}|\Phi_{i}\rangle
\end{equation}

But in practice, the number of N-electron basis functions are finite. 
A Hamiltonian matrix is constructed as $H_{ij}=\left\langle \Phi_{i}|H|\Phi_{j}\right\rangle $,
and is diagonalized for obtaining eigenstates and eigenvalues. The
above wavefunction can also be written in terms of the N-electron
basis functions expressed as excitations or substitutions from the
reference Hartree-Fock Determinant, \emph{i.e.}
\begin{equation} 
\label{eq:ci-expansion}
|\Psi\rangle=c_{0}|\Phi_{0}\rangle+\sum_{ra}c_{a}^{r}|\Phi_{a}^{r}\rangle+\sum_{r<s,a<b}c_{ab}^{rs}|\Phi_{ab}^{rs}\rangle+\sum_{r<s<t,a<b<c,}c_{abc}^{rst}|\Phi_{abc}^{rst}\rangle+\cdots
\end{equation}

In $|\Phi_{a}^{r}\rangle$, a spin orbital $a$ is replaced by spin orbital $r$ in reference configuration $|\Phi_{0}\rangle$. \cite{szabo_book}
So all such configurations made up of substitutions from reference Slater determinant constitutes ``Configuration Interaction''. Since we
are making the basis set bigger by considering all possible configurations, the system can now be accurately described. It should give exact results
if we take into account all the terms in the above expansion. This formalism is also applicable to excited states, open shell systems
and systems far from equilibrium geometries, in contrast to the other non-variational single reference approaches. 
  
 \subsubsection{Full Configuration Interaction}
If we take into account all possible N-electron basis functions \{$\Phi_{i}$\}
with given set of one-particle functions \{$\chi_{i}(\mathbf{x})$\},
then the procedure is called \ac{FCI}. It will be an exact solution
to the Schrodinger's equation within the space spanned by the specified
one-electron basis. And if that one electron basis forms a complete
set (not possible practically), then the method is called Complete-CI.
Even after restricting the one particle basis set to a small number, the
number of possible determinants in \ac{FCI} are astronomically large. 
The number of determinants (considering the spin symmetry and ignoring the spatial symmetry) to be included
are given by, 
\begin{equation} \label{eq:fullci}
D_{mN_{\alpha}N_{\beta}}={}^{m}C_{N_{\alpha}}  \times  {}^{m}C_{N_{\beta}} 
\end{equation}
So for Boron dimer (10 electrons) with single electron wavefunction
expanded in AUG-CC-PVDZ basis set, this number comes out to be of
the order $10^{12}$. Hence the Hamiltonian matrix to be diagonalized
is of the order $10^{12}\times10^{12}$. The situation worsens for
bigger systems. Hence, the applicability of this method is very limited. 

 \subsubsection{Configuration Interaction Singles and Doubles}
One can terminate the Eqn. \ref{eq:ci-expansion} according to the excitation level. Suppose we only allow only single or double
virtual excitations or substitutions from reference determinant, then the corresponding \ac{CI} is called \ac{CIS} or \ac{CISD} respectively.
Owing to Brillouin's theorem, $\langle \Phi_{a}^{a} | \mathscr{H} | \Phi_{0} \rangle  = 0$, the inclusion of singly-substituted Slater
determinants cannot improve the ground state energy of the system. However, this approach can be 
used to describe excited states. \cite{chem-review-cis}

The only type of excited state Slater determinant that interacts with \ac{HF} ground state is doubly-substituted one. When, Eqn. \ref{eq:ci-expansion} 
is terminated at 2$^{nd}$ place (from Slater reference determinant), we get what is known as \ac{CISD}. Since, the many-body 
Hamiltonian contains only one- and two-electron operators, terminating the \ac{CI} expansion series at 2$^{nd}$ place may be 
a very good first approximation for the ground state calculations. Indeed, in most of the cases \ac{CISD} accounts for the 95\% of electron 
correlation for the ground state.\cite{harrison95}

\subsubsection{Multi-Reference Singles Doubles Configuration Interaction}\label{subsection-mrsdci}
Including singly- and doubly-substituted determinants to describe the molecular ground state may not be enough if the ground state itself
is near degenerate. So choosing only one ``reference'' Slater determinant would describe system inadequately. The same is true for excited states calculations. 
To overcome this issue, many such ``references'' can be included in the \ac{CI} reference space, and subsequently singly- and doubly- substituted 
determinants can be formed. Such an inclusion of many references is called as ``\ac{MRSDCI}''.

Once the method to shortlist configurations in the reference space is known, \ac{MRSDCI} is the most accurate and efficient method 
for electronic structure calculations. Choosing right configurations for reference space for \ac{MRSDCI} calculations is a 
difficult task. Picking configurations before doing any calculations requires deep intuition about the system. 

Having more than one configurations as reference for \ac{CI} calculations, would naturally address the issue of static electron correlation.
In the case of systems with nearly degenerate ground states, including more references would allow in describing electrons arranged 
in molecular states. We have mainly used this approach to get ground as well as excited state energies of various clusters.

To calculate \emph{ab initio} photoabsorption using this method, a number of singly- and doubly-excited configurations from a set of 
reference configurations are considered in obtaining both ground as well as excited state energies of various geometries of clusters, 
as implemented in the computer program MELD. \cite{meld}

Using the ground- and excited-state wavefunctions obtained from these \ac{MRSDCI} calculations, electric dipole matrix
elements are computed. For finite systems, such as clusters or quantum dots, the ratio $\frac{a}{\lambda}$ $\ll$ 1, where $a$ is system size 
and $\lambda$ is incident wavelength of light. In this case, the optical absorption cross section of the system is given by,
 \begin{equation}  
\label{eq:opt}
\sigma_{n}(\omega) = 4\pi\alpha\sum_{i} \frac{\omega_{in} \left| \langle i|\mathbf{\hat{e}}.\mathbf{r} | n\rangle  \right|^2 \gamma}{(\omega_{in}-\omega)^2+\gamma^2}
 \end{equation}
where, $\omega_{in}=\frac{\epsilon_i - \epsilon_n}{\hbar}$, $\alpha$ is fine structure constant, $\omega$ is frequency of incident 
radiation, $|i\rangle$ is final state, $|n\rangle$ is initial state and $\gamma$ is the assumed linewidth of absorption. In our calculations, 
we have fixed initial state as ground state of the system, and assumed that no multi-photon absorption takes place.

 By analyzing the wavefunctions of the excited states contributing to the peaks of the computed spectrum obtained from a given calculation, bigger
\ac{MRSDCI} calculations were performed by including a larger number of reference states. The choice of the reference states to be included
in a given calculation was based upon the magnitude of the corresponding coefficients in the \ac{CI} wavefunction of the excited state (or states)
contributing to a peak in the spectrum. This procedure was repeated until the computed spectrum converged within an acceptable tolerance,
and all the configurations contributing significantly to various excited states were included in the list of the reference states. We used this approach
for calculating linear optical absorption spectra of various isomers of B$_{n}$, Al$_{n}$ and Mg$_{n}$ (n = 2 -- 5) clusters. 
Such an iterative \ac{MRSDCI} approach has been used to perform large-scale correlated calculations 
of linear and nonlinear optical properties of a number of conjugated polymers.\cite{mrsd_jcp_09,mrsd_prb_07,mrsd_prb_05,mrsd_prb_02}

The number of molecular orbitals, and thus the size of the CI expansion, increases rapidly with the increasing number of atoms in the clusters.
Such a proliferation in the size of calculations can essentially make high-quality MRSDCI calculations impossible even for clusters of the
sizes discussed in this work. Therefore, wherever possible, we have used the point-group symmetries corresponding to $D_{2h}$, and its
subgroups, at all levels of calculations to reduce the size of the CI expansions. During the MRSDCI calculations, the frozen-core approximation
was employed, \emph{i.e.}, while constructing the CI expansion, no virtual excitations from the chemical core ($1s^{2}$ for boron and 
$1s^{2} 2s^{2} 2p^{6}$ for aluminum and magnesium) of the atoms of the cluster were considered. Similarly, excitations into very high energy 
virtual orbitals were not considered with the purpose of keeping the calculations manageable. The impact of both the frozen-core 
approximation, and the deletion of high-energy virtual orbitals, along with the influence of the choice of the basis sets on our 
calculations is examined carefully. 

\subsection{Coupled Cluster Method}
 Coupled cluster method is another popular and one of the most successful methods in quantum chemistry for electronic structure calculations.\cite{coupled-cluster-rmp07}
 Unlike truncated \ac{CI}, this method is size-extensive, which means that the correlation energy scales with number of electrons in the system. 
The coupled-cluster wavefunction is written as,\cite{coupled-cluster-rmp07}
 \begin{equation} 
 \label{eq:ccsd-equation}
 | \Psi_{cc} \rangle =  e^{\hat T} | \Phi_0 \rangle 
\end{equation} 
\begin{equation} 
| \Psi_{cc} \rangle = \left( 1 + \hat T + \frac{\hat T^2}{2!} + \frac{\hat T^3}{3!} + ... \right) | \Phi_0 \rangle 
 \end{equation}
where, the operator $\hat T$ is a series of connected operators, and $\Phi_0$ is a reference state.
 \begin{equation}
  \hat T = \hat T_{1} + \hat T_{2} + \hat T_{3} + ... \hat T_{n}
 \end{equation}
The constituent operators in the above summation are given by,
 \begin{equation}
  \hat T_{1}  | \Phi_0 \rangle = \sum_{i,a} t_{i}^{a} |\Phi_{i}^{a} \rangle
 \end{equation}
 \begin{equation}
  \hat T_{2}  | \Phi_0 \rangle = \sum_{i>j,a>b} t_{ij}^{ab} |\Phi_{ij}^{ab} \rangle
 \end{equation}
 \begin{equation}
  \hat T_{3}  | \Phi_0 \rangle = \sum_{i>j>k,a>b>c} t_{ijk}^{abc} |\Phi_{ijk}^{abc} \rangle
 \end{equation}
The terms $t_{ij...}^{abc...}$ are called as cluster amplitudes. The operators are connected to each other, evident from the following relations.
 \begin{equation} 
 \label{eq:cc-t1square}
  \frac{1}{2} \hat T_{1}^{2}  | \Phi_0 \rangle = \sum_{ia, jb} t_{i}^{a} t_{j}^{b} | \Phi_{ij}^{ab} \rangle 
 \end{equation}
 \begin{equation}
  \hat T_{1}\hat T_{2}  | \Phi_0 \rangle =  \sum_{ia, k>l, c>d} t_{i}^{a} t_{kl}^{cd} | \Phi_{ijk}^{abc} \rangle 
 \end{equation}
In the Eqn. \ref{eq:cc-t1square}, the operator introduces quadruple excitations into the reference state, still they can be greatly simplified
as their coefficients are composed of products of just double excitation coefficients. In a shorthand notation, the wavefunctions of couple-cluster 
and configuration interaction can be written as,
 \begin{equation}
  | CC \rangle = \left[ \prod_{i} (1 + t_{i}\hat{T}_{i}) \right] | \Phi_0 \rangle 
 \end{equation}
 \begin{equation}
  | CI \rangle = \left[ \sum_{i} (1 + c_{i}\hat{T}_{i}) \right] | \Phi_0 \rangle 
 \end{equation}
Consider the non-relativistic time-independent Schr\"{o}dinger equation.
 \begin{equation} 
 \label{eq:coupled-cluster} 
 \hat{H} | \Psi_{cc} \rangle  = E | \Psi_{cc} \rangle 
 \end{equation}
Since $ | \Psi_{cc} \rangle = e^{\hat T} | \Phi_0 \rangle$, substituting it in the above equation gives,
 \begin{equation} 
  \hat{H}  e^{\hat T} | \Phi_0 \rangle  =  E e^{\hat T} | \Phi_0 \rangle 
 \end{equation}
 \begin{equation} 
  e^{-\hat{T}} \hat{H}  e^{\hat T} | \Phi_0 \rangle	 =  E e^{-\hat T} e^{\hat T} | \Phi_0 \rangle 
 \end{equation}
 \begin{equation} 
 \label{eq:coupled-cluster-energy}
 e^{-\hat{T}} \hat{H}  e^{\hat T} | \Phi_0 \rangle	 =  E | \Phi_0 \rangle 
 \end{equation}
The energy and cluster amplitude equations can be obtained from Eqn. \ref{eq:coupled-cluster-energy} by left multiplying by 
the reference and any excited state determinant, respectively, and integrating over all space.
 \begin{equation} 
 \langle \Phi_0 |  e^{-\hat{T}} \hat{H}  e^{\hat T} | \Phi_0 \rangle	 =  E 
 \end{equation}
 \begin{equation} 
 \label{eq:coupled-cluster-amplitude-equation}
 \langle \Phi_{ij..}^{ab..} |  e^{-\hat{T}} \hat{H}  e^{\hat T} | \Phi_0 \rangle	=  0 
 \end{equation}
These equations are true only if the cluster expansion includes all possible excitations in the summation. The Eqn. \ref{eq:coupled-cluster-energy} 
can further be written in a simplified expression after Hausdorff expansion,\cite{coupled-cluster-rmp07}
 \begin{equation}
  E = E_{0}  + \langle \Phi_0  | [\hat H, \hat T_2 ] |  \Phi_0  \rangle + \frac{1}{2} \langle \Phi_0  | [[\hat H, \hat T_1 ], \hat T_1|  \Phi_0  \rangle 
 \end{equation}
where, $\Phi_0$ is reference state and $E_{0} $ is corresponding energy. Clearly, only singles and doubles amplitude contribute directly to the 
energy, but the singles and doubles are indirectly connected to all other remaining amplitudes. To a first approximation, we can terminate the 
series of coupled cluster expansion after doubles. The resultant method is known as \ac{CCSD}. 

\subsubsection{Equation-of-Motion Coupled Cluster Singles Doubles}
 \ac{EOM-CCSD} is one of the most accurate and compact single-reference electronic structure calculation methods. \cite{eom-ccsd-ann-rev, eomccsd-jctc}
It is conceptually very similar to the method \ac{CI}. Not just excitation energies, but also ionization potential, electron affinities, 
charge-transfer effects can be obtained using this approach. 
In equation-of-motion theory, one attempts to find excitation operators R$_{EE}$, which acting upon molecular ground state give
wavefunctions of excited electronic states of the molecules or clusters.\cite{eomccsd-smith-jcp} 
\begin{equation}
 | \Phi_{ex} \rangle = R_{EE} | \Phi_g \rangle
\end{equation}
Since, both $\Phi_g$ and $\Phi_{ex}$  are eigenfunctions of the Born-Oppenheimer Hamiltonian H,
\begin{equation}
 H | \Phi_g \rangle = E_g | \Phi_g \rangle 
\end{equation}
\begin{equation}
 H | \Phi_{ex} \rangle = E_{ex} | \Phi_{ex} \rangle 
\end{equation}
we can write equations of motion in the following form.
\begin{equation}
\label{eq:eom-ccsd-equation}
 [H, R_{EE}] | \Phi_g \rangle  = (E_{ex} - E_{g}) | \Phi_g \rangle 
\end{equation}
So, the application of commutator on ground state of the system gives the excitation energy. Thus, to apply this method one seeks a set of 
operators $R_{EE}$ whose commutator with electronic Hamiltonian operating on ground state of the system gives a constant times 
R$_{EE}$ $| \Phi_g \rangle$. Such resulting operators are called as excitation operators and the constant multipliers are the excitation energies.

It is difficult to solve Eqn. \ref{eq:eom-ccsd-equation} exactly for multi-electron systems, hence approximate solutions are used in practice.
It involves calculations of ground state and excitation operators by self-consistent method.

The computational scaling of EOM-CCSD (i.e. equation-of-motion approach applied for only singles and doubles excitation in coupled cluster)
 is N$^{6}$. If the wavefunction $|\Phi_g\rangle$ corresponds to ground state and operator $R$ conserves the number and total spin of electrons, 
then \ac{EOM-CCSD} gives excited state description of the system. \cite{eom-ccsd-ann-rev}
\begin{equation}
 R_{EE} | \Phi_g \rangle  = \sum_{i,a} t_{i}^{a} |\Phi_{i}^{a} \rangle + \frac{1}{(2!)^2} \sum_{ij,ab} t_{ij}^{ab}  |\Phi_{ij}^{ab} \rangle + \frac{1}{(3!)^2}\sum_{ijk,abc} t_{ijk}^{abc}  |\Phi_{ijk}^{abc} \rangle + ...
\end{equation}
The operator can be made to yield only $\alpha \rightarrow \alpha$ and $\beta \rightarrow \beta$ transitions, thereby conserving the total spin.
The error in excitation energy obtained using \ac{EOM-CCSD} is generally in the range of 0.1 -- 0.3 eV with accurate reproduction of
 relative spacing between excited states .\cite{eom-ccsd-ann-rev, fci-eomccsd-jcp}

    \subsection{Density Functional Theory}
All methods discussed above require wavefunction of electron. Sometimes working with wavefunction based methods becomes clumsy for 
large systems. Ignoring spin degree of freedom, for system of $N$ electrons, the many-body wavefunction is a complex function with $3N$ coordinates.
For a simple carbon atom, this amounts to 18 coordinates.  In a self-consistent method, evaluating and storing such a complex function is 
computationally extensive and time consuming. An altogether different approach was proposed by Hohenberg and Kohn, which does not use 
many-body wavefunction for description of system.\cite{hohenberg_kohn} Rather it uses electronic density $n(\textbf{r})$ as primary variable. 
The theorem says, (a) The electronic density of an interacting  system of electrons completely and uniquely determines the 
external potential $v(\textbf{r})$, that these electrons experience. Hence it also determines the Hamiltonian, the many-body 
wavefunction, and all the observables of the system. (b) The ground state energy of the system can be obtained by variationally 
minimizing the total energy with respect to density, and (c) There exists an universal functional $F[n]$ such that the total energy, $E[n]$ can
be written in the form 
\begin{equation}
 E[n] = F[n] + \int n(\mathbf{r}) v(\mathbf{r}) d^{3}r 
\end{equation}
The same formalism is also applicable to spin-dependent version of \ac{DFT}, in which total energy $E$ and universal functional $F$ are explicit
functionals of the spin-up or spin-down electron densities.

The Hohenberg-Kohn theorem is exact, but it does not talk about the prescription to obtain energy from the density alone. Kohn and Sham
proposed a method by constructing an auxiliary system of non-interacting electrons with same density as that of an interacting electron system.\cite{kohn_sham}
This way, a non-interacting electrons problem can be solved by one-particle Schr\"{o}dinger equation.
\begin{equation} 
\label{eq:schrodinger-kohn-sham}
 \left(  -\frac{\nabla^2}{2} + v_{KS}[n(\mathbf{r})] \right) \phi_i (\mathbf{r}) = \epsilon_i \phi_i (\mathbf{r})
\end{equation}
This self-consistent equation can be solved for $\phi_i (\mathbf{r})$, which in turn gives ground state electronic density of $N$ electrons as,
\begin{equation}
 n(\mathbf{r}) = \sum_{i}^{N} {\left | \phi_i (\mathbf{r}) \right |}^{2} 
\end{equation}
The Kohn-Sham potential term in the Eqn. \ref{eq:schrodinger-kohn-sham} is conventionally split into three parts.
\begin{equation}
 v_{KS}[n(\mathbf{r})] = v_{ext}(\mathbf{r}) + \int \frac{n(\mathbf{r^{'}})}{\left|\mathbf{r}-\mathbf{r^{'}}\right|} d^{3}\mathbf{r^{'}} + v_{xc}[n(\mathbf{r})]
\end{equation}
The term $v_{ext}(\mathbf{r})$ is an external potential usually caused by nuclei in the system. Second term is electron-electron Coulomb 
interaction. The final term, $v_{xc}[n(\mathbf{r})]$ accounts for all other non-trivial many-body electron interactions, which are historically known as
exchange-correlations. This is also written in the form of a functional derivative of so-called exchange-energy $E_{xc}$ as,
\begin{equation}
 v_{xc}[n(\mathbf{r})] = \frac{\delta E_{xc}[n]}{\delta n (\mathbf{r})}
\end{equation}
If the exchange-correlation energy functional $E_{xc}$ is exactly known, the Kohn-Sham equation would provide the exact density of the
interacting many-body system. However, in practice, it is not known, and must be approximated to solve Kohn-Sham equations. Following are the 
two most popular approximations.

\textbf{\ac{LDA}} This approximation was proposed by Kohn and Sham \cite{kohn_sham}. The exchange-
correlation energy is approximated by using $\epsilon_{xc}^{uniform}$, which is energy per particle of the homogeneous (uniform) electron gas
with constant density. So $E_{xc}$ is given by,
\begin{equation}
 E_{xc}^{LDA} = \int n (\mathbf{r}) \epsilon_{xc}^{uniform} (n (\mathbf{r})) d^{3} r
\end{equation}
Since the energy density at point $\mathbf{r}$ depends on the electron density at the same point $\mathbf{r}$, this approach is called as 
\emph{local}. Owing to a crude approximation of uniform density, this approach fails to describe highly inhomogeneous systems such as atoms or 
molecules. However, for calculations of some properties, it is well suited for inhomogeneous systems as well as systems with slowly varying 
densities for which it was originally proposed.

\textbf{\ac{GGA}} The functional in this approximation is defined as,
\begin{equation}
 E_{xc}^{GGA} = \int f \left( n (\mathbf{r}), \nabla n (\mathbf{r}) \right) d^{3} r
\end{equation}
The function $f$ now depends on both local density as well as gradient of the density. Hence this approach is also called as semi-local. 
Contrary to the exact form of $\epsilon_{xc}^{uniform}$ in \ac{LDA}, the function $f$ must be parametrized to obtain the exchange-correlation energy,
except in the limiting case of weakly inhomogeneous system, in which it is uniquely defined.\cite{hohenberg_kohn} The parametrization is 
achieved by studying well known systems such as, uniform electron gas, or using sum rules or other features of the exact exchange functional, or
fitting properties of well known smaller systems.

  \subsubsection{Time-Dependent Density Functional Theory}
  \ac{TDDFT} is an extension of Hohenberg-Kohn density functional theory. It was formally derived by Gross
and Runge. \cite{runge-gross} Suppose a system is subjected to a time-dependent perturbation (a laser pulse) to the external potential $v_{ext}(\mathbf{r},t)$ 
\begin{equation}
v_{ext}(\mathbf{r},t) = E f(t)  sin(\omega t)\mathbf{r}.\boldsymbol\alpha - \sum_{n=1}^{N} \frac{Z_{n}}{|\mathbf{r} - \mathbf{R_{n}}|}
\end{equation}
where $\boldsymbol\alpha$, $\omega$ and E are polarization, frequency and amplitude of the laser pulse. The function $f(t)$ defines the 
envelope of the laser pulse. The Runge-Gross theorem in the \ac{TDDFT} states that there is one-to-one 
correspondence between the external time-dependent potential, $v_{ext}(\mathbf{r},t)$, and time-dependent electron density, $n(\mathbf{r},t)$.
\cite{runge-gross} This implies, if we know only the time-dependent density of the system, evolving from a given initial state, then 
the external potential that caused this density can be known up to a time-dependent constant.\cite{miguel-fundamental-tddft-book}
The external potential, in turn, can identify the Hamiltonian of the system and all the operators of the observable quantities. 
The time-dependent constant does not alter the physics, as it only introduces a purely time-dependent phase factor to the wavefunction, and 
that gets nullified in the expectation values of Hermitian operators. This treatment is very much analogous to the 
Hohenberg-Kohn ground state \ac{DFT}. However, the variational principle does not hold good for time-dependent case, as the total energy
is not a conserved quantity. An analogous quantity to total energy called as quantum mechanical action is varied instead. \cite{miguel-fundamental-tddft-book}
\begin{equation}
 \mathscr{A}[\Phi] = \int_{t_1}^{t_2} dt  \langle \Phi(t) | i\frac{\partial}{\partial t} - \hat{H}(t) | \Phi(t) \rangle
\end{equation}
where, $\Phi$ is some $N$-body function. The action $\mathscr{A}$ is varied and equated to zero to get stationary point of the functional $\mathscr{A}$.
The function, say $\Psi(t)$ which makes the functional stationary will be the solution of the time-dependent Schr\"{o}dinger equation.

Like in the ground-state theory, the problem of finding correct energy functional also arises here. To approximate the ground state
electron density, Kohn -- Sham proposed an equivalent density due to non-interacting electrons. Owing to one-to-one correspondence 
between external potential and electron density, the formalism of using non-interacting electron density can also be applied to 
time-dependent systems. In this auxiliary system, the Kohn-Sham electrons obey time-dependent Schr\"{o}dinger equation, 
 very similar to Eqn. \ref{eq:schrodinger-kohn-sham}, 
\begin{equation} 
\label{time-dependent-kohn-sham}
 \left(  -\frac{\nabla^2}{2} + v_{KS}[n(\mathbf{r},t)] \right) \phi_i (\mathbf{r},t) =  i \frac{\partial}{\partial t}\phi_i (\mathbf{r},t)
\end{equation}
Along with the density of the interacting system can now be written in terms of time-dependent Kohn-Sham orbitals, given by,
\begin{equation}
 n(\mathbf{r},t) = \sum_{i}^{N} {\left | \phi_i (\mathbf{r},t) \right |}^{2} 
\end{equation}
The Kohn-Sham potential for the time-dependent case takes the following form,
\begin{equation}
 v_{KS}[n(\mathbf{r},t)] = v_{ext}(\mathbf{r},t) + \int \frac{n(\mathbf{r^{'}},t)}{\left|\mathbf{r}-\mathbf{r^{'}}\right|} d^{3}\mathbf{r^{'}} + v_{xc}[n(\mathbf{r},t)]
\end{equation}
where, exchange-correlation potential can conventionally be written as,\cite{leeuwen_causality_tddft}
\begin{equation}
 v_{xc}[n(\mathbf{r},t)] = \frac{\delta \mathscr{A'}_{xc}[n]}{\delta n (\mathbf{r},\tau)}  \Bigg|_{n(\mathbf{r},t)}
\end{equation}
with modified action to avoid problems of causality. $\tau$ stands for Keldish pseudo-time. \cite{leeuwen_causality_tddft}

Since the exact form of $v_{xc}$ is not known, it is approximated on the similar lines of ground state \ac{DFT}. However, for ground state
\ac{DFT}, there exists a vast number of energy functionals, as compared to a very few for the \ac{TDDFT}. This is the only approximation made
in the formalism of \ac{TDDFT}. \ac{ALDA}, time-dependent exact exchange (EXX) functionals are the very few approximate methods
 developed for \ac{TDDFT} functionals. In the \ac{ALDA}, existing ground-state xc-functionals are used in adiabatic limit, \emph{i.e.}, the 
same functional form is used to evaluate $v_{xc}$ at each time with density $n(\mathbf{r},t)$. This local approximation in time appears to work 
exceedingly well for low-lying excited states of valence types when used 
with conventional \ac{DFT} functionals. \cite{adiabatic-lda-casida-jcp,frisch-implementation-tddft-jcp}

    The optical absorption in the clusters and other nanostructures is characterized by an weak external potential $v_{ext}(\mathbf{r},\omega)$ 
of time-dependent electric field. This perturbation causes instantaneous change in the electron density $\delta n (\mathbf{r},\omega)$. In the
linear response regime, we can safely ignore contribution from magnetic field, giving a relation between change in the electron density to the 
external potential.
\begin{equation}
 \delta n (\mathbf{r},\omega) = \int \chi(\mathbf{r,r'},\omega) v_{ext}(\mathbf{r'},\omega) d^{3} r'
\end{equation}
The term $\chi(\mathbf{r,r'},\omega)$ is called dynamic susceptibility. From the change in electron density  $\delta n (\mathbf{r},\omega)$,
the dynamic polarizability can be obtained by taking ratio between induced dipole moment and magnitude of the applied electric field $E_{0}$.
It is given by,
\begin{equation}
 \alpha(\omega) = \frac{e}{E_0}  \int v_{ext}(\mathbf{r},\omega) \delta n (\mathbf{r},\omega) d^{3}r
\end{equation}
By applying Fermi's golden rule, one can obtain the photoabsorption cross section as,
\begin{equation}
 \sigma(\omega) = \frac{4\pi\omega}{c} \emph{Im}(\alpha(\omega))
\end{equation}
where $\emph{Im}(\alpha(\omega))$ denotes the imaginary part of dynamical polarizability. The cross-section integrated over entire space 
should be equal to the number of electrons multiplied by a constant. 

  \lhead{{\chaptername\ \thechapter.}{  Boron Clusters B$_{n}$ ($n$=2 -- 5)}}
  \chapter{\label{chap:main_smallboron}Theory of Linear Optical Absorption in Various Isomers of Boron Clusters B$_{n}$ ($n$=2 -- 5)}
\emph{This chapter is based on a published paper, Nano LIFE, \textbf{2}, 1240004 (2012) \\ by Ravindra Shinde and Alok Shukla.}
\par

Boron clusters are attracting great attention because of their novel properties, and potential applications in nanotechnology
and hydrogen storage related capabilities.\cite{boron_age2004,zhao_prl2005,cabria_alonso,shelvin_guo}
Boron atom, having $s^{2}p^{1}$ valence electronic configuration, has short covalent radius and tends to form strong directional bonds
producing clusters of covalent nature. Because of this strong covalent bonding, it has hardness close to that of diamond. The ability of
boron to form structures of any size due to catenation is only comparable to its neighbor carbon.\cite{boron_age2004} Planar boron clusters
exhibit aromaticity\cite{borozene_sahu} due to the presence of itinerant $\pi$ electrons, and some of them are analogous to aromatic 
hydrocarbons.\cite{kiran_wang} Boron fullerenes, boron sheets and single-sheet boron nitride---a
graphene analogue---are the other examples of boron-based clusters.

As far as the studies of boron-based clusters are concerned, small ionic boron clusters B$_{n}^{+}$ (n $\leqslant$ 20) were experimentally
studied by Hanley, Whitten and Anderson.\cite{hanley_whitten} Wang and coworkers have reported joint theoretical and experimental studies
of the electronic structure of bare boron wheels, rings, tubes and large quasi-planar clusters.\cite{kiran_wang, wheel_wang,ring_wang,tube_wang}
Using the photoelectron spectroscopy, they predicted that tubular B$_{20}$ can act as the smallest boron single walled nanotube. Transition
metal-centered boron ionic ring clusters were studied by Constantin \emph{et. al.} \cite{ring_wang}, in a photo-electron spectroscopy
experiment, supported by first-principles calculations. The abundance spectrum of boron clusters generated by laser ablation of hexagonal
boron nitride was studied by time of flight measurements performed by La Placa, Roland and Wynne.\cite{la_placa} They also postulated
the existence of B$_{36}$N$_{24}$ cluster having a structure similar to that of C$_{60}$ fullerene. Lauret \emph{et al.}\cite{prl_lauret}
probed the optical transitions in single walled boron nitride nanotubes by means of optical absorption spectroscopy.

Larger pure boron clusters have also been investigated extensively. Cage-like structure of B$_{80}$---similar to C$_{60}$ fullerene---has
been proposed theoretically.\cite{prl_szwacky} A density functional theory (DFT) study of pure boron sheets and nanotubes was carried
out by Cabria, Lopez and Alonso to explore their potential hydrogen storage materials.\cite{cabria_alonso} Chacko, Kanhere and Boustani
investigated different equilibrium geometries of B$_{24}$ cluster using Born-Oppenheimer molecular dynamics within the framework of
DFT.\cite{chacko_kanhere} Abdurahman \emph{et al.}\cite{ayjamal} studied the ladder-like planar boron chains B$_{n}$ ($n$=4-14),
and computed their static dipole polarizabilities using the \emph{ab initio} CI method. Johansson discussed strong toroidal ring currents
in B$_{20}$ and other toroidal boron clusters.\cite{toroid_current} Double aromaticity was proposed in toroidal boron clusters B$_{2n}$
(n = 6,14) by Bean and Fowler.\cite{aromatic_toroid}

Regarding the smaller sized boron clusters, an early theoretical study of boron dimer was carried out by Langhoff and Bauschlicher,\cite{langhoff}
who performed an extensive calculations using the complete-active-space self-consistent-field (CASSCF) multireference configuration interaction
(MRCI) with a large basis set. A similar study was carried out by Bruna and Wright\cite{bruna_wright} for the excited states of B$_{2}$,
and by Howard and Ray\cite{howard_ray} using the many-body perturbation theory. A systematic geometry and electronic structure calculations
of bare boron clusters was reported by Boustani.\cite{boustani_prb97} He performed all-electron calculations at the SDCI level, but the
contracted Gaussian basis sets used were small. Niu, Rao and Jena,\cite{rao_jena} using DFT and quantum chemical methods, presented an account of electronic
structures of neutral and charged boron clusters. In their study on small clusters, M\"{o}ller-Plesset perturbation theory of fourth order
(MP4) was used to account for the electron correlation effects. More recently, Ati\c{s}, \"{O}zdogan, and G\"{u}ven\c{c} investigated structure and
energetics of boron clusters using the DFT.\cite{turkish_boron} Aromaticity in planar boron clusters was addressed by Aihara, Kanno and Ishida.\cite{ishida_jacs}

In spite of many theoretical studies of boron clusters of various shapes and sizes, very little experimental information about their
ground and excited states is available. Conventional mass spectrometry can distinguish between different clusters only according to their
mass, but not according to their geometry. One has to rely on other theoretical or experimental data to be able to differentiate one isomer
from another. For example, using first principles calculations of vibrionic fine structure in C$_{20}^{-}$, and comparing it with experimentally
available data, Saito and Y. Miyamoto\cite{saito_fullerene} identified the cage and bowl structures. Optical absorption spectroscopy, coupled
with extensive theoretical calculations of the optical absorption spectra, can be used to distinguish between distinct isomers of clusters
produced experimentally, because normally optical absorption spectra are sensitive to the geometries of the clusters. The optical absorption
of alkali metal clusters has been extensively studied both experimentally 
and theoretically.\cite{na_dehaar_prl, na_opt_kappes_jcp, na_opt_bethe_salp,li_boustani_prb,li_na_tddft}
However, a very few such studies exist for the case of boron clusters. Marques and Botti\cite{marques_botti} calculated optical absorption
on different B$_{20}$ isomers using time-dependent (TD) DFT. Boron fullerenes such as B$_{38}$, B$_{44}$, B$_{80}$ and B$_{92}$ were
also studied by Botti and coworkers\cite{marques_fullerene} using the same technique. However, to the best of out knowledge, there are
no experimental and theoretical study of optical absorption on other bare boron clusters, particularly the smaller ones. It is with the
aim of filling this void that we undertake a systematic study of the optical absorption in small boron clusters B$_{n}$(n=2--5), employing
the MRSDCI method, and high-quality Gaussian basis functions. We perform careful geometry optimization for each possible isomer, and compute
the optical absorption spectra of various structures. We also analyze the many-body wavefunctions of various excited states contributing
to the peaks in the computed spectra, and conclude that most of the excitations are collective in nature, signaling the presence of plasmons.

The remainder of this chapter is organized as follows. Next section describes the theoretical and computational details of the work, followed
by section \ref{sec:results-boron1} in which our results are presented and discussed. In section  \ref{sec:conclusions-nanolife} we present 
our conclusions and discuss possibilities for future work. Detailed information about various excited states contributing to optical 
absorption is presented in the Appendix \ref{app:wavefunction-boron2-5}.

\section{\label{sec:theory-boron1}Theoretical and Computational Details}

\begin{figure*}
 \begin{center}
 \subfloat[B$_{2}$, D$_{\infty h}$, $^{3}\Sigma_{g}^{-}$ \label{fig:b2geom}]{\includegraphics[width=2.7cm]{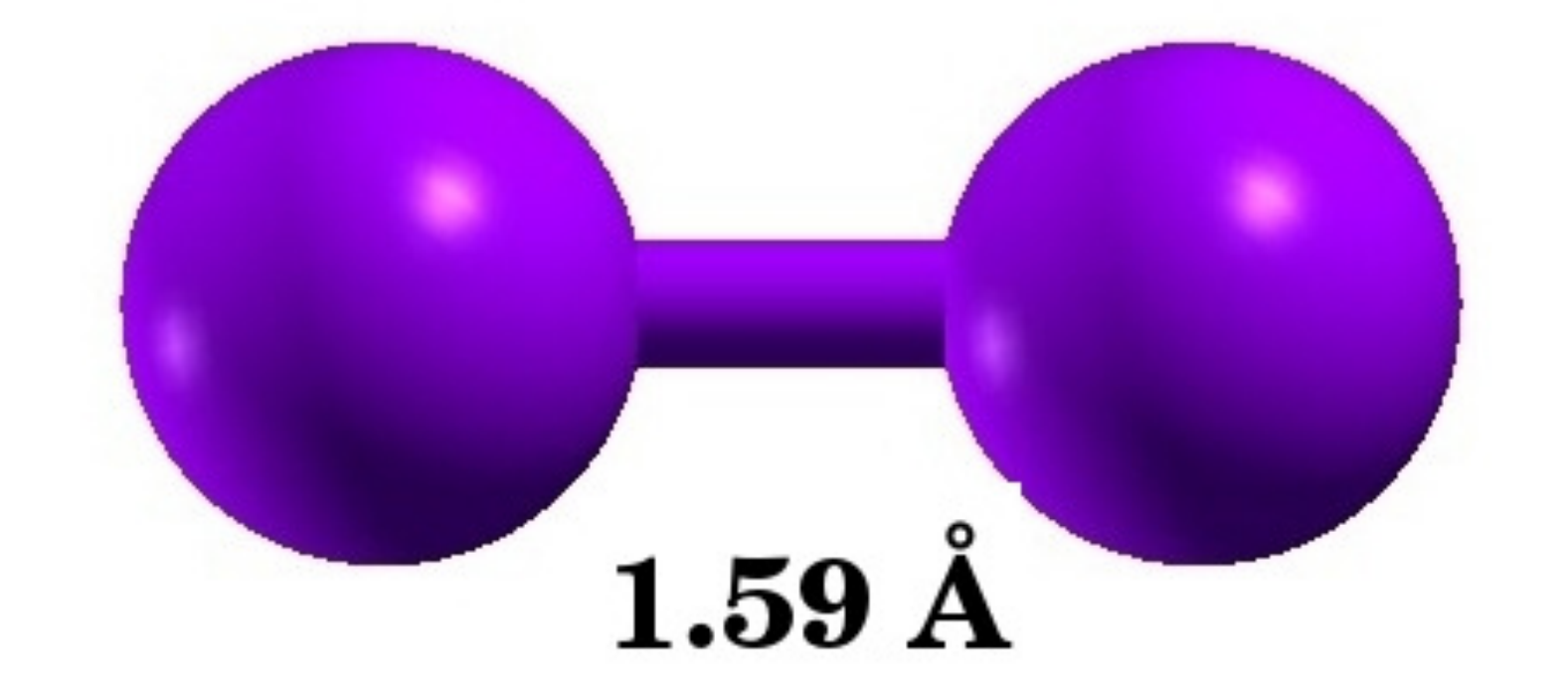}}
 \hspace{0.25cm} \subfloat[B$_{3}$, D$_{3h}$, $^{2}$A$_{1}^{'}$ \label{fig:b3trgeom} ]{\includegraphics[scale=0.135]{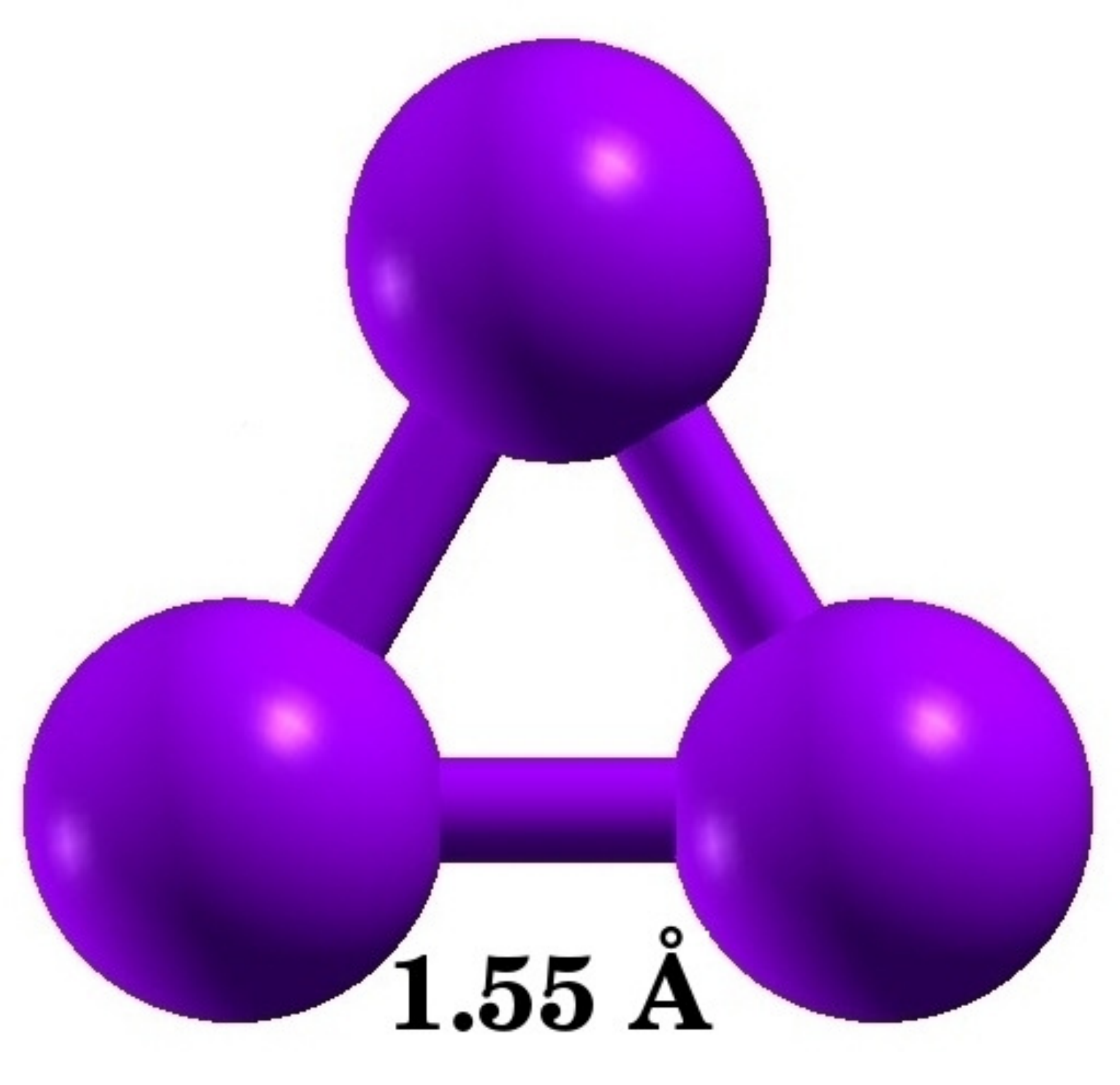}}
 \hspace{0.25cm} \subfloat[B$_{3}$, D$_{\infty h}$, $^{2}\Sigma_{g}^{-}$ \label{fig:b3lingeom}]{\includegraphics[scale=0.155]{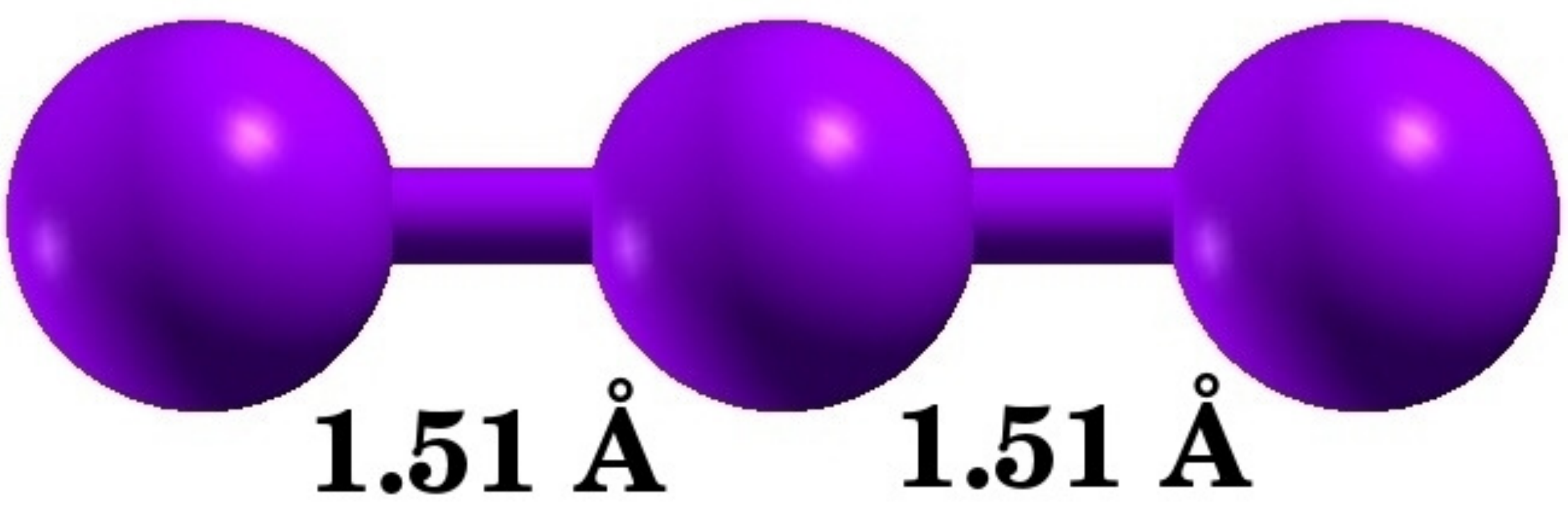}}  
 \hspace{0.25cm} \subfloat[B$_{4}$, D$_{2h}$, $^{1}$A$_{g}$ \label{fig:b4rhogeom}]{\includegraphics[width=2.6cm]{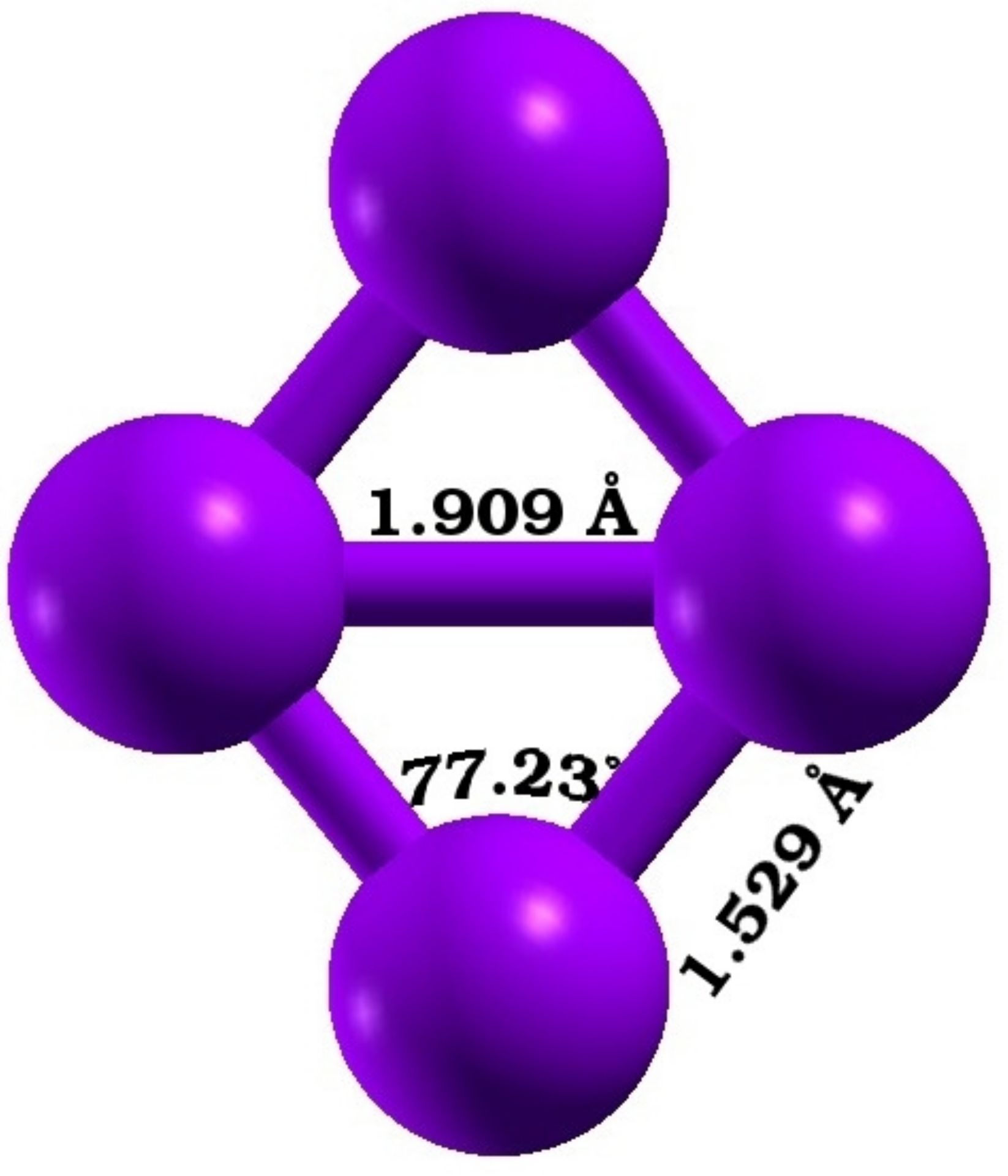}} 
  \hspace{0.25cm}\subfloat[B$_{4}$, D$_{4h}$, $^{1}$A$_{1g}$ \label{fig:b4sqrgeom}]{\includegraphics[width=2.4cm]{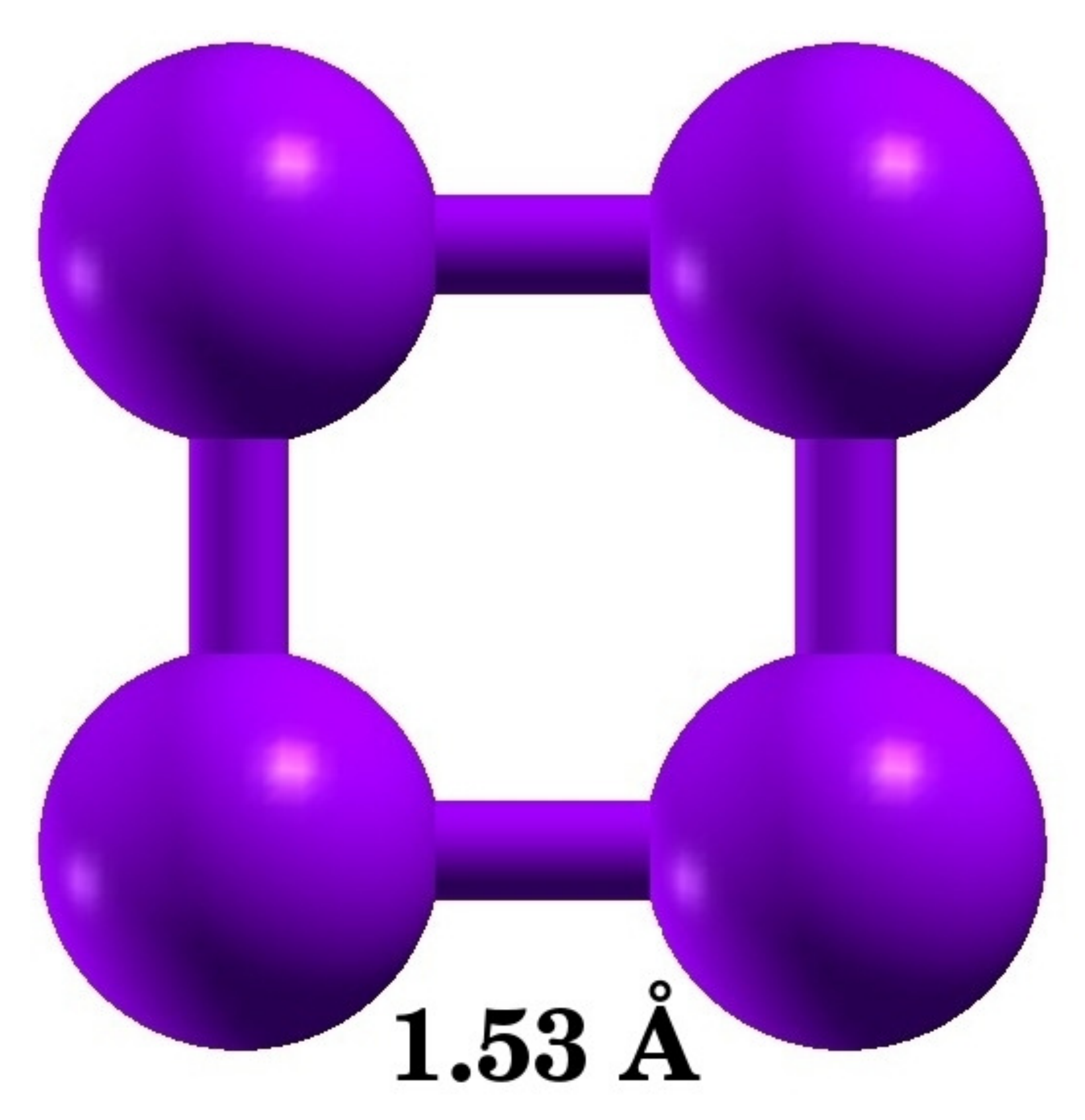}} 
  \hspace{0.25cm}\subfloat[B$_{4}$, D$_{\infty h}$, $^{1}\Sigma_{g}^{-}$ \label{fig:b4lingeom}]{\includegraphics[scale=0.21]{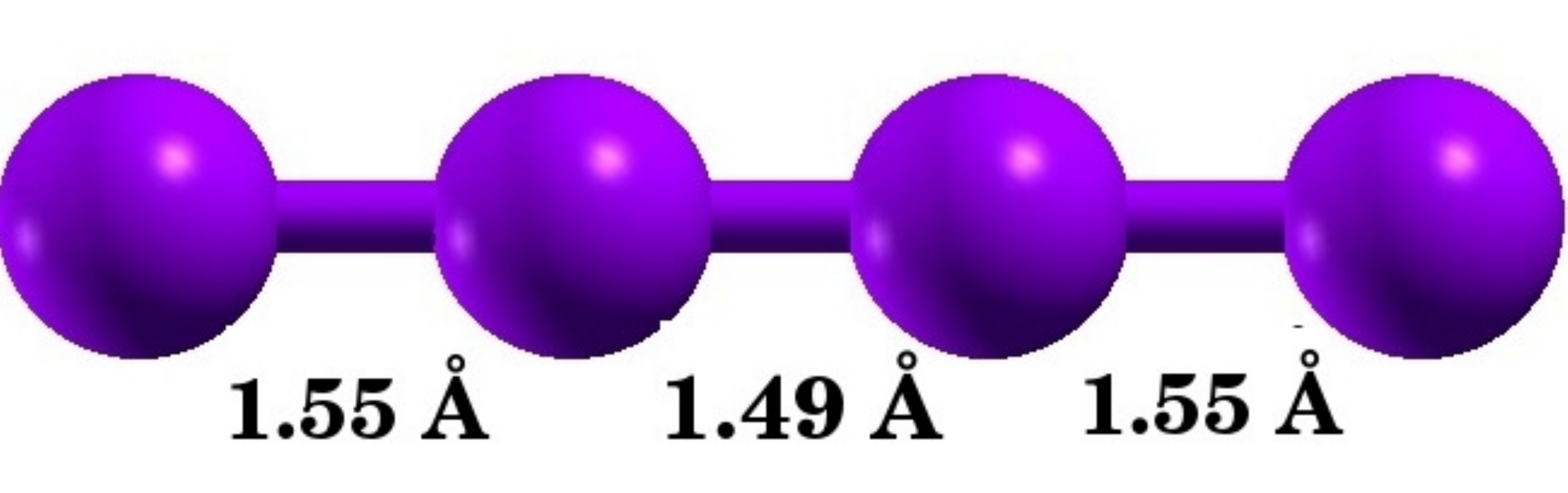}} 
  \hspace{0.25cm}\subfloat[B$_{4}$, C$_{2v}$, $^{1}$A$_{1}$ \label{fig:b4tetgeom}]{\includegraphics[width=2.6cm]{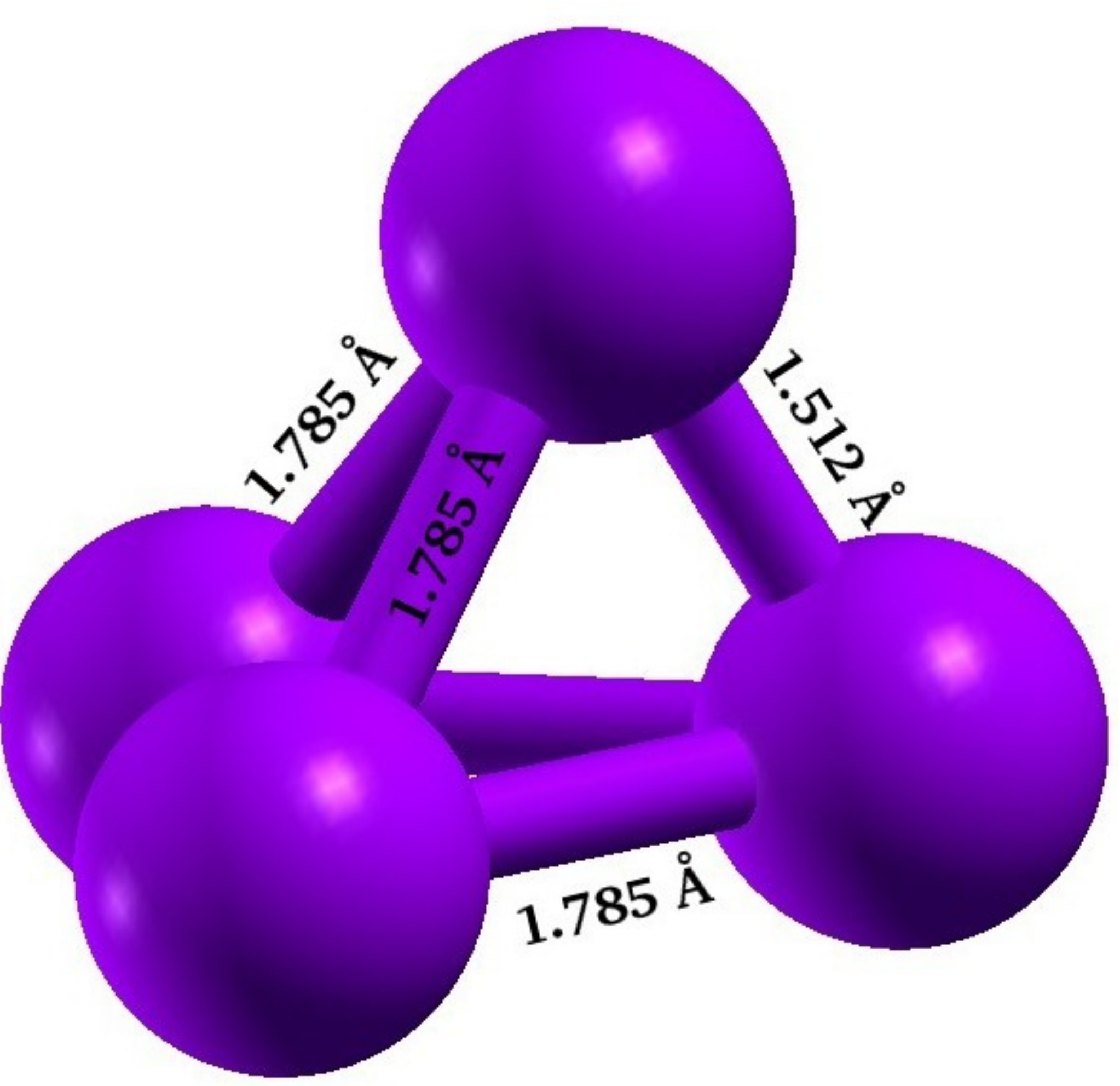}}
  \hspace{0.25cm}\subfloat[B$_{5}$, C$_{2v}$, $^{2}$B$_{2}$ \label{fig:b5pengeom}]{\includegraphics[width=3.8cm]{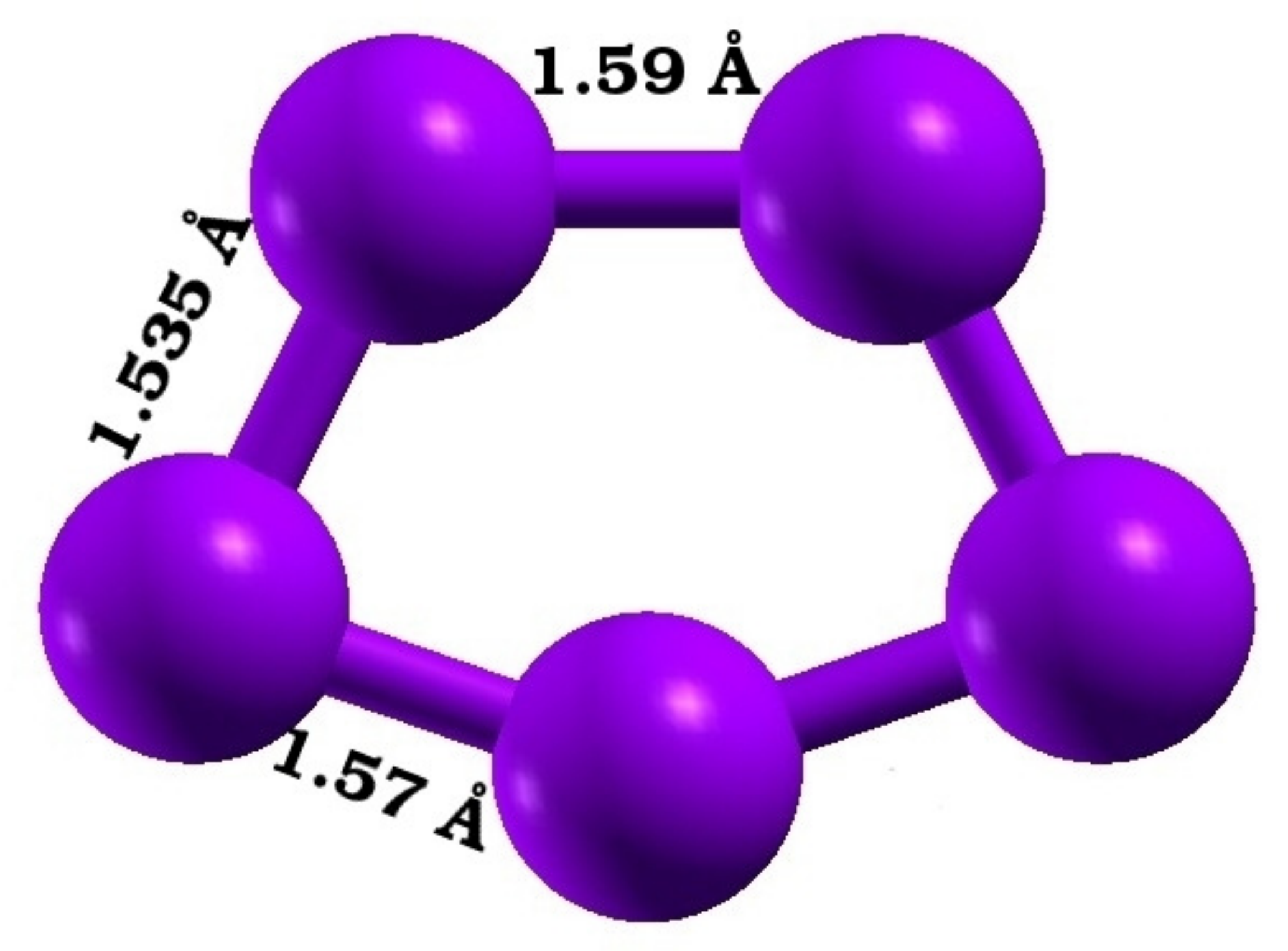}}
  \hspace{0.25cm}\subfloat[B$_{5}$, C$_{s}$, $^{2}$A \label{fig:b5tetgeom}]{\includegraphics[width=2.6cm]{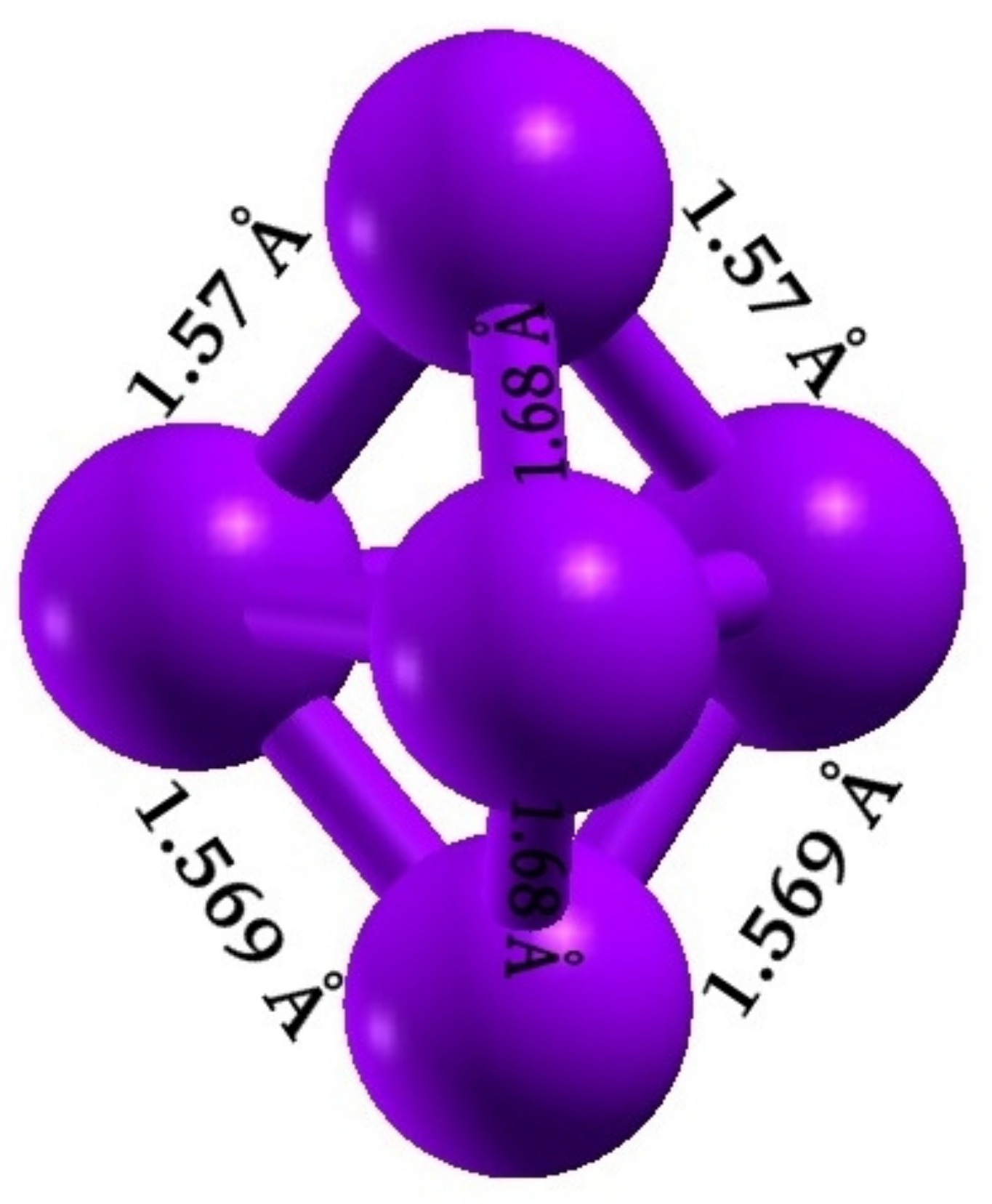}}
  \caption{Geometry optimized structures of boron clusters with point group symmetry and the electronic ground 
state at the configuration interaction level.}
 \label{fig:geometries-nanolife} 
\end{center}
 \end{figure*}

The geometry optimization of various isomers was done using the size-consistent
coupled-cluster singles doubles (CCSD) based analytical gradient approach,
as implemented in the package \textsc{gamess-us}.\cite{gamess}
For the purpose, we used the 6-311G(d,p) basis set included in the
program library,\cite{gamess} which is known to be well-suited for
this task.  The process of optimization was initiated by using the
geometries reported by Ati\c{s} \emph{et al.}\cite{turkish_boron},
based upon first principles DFT based calculations. For some simple
geometries such as B$_{2}$, B$_{3}$ (D$_{3h}$ symmetry), the optimization
was carried out manually, by performing the MRSDCI calculations at
different geometries, and locating the energy minima. Figure \ref{fig:geometries-nanolife}
shows the final optimized geometries of the isomers studied in this chapter.

The linear photoabsorption spectra of various isomers of the boron clusters were computed using MRSDCI method, as described in 
subsection \ref{subsection-mrsdci}.



\section{\label{sec:results-boron1}Results and Discussion}

In this section, first we discuss the convergence of our calculations
with respect to various approximations and truncation schemes. Thereafter,
we present and discuss the results of our calculations for various
clusters.

\subsection{Convergence of Calculations}

Here, we carefully examine the convergence of the calculated absorption
spectra with respect to the size and quality of the basis set, along
with various truncation schemes in the \ac{CI} calculations.

\subsubsection{Choice of the basis set}

\begin{figure}
\centering
\includegraphics[width=8.3cm]{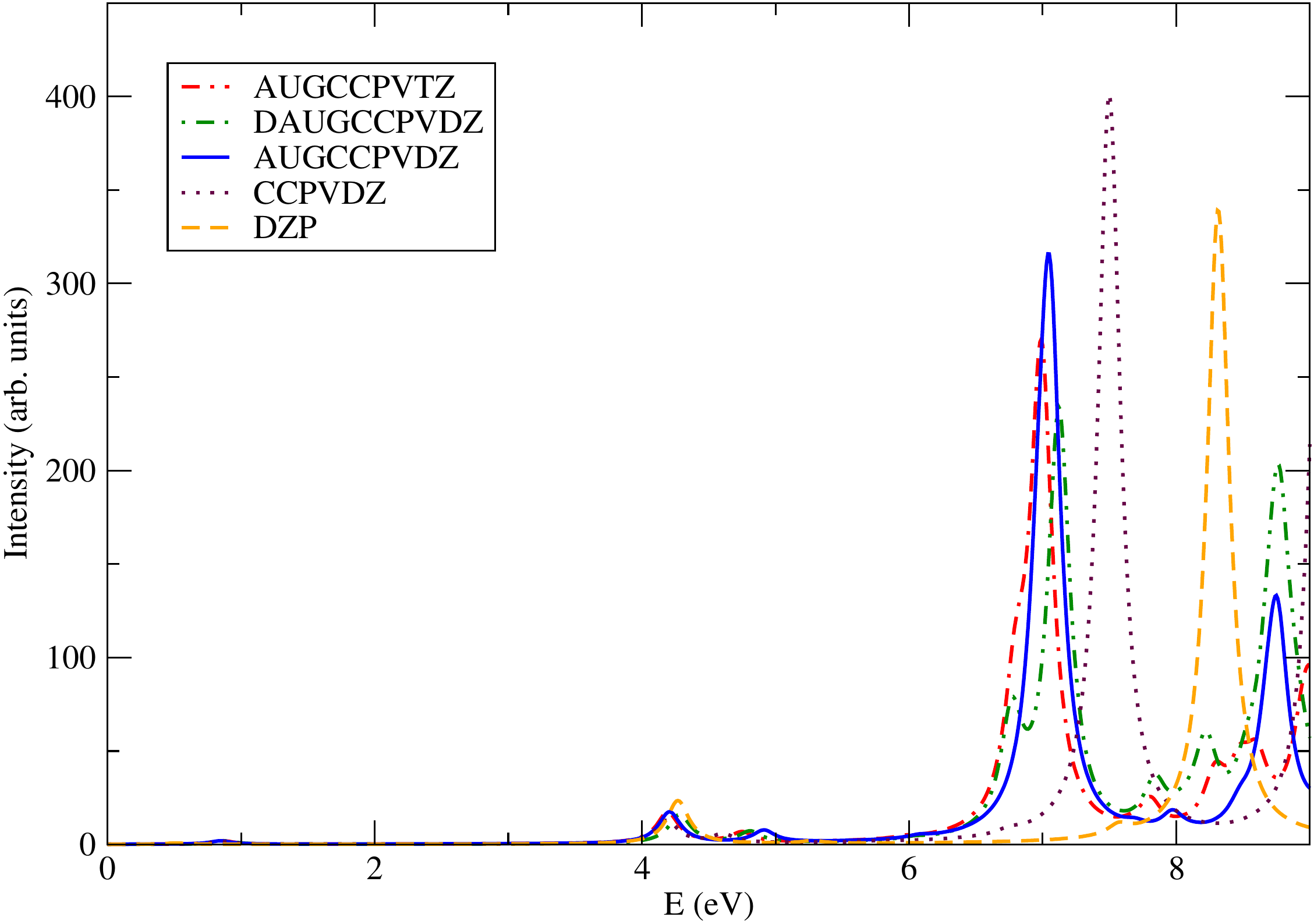}
\caption{Optical absorption in B$_{2}$ calculated using various
Gaussian contracted basis sets. Increasing more and more diffuse \textit{d}
type Gaussians shows negligible effect on optical spectra. }
\label{fig:basis-nanolife} 
\end{figure}

In general, the results of electronic structure calculations depend
upon the quality and the size of the basis set employed. While several
contracted Gaussian basis functions have been devised which can deliver
high-quality results on various quantities such as the total energy,
correlation energy, and the static polarizabilities of molecules,
to the best of our knowledge the basis set dependence of linear optical
absorption has not been explored. Since boron shows strong covalent
bondings, the basis set used for calculations should have diffuse
Gaussian contractions. Therefore, to explore the basis set dependence
of computed spectra, we used several basis sets\cite{emsl_bas1,emsl_bas2}
to compute the optical absorption spectrum of the smallest cluster,
\emph{i.e.}, B$_{2}$. For the purpose, we used correlation-consistent
basis sets named AUG-CC-PVTZ, DAUG-CC-PVDZ, AUG-CC-PVDZ, CC-PVDZ,
and DZP, which consist of polarization functions along with diffuse
exponents, and were designed specifically for post Hartree-Fock correlation
calculations\cite{emsl_bas1,emsl_bas2}. From the calculated spectra
presented in Fig. \ref{fig:basis-nanolife} the following trends emerge: the
spectra computed by various augmented basis sets (AUG-CC-PVTZ, DAUG-CC-PVDZ,
AUG-CC-PVDZ) are in good agreement with each other in the energy range
up to 8 eV, while those obtained using the nonaugmented sets (CC-PVDZ
and DZP) disagree with them substantially, particularly in the higher
energy range. Given the fact that augmented basis sets are considered
superior for molecular calculations, we decided to perform calculations
on the all the clusters using the AUG-CC-PVDZ basis set. This is the
smallest of the augmented basis sets considered by us, and, therefore,
does not cause excessive computational burden when used for larger
clusters.

\subsubsection{Orbital Truncation Schemes}

If the total number of orbitals used in a \ac{CI} expansion is $N$, the
number of configurations in the calculation proliferates as $\approx N^{6}$,
which can become intractable for large values of $N$. Therefore,
it is very important to reduce the number of orbitals used in the
\ac{CI} calculations. The occupied orbitals are reduced by employing the
so-called ``frozen-core approximation'' described earlier, while
the unoccupied (virtual) set is reduced by removing very high-energy
orbitals.

\begin{figure}
\centering
\includegraphics[width=8.3cm]{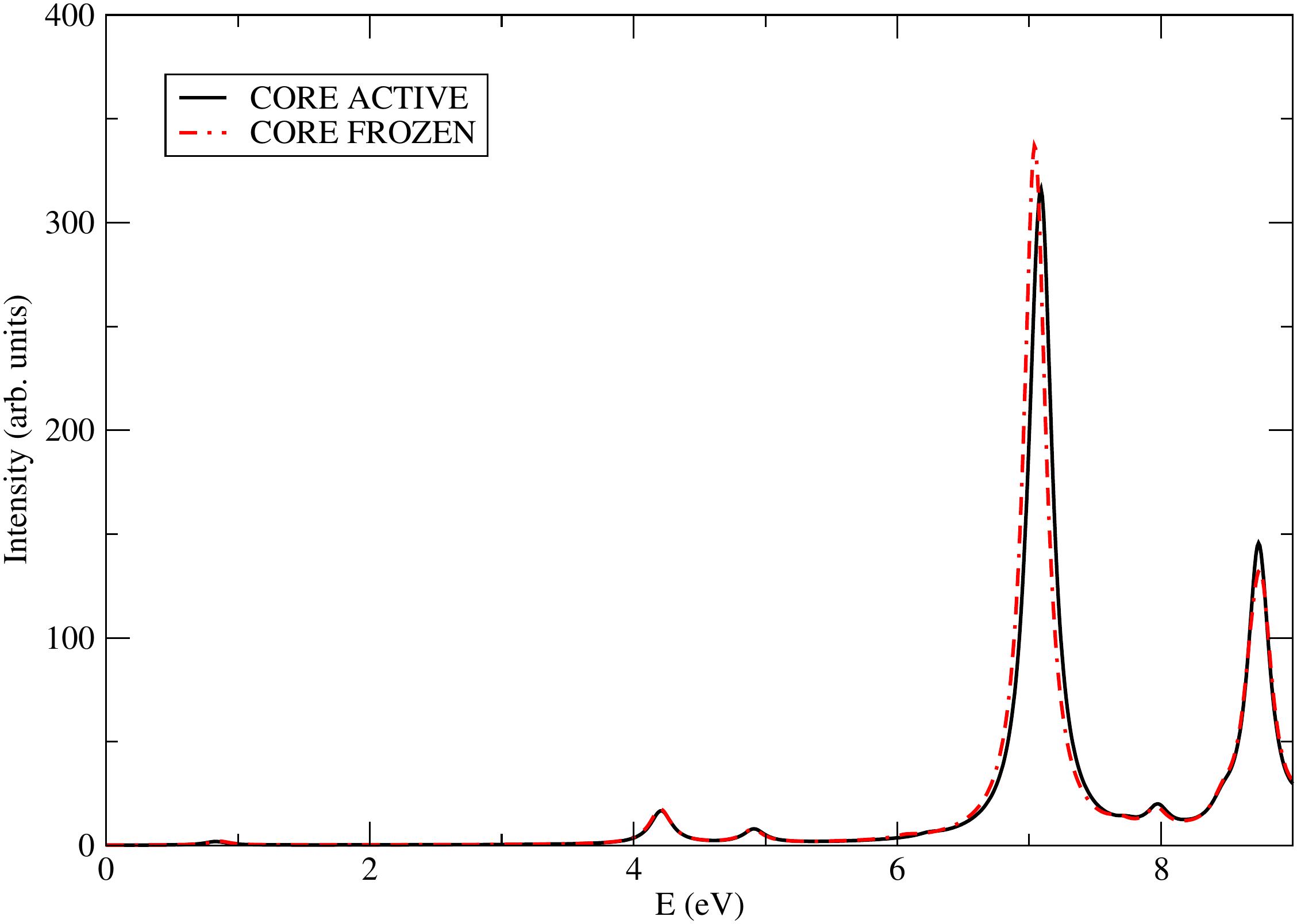}
\caption{  The effect of freezing the core orbitals ($1s$) of
boron atoms on optical absorption spectrum of B$_{2}$. It renders
almost no effect on optical absorption spectrum.}
\label{fig:core-nanolife} 
\end{figure}

\begin{figure}
\centering
\includegraphics[width=8.3cm]{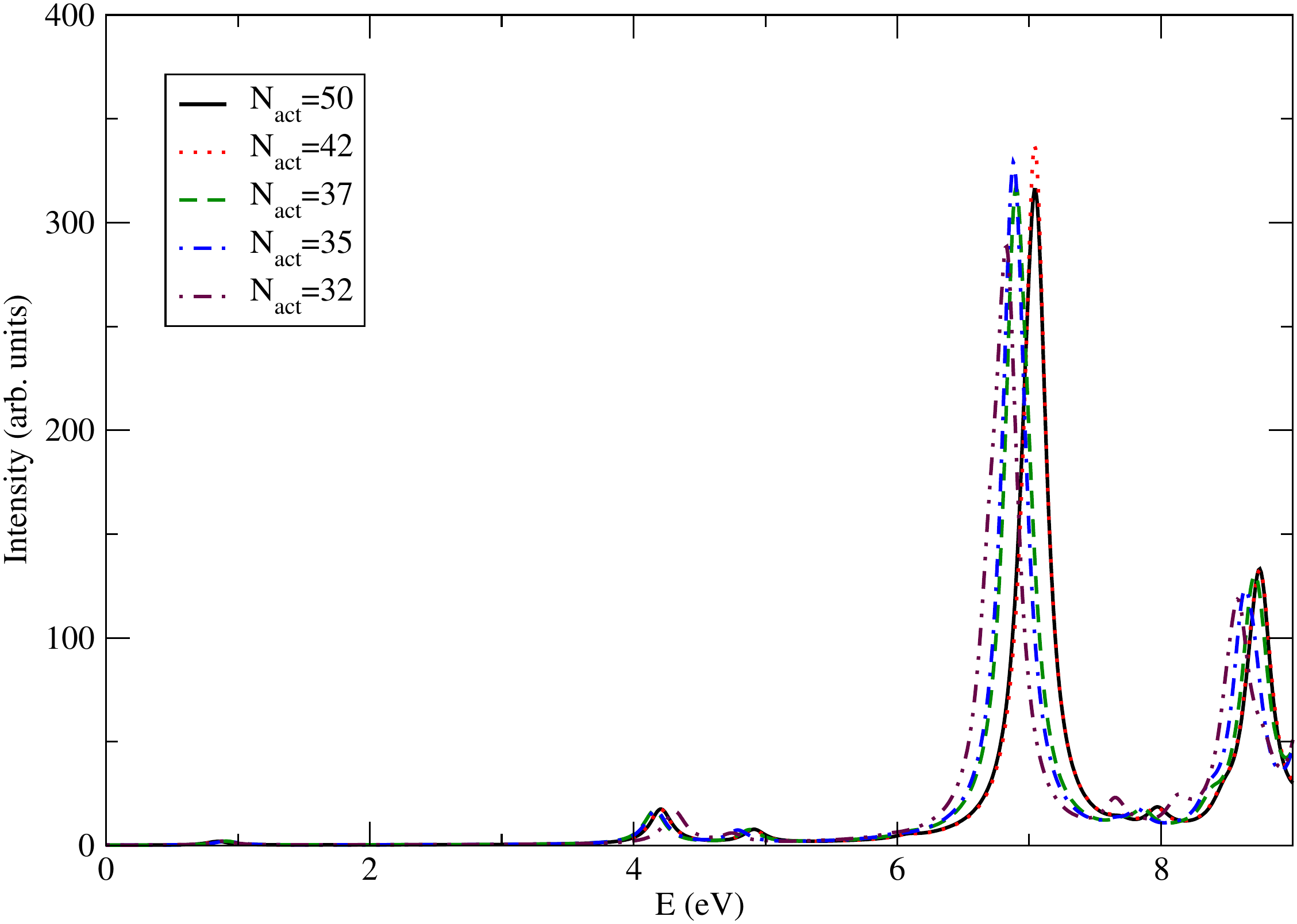}
\caption{  The effect of the number of active orbitals (N$_{act}$)
on the optical absorption spectrum of B$_{2}$. Until N$_{act}$=42,
the optical spectrum does not exhibit any significant change. It corresponds
to 1.0 Hartree ($\approx27.2$ eV) virtual orbital energy. }
\label{fig:nref-nanolife} 
\end{figure}

The influence of freezing the $1s$ core orbitals on the optical absorption
spectrum of B$_{2}$ cluster is displayed in Fig. \ref{fig:core-nanolife},
from which it is obvious that it makes virtually no difference to
the results whether or not the core orbitals are frozen. The effect
of removing the high-energy virtual orbitals on the absorption spectrum
of B$_{2}$ is examined in Fig. \ref{fig:nref-nanolife}. From the figure it
is obvious that if all the orbitals above the energy of 1 Hartree
are removed, the absorption spectrum stays unaffected. Therefore,
in rest of the calculations, wherever needed, orbitals above this
energy cutoff were removed from the list of active orbitals. Theoretically
speaking this cutoff is sound, because we are looking for absorption
features in the energy range much smaller than 1 Hartree.

\subsubsection{Size of the CI expansion}

As mentioned earlier that the electron correlation effects in both
the ground and the excited states were accounted in our calculations
by including the relevant configurations in the reference list of
the \ac{MRSDCI} expansion. The greater numerical accuracy demands the inclusion
of a large number of configurations in the reference list, but that
leads to a rapid growth in the size of the \ac{CI} expansion, making the
calculations numerically prohibitive. However, here we are interested
in computing the energy differences rather than the absolute energies
of various states, for which good accuracy can be achieved even with
moderately large \ac{CI} expansions. In Table \ref{tab:energies-irrep-nanolife} we present
the average number of reference states (N$_{ref}$) included in the
\ac{MRSDCI} expansion and average number of configurations (N$_{total}$)
for different isomers. For a given isomer, the average has been calculated
across different irreducible representations which were needed in
these symmetry adapted calculations in order to compute the ground
and various excited states. The extensiveness of our calculations
can be seen from the number N$_{total}$, which is $\approx$ 77000
for the simplest cluster, and around four million for each symmetry
subspace of B$_{5}$. This makes us believe that our results are fairly
accurate.

Before we discuss the absorption spectrum for each isomer, we present
the ground state energies along with the relative energies of each
isomer are given in Table \ref{tab:energies-irrep-nanolife}. The \ac{MRSDCI} energy convergence
threshold was 10$^{-5}$ for all the isomers, with 10$^{-4}$ as convergence
threshold for configuration coefficients. From the results it is obvious
that as far as the energetics are concerned, for the B$_{3}$ the
triangular structure is most stable, while for B$_{4}$ and B$_{5}$
the rhombus and pentagonal structures, respectively, are favorable.

\begin{table*}
\small
\centering
\begin{threeparttable}
  \caption{The average number of reference configurations (N$_{ref}$), and average
number of total configurations (N$_{total}$) involved in \ac{MRSDCI} calculations,
ground state (GS) energies (in Hartree) at the \ac{MRSDCI} level, relative
energies and correlation energies (in eV) of various isomers of boron clusters.\label{tab:energies-irrep-nanolife}}
\par

{\begin{tabular}{cccccc}
\hline
Cluster  	& Isomer  & N$_{ref}$  	& N$_{total}$  & GS energy 	& Relative \tabularnewline
		&  		 &  			&  			& (Ha) 		& energy (eV) \tabularnewline
\hline 
B$_{2}$  & Linear  	& 24 		& 77245  		& -49.27844 	& 0.0 \tabularnewline
	      &  		&  			&  			&  			&  \tabularnewline
B$_{3}$  & Triangular  &  36 		& 596798  	& -73.98998 	& 0.00 \tabularnewline
	      & Linear 	&  41 		& 671334 		& -73.92906 	&  1.66 \tabularnewline
	      &  		&  			&  			&  			&  \tabularnewline
B$_{4}$  & Rhombus &  37 		& 1127918  	& -98.74004 	& 0.00 \tabularnewline
	      & Square 	&  40 		& 1070380  	& -98.73785 	&  0.06 \tabularnewline
	      & Linear 	& 34 		& 1232803  	&  -98.66575 	&  2.02 \tabularnewline
	      & Distorted Tetrahedron & 28 & 1253346  & -98.63213 	&  2.94 \tabularnewline
	      &  		&  			&  			&  			&  \tabularnewline
B$_{5}$  & Pentagon & 22 		& 3936612 	&  -123.42652  & 0.00 \tabularnewline
	      & Distorted Tri. bipyramid\tnote{1}  & 7 & 3927508 & -123.31485 &  3.04 \tabularnewline
\hline
\end{tabular}}
  \begin{tablenotes}
    \item[1]C$_{s}$ symmetry of isomer converted to C$_{1}$ in calculations.
  \end{tablenotes}
\end{threeparttable}
\end{table*}

\subsection{MRSDCI Photoabsorption Spectra of Boron Clusters}

Next we present and discuss the results of our photoabsorption calculations
for each isomer.

\subsubsection{B$_{2}$}

The simplest and most widely studied cluster of boron is B$_{2}$
with D$_{\infty h}$ point group symmetry. Using the \ac{CISD} method,
we obtained its optimized bond length to be 1.59 \AA{} (\emph{cf.}
Fig. \ref{fig:geometries-nanolife}\subref{fig:b2geom}), which is in excellent
agreement with the experimental value 1.589 \AA{}.\cite{herzberg_book}.
Using a DFT based methodology, Ati\c{s} \emph{et al.}\cite{turkish_boron},
obtained a bond length of 1.571 \AA{}, while Howard and Ray calculated
it to be 1.61 \AA{}, using the fourth-order perturbation theory (MP4).\cite{howard_ray}

\begin{figure}
\centering
\includegraphics[width=8.3cm]{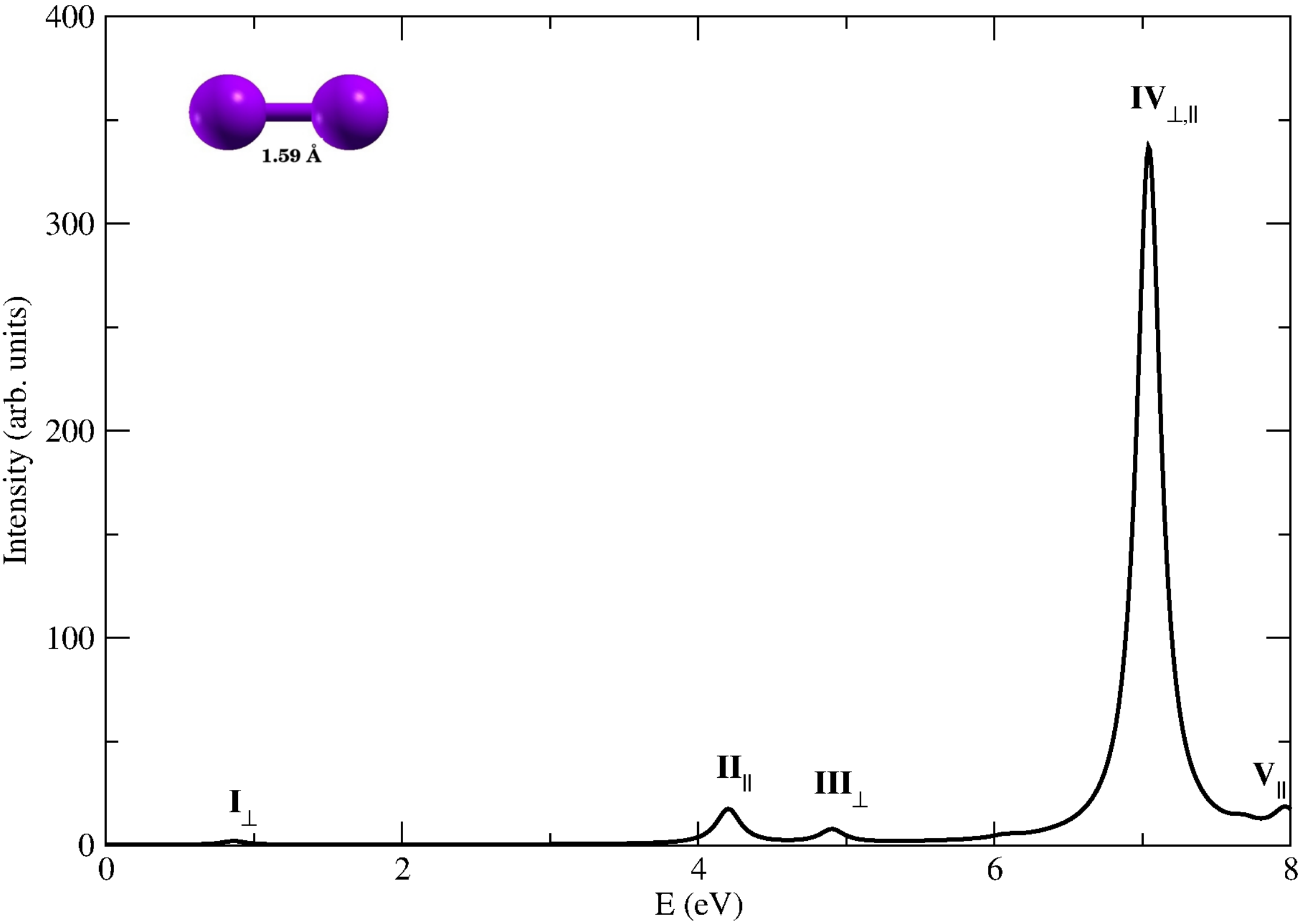}
\caption{The linear optical absorption spectrum of B$_{2}$, calculated using
the \ac{MRSDCI} approach. The peaks corresponding to the light polarized
along the molecular axis are labeled with the subscript $\parallel$,
while those polarized perpendicular to it are denoted by the subscript
$\perp$. For plotting the spectrum, a uniform linewidth of 0.1 eV
was used.}
\label{fig:b2_lin_final} 
\end{figure}

Because the ground state of B$_{2}$ is a spin triplet, its many-particle
wavefunction predominantly consists of a configuration with two degenerate \ac{SOMO} referred to as $H_{1}$
and $H_{2}$ in rest of the discussion. The excited state wavefunctions
will naturally consist of configurations involving electronic excitations
from the occupied MOs to the unoccupied MOs starting from \ac{LUMO} ($L$ for short). Our calculated photoabsorption
spectrum shown in Fig. \ref{fig:b2_lin_final} is characterized by
weaker absorptions at low energies, and a very intense one at high
energy. The many-particle wavefunctions of excited states contributing
to various peaks are presented in Table \ref{Tab:table_b2_lin}. A
feeble peak appears near 0.85 eV, dominated by $H_{2}\rightarrow L$
and $H_{2}\rightarrow L+4$ excitations compared to the \ac{HF} reference
configuration. It is followed by a couple of smaller peaks at 4.20
and 4.91 eV. The most intense peak is found at 7.05 eV, to which two
closely spaced states contribute. Transition to the state near 6.97
eV is polarized transverse to the bond length, while the one close
to 7.05 eV carries the bulk of oscillator strength, and is reached
by longitudinally polarized photons. All these states exhibit strong
mixing of singly-excited configurations. Near 8 eV, a smaller peak
appears which has strong contributions from doubly-excited configurations
$H-1\rightarrow L$; $H_{1}\rightarrow L+2$ and $H-1\rightarrow L$;
$H_{2}\rightarrow L+2$. The wavefunctions of the excited states
contributing to all the peaks exhibit strong configuration mixing,
instead of being dominated by single configurations, pointing to the
plasmonic nature of the optical excitations.\cite{plasmon}

\subsubsection{B$_{3}$}

Boron trimer has two possible isomers, triangular and the linear one
shown in Figs. \ref{fig:geometries-nanolife}\subref{fig:b3trgeom} and \ref{fig:geometries-nanolife}\subref{fig:b3lingeom}.
We found equilateral triangle with D$_{3h}$ symmetry to be the most
stable isomer. The optimized bond length for triangular isomer is
1.55 \AA{}, with the ground state ($^{2}A_{1}^{'}$) energy 1.66 eV
lower than that of its linear counterpart. We also explored the possibility
of isosceles triangular structure as a favorable one, because B$_{3}$
is an open-shell system, making it a possible candidate for Jann-Teller
distortion. However, the \ac{CCSD} optimized geometry corresponding to
the isosceles structure is so slightly different compared to the equilateral
one, that it is unlikely to affect the optical absorption spectrum
in a significant manner. Our calculated bond length is in good agreement
with experimental value 1.57 \AA{}\cite{hanley_whitten}, as well
as with other reported theoretical values of 1.553 \AA{},\cite{boustani_prb97}
1.56 \AA{} \cite{howard_ray} and 1.548 \AA{}.\cite{turkish_boron} 

The linear B$_{3}$ isomer with the D$_{\infty h}$ symmetry, and
the $^{2}\Sigma_{g}^{-}$ as ground state, was found to have equal
bond lengths. Our \ac{CISD} optimized bond length of 1.51 \AA{} agrees
well with the value 1.518 \AA{} reported by Ati\c{s} \emph{et al}.\cite{turkish_boron}

\begin{figure}
\centering
\includegraphics[width=8.3cm]{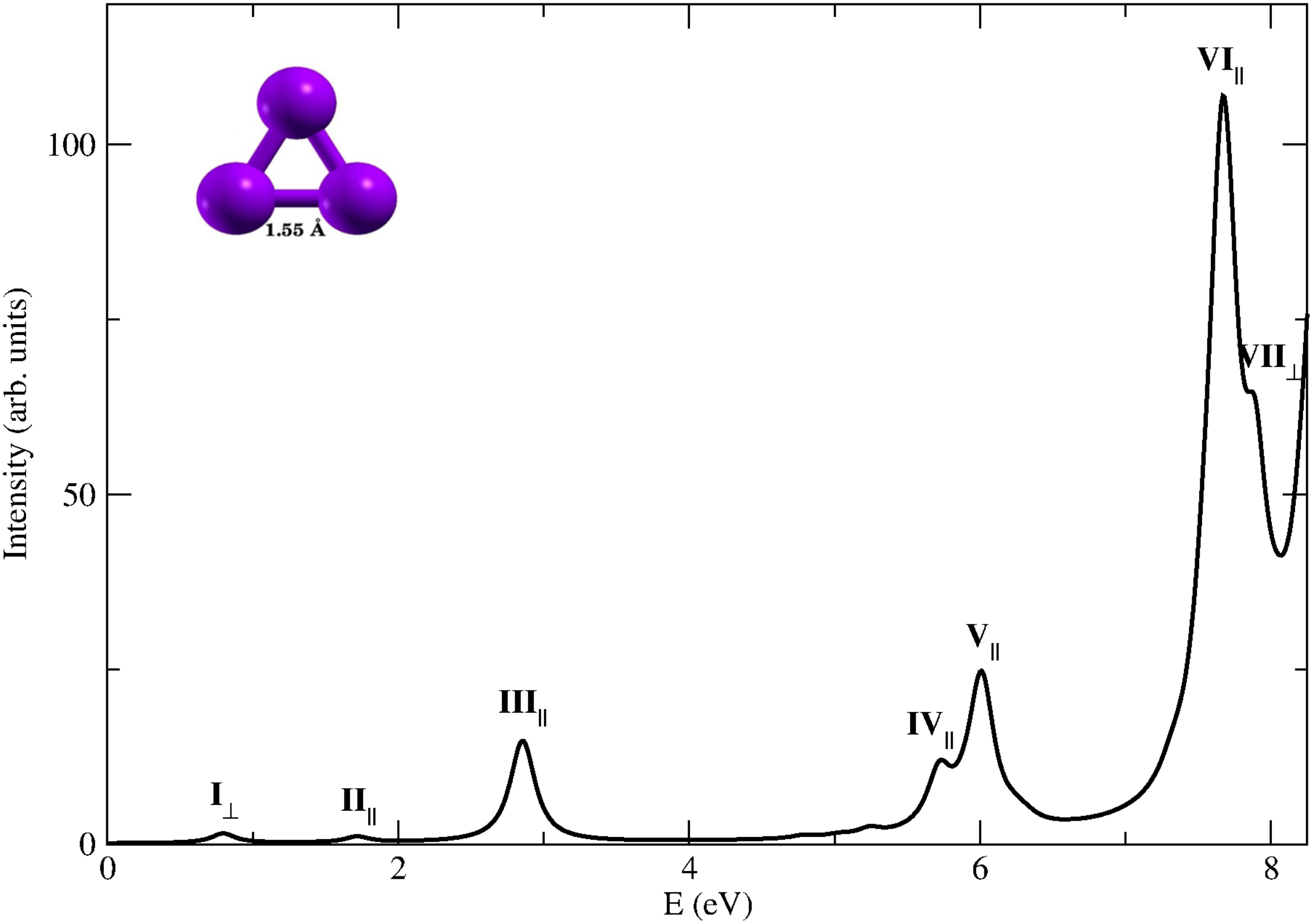}
\caption{  The linear optical absorption spectrum of triangular
B$_{3}$ calculated using the \ac{MRSDCI} approach. Peaks corresponding
to light polarized in the plane of the molecule are labeled with subscript
$\parallel$, while those polarized perpendicular to the plane are
denoted by the subscript $\perp$. For plotting the spectrum, a uniform
linewidth of 0.1 eV was used.}
\label{fig:b3_tri_combined} 
\end{figure}

\begin{figure}
\centering
\includegraphics[width=8.3cm]{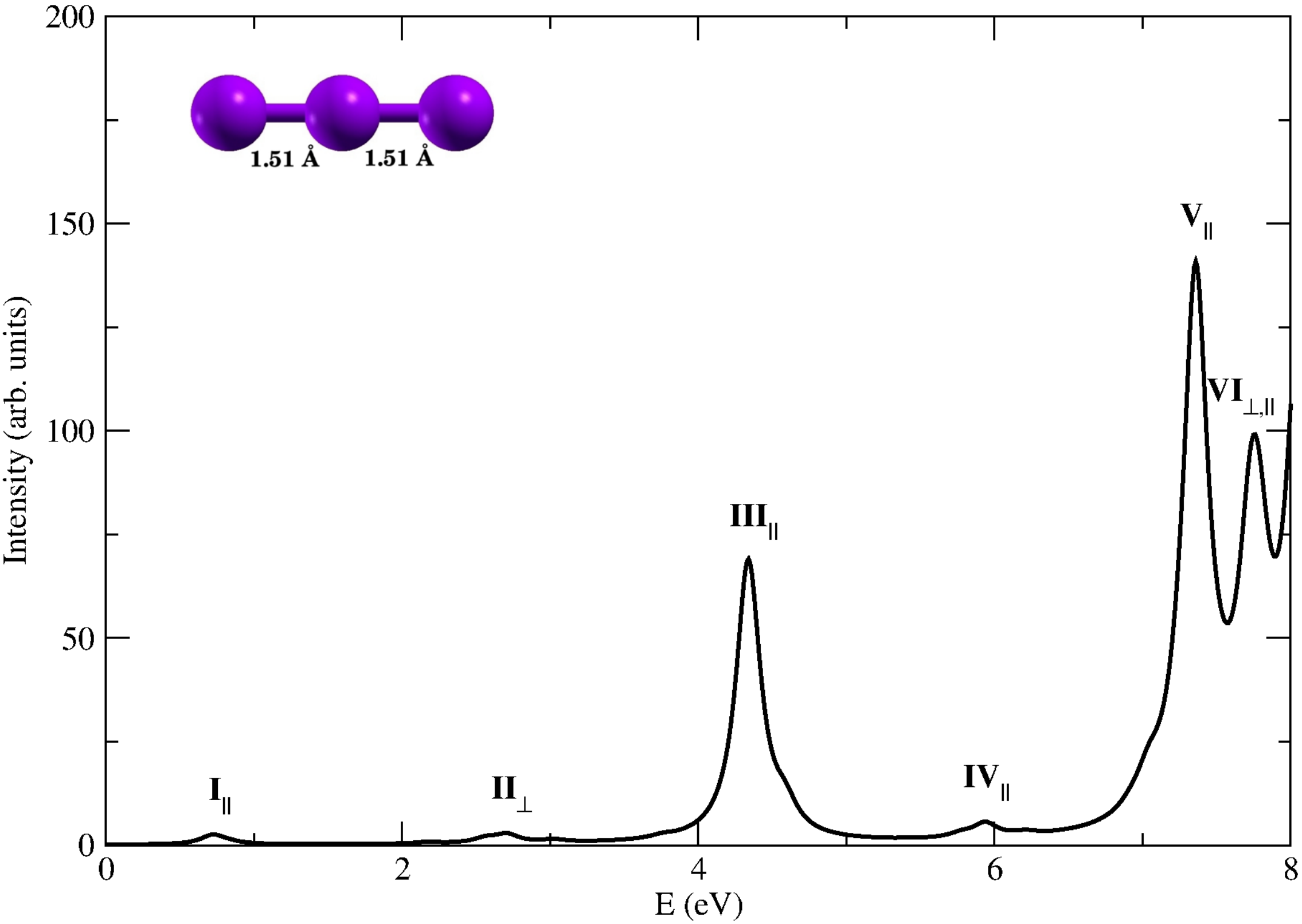}
\caption{The linear optical absorption spectrum of linear B$_{3}$, calculated
using the \ac{MRSDCI} approach. The peaks corresponding to the light polarized
along the molecular axis are labeled with subscript $\parallel$,
while those polarized perpendicular to it are denoted by the subscript
$\perp$. For plotting the spectrum, a uniform linewidth of 0.1 eV
was used.}
\label{fig:b3_lin_final} 
\end{figure}

The photoabsorption spectra of two isomers of B$_{3}$ are presented
in Figs. \ref{fig:b3_tri_combined} and \ref{fig:b3_lin_final}. The
corresponding many-particle wavefunctions of excited states contributing
to various peaks are presented in Table  \ref{Tab:table_b3_tri} and 
\ref{Tab:table_b3_lin}. It is obvious that in the linear structure,
absorption begins at a lower energy as compared to the triangular
one, although the intensity of its low-energy peaks is very small.
In the triangular isomer on the other hand, most of the intensity
is concentrated at rather high energies, except for a weaker peak
close to 3 eV. The optical spectra of linear isomer begins with very
weak peaks at 0.7 eV (longitudinal polarization) and 2.7 eV (transverse
polarization), with their many-particle wavefunctions dominated by
singly-excited configurations. The relatively intense peak at 4.3
eV corresponding to a longitudinally polarized transition, is dominated
by doubly-excited configurations. It is followed by a small peak mainly
due to single excitation $H-2\rightarrow L$, near 5.9 eV. The most
intense peak of the spectrum occurs at 7.4 eV, followed by another
strong peak close to 7.7 eV. Both the features correspond to longitudinally
polarized transitions, with the many particle wavefunctions of the
concerned states being strong mixtures of single and double excitations
with respect to the HF reference state. We note that in the absorption
spectrum of the linear cluster, quite expectedly, the bulk of the
oscillator strength is carried by longitudinally polarized transitions.

Because the triangular cluster is a planar cluster, its orbitals can
be classified as in-plane $\sigma$ orbitals, and the out-of-plane
$\pi$ orbitals. Both the \ac{HOMO} (a singly occupied orbital, in this
case) and the \ac{LUMO} for this isomer are $\sigma$-type orbitals. For
this system, two types of optical absorptions are possible: (a) those
polarized in the plane of the cluster, and (b) the ones polarized
perpendicular to that plane. Our calculations reveal that the transitions
corresponding to perpendicular polarization ($z$ direction), except
for a couple of peaks, have negligible intensities. From Fig. \ref{fig:b3_tri_combined}
it is obvious that the optical absorption in the triangular isomer
starts with a very weak $z-$polarized feature near 0.8 eV (peak I),
corresponding  to a state with the wavefunction dominated by single
excitations ($\pi\rightarrow\sigma^{*}$).  This is followed by a
series of peaks ranging from II to VI which correspond to the photons
polarized in the plane of the cluster. All these peaks are dominated
by states consisting primarily of singly-excited configurations of
the $\sigma\rightarrow\sigma^{*}$ type. The most intense peak VI
is followed by a shoulder-like feature (VII) corresponding to a $z$-polarized
absorption. 

If we compare the absorption spectra of the linear and the triangular
B$_{3}$, the peak at 4.34 eV in the spectrum of the linear
cluster is the distinguishing feature, and can be used to differentiate
between the two isomers.

\subsubsection{B$_{4}$}

For the B$_{4}$ cluster, we investigated the rhombus, square, linear
and tetrahedral structures. While the rhombus shaped isomer was found
to have the lowest energy, but the square isomer is higher in energy
only by a small amount. As a matter of fact, at the \ac{HF} level the energies
of the two isomers were found to be almost degenerate. It was only
after the electron correlation effects were included at the \ac{CI} level
that the rhombus stabilized by $\approx$ 0.06 eV (\emph{cf.} Table
\ref{tab:energies-irrep-nanolife})with respect to the square. For the rhombus, the
ground state was $^{1}A_{g}$, with the optimized bond length 1.529
\AA{}, and the short diagonal length 1.909 \AA{}. These results are
in good agreement with with the corresponding lengths of 1.528 \AA{}
and 1.885 \AA{} reported by Boustani,\cite{boustani_prb97}, and 1.523
\AA{} and 1.884 \AA{} computed by Ati\c{s} \emph{et al.}\cite{turkish_boron}
Both \ac{HOMO} and \ac{LUMO} of rhombus isomer are $\sigma$ orbitals.

For the square isomer, with D$_{4h}$ symmetry, the electronic ground
state is expectedly $^{1}$A$_{1g}$. As shown in Fig. \ref{fig:geometries-nanolife}\subref{fig:b4sqrgeom},
our optimized bound length is 1.53 \AA{}, which agrees well with the
values 1.527 \AA{} and 1.518 \AA{} as reported in Refs. \citep{boustani_prb97} and \citep{turkish_boron}.
In this isomer, \ac{HOMO} is a $\sigma$ orbital while \ac{LUMO} is a $\pi$
orbital.

Linear B$_{4}$, with the D$_{\infty h}$ symmetry, has the electronic
ground state of $^{1}\Sigma_{g}^{-}$. However, energetically linear
structure is 2.02 eV higher than the rhombus one (\emph{cf.} Table
\ref{tab:energies-irrep-nanolife}) which rules out its existence at the room temperatures.
As per Fig. \ref{fig:geometries-nanolife}\subref{fig:b4lingeom}
, the central
bond length was found to be 1.49 \AA{}, with the two outer bonds being
1.55 \AA{} in length. For the same bonds, Ati\c{s} \emph{et al.} reported
these lengths to be 1.487 \AA{} and 1.568 \AA{}, respectively.\cite{turkish_boron}

The distorted tetrahedral structure having C$_{3v}$ symmetry, made
up of four isosceles triangular faces with lengths 1.785 \AA{}, 1.785
\AA{} and 1.512 \AA{}. This isomer also lies much higher in energy
as compared to the most stable rhombus structure.

\begin{figure}
\centering
\includegraphics[width=8.3cm]{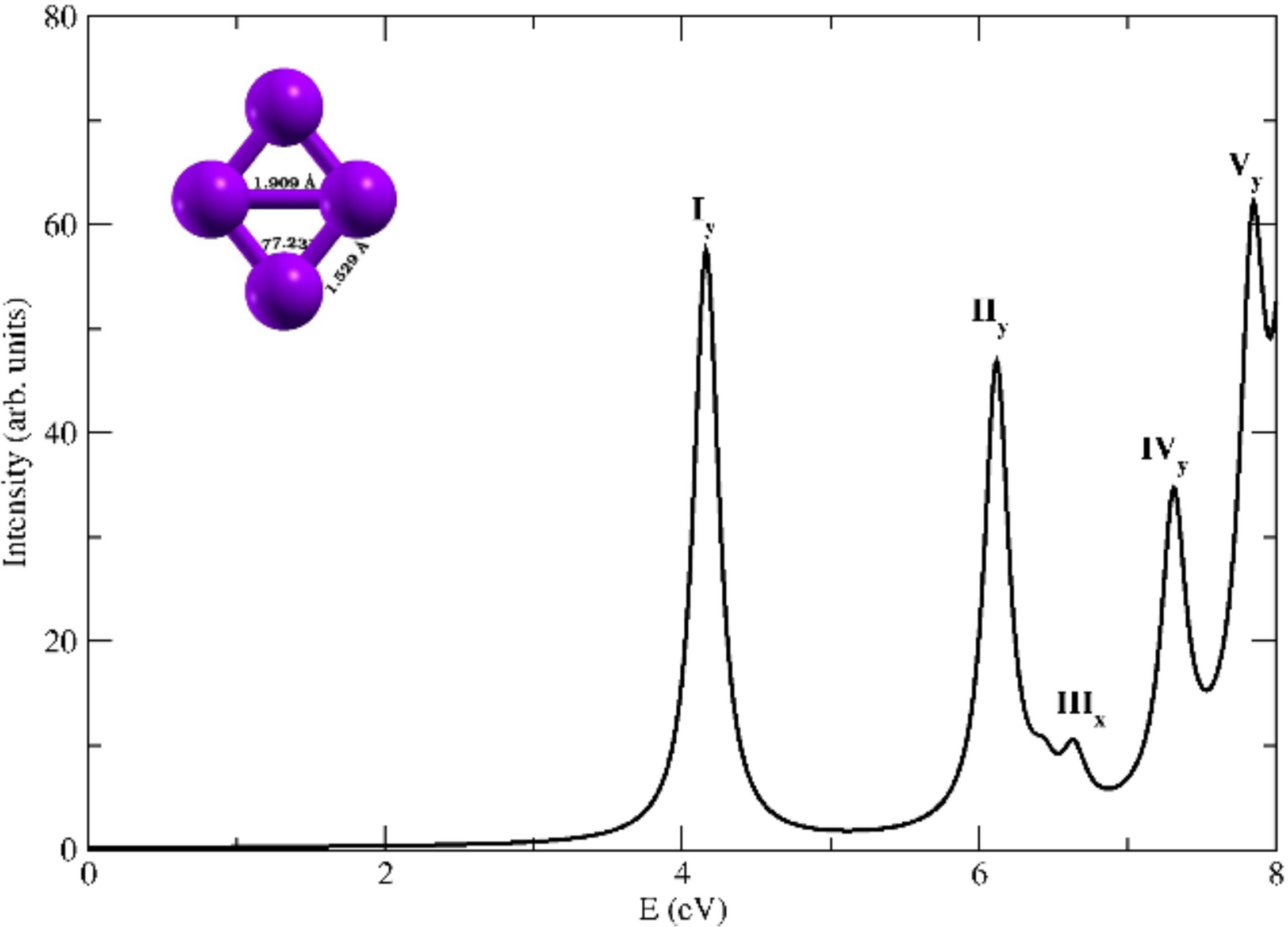} 
\caption{  The linear optical absorption spectrum of B$_{4}$
rhombus geometry using the \ac{MRSDCI} approach. Isomer is aligned in $x-y$
plane with short diagonal along $x$-axis. Peaks corresponding to
light polarized along $x$ and $y$-axis are labeled with subscript
$x$ and $y$. For plotting the spectrum, a uniform linewidth of 0.1
eV was used.}
\label{fig:b4_rho_combined} 
\end{figure}

\begin{figure}
\centering
\includegraphics[width=8.3cm]{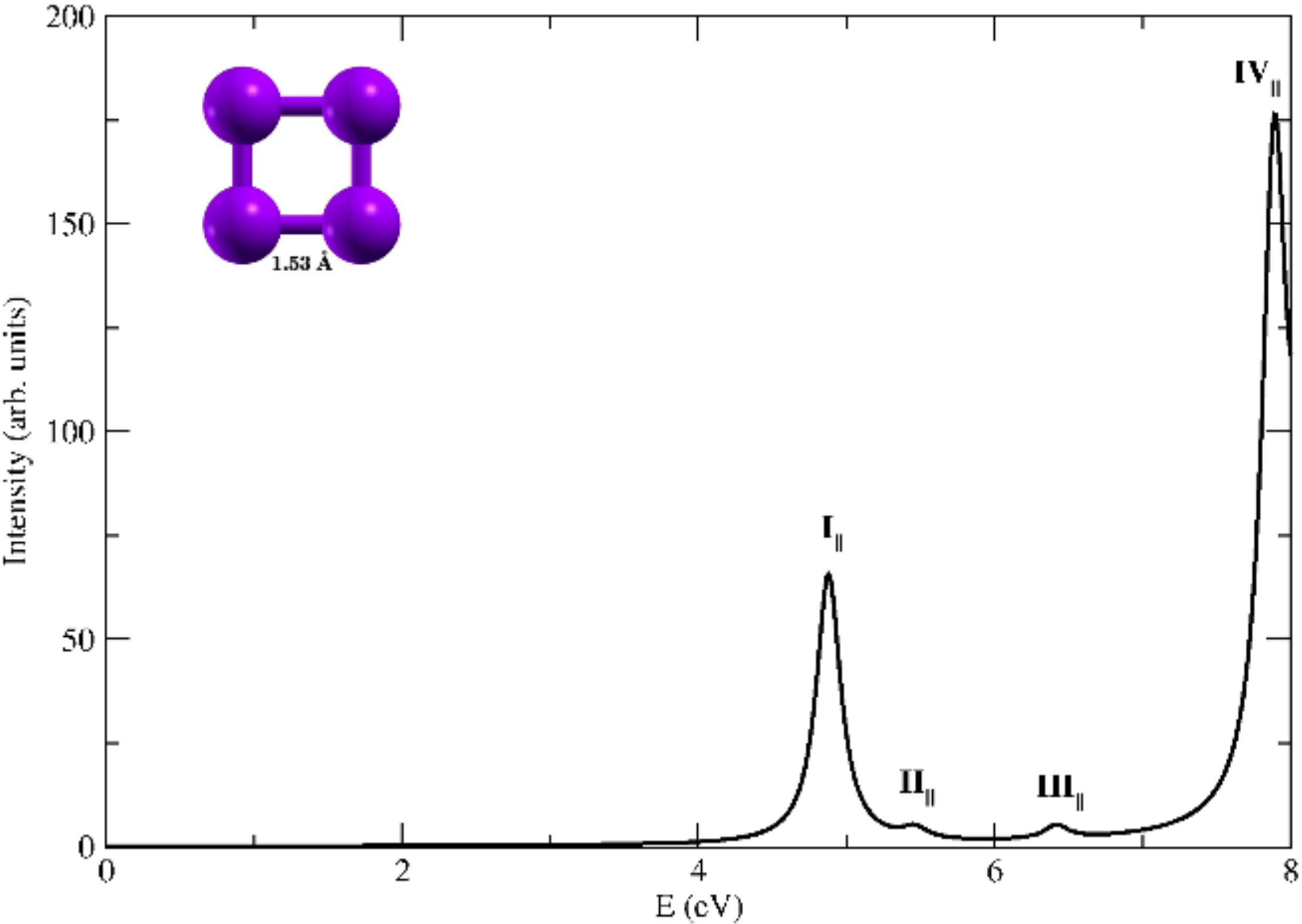}
\caption{  The linear optical absorption spectrum of B$_{4}$
square geometry using \ac{MRSDCI} approach. Isomer is aligned in $x-y$
plane. Spectrum represents the equal contribution from light polarized
in $x$ and $y$ direction. Peaks corresponding to light polarized
in the plane of the molecule are labeled with subscript $\parallel$.
For plotting the spectrum, a uniform linewidth of 0.1 eV was used.}
\label{fig:b4_sqr_combined} 
\end{figure}

\begin{figure}
\centering
\includegraphics[width=8.3cm]{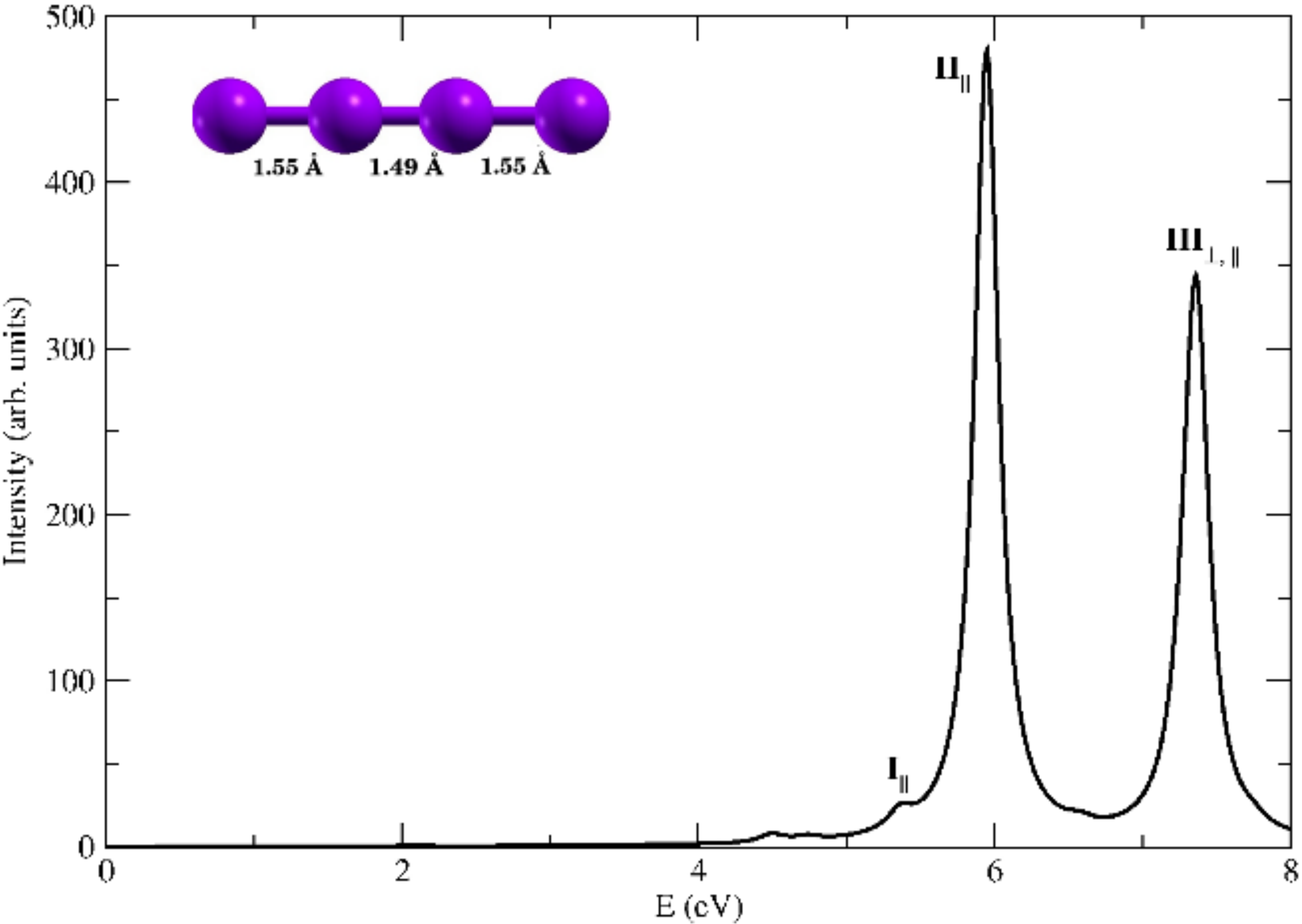}
\caption{The optical absorption spectrum of linear B$_{4}$, calculated using
the \ac{MRSDCI} approach. The peaks corresponding to the light polarized
along the molecular axis are labeled with subscript $\parallel$,
while those polarized perpendicular to it are denoted by the subscript
$\perp$. For plotting the spectrum, a uniform linewidth of 0.1 eV
was used.}
\label{fig:b4_lin_final} 
\end{figure}

\begin{figure}
\centering
\includegraphics[width=8.3cm]{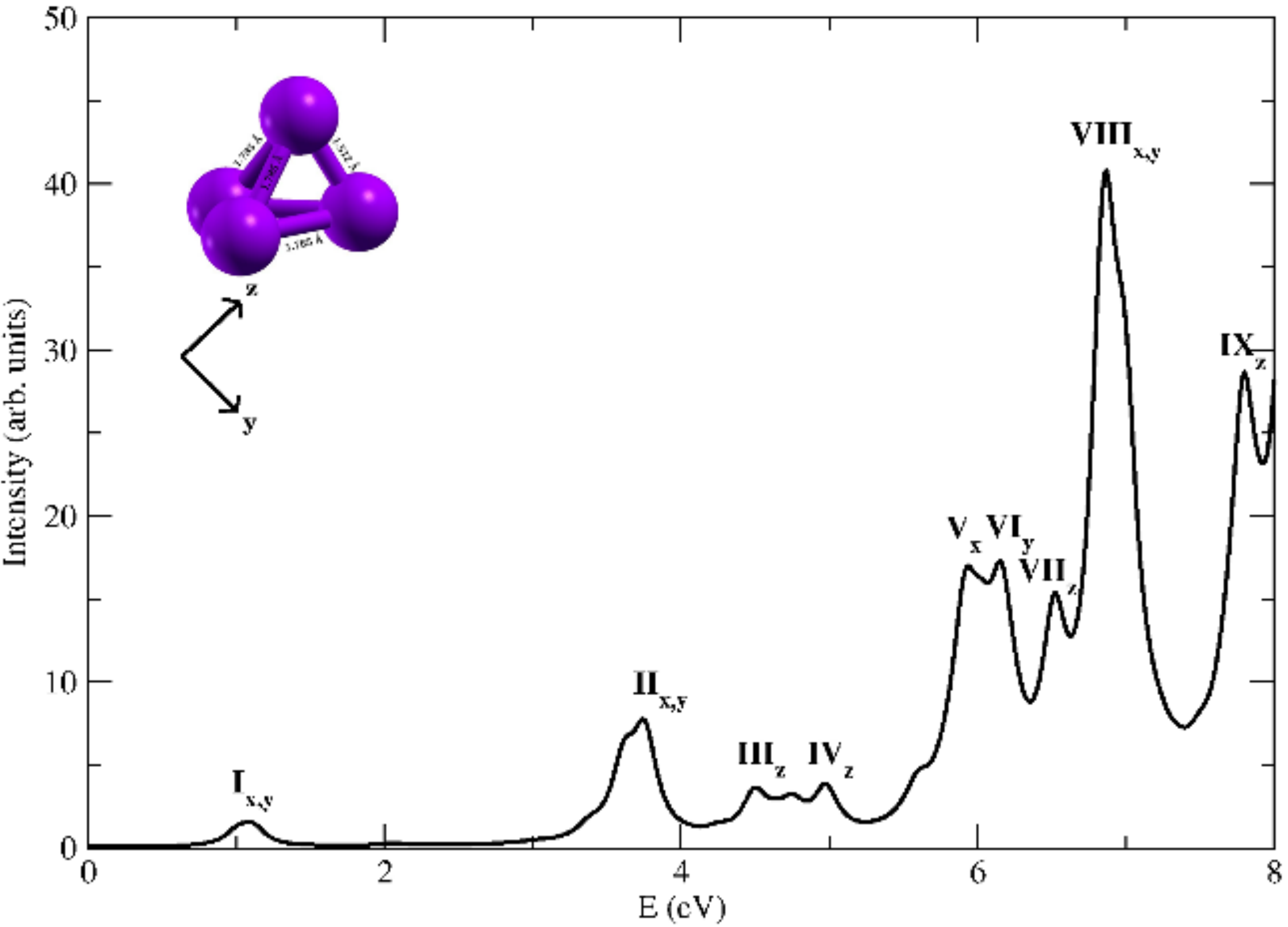}
\caption{  The linear optical absorption spectrum of B$_{4}$
distorted tetrahedral geometry using the \ac{MRSDCI} approach. Peaks corresponding
to light polarized along $x$, $y$ and $z$-axis are labeled with
subscript $x$, $y$ and $z$ respectively. For plotting the spectrum,
a uniform linewidth of 0.1 eV was used.}
\label{fig:b4_tetra_combined} 
\end{figure}

The absorption spectra of rhombus, square, linear, and tetrahedral
isomers are presented in Figs. \ref{fig:b4_rho_combined}, \ref{fig:b4_sqr_combined},
\ref{fig:b4_lin_final}, and \ref{fig:b4_tetra_combined} 
respectively.
From the figures it is obvious that the general features of the absorption
spectra of rhombus and square isomers are similar, except that the
rhombus spectrum, with the onset of the absorption near 4 eV, is red-shifted
by about 1 eV as compared to the square. The absorption spectrum of
the linear structure is blue-shifted as compared to the rhombus
and square shaped isomers, with the majority of absorption occurring
in the energy range 5--8 eV. This aspect of the photoabsorption in
B$_{4}$ is similar to the case of B$_{3}$ for which also the linear
structure exhibited a red-shifted absorption compared to the triangular
one.

Since B$_{4}$ rhombus isomer has $D_{2h}$ symmetry, we can represent
the absorption due to light polarized in different directions in terms
of irreducible representations of $D_{2h}$. So absorption due to
in-plane polarized light corresponds to $B_{1u}$ and $B_{2u}$, while
$B_{3u}$ corresponds to light polarized in the direction perpendicular
to the plane of the isomer. 

The polarization resolved absorption spectrum of rhombus B$_{4}$,
as shown in Fig. \ref{fig:b4_rho_combined}, exhibits a rather red-shifted
nature as compared to the linear isomer. The many-particle wavefunctions
of excited states contributing to various peaks are presented in Table
\ref{Tab:table_b4_rho}. The onset of spectrum is seen at 4.15 eV
followed by a peak at around 6.12 eV. Both of them are due to $y-$polarized
component, i.e. along the larger diagonal. The dominant contribution
to these peaks come from $\sigma\rightarrow\pi^{*}$ for former, and
$\pi\rightarrow\pi^{*}$ for latter. The $x$-component does not contribute
much in the whole spectrum, except for minor peaks at 4.2 eV and 6.6
eV. It is characterized by mainly $\pi\rightarrow\pi^{*}$ type transitions.
It is followed by a relatively low intensity peak at 7.3 eV due to
$y$-polarized component with leading contribution from $\sigma\rightarrow\pi^{*}$
transitions. The most intense peak, at 7.84 eV, having $y$-polarization
component, is characterized by $\sigma\rightarrow\pi^{*}$ type of
transitions. There are no direct $H\rightarrow L$ transitions for
this isomer, because they are dipole forbidden. The absorption due
to light polarized in the direction perpendicular to the plane of
isomer is negligible.

The square B$_{4}$ isomer, because of its symmetry, gets equal contribution
to absorption spectrum from both $x-$ and $y-$component. It corresponds
to in-plane polarization due to $B_{1u}$ and $B_{2u}$ irreducible
representation, while $B_{3u}$ corresponds to light polarized in
the direction perpendicular to the plane of the isomer. However, in
this isomer also, the contribution due to latter is quite negligible.
The many-particle wavefunctions of excited states contributing to
various peaks are presented in Table \ref{Tab:table_b4_sqr}. It shows
just one major peak at 4.88 eV below 7 eV, characterized by $\sigma\rightarrow\pi^{*}$;$\sigma\rightarrow\pi^{*}$
double excitation. Two smaller peaks appear in this range at 5.5 eV
and 6.4 eV, with leading contributions from $\sigma\rightarrow\pi^{*}$;$\sigma\rightarrow\pi^{*}$
and $\sigma\rightarrow\pi^{*}$;$\pi\rightarrow\pi^{*}$ excitations
respectively. Beyond 7 eV, there are many closely spaced peaks including
the most intense one at 7.89 eV. It is characterized by double excitation
$\sigma\rightarrow\pi^{*}$;$\pi\rightarrow\pi^{*}$. In this isomer
also, a direct $H\rightarrow L$ transition is forbidden. Though,
there is very little difference in total energies of rhombus and square
isomers of B$_{4}$, their optical absorption spectra are completely
different. They can be easily identified from each other by looking
at number of peaks below 7 eV energy. Rhombus exhibits two major peaks,
while square has just one. 

Linear B$_{4}$ isomer exhibits absorption with few, but sharp peaks.
The many-particle wavefunctions of excited states contributing to
various peaks are presented in Table \ref{Tab:table_b4_lin}. The
onset of optical absorption occurs near 4.5 eV, due to absorption
of long-axis polarized light, followed by two major peaks at 5.95
eV, and 7.36 eV. The first of these two intense peaks, peak II is
dominated by singly-excited configurations, while the second one (peak
III) is a strong mixture of both singly- and doubly-excited configurations
with respect to the HF reference configuration. 

The 3D structure, a distorted tetrahedral isomer, exhibits an absorption
spectrum very different from other isomers, as displayed in Fig. \ref{fig:b4_tetra_combined}.
The many-particle wavefunctions of excited states contributing to
various peaks are presented in Table \ref{Tab:table_b4_tetra}. It
is the only B$_{4}$ isomer to exhibit peaks below 4 eV. The absorption
spectrum is spread over a much larger energy range, and is almost
continuous. The oscillator strengths associated with various peaks
are much smaller than in other isomers, and most of the peaks appear
pairwise. The onset of absorption spectrum is seen at around 1.1 eV,
characterized mainly by an excited dominated by single-excitation
$H\rightarrow L$ (\emph{cf.} Table \ref{Tab:table_b4_tetra}). In
this isomer, in contrast to other B$_{4}$ isomers, direct $H\rightarrow L$
transitions are allowed. Higher energy peaks in this isomer are dominated
by doubly-excited configurations, and, are, therefore, sensitive to
the electron-correlation effects.

\subsubsection{B$_{5}$}

We investigated two isomers of B$_{5}$: a Jahn-Teller distorted pentagon
with the C$_{2v}$ symmetry, and (b) a triangular bipyramid with the
C$_{s}$ point group symmetry. The latter one is the second 3-D structure
of the boron clusters probed in this work. The lowest lying pentagon
isomer, has $^{2}$B$_{2}$ electronic ground state, and is 3.04 eV
lower in energy as compared to the bipyramid structure. For the pentagon,
the symmetry of ground state at the \ac{SCF} level was A$_{1}$, however,
at the \ac{MRSDCI} level the B$_{2}$ state became lower in energy, in
agreement with the previous calculations of Boustani.\cite{boustani_prb97}
Our optimized geometry for the pentagon (Fig. \ref{fig:geometries-nanolife}\subref{fig:b5pengeom})
corresponds to an average bond length of 1.56 \AA{}, as against 1.57
\AA{} reported by Boustani\cite{boustani_prb97}, and 1.644 \AA{}
reported by Ati\c{s} \emph{et al}.\cite{turkish_boron}. The singly
occupied molecular orbital (denoted by $H$) and \ac{LUMO} of pentagon
isomers are of $\pi$ and $\sigma$ type, respectively. The bond lengths
for the bipyramid structure are shown in Fig. \ref{fig:geometries-nanolife}\subref{fig:b5tetgeom},
with an average bond length of 1.704 \AA{}. The triangular base was
found to be isosceles with 1.97 \AA{} as equal sides, and 1.75 \AA{}
as the other side. 

\begin{figure}
\centering
\includegraphics[width=8.3cm]{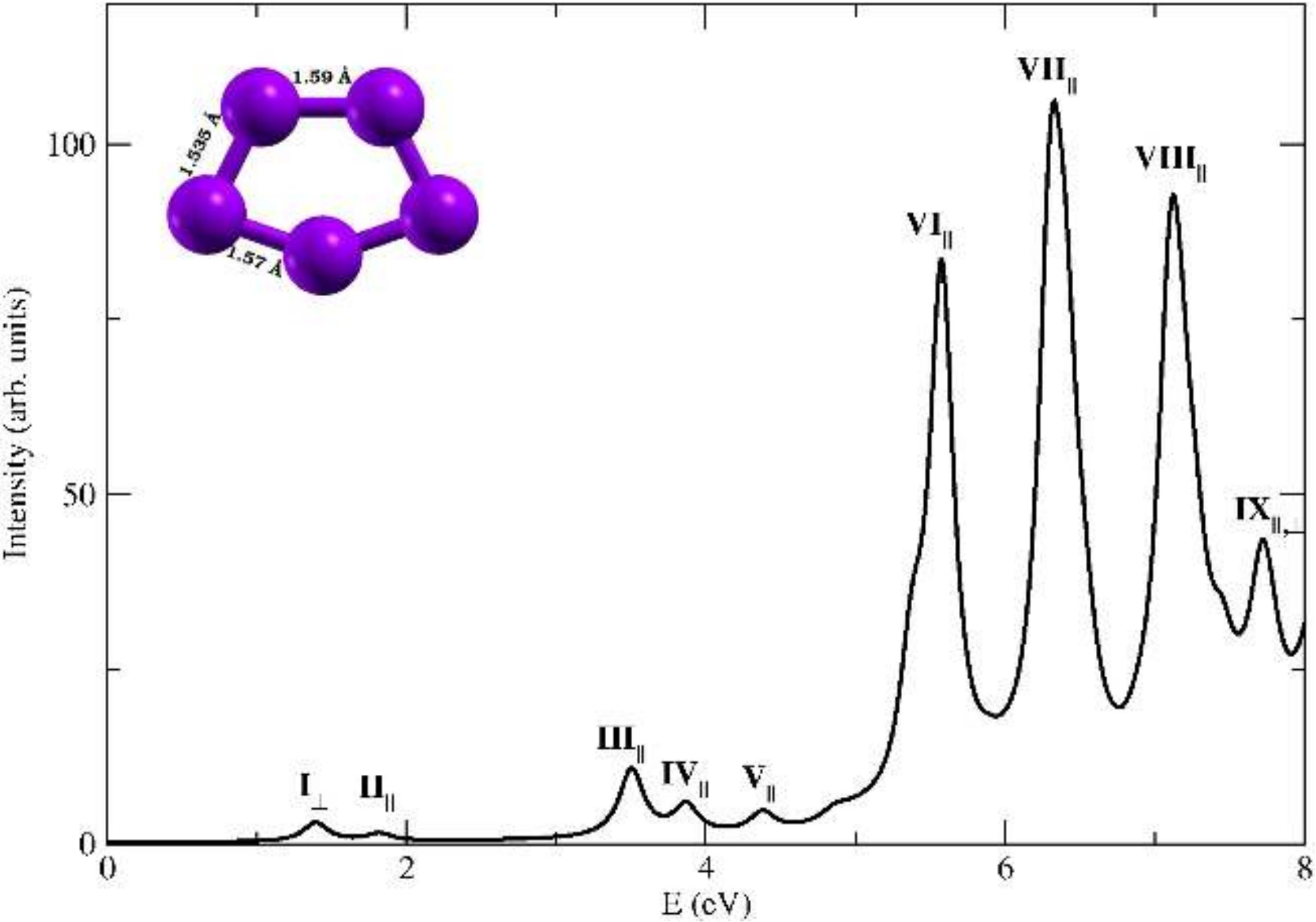}
\caption{  The linear photo-absorption spectrum of pentagon B$_{5}$,
calculated using the \ac{MRSDCI} approach. The peaks corresponding to the
light polarized in the plane of the molecule are labeled with subscript
$\parallel$, while those polarized perpendicular to it are denoted
by the subscript $\perp$. For plotting the spectrum, a uniform linewidth
of 0.1 eV was used.}
\label{fig:b5_pen_combined} 
\end{figure}

\begin{figure}
\centering
\includegraphics[width=8.3cm]{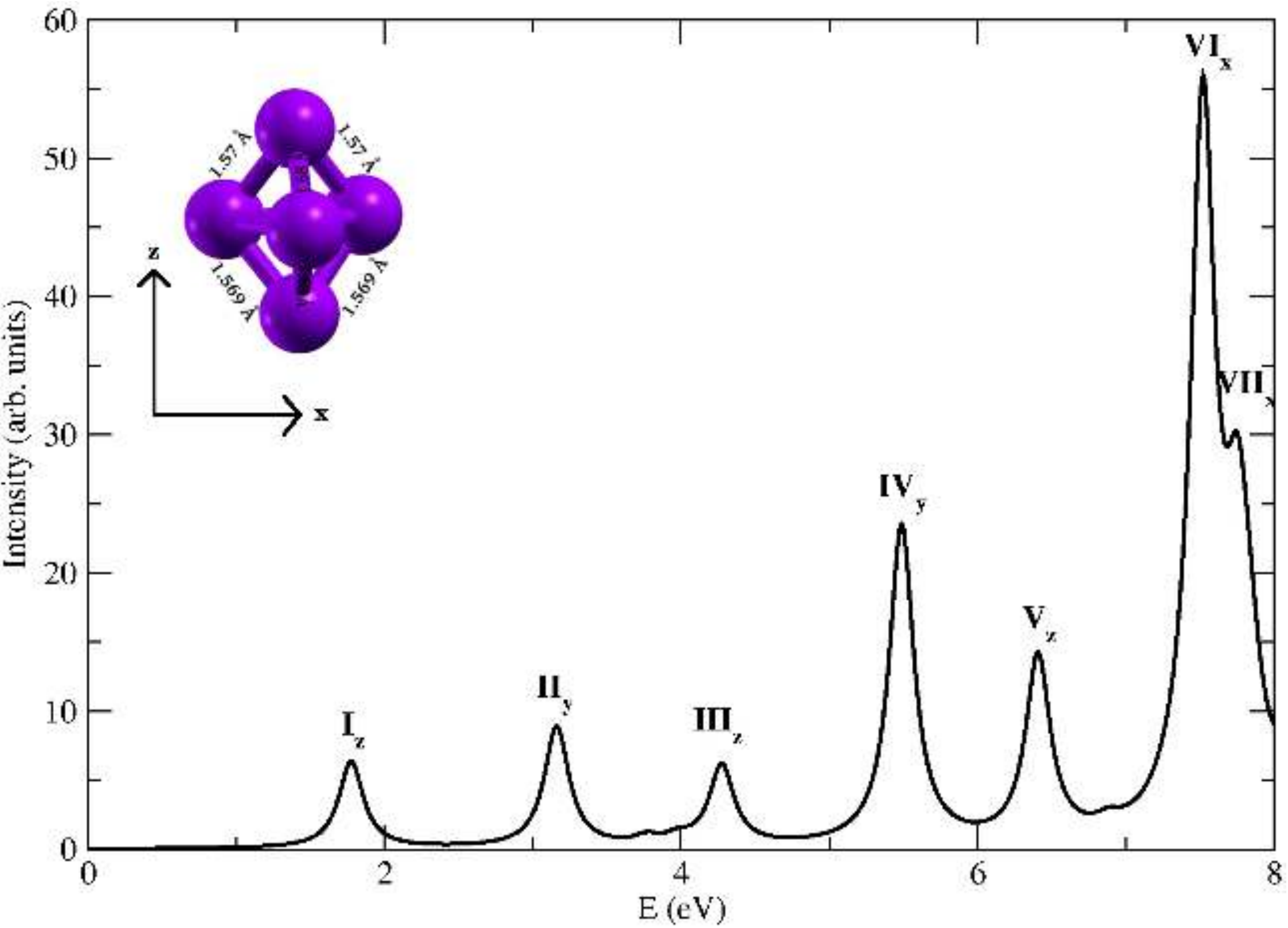}
\caption{  The linear photo-absorption spectrum of distorted triangular
bipyramid B$_{5}$, calculated using the \ac{MRSDCI} approach. Peaks corresponding
to light polarized along $x$, $y$ and $z$-axis are labeled with
subscript $x$, $y$ and $z$ respectively. For plotting the spectrum,
a uniform linewidth of 0.1 eV was used.}
\label{fig:b5_tet_combined} 
\end{figure}

The absorption spectra of the two isomers are presented in Figs. \ref{fig:b5_pen_combined}
and \ref{fig:b5_tet_combined}. The many-particle wavefunctions of
excited states contributing to various peaks are presented in Table
\ref{Tab:table_b5_pen} and \ref{Tab:table_b5_tet} respectively.
From the figures it is obvious that the intense absorption in the bipyramid
starts at much lower energies as compared to the pentagonal isomer.
Intense absorption peaks in pentagon B$_{5}$ are located at energies
higher than 5 eV, with three equally intense peaks at 5.58 eV, 6.30
eV and 7.16 eV, with the photons polarized along the plane of the
molecule direction. It has an underlying low intensity absorption
contribution from photons polarized along $z-$ direction, which is
perpendicular to the molecular plane. The major contribution to the
peak at 5.58 eV comes from $\pi\rightarrow\sigma^{*}$ and $\sigma\rightarrow\sigma^{*}$,
single excitations. The latter configuration also contributes to the
most intense peak at 6.30 eV. The peak at 7.16 eV is mainly due to
$\sigma\rightarrow\sigma^{*}$ type transitions.

Since the B$_{5}$ triagonal bipyramid isomer is not a symmetric one,
the calculations were done using C$_{1}$ symmetry, thereby increasing
the difficulty in diagonalizing the Hamiltonian. Hence, in order to
reduce the matrix size, we have used a smaller number of reference
configurations, and also relaxed the energy convergence threshold
criterion a little.

The optical absorption spectrum of B$_{5}$ triangular bipyramid isomer
is exhibited by almost equally spaced peaks at relatively lower energies.
The optical absorption starts at 1.74 eV characterized by $H-1\rightarrow L+4$
configuration. It is followed by two equal intensity peaks at 3.16
eV and 4.27 eV with contributions from single excitations $H-2\rightarrow H\mbox{ and }H-1\rightarrow L+2$,
respectively. The most intense peak is found at 7.52 eV dominated
by the doubly excited configuration $H-2\rightarrow H\:;\: H-1\rightarrow L+2$.
There are two distinguishing features as far as the optical absorption
in the two isomers is concerned: (a) presence of three intense peaks
in the higher energy region of the absorption spectrum of the pentagonal
isomer, and (b) occurrence of equally spaced absorption peaks at lower
energies in the spectrum of bipyramidal isomer.

\section{Summary}
\label{sec:conclusions-nanolife}

We presented systematic large-scale all-electron correlated calculations of photoabsorption spectra of boron clusters B$_{n}$, ($n=$2--5)
with several possible isomers of each cluster. The calculations were performed using the \ac{MRSDCI} method which takes electron correlations
into account at a sophisticated level, both for the ground and the excited states. For a cluster consisting of a given number of atoms,
significant changes were observed in absorption spectra for different isomers, indicating a strong structure-property relationship. Therefore,
our computed spectra can be used in the future photoabsorption experiments to distinguish between different isomers of a cluster, something which
is not possible with the conventional mass spectrometry. We also analyzed the many-particle wavefunctions of various excited states and found
them to be a mixture of a large number of configurations, indicating the nature of photoexcited states in these clusters to be plasmonic.\cite{plasmon}
A noteworthy aspect of the ground state photoabsorption of various clusters was the absence of high-intensity peaks in the low-energy
region of the spectrum. The most intense peaks occurred at higher energies involving orbitals away from the Fermi level, consistent
with the fact that the bulk boron is an indirect bandgap semiconductor, with no optical absorption at the gap. In other words, optical absorption
features of bulk boron were already evident in smaller clusters. 

  \lhead{{\chaptername\ \thechapter.}{  Boron Clusters B$_{6}$ and B$_{6}^{+}$}}
   \chapter{\label{chap:main_boron6}Theory of Linear Optical Absorption in Various Isomers of Boron Clusters B$_{6}$ and B$_{6}^{+}$}
\emph{This chapter is based on a published paper, Eur. Phys. J. D, \textbf{67}, 98 (2013) \\ by Ravindra Shinde and Alok Shukla.}
\par

Boron clusters exhibit a number of interesting properties, comparable only to its neighbor, carbon. The study of smaller boron clusters has
unraveled a great potential for applications in nanotechnology. For example, some planar boron structures are also found to be 
analogous to hydrocarbons \cite{kiran_wang}. The all-boron clusters are found promising candidates as inorganic 
ligands \cite{coord-chem-review,boron-8-ligand}.  A circular B$_{19}^{-}$ cluster, with a unit of B$_{6}$ wheel in the center 
behaves as a Wankel motor, \emph{i.e.} the inner B$_{6}$ wheel rotating opposite to the outer B$_{13}$ 
ring \cite{boron-19-rotor,boron19-natchem}. %
%
 As far as optical properties of boron clusters are
 concerned, only a small number of reports available. Marques
 and Botti studied the optical absorption spectra of B$_{20}$, B$_{38}$,
 B$_{44}$, B$_{80}$ and B$_{92}$ using time-dependent DFT \cite{marques_fullerene}. 
 To the best of our knowledge, no other experimental results are available
 on the optical absorption of boron clusters.

 In the previous chapter, we studied the optical absorption in boron clusters B$_{n}$ (n=2 -- 5) using a large-scale multi-reference 
 configuration interaction method \cite{nano_life}. This method is quite expensive and scales as N$^6$, where N is the
number of orbitals used in the calculations, it becomes more and more computationally 
demanding for large clusters, or clusters with no symmetry. However good insights can be achieved with much
less expensive method known as \ac{CIS}, containing only one electron excitations
from the Hartree-Fock ground state \cite{chem-review-cis}. This method has been 
extensively used for the study of the excited states and optical absorption in
various other systems \cite{sci-oligofluorenes,sci-phenylene,sci-spectra-jcp,cpl-indo-sci,sci-si29-apl,sci-c60-prb}. 
Since optical absorption spectra is very sensitive to the structural geometry, the optical absorption spectroscopy along with the extensive
calculations of optical absorption spectra, can be used to distinguish between distinct isomers of a cluster. In this
report, we present extensive calculations of the linear optical absorption spectrum of low-lying isomers of
B$_{6}$ and B$_{6}^{+}$ clusters with different structures. This study, along with
the experimental absorption spectra, can lead to identification of these distinct isomers. Also, in the interpretation of the measured spectra,
the theoretical understanding of the excited states of clusters plays an important role \cite{vlasta-cis-excited}.

The remainder of this chapter is organized as follows. Next section describes the theoretical and computational
details of the work, followed by section \ref{sec:results-epjd}, in which results are presented and discussed.
In the last section we summarize our findings. 
A Detailed information regarding the excited states contributing to the optical absorption is presented in the Appendix \ref{app:wavefunction-epjd}.

\section{Theoretical and Computational Details}
\label{sec:theory-epjd}

Different possible arrangements and orientations of atoms of the B$_{6}$ cluster 
(both neutral and cationic) were randomly selected for the initial configurational 
search of geometries of isomers. For a given spin multiplicity, the geometry optimization was done at a 
correlated level, i.e., at the \ac{CCSD} level \cite{ccsd-book-shavitt} with 6-311G(d,p) basis set 
as implemented in \textsc{gaussian09}\cite{gaussian09}. 
Since neutral cluster can have singlet or higher spin multiplicity, the optimization was repeated
for different spin configurations to get the lowest energy isomer. Similarly for cationic clusters with
odd number of electrons, spin multiplicities of 2 and 4 were considered in the optimization.
In total, we have obtained 11 neutral B$_{6}$  and 8 cationic B$_{6}^{+}$ low-lying isomers.
These optimized geometries of neutral B$_{6}$ cluster, as shown in Fig. \ref{fig:geometries-neutral}, 
are found to be in good agreement with other available reports. Figure \ref{fig:geometries-cationic}
shows the corresponding geometries of cationic B$_{6}^{+}$ cluster. 
The unique bond lengths, point group symmetry and the electronic ground states are given 
in respective sub-figures.

The excited state energies of isomers are obtained using the \emph{ab initio} \ac{CIS} approach. 
In this method, different configurations are constructed by promoting an electron from an occupied
orbital to a virtual orbital. For open-shell systems, we have used unrestricted 
 Hartree-Fock formalism for constructing \ac{CIS} configurations.
Excited states of the system will have a linear combination of all such substituted configurations,
with corresponding variational coefficients. The energies of the excited states will then be obtained 
by diagonalizing the Hamiltonian in this configurational space \cite{meld}.
 The dipole matrix elements are calculated using the ground state and the excited state wavefunctions.
This is subsequently used for calculating the optical absorption 
cross section assuming Lorentzian lineshape, with some artificial finite linewidth. 

\ac{EOM-CCSD} calculations were done on few representative clusters in order to
justify the use of \ac{CIS} method for optical absorption calculations \cite{ eom-ccsd-ann-rev, eom-ccsd-jcp}. 
Details are discussed in the next section.
The contribution of wavefunction of the excited states to the absorption peaks as well as an analysis based on
natural transition orbitals gives an insight into the nature of optical excitation.

As discussed in the chapter \ref{chap:main_smallboron}, 
we have extensively studied the dependence of basis sets, freezing of 1s$^{2}$ chemical core on the 
computed photoabsorption spectra of neutral boron clusters \cite{nano_life}.
We have shown that the optical absorption spectra of small boron clusters do not change even if we freeze the chemical core of 
boron atoms. Therefore, in all these calculations $1s^2$ chemical core of each boron atom has been frozen.

\section{Results and Discussion}
\label{sec:results-epjd}
In this section, we discuss the structure and energetics of 
various isomers of neutral and cationic B$_{6}$ cluster, followed by
discussion of results of computed absorption spectra and nature of 
photo-excitations.

In the many-particle wavefunction analysis of excited states contributing to the various peaks
, we have used following convention. For doublet systems $H_{1\alpha}$ denotes
the singly occupied molecular orbital. For triplet systems, two \ac{SOMO}s
are denoted by $H_{1\alpha}$ and $H_{2\alpha}$, while $H$ and $L$ stands for highest occupied molecular 
orbital and lowest unoccupied molecular orbital respectively. For quartets, the third singly
occupied molecular orbital is denoted by $H_{3\alpha}$.

\subsection{B$_{6}$}

\begin{figure*}
\begin{center}
\subfloat[C$_{2h}$, $^{3}A_{u}$ \newline Planar ring (Tp)]
{\includegraphics[scale=0.12]{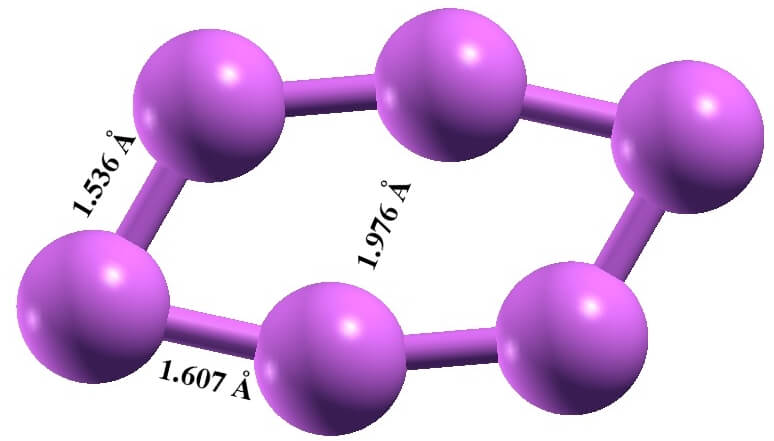}} \hfill
\subfloat[C$_{2v}$, $^{3}B_{1}$ \newline Incomplete Wheel]
{\includegraphics[scale=0.085]{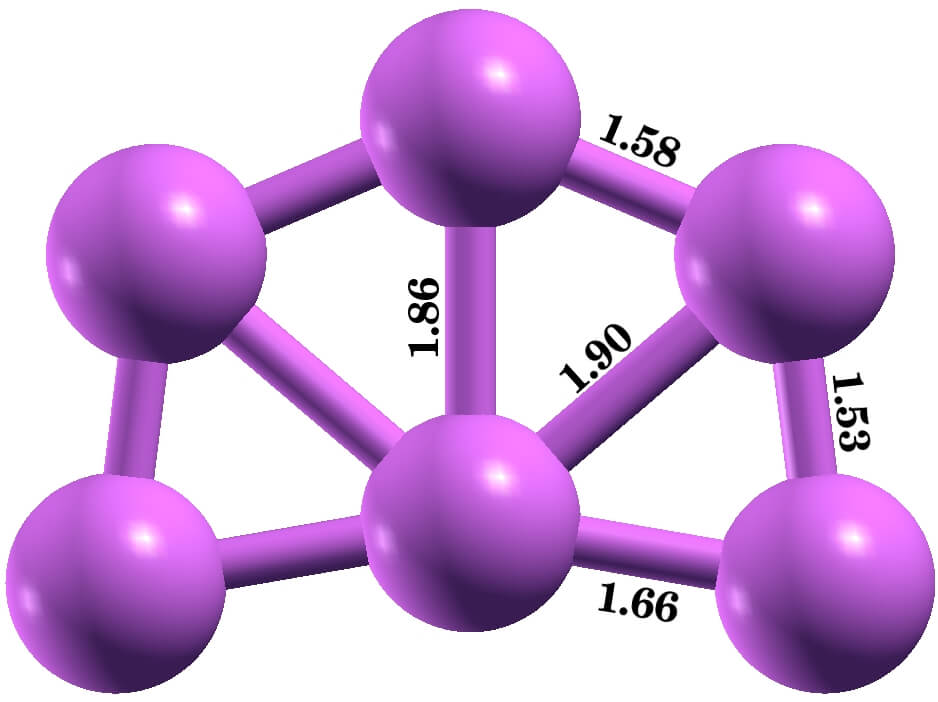}} \hfill
\subfloat[C$_{5v}$, $^{1}A_{1}$ Bulged wheel]
{\includegraphics[scale=0.15]{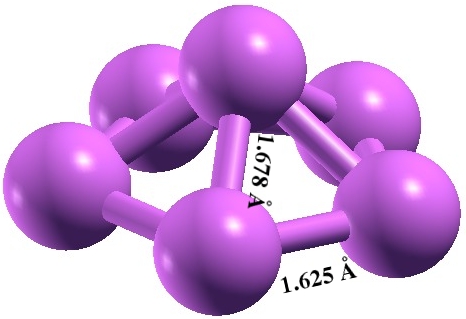}} \hfill
\subfloat[C$_{s}$, $^{1}A^{'}$ \newline Planar ring (Sg)]
{\includegraphics[scale=0.09]{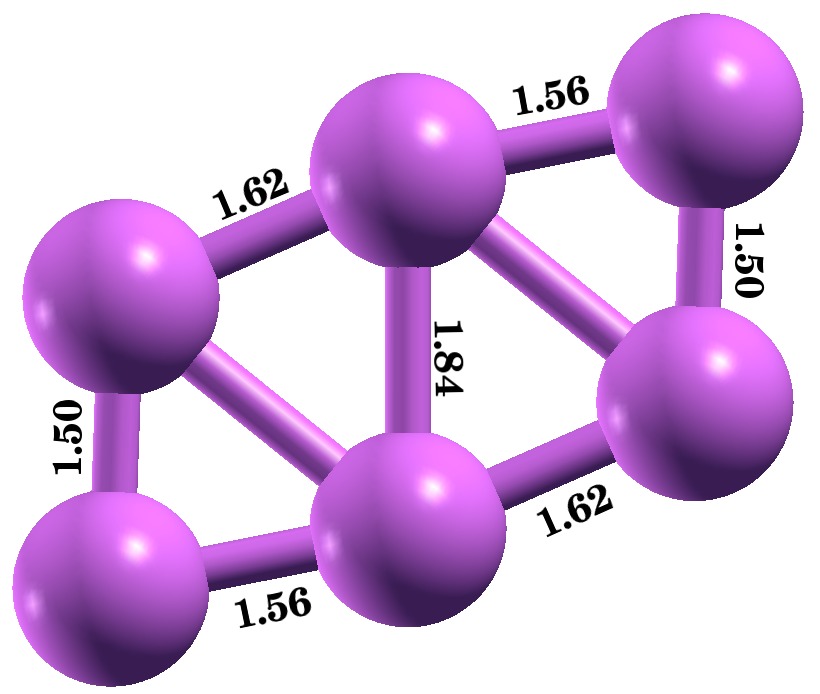}} \\
\subfloat[O$_{h}$, $^{3}A_{1g}$ Octahedron]
{\includegraphics[scale=0.15]{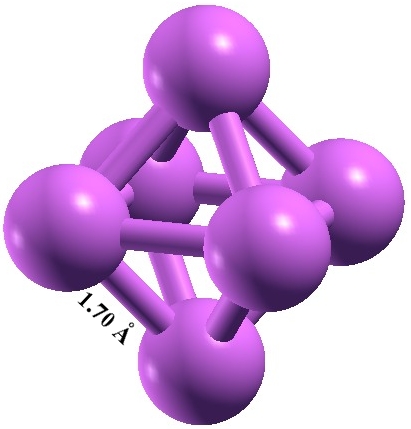}} \hfill
\subfloat[C$_{1}$, $^{3}A$ Threaded tetramer]
{\includegraphics[scale=0.15]{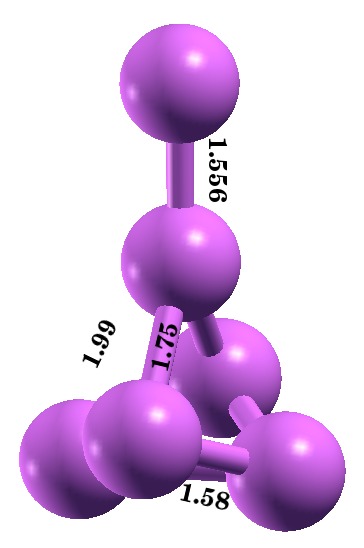}} \hfill
\subfloat[C$_{2v}$, $^{3}B_{1}$ Threaded trimer]
{\includegraphics[scale=0.13]{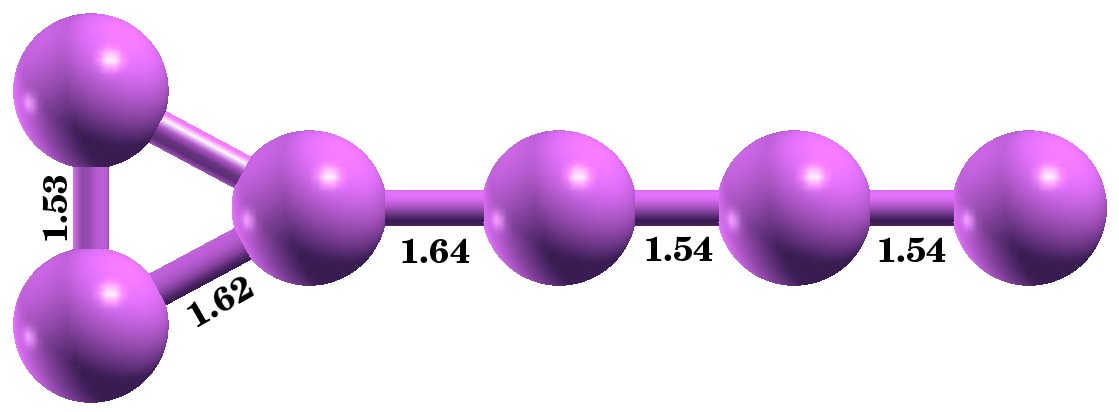}}  \hfill
\subfloat[C$_{1}$, $^{1}A$ \newline Twisted trimers]
{\includegraphics[scale=0.14]{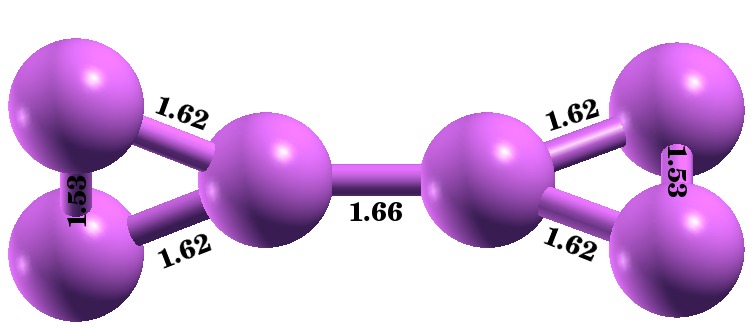}}  \\
\subfloat[D$_{2h}$, $^{1}A_{g}$ Planar trimers]
{\includegraphics[scale=0.17]{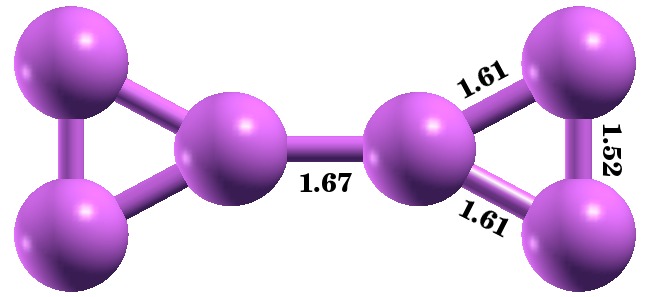}} \hfill
\subfloat[C$_{1}$, $^{1}A$ Convex bowl]
{\includegraphics[scale=0.16]{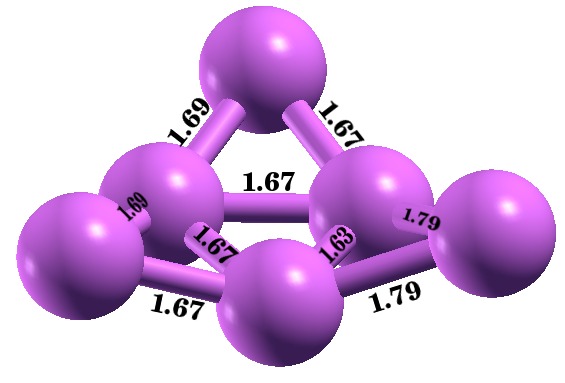}} \hfill
\subfloat[D$_{\infty h}$, $^{1}\Sigma_{g}$ Linear]
{\includegraphics[width=6cm]{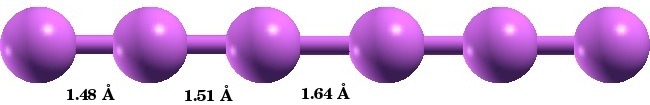}}  
\end{center}
\caption{\label{fig:geometries-neutral} Geometry optimized ground state structures
of different isomers of neutral B$_{6}$ clusters, along with the point group symmetries
obtained at the \ac{CCSD} level. }
\end{figure*}

We have found a total of 11 isomers of neutral B$_{6}$ cluster with stable geometries 
as shown in the Fig. \ref{fig:geometries-neutral}. \cite{gabedit} The relative standings in energy 
are presented in the Table \ref{tab:energies-neutral}, along with point group symmetries and
electronic states.

\begin{table*}
\centering
\caption{Point group, electronic state, total energies and values of $\langle S^2 \rangle$ 
before and after ($\langle S^2_a \rangle$) spin annihilation operation for different isomers of of B$_{6}$ cluster.}
\label{tab:energies-neutral}       
\begin{tabular}{clllccc}
\hline\noalign{\smallskip}
Sr.    	& Isomer		& Point 	& Elect. 	& Total  	& $\langle S^2 \rangle$ & $\langle S^2_a \rangle$\\
no. 	& 			& group 	& State      	& Energy (Ha) 	&			    & 				\\  
\noalign{\smallskip}\hline\noalign{\smallskip}
1	& Planar ring (triplet)	& C$_{2h}$	& ${}^3 A_{u}$	&	-147.795051	&2.720	&2.180	\\
2	& Incomplete wheel 	& C$_{2v}$	& ${}^3 B_{1}$	& 	-147.774166	&2.750	&2.217	\\
3	& Bulged wheel 		& C$_{5v}$	& ${}^1 A_{1}$	&	-147.764477	&0.000	&0.000	\\
4	& Planar ring (singlet)	& C$_{s}$	& ${}^1 A^{'}$	&	-147.720277	&0.000	&0.000	\\
5	& Octahedron	 	& O$_{h}$	& ${}^3 A_{1g}$	&	-147.678302	&2.082	&2.003	\\
6	& Threaded tetramer 	& C$_{1}$	& ${}^3 A$	&	-147.676776	&2.019	&2.000	\\
7	& Threaded trimer 	& C$_{2v}$	& ${}^3 B_{1}$	&	-147.667709	&2.087	&2.001	\\
8	& Twisted trimers 	& C$_{1}$	& ${}^1 A$	&	-147.645847	&0.000	&0.000	\\
9	& Planar trimers 	& D$_{2h}$	& ${}^1 A_{g}$	&	-147.645522	&0.000	&0.000	\\
10	& Convex bowl	 	& C$_{1}$	& ${}^1 A$	&	-147.612607	&0.000	&0.000	\\
11	& Linear 		& D$_{\infty h}$& ${}^1 \Sigma_{g}$&	-147.449013	&0.000	&0.000	\\
\noalign{\smallskip}\hline
\end{tabular}
\end{table*}

\FloatBarrier

\begin{figure}[h!]
\centering
\includegraphics[width=8.3cm]{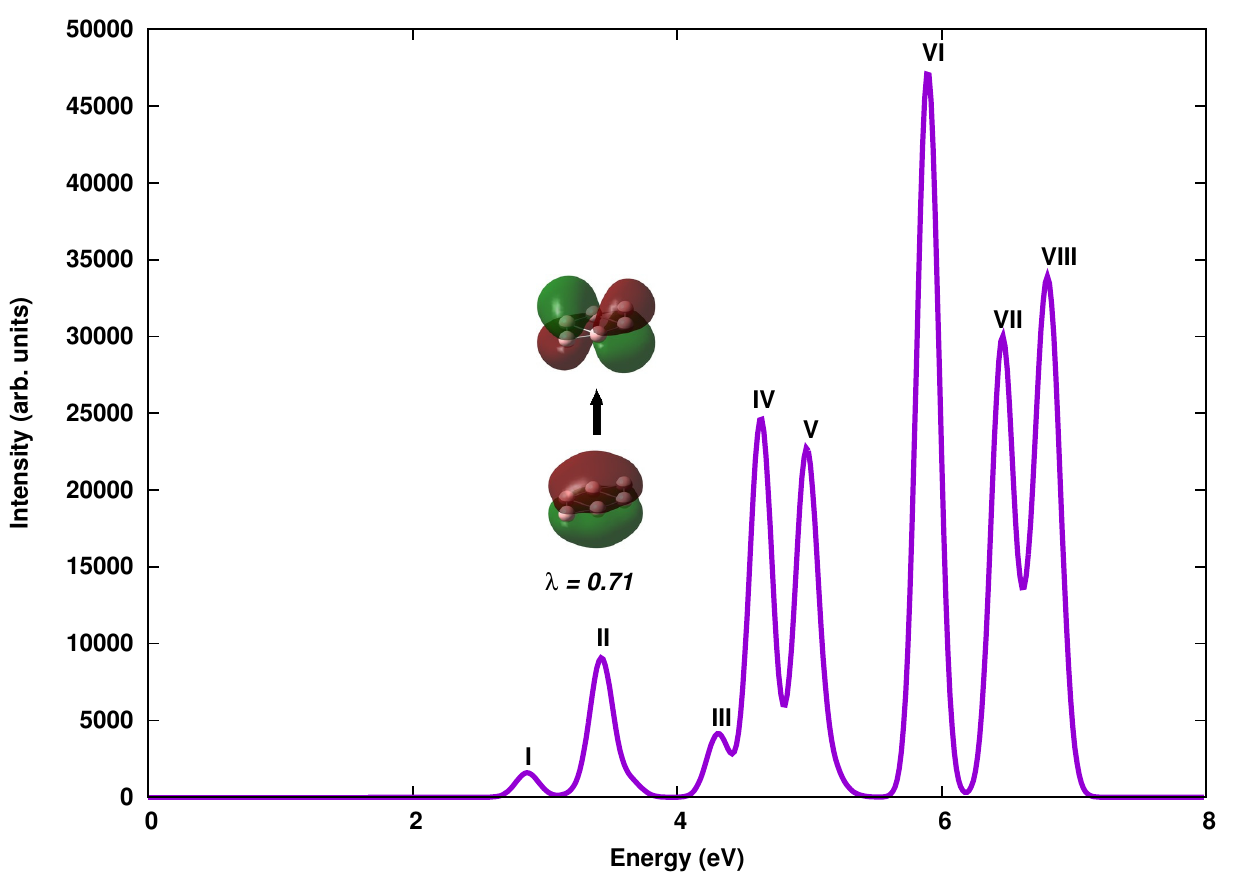}
\caption{\label{fig:neutral-plot-planar-ring-triplet}  The linear optical absorption spectrum
of planar ring (triplet) B$_{6}$ isomer, calculated using the \ac{CIS} approach, along with the natural transition orbitals
involved in the excited states corresponding to the peak II (3.42 eV). Parameter $\lambda$ refers to a fraction of the
\ac{NTO} pair contribution to a given electronic excitation.}
\end{figure}

The most stable isomer of B$_{6}$ cluster has ring-like
planar structure, with C$_{2h}$ point group symmetry. Although
B$_{6}$ has an even number of electrons, the electronic ground
state of this isomer is a triplet -- an open shell system. The
equilibrium geometry obtained in our calculation is in good
agreement with the recently reported values.\cite{turkish_boron, structure-bonding-b6, b6-isomerization,vlasta-chem-review}
 The optical absorption spectrum calculated using the \ac{CIS} approach
is as shown in the Fig. \ref{fig:neutral-plot-planar-ring-triplet}.
It is mainly characterized by feeble absorption in the visible range, but much stronger absorption 
at higher energies. The many particle wave-functions of excited states contributing to various peaks 
are presented in Table \ref{Tab:table-neutral-planar-ring-triplet}. The first absorption peak at 2.85 eV
with very low intensity is characterized by $H_{\alpha} - 2 \rightarrow L_{\alpha}$ and 
$H_{1\alpha} \rightarrow L_{\alpha}$ transitions. The natural transition orbital analysis of the peak at 3.42 eV shows 
that this is dominated by a $\pi \rightarrow \pi^{*}$ transition.
Due to planar nature of the isomer, we can classify the absorption into two categories: (a) those 
with polarization along the direction of the plane and (b) polarization perpendicular to the plane.
In this case, it is seen that, in most of the cases, the absorption is due to polarizations along the plane of the isomer.
Also, instead of being dominated by single configurations, the wavefunctions of the excited states 
contributing to all the peaks exhibit strong configuration mixing. This is an indicator of 
plasmonic nature of the optical excitations.\cite{plasmon}

\begin{figure}[h!]
\begin{center}
 \includegraphics[width=8.3cm]{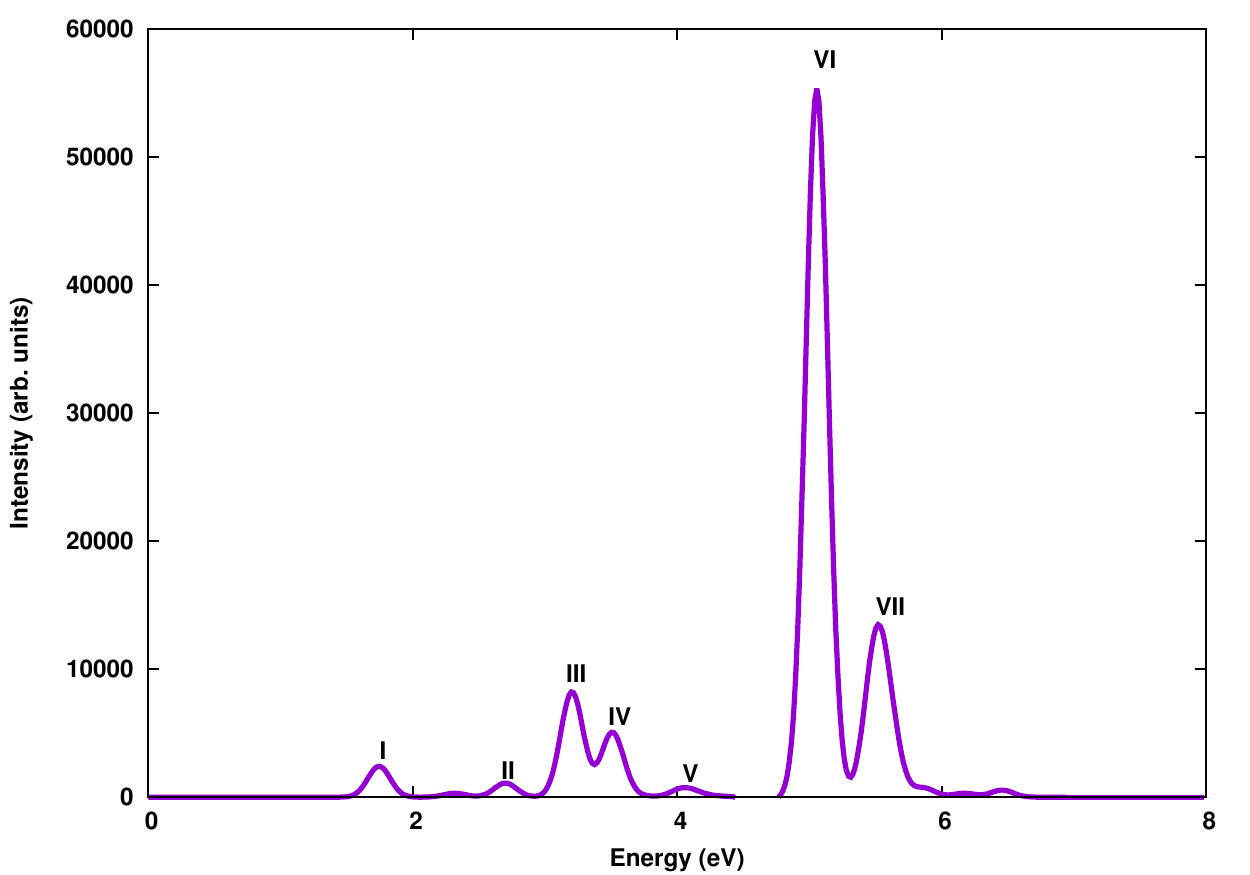} \\
\end{center}
\caption{\label{fig:neutral-plot-planar-ring-triplet-eomccsd}  The linear optical absorption spectrum
of planar ring (triplet) B$_{6}$ isomer, calculated using the \ac{EOM-CCSD} approach.}
\end{figure}

The optical absorption spectrum for the same isomer is calculated using a sophisticated \ac{EOM-CCSD} method, as shown in 
Fig. \ref{fig:neutral-plot-planar-ring-triplet-eomccsd}. A complete one-to-one mapping of configurations involved in excited
states of \ac{CIS} and \ac{EOM-CCSD} calculations is observed, along with some double excitations with minor contribution. The spectrum
of \ac{EOM-CCSD} is red-shifted, as expected, because this method takes electron correlations into account at a high level.

\begin{figure}[h!]
\centering
\includegraphics[width=8.3cm]{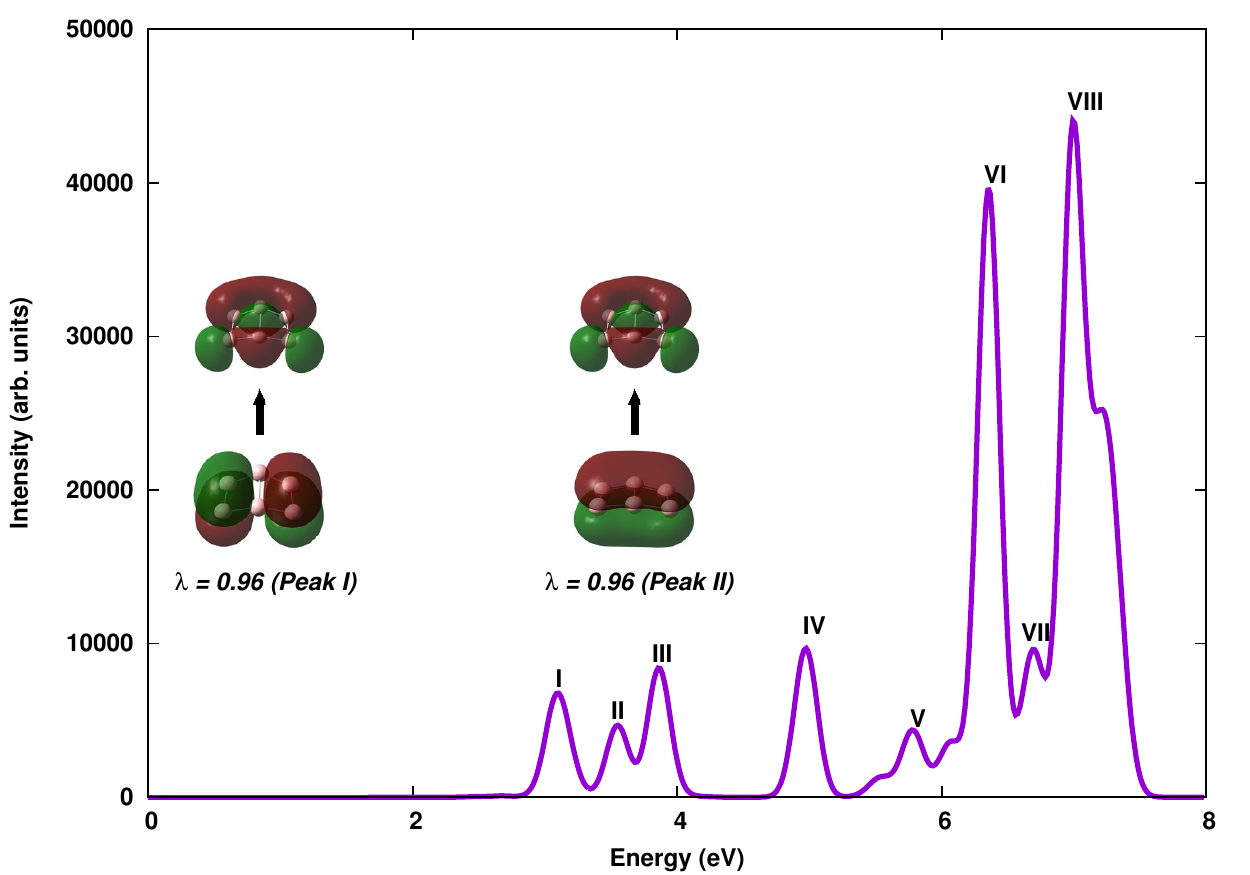}
\caption{\label{fig:neutral-plot-incomplete-wheel}  The linear optical absorption spectrum
of incomplete-wheel B$_{6}$ isomer, calculated using the \ac{CIS} approach, along with the natural transition orbitals
involved in the excited states corresponding to the peaks I (3.09 eV) and II (3.55 eV) respectively. 
Parameter $\lambda$ refers to a fraction of the \ac{NTO} pair contribution to a given electronic excitation.}
\end{figure}

The second low lying isomer of B$_{6}$ is another planar structure resembling an incomplete wheel, \emph{i.e.}
one outer atom removed from B$_{7}$ wheel cluster.
This isomer is also a triplet system with C$_{2v}$ symmetry, lying 0.56 eV above the global minimum structure.
The optimized geometry is in good agreement with the other previously available reports \cite{b6-isomerization,b6-dft}. 
This is one of the isomers showing feeble optical absorption at lower energies 
(\emph{cf.} Fig. \ref{fig:neutral-plot-incomplete-wheel}). 
The many particle wave-functions of excited states contributing to various peaks 
are presented in Table \ref{Tab:table-neutral-incomplete-wheel}.
Degenerate $\pi$ orbitals are involved in the excitations at peak I and II as evident from the \ac{NTO}s shown 
in Fig. \ref{fig:neutral-plot-incomplete-wheel}.

\begin{figure}[h!]
\centering
\includegraphics[width=8.3cm]{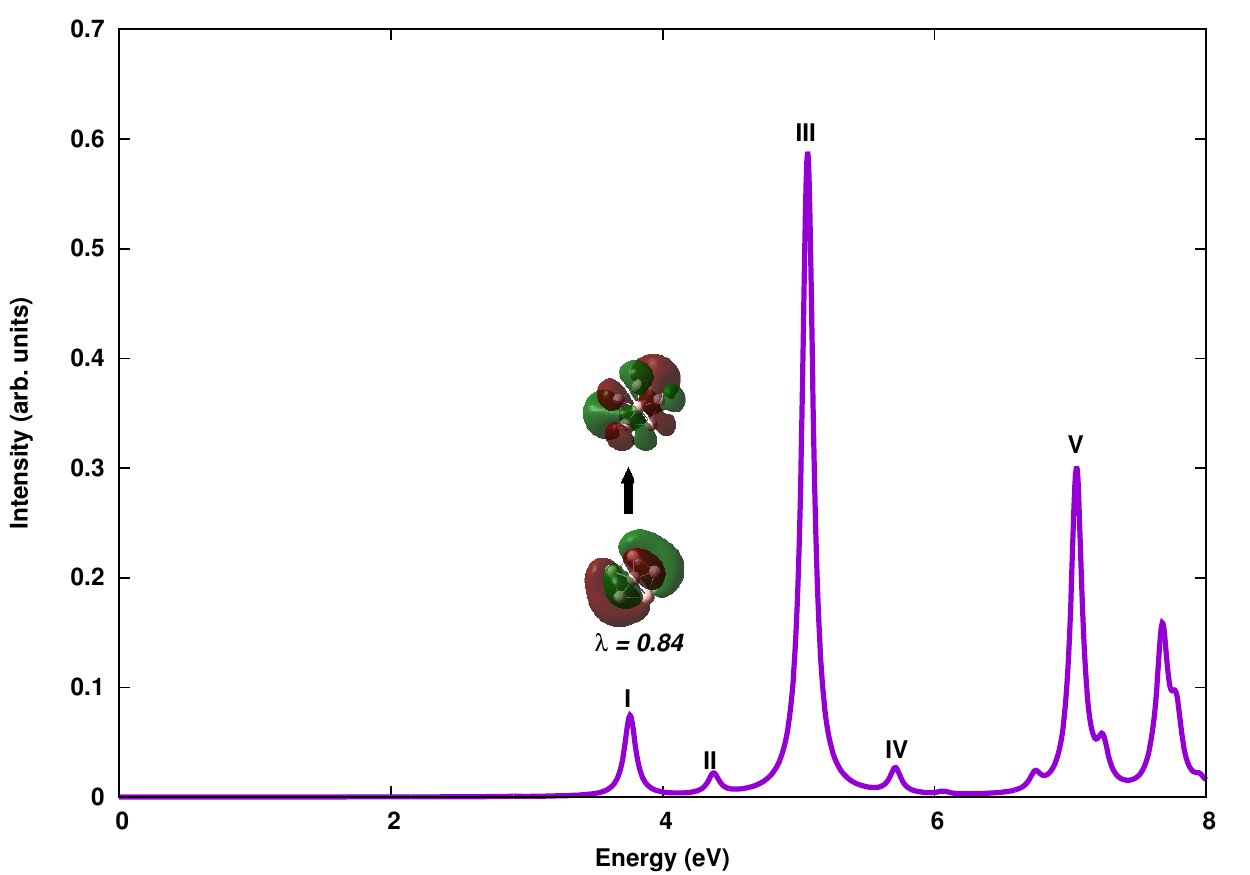} \label{subfig:neutral-plot-bulged-wheel} 
\caption{\label{fig:plots-neutral-bulged-wheel}  The linear optical absorption spectrum
of bulged-wheel B$_{6}$ isomer, calculated using the \ac{CIS} approach, along with the natural transition orbitals
involved in the excited states corresponding to the peak I (3.76 eV). 
Parameter $\lambda$ refers to a fraction of the \ac{NTO} pair contribution to a given electronic excitation.}
\end{figure}

A wheel kind of structure, with its center slightly bulged out, is found to be the next stable isomer
of B$_{6}$. A singlet system with C$_{5v}$ point group symmetry, lies just 0.83 eV above in energy 
as compared to the most stable isomer. The pentagonal base has bond length of 1.625 \AA{} and the vertex 
atom is 1.678 \AA{} away from the corners of the pentagon. Other reported values for those bond lengths are
1.61 \AA{}, 1.66 \AA{} \cite{turkish_boron, b6-dft} and 1.61 \AA{}, 1.659 \AA{} \cite{structure-bonding-b6,b6-isomerization} 
respectively. The optical absorption spectrum calculated using CIS approach is presented in Fig. \ref{fig:plots-neutral-bulged-wheel}.
The many-particle wavefunctions of excited states contributing to various peaks are presented in Table
\ref{Tab:table-neutral-bulged-wheel}. The onset of the spectrum occurs near 3.76 eV, with polarization in the plane
of the pentagonal base, characterized by excitations $H - 1 \rightarrow L + 6$ and  $H \rightarrow L + 6$ with equal 
contribution.

The spectrum of this singlet isomer is calculated again using \ac{EOM-CCSD} approach to investigate any involvement 
of double excitation. The spectrum appears to be red-shifted but the configurations contributing to the excited states
corresponding to the peaks of the spectrum are the same as observed in the case of \ac{CIS}. This makes us confident to use \ac{CIS} 
method only to investigate optical absorption spectrum of other isomers.

\begin{figure}[h!]
\begin{center}
\includegraphics[width=8.3cm]{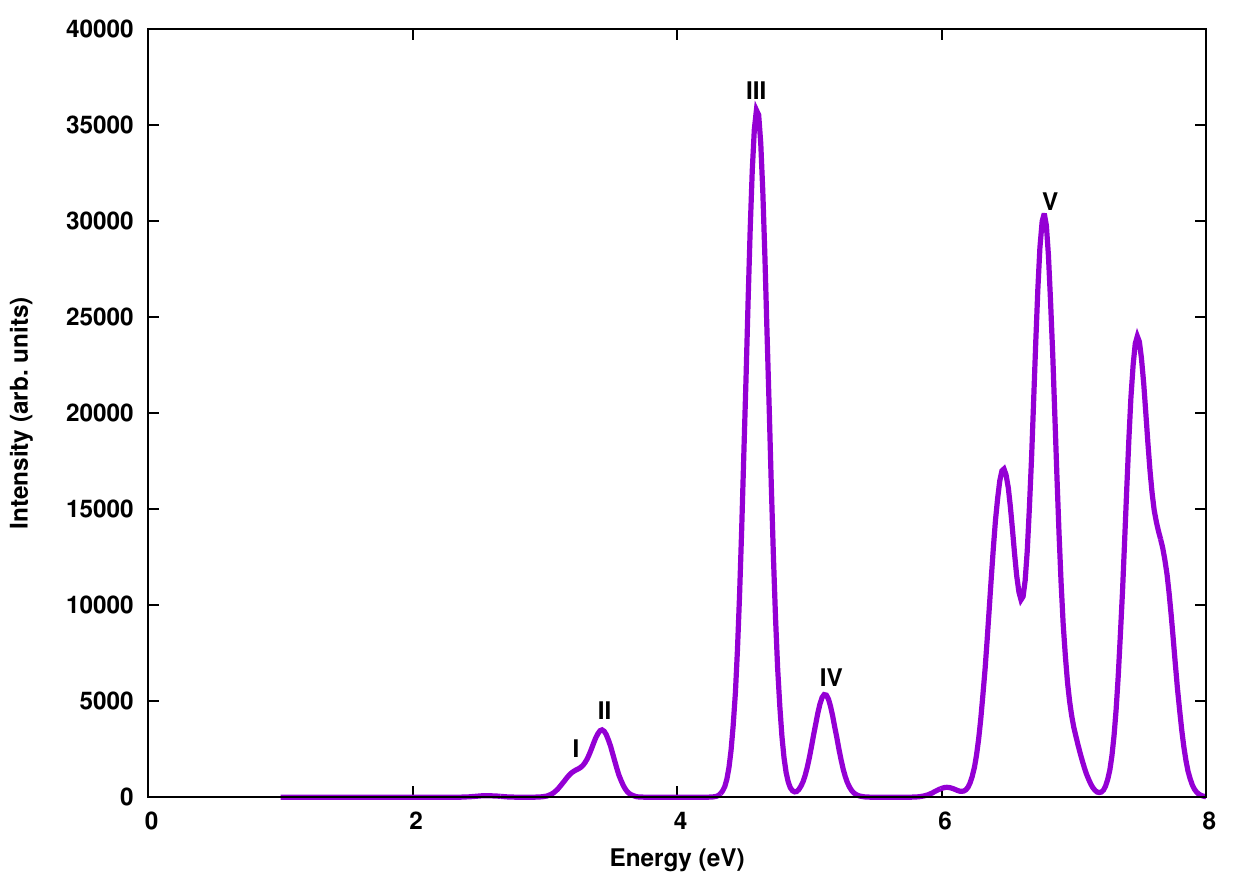} \\
\end{center}
\caption{\label{fig:neutral-plot-bulged-wheel-eomccsd}  The linear optical absorption spectrum
of bulged-wheel B$_{6}$ isomer, calculated using the \ac{EOM-CCSD} approach.}
\end{figure}

\begin{figure}[h!]
\centering
\includegraphics[width=8.3cm]{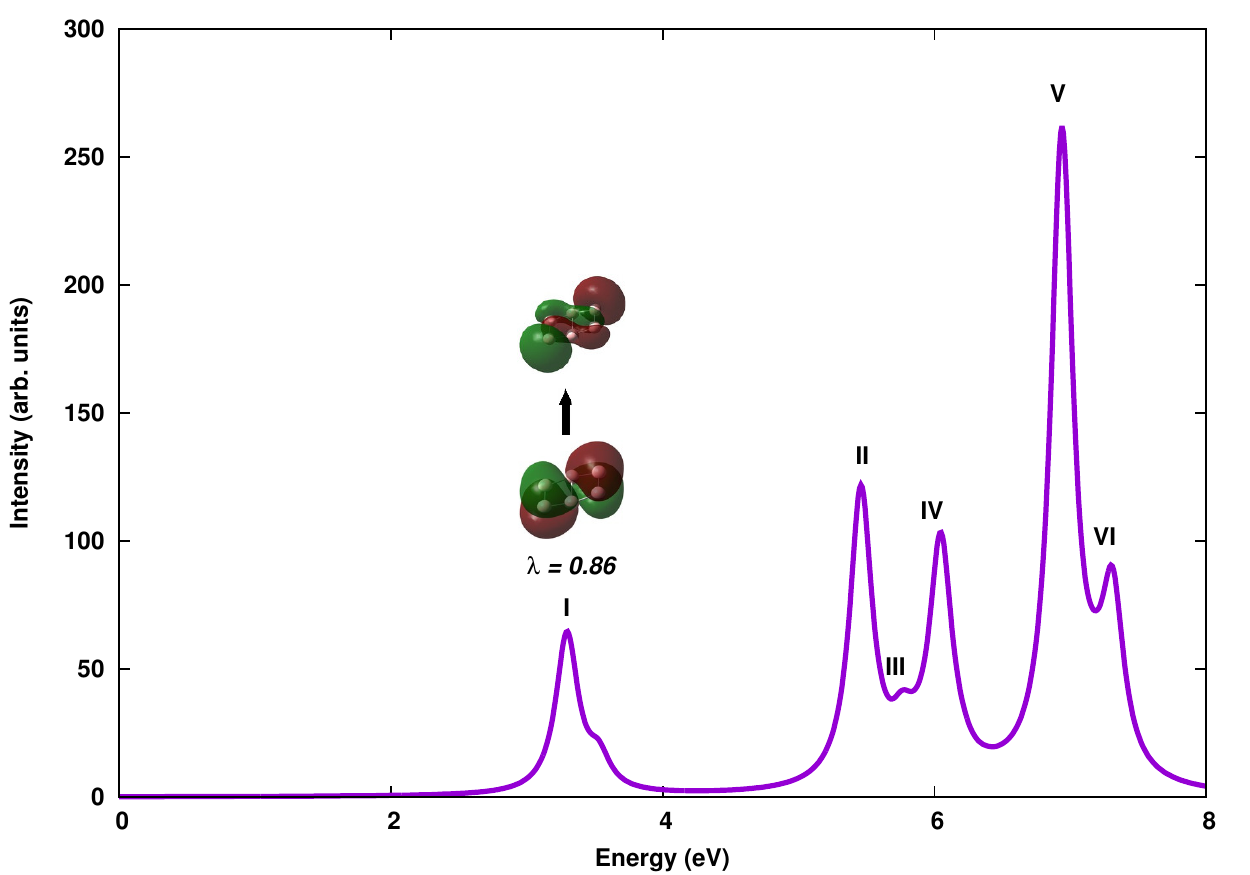}
\caption{\label{fig:neutral-plot-planar-ring-singlet}  The linear optical absorption spectrum
of planar-ring B$_{6}$ isomer, calculated using the \ac{CIS} approach, along with the natural transition orbitals
involved in the excited states corresponding to the peak I (3.13 eV).
Parameter $\lambda$ refers to a fraction of the \ac{NTO} pair contribution to a given electronic excitation. }
\end{figure}

A planar ring like structure, resembling the global minimum one; however, with singlet state and C$_{s}$ symmetry, 
is the next low lying isomer of B$_{6}$ cluster. The optimized geometry is in good agreement with Ref. \citenst{b6-dft}.
The absorption spectrum is presented in Fig. \ref{fig:neutral-plot-planar-ring-singlet} and
corresponding many-particle wavefunctions of various excited states are presented in Table 
\ref{Tab:table-neutral-planar-ring-singlet}. The absorption onset occurs at 3.1 eV, characterized by 
delocalized orbitals with $H - 1 \rightarrow L $ and $H \rightarrow L + 23$ configurations.  

\begin{figure}[h!]
\centering
\includegraphics[width=8.3cm]{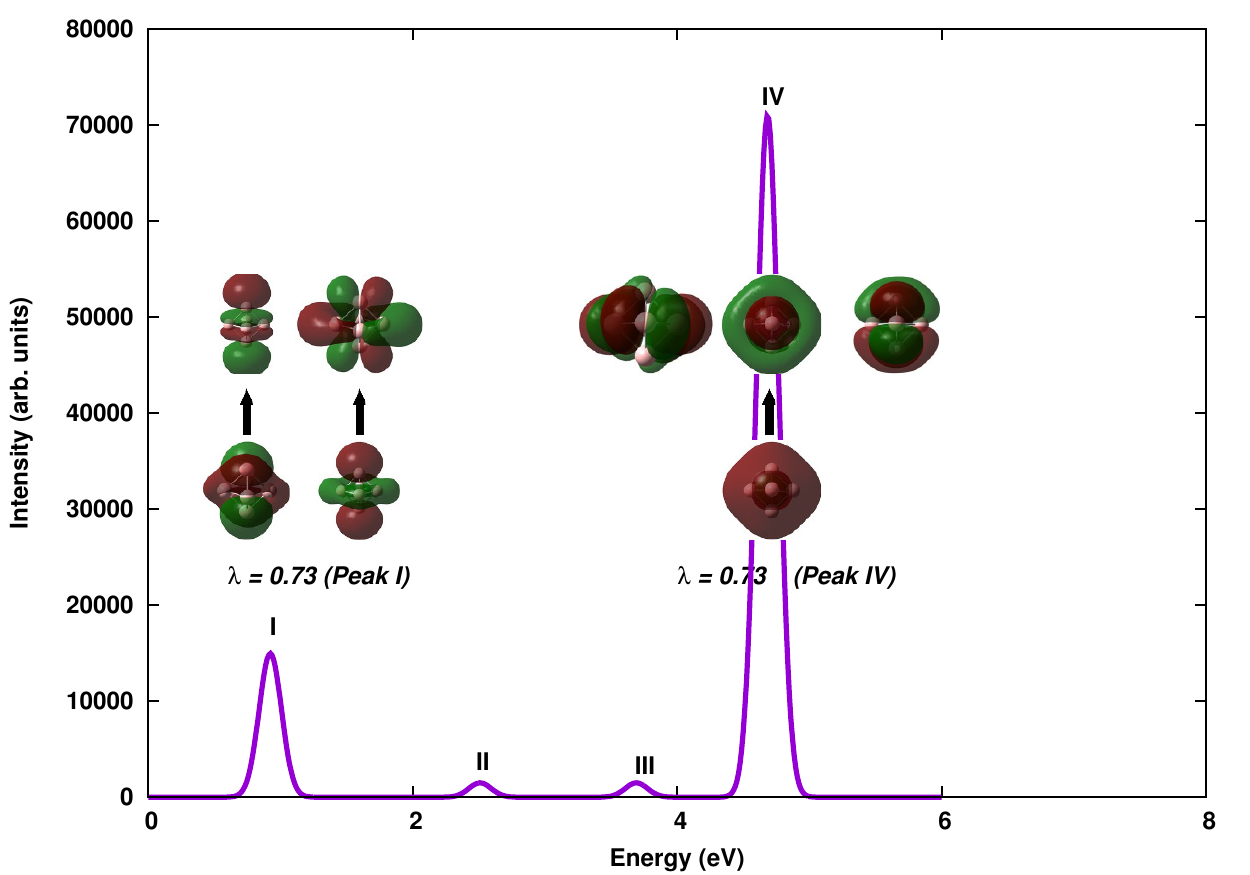}
\caption{\label{fig:neutral-plot-octahedron}  The linear optical absorption spectrum
of octahedron B$_{6}$ isomer, calculated using the \ac{CIS} approach, along with the natural transition orbitals
involved in the excited states corresponding to the peaks I (0.92 eV) and IV (4.68 eV) respectively.
Parameter $\lambda$ refers to a fraction of the \ac{NTO} pair contribution to a given electronic excitation. }
\end{figure}

An octahedron structure with O$_{h}$ point group symmetry is the next stable isomer of neutral B$_{6}$. Each side of the
octahedron is found to be 1.7 \AA{} as compared to the 1.68 \AA{} reported by Ati\c{s} \emph{et al}.\cite{turkish_boron} and Jun \emph{et al}.\cite{b6-dft}
The many-particle wavefunctions of the excited states corresponding to various peaks 
(\emph{cf.} Fig. \ref{fig:neutral-plot-octahedron}) are presented in Table
\ref{Tab:table-neutral-octahedron}. A very feeble absorption at 0.92 eV opens the spectrum, mainly 
characterized by excitations $H_{1\alpha} \rightarrow L_{\alpha}$ and  $H_{2\alpha} \rightarrow L_{\alpha} + 1$ 
with equal contribution. The transition orbitals corresponding to peaks at 0.92 eV and 2.50 eV
are as shown in Fig. \ref{fig:neutral-plot-octahedron}.

\begin{figure}[h!]
\centering
\includegraphics[width=8.3cm]{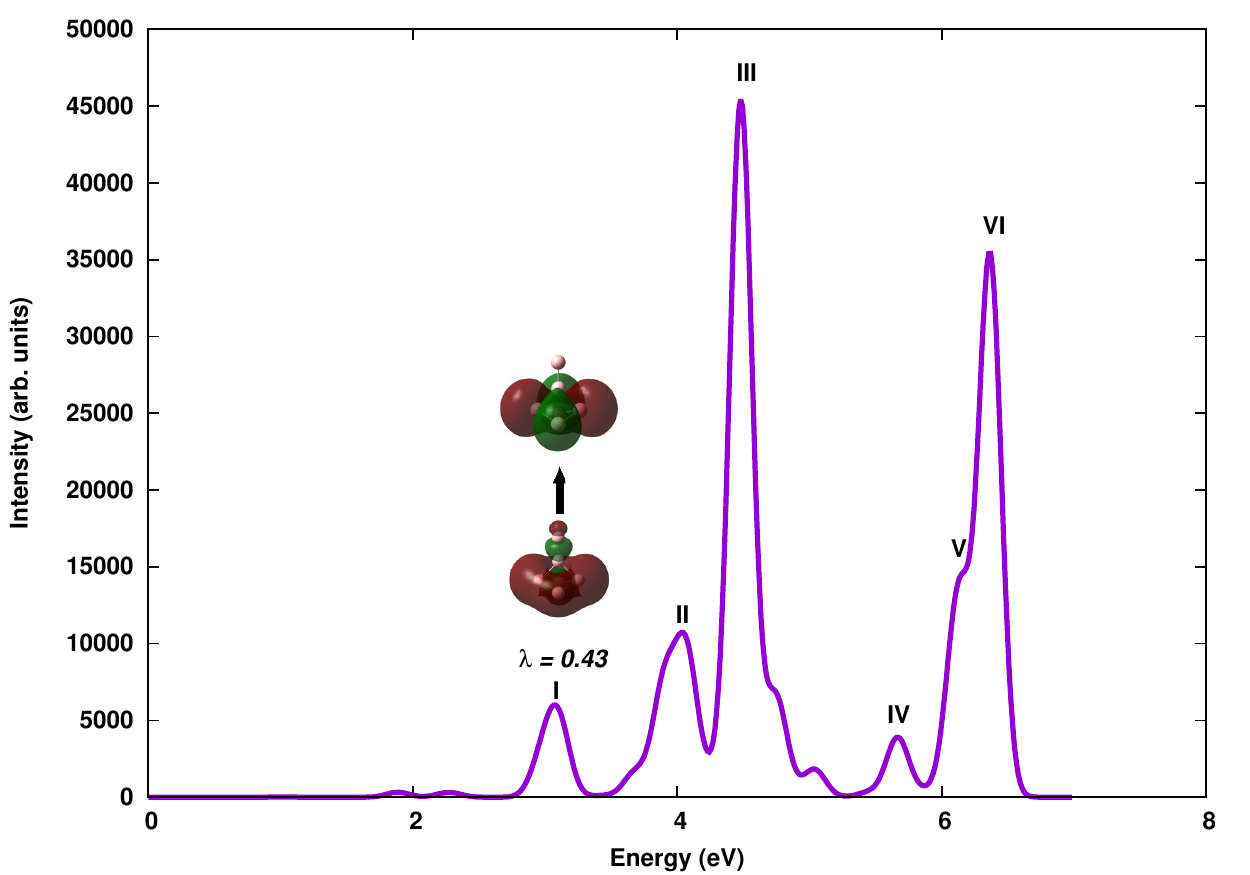}
\caption{\label{fig:neutral-plot-threaded-tetramer}  The linear optical absorption spectrum
of threaded-tetramer B$_{6}$ isomer, calculated using the \ac{CIS} approach, along with the natural transition orbitals
involved in the excited states corresponding to the peak I (3.06 eV). 
Parameter $\lambda$ refers to a fraction of the \ac{NTO} pair contribution to a given electronic excitation. }
\end{figure}

Next isomer is previously unreported, with structure of a saddle threaded with dimer from top. It lies just 0.04 eV above 
the previous octahedron isomer. However the optical absorption spectrum 
(\emph{cf.} Fig. \ref{fig:neutral-plot-threaded-tetramer}) is completely different. 
A narrow energy range hosts all the peaks. The onset of intense absorption occurs near 3.06 eV, with major contribution from
H$_{2\alpha} \rightarrow L_{\alpha} + 2$ (\emph{cf.} Table \ref{Tab:table-neutral-threaded-tetramer}). The \ac{NTO} analysis shows that 
excitation occurs from the tail end to the saddle section of the isomer.

\begin{figure}[h!]
\centering
\includegraphics[width=8.3cm]{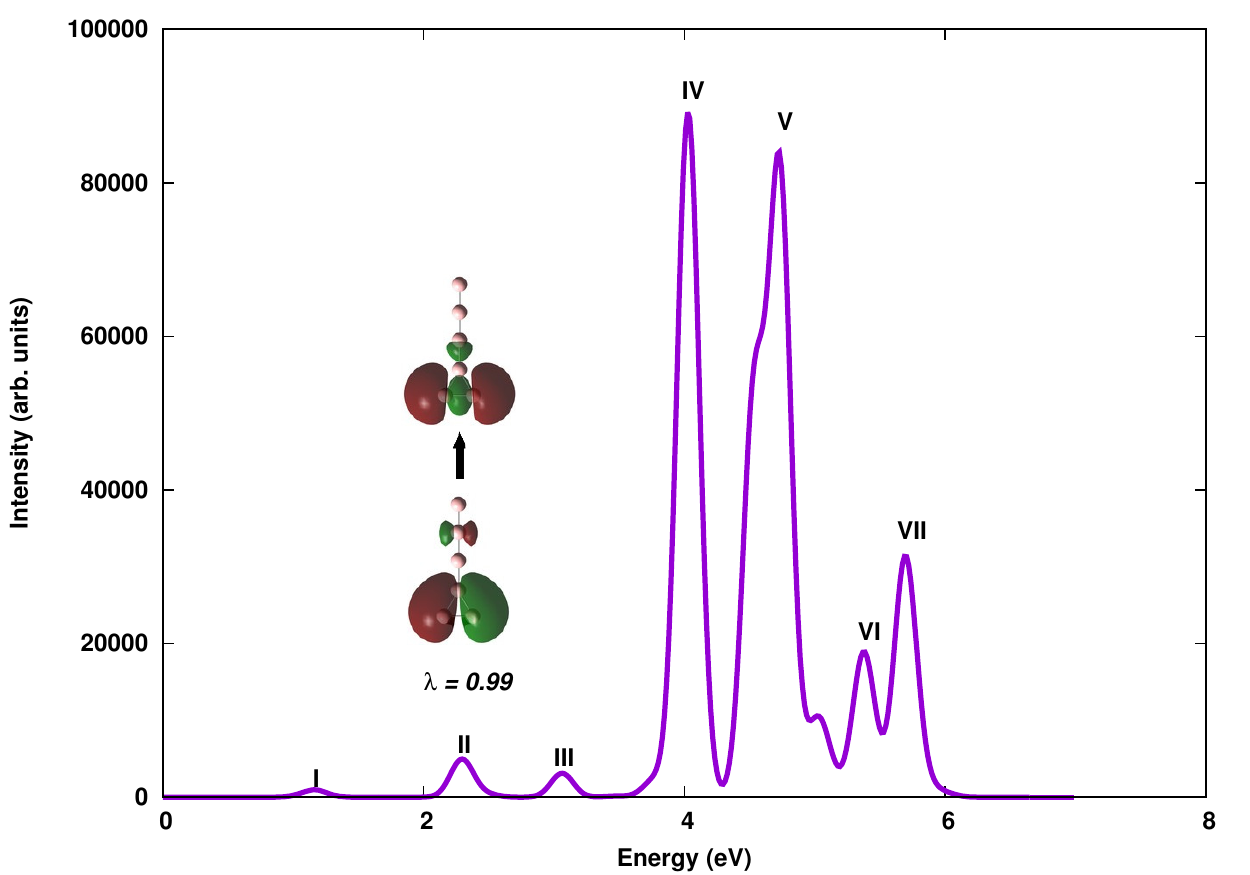}
\caption{\label{fig:neutral-plot-threaded-trimer}  The linear optical absorption spectrum
of threaded-trimer B$_{6}$ isomer, calculated using the \ac{CIS} approach, along with the natural transition orbitals
involved in the excited states corresponding to the peak II (2.28 eV). 
Parameter $\lambda$ refers to a fraction of the \ac{NTO} pair contribution to a given electronic excitation. }
\end{figure}

An isosceles triangle connected to a linear chain of boron atoms forms the next isomer. This structure with C$_{2v}$
symmetry and a triplet electronic state has been reported in Ref. \citenst{b6-dft}, which is in close agreement with our results.
The optical absorption spectrum (\emph{cf.} Fig. \ref{fig:neutral-plot-threaded-trimer})
has distinctive closely lying peaks at 4.03 eV and 4.73 eV. The many-particle wavefunctions of excited states corresponding
to various peaks are presented in Table \ref{Tab:table-neutral-threaded-trimer}. As evident from the \ac{NTO}s involved in 
transitions at peak I, the electrons tend to be localized at the triangular end of the isomer.

\begin{figure}[h!]
\centering
\includegraphics[width=8.3cm]{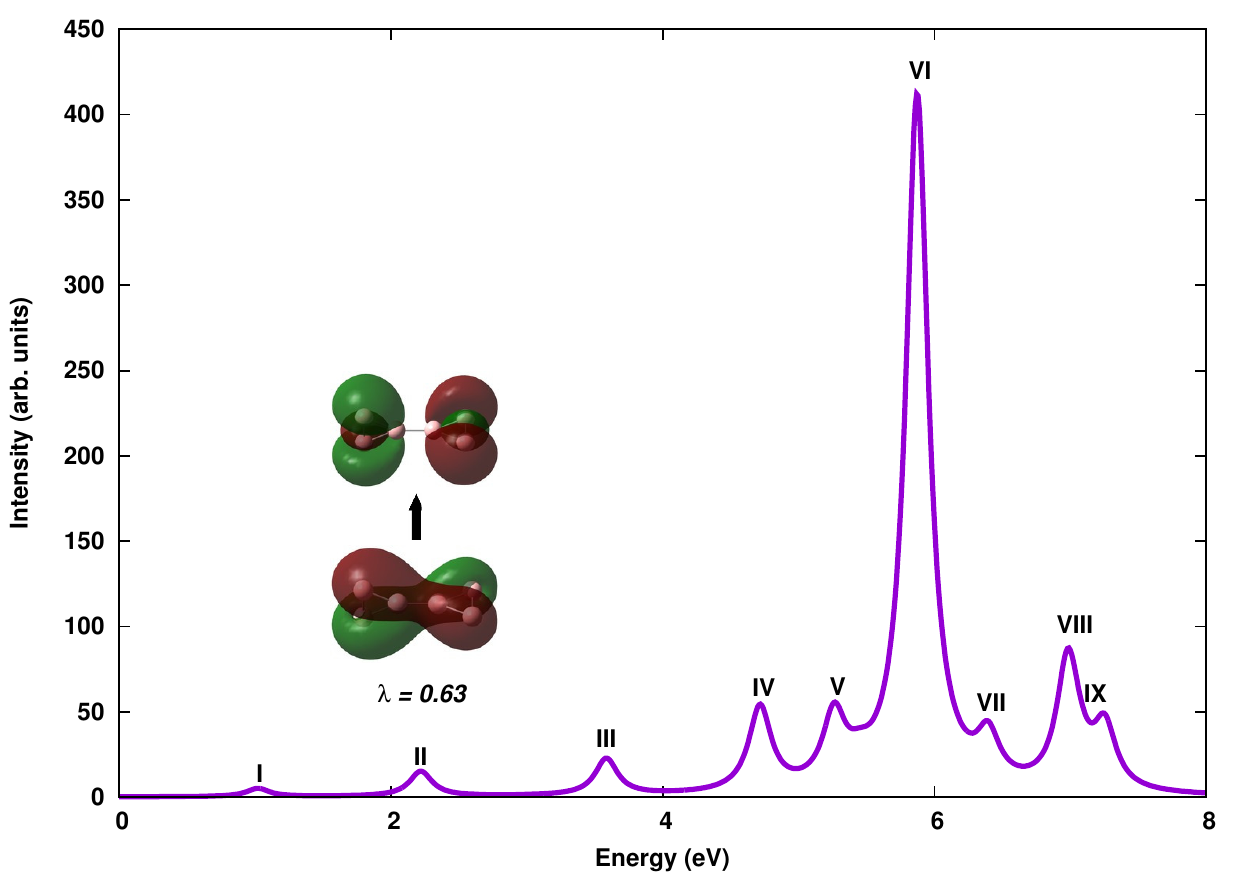}
\caption{\label{fig:neutral-plot-twisted-trimers}  The linear optical absorption spectrum
of twisted trimers B$_{6}$ isomer, calculated using the \ac{CIS} approach, along with the natural transition orbitals
involved in the excited states corresponding to the peak I (2.22 eV).
Parameter $\lambda$ refers to a fraction of the \ac{NTO} pair contribution to a given electronic excitation. }
\end{figure}

A structure with two out of plane isosceles triangles joined together is found to be one of the isomers. 
The geometry has isosceles triangle with lengths 1.62 \AA{}, 1.62 \AA{} and 1.53 \AA{}, while two such triangles are
joined by a bond of length 1.66 \AA{}. The respective numbers reported by Ref. \citenst{b6-dft} are 1.60 \AA{}, 1.60 \AA{},
1.50 \AA{} and 1.647 \AA{} respectively. The optical absorption spectrum contains many low intensity peaks except for 
strongest one at 5.87 eV, as presented in Fig. \ref{fig:neutral-plot-twisted-trimers}. 
A $\pi  \rightarrow \pi^*$ transition is observed at 2.22 eV. (\emph{cf.} Table \ref{Tab:table-neutral-twisted-trimers}). 

\begin{figure}[h!]
\centering
\includegraphics[width=8.3cm]{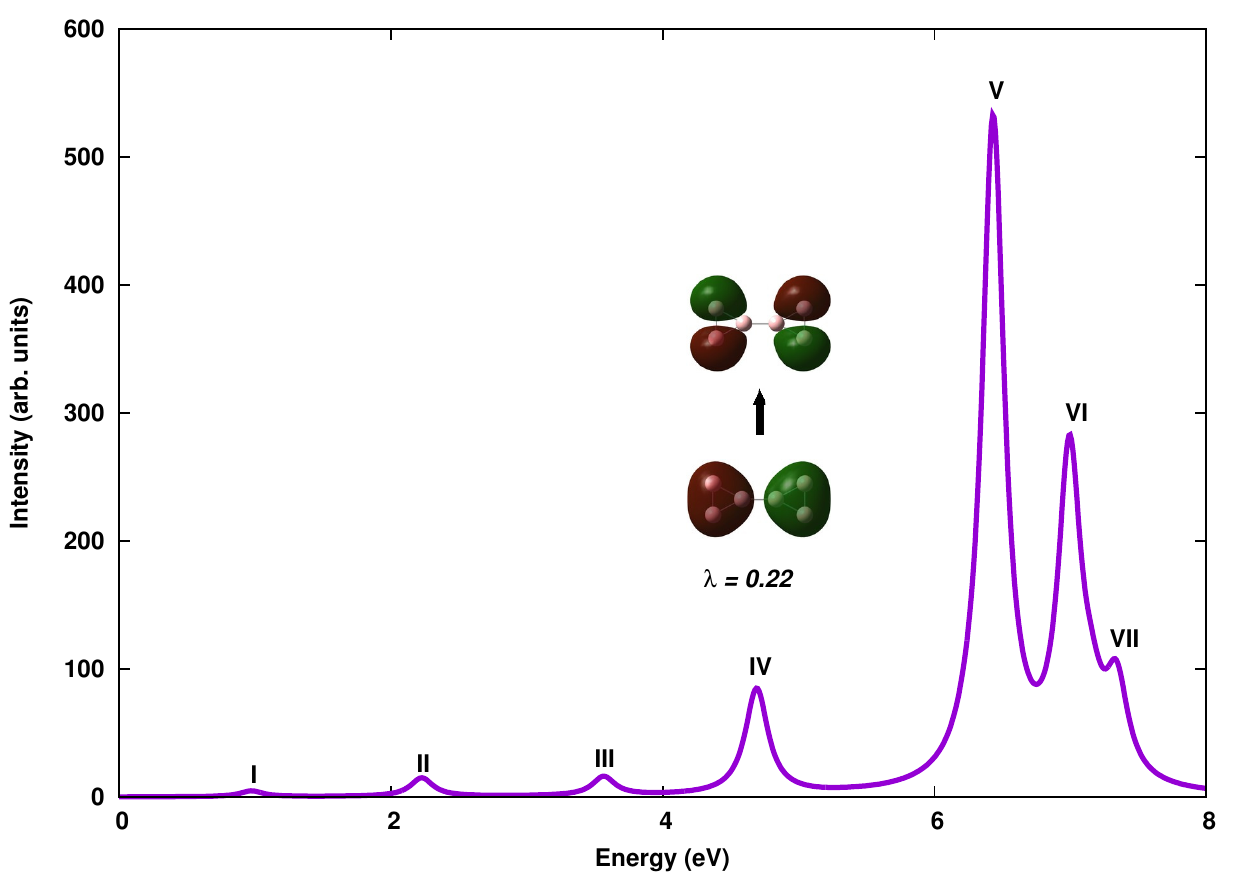}
\caption{\label{fig:neutral-plot-planar-trimers}  The linear optical absorption spectrum
of planar trimers B$_{6}$ isomer, calculated using the \ac{CIS} approach, along with the natural transition orbitals
involved in the excited states corresponding to the peak IV (4.69 eV).
 Parameter $\lambda$ refers to a fraction of the \ac{NTO} pair contribution to a given electronic excitation. }
\end{figure}

An almost degenerate structure forms the next isomer, lying just 0.009 eV above the previous isomer. Contrary to
the previous one, this geometry is completely planar and is a triplet system, with C$_{2v}$ point group symmetry. 
Probably because of such a strong near-degeneracy, this isomer has not been reported in the literature before.
The many-particle wavefunctions of the excited states corresponding to various peaks 
(\emph{cf.} Fig. \ref{fig:neutral-plot-planar-trimers}) are presented in Table
\ref{Tab:table-neutral-planar-trimers}. The spectrum opens with feeble peak. First major peak at 4.69 eV
is characterized by $\pi  \rightarrow \pi^*$ transition, as is evident from the natural transition orbital analysis. 
First four peaks in the absorption spectra of this isomer are identical to that of twisted trimers isomer.
The effect of twisting has effect only on the high energy excitations.

\begin{figure}[h!]
\centering
\includegraphics[width=8.3cm]{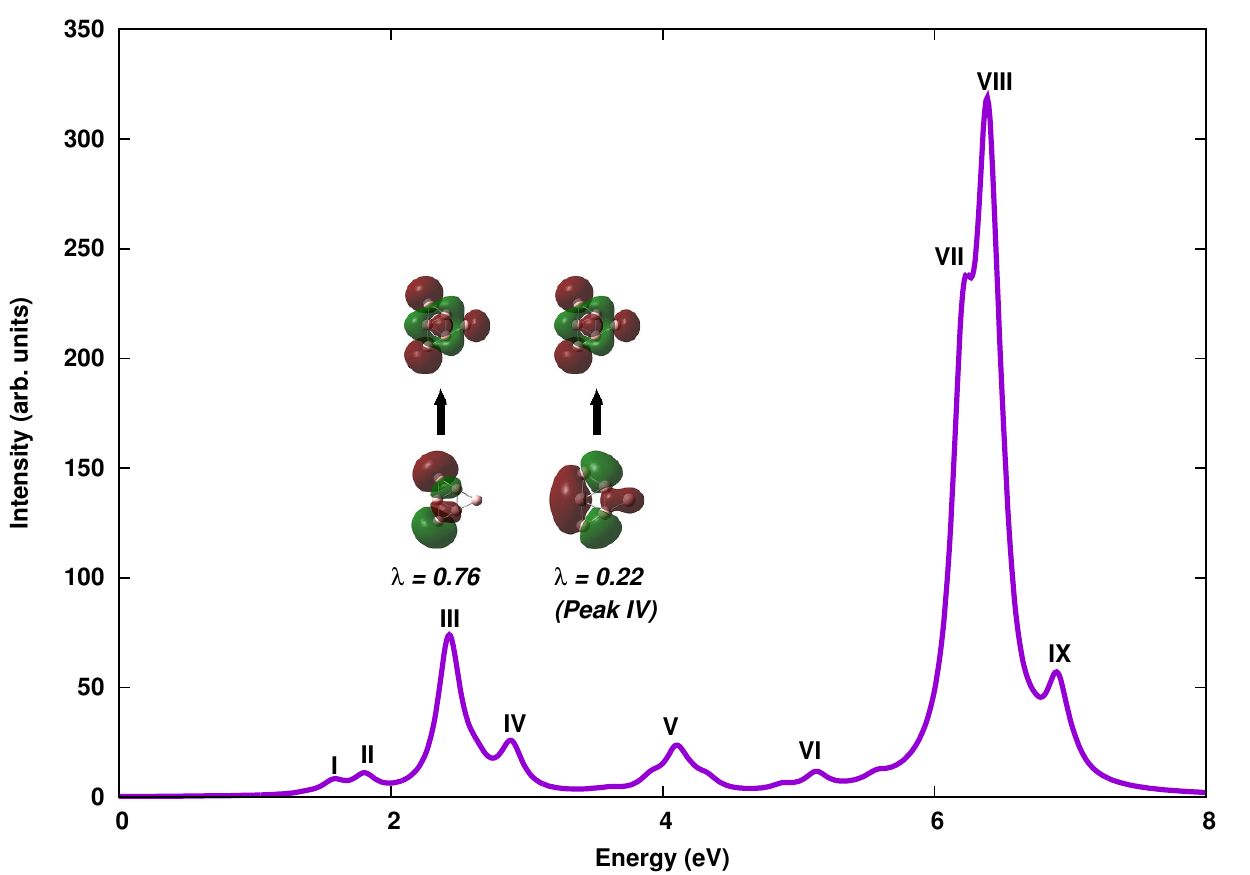}
\caption{\label{fig:neutral-plot-convex-bowl}  The linear optical absorption spectrum
of convex-bowl B$_{6}$ isomer, calculated using the \ac{CIS} approach, along with the natural transition orbitals
involved in the excited states corresponding to the peaks III (2.43 eV) and IV (2.88 eV) respectively.
Parameter $\lambda$ refers to a fraction of the \ac{NTO} pair contribution to a given electronic excitation. }
\end{figure}

\begin{figure}[h!]
\centering
\includegraphics[width=8.3cm]{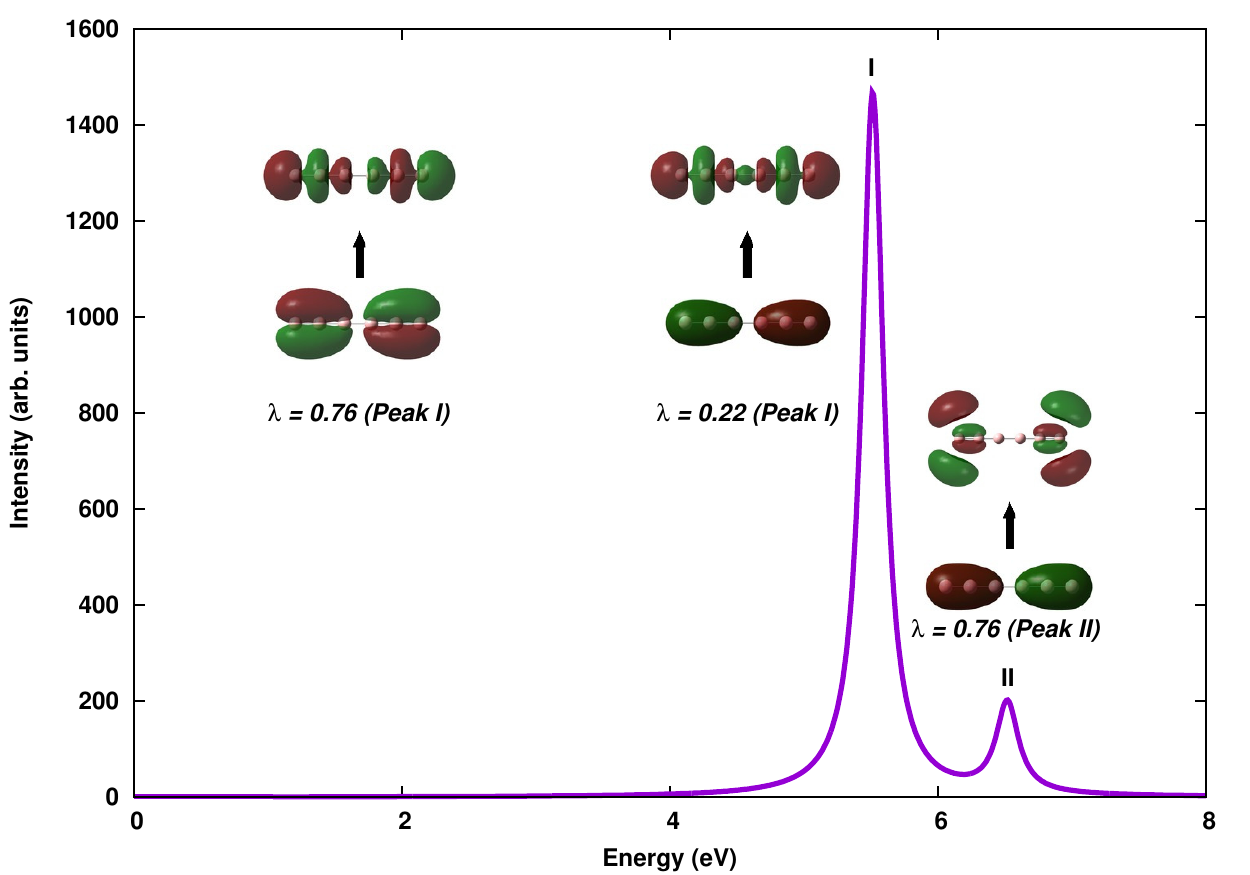}
\caption{\label{fig:neutral-plot-linear}  The linear optical absorption spectrum
of linear B$_{6}$ isomer, calculated using the \ac{CIS} approach, along with the natural transition orbitals
involved in the excited states corresponding to the peaks I (5.51 eV) and II (6.51 eV) respectively. 
Parameter $\lambda$ refers to a fraction of the \ac{NTO} pair contribution to a given electronic excitation. }
\end{figure}

Convex bowl shaped isomer and a perfect linear chain are found very high in energy, ruling out their existence at room
temperature. The optical spectra are presented in Figs. \ref{fig:neutral-plot-convex-bowl}
and \ref{fig:neutral-plot-linear} respectively. The corresponding many-particle wavefunctions 
of excited states of various peaks are presented in Table \ref{Tab:table-neutral-convex-bowl} and 
\ref{Tab:table-neutral-linear}. Peak at 2.43 eV in the absorption spectrum of convex bowl isomer shows partially 
delocalized to fully delocalized nature of transition. The bulk of the oscillator strength of the spectrum of linear 
isomer is carried by $ H - 1 \rightarrow L + 3$ and $H \rightarrow L + 2$ having equal contributions. The \ac{NTO}s 
corresponding to the excitations involved in the spectrum are shown in Fig. \ref{fig:neutral-plot-linear}.

 \FloatBarrier

\subsection{B$_{6}^{+}$}

\begin{figure*}
\begin{center}
\subfloat[C$_{s}$, $^{2}A^{''}$ \newline Planar ring (I)]
{\includegraphics[scale=0.12]{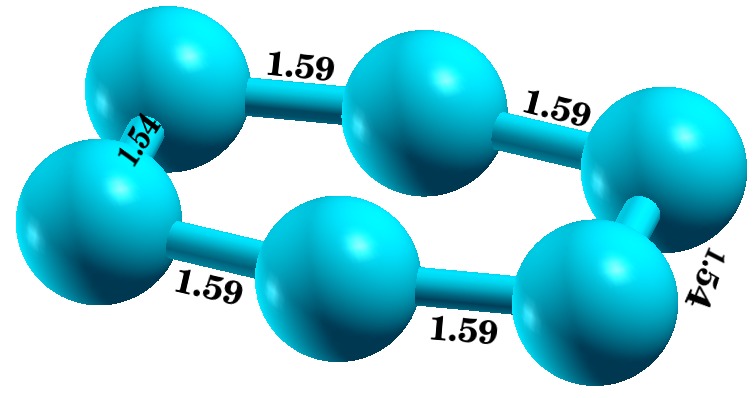}} \hfill
\subfloat[C$_{1}$, $^{2}A$ \newline Bulged wheel]
{\includegraphics[scale=0.15]{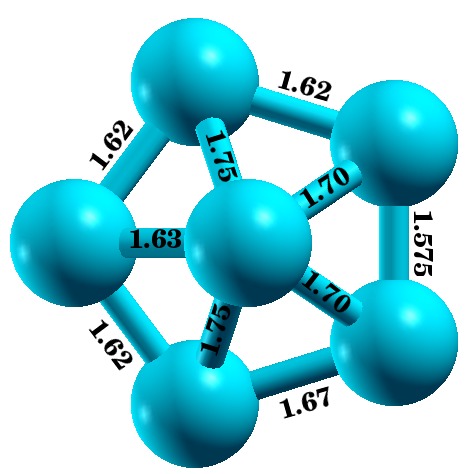}} \hfill
\subfloat[D$_{2h}$, $^{2}A_{g}$ \newline Planar ring (II)]
{\includegraphics[scale=0.10]{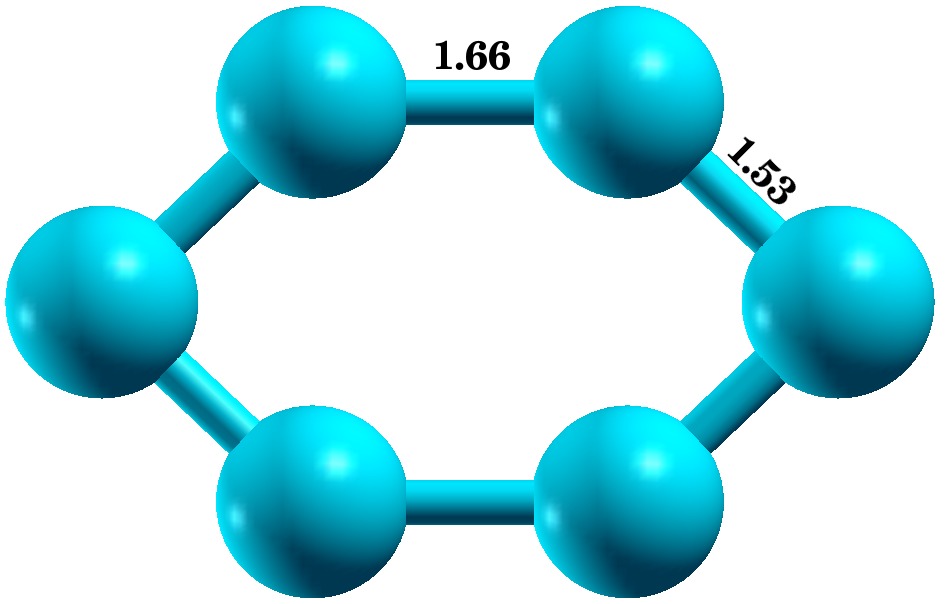}} \hfill
\subfloat[C$_{2v}$, $^{4}A_{2}$ \newline Incomplete wheel]
{\includegraphics[scale=0.10]{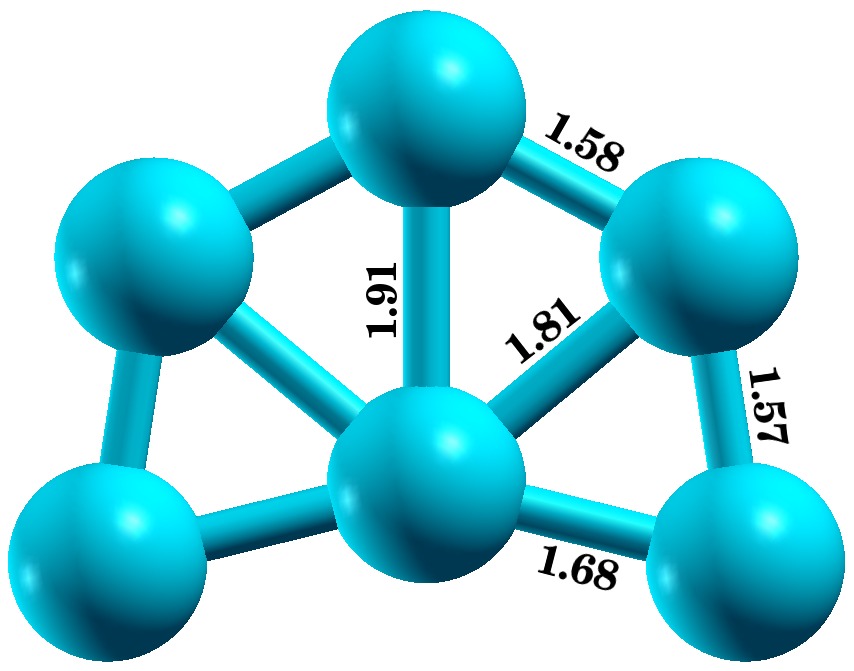}} \\
\subfloat[C$_{2v}$, $^{4}A_{2}$ Threaded trimer]
{\includegraphics[scale=0.13]{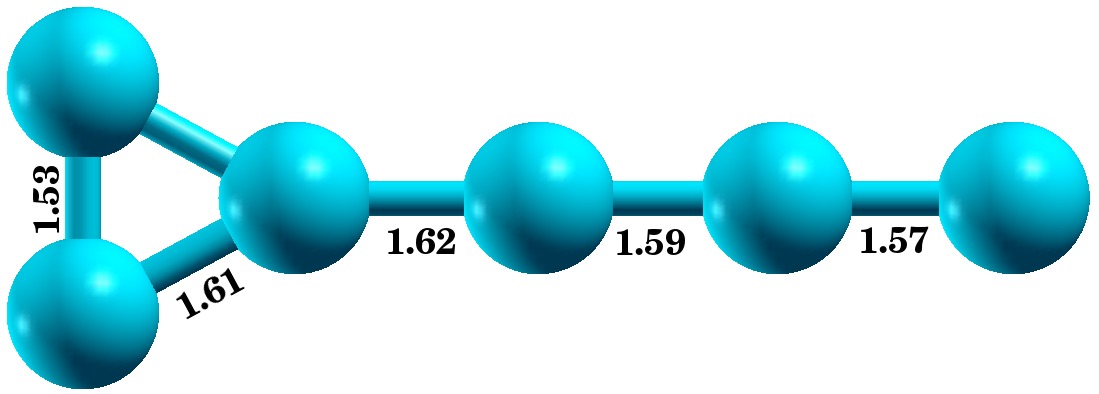}} \hfill
\subfloat[D$_{4h}$, $^{2}B_{1g}$ Bipyramid]
{\includegraphics[scale=0.15]{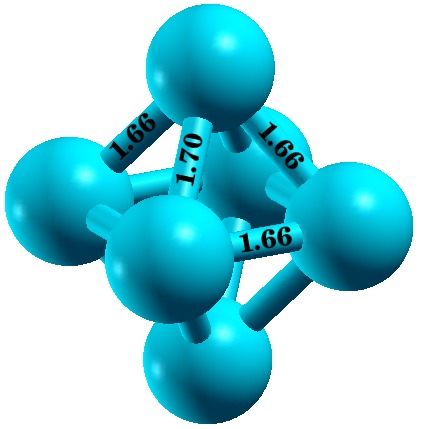}} \hfill
\subfloat[D$_{\infty h}$, $^{4}\Sigma_{u}$ Linear]
{\includegraphics[width=6.4cm]{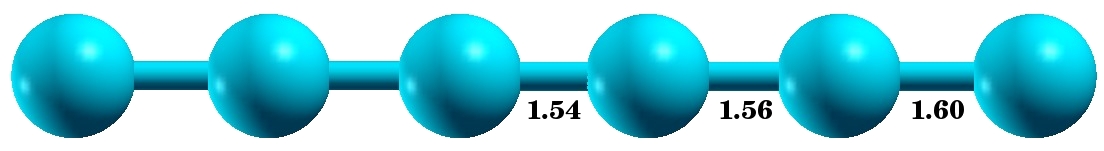}}   \hfill
\subfloat[D$_{2h}$, $^{2}B_{2g}$ Planar trimers]
{\includegraphics[scale=0.15]{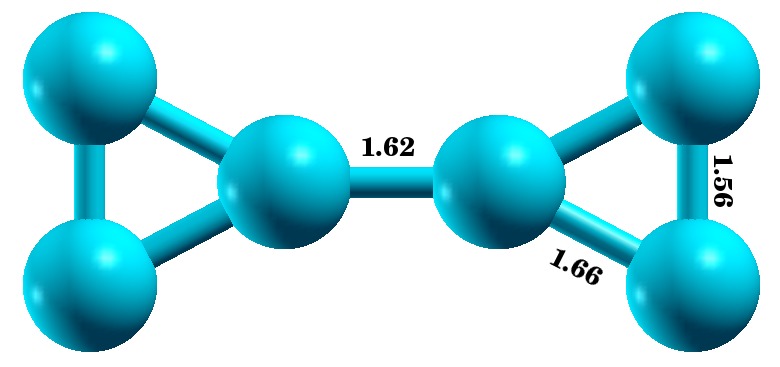}} \\
\end{center}
\caption{\label{fig:geometries-cationic}   Geometry optimized ground state structures
of different isomers of cationic B$_{6}^{+}$ clusters, along with the point group symmetries
obtained at the \ac{CCSD} level. }
\end{figure*}

We have found a total of 8 isomers of cationic (B$_{6}^{+}$) cluster with stable geometries 
as shown in the Fig. \ref{fig:geometries-cationic}. The relative standings in energy 
are presented in the Table \ref{tab:energies-cationic}, along with point group symmetries,
electronic states and expectation value of $S^2$ operator. Since this is a case of an open-shell system,
the spin contamination may induce large errors in he computed absorption spectra. We have reported 
$\langle S^2 \rangle$ values for excited states corresponding to each peak in the spectra.
In most of the cases the geometry of the neutral isomer is retained, reflected in the fact that some peaks
show up in the optical absorption spectra at the same energies as those in the neutral cluster.

\begin{table*}
\caption{Point group, electronic state, total energies and values of $\langle$ S$^2$ $\rangle$ before and after spin 
annihilation operation for different isomers of B$_{6}^{+}$ cluster.}
\label{tab:energies-cationic}       
\centering
\begin{tabular}{clllccc}
\hline\noalign{\smallskip}
Sr.    	& Isomer		& Point 	& Elect. 	& Total  	& $\langle$ S$^2$ $\rangle$ & $\langle$ S$^2_a$ $\rangle$\\
no. 	& 			& group 	& State      	& Energy (Ha) 	&			    &				\\  
\noalign{\smallskip}\hline\noalign{\smallskip}
1	& Planar ring (I)	& C$_{s}$	& ${}^2 A^{''}$	& -147.492831	&0.8410	&0.7524	\\
2	& Bulged wheel 		& C$_{1}$	& ${}^2 A$	& -147.491994	&1.0450	&0.7909	\\
3	& Planar ring (II)	& D$_{2h}$	& ${}^2 A_{g}$	& -147.480796	&0.8503	&0.7531	\\
4	& Incomplete wheel 	& C$_{2v}$	& ${}^4 A_{2}$	& -147.454234	&4.6090	&3.9490	\\
5	& Threaded trimer 	& C$_{2v}$	& ${}^4 A_{2}$	& -147.429627	&3.7671	&3.7501	\\
6	& Tetra. bipyramid 	& D$_{4h}$	& ${}^2 B_{1g}$	& -147.413145	&1.2565	&0.8742	\\
7	& Linear 		& D$_{\infty h}$& ${}^4\Sigma_{u}$&-147.392263	&4.8354	&4.0712	\\
8	& Planar trimers	& D$_{2h}$	& ${}^2 B_{2g}$	& -147.358494	&1.0000	&0.7808	\\
\noalign{\smallskip}\hline
\end{tabular}
\end{table*}


\begin{figure}[h!]
\centering
\includegraphics[width=8.3cm]{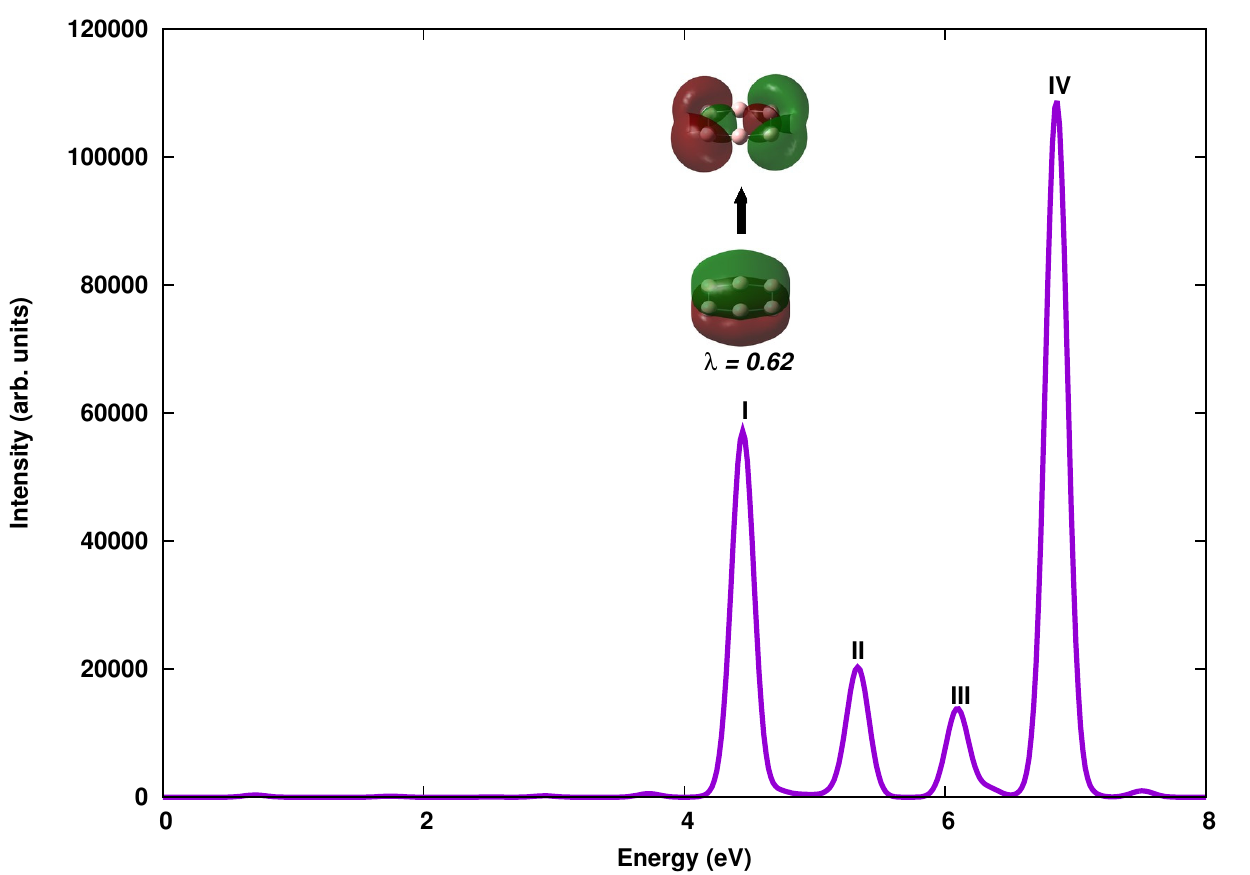}
\caption{\label{fig:cationic-plot-planar-ring}  The linear optical absorption spectrum
of planar ring B$_{6}^{+}$ isomer (I), calculated using the \ac{CIS} approach, along with the natural transition orbitals
involved in the excited states corresponding to the peak I (4.44 eV). 
Parameter $\lambda$ refers to a fraction of the \ac{NTO} pair contribution to a given electronic excitation. }
\end{figure}

The most stable isomer of B$_{6}^{+}$ cluster is a planar ring-type of structure, with C$_{s}$ point group symmetry.
This is in contrast to the other reported geometries which have D$_{2h}$ symmetry \cite{b6-isomerization,vlasta-chem-review,b6-dft}.
A slight difference in the orientation makes it less symmetric.
However, the bonds lengths obtained are in good agreement with those with D$_{2h}$ symmetric geometry cited above.
 The optical absorption spectrum calculated using \ac{CIS} approach is as shown in the 
Fig. \ref{fig:cationic-plot-planar-ring} and corresponding many-particle 
wavefunctions of excited states contributing to the various peaks are presented in 
Table \ref{Tab:table-cationic-planar-ring}.
 Similar to the neutral counterpart, this isomer also has very feeble absorption in the 
visible range, with polarization perpendicular to the plane of the isomer. Transitions involved corresponding to
peak 4.44 eV are from completely delocalized orbitals to the localized ones on each corner of the isomer.

\begin{figure}[h!]
\centering
\includegraphics[width=8.3cm]{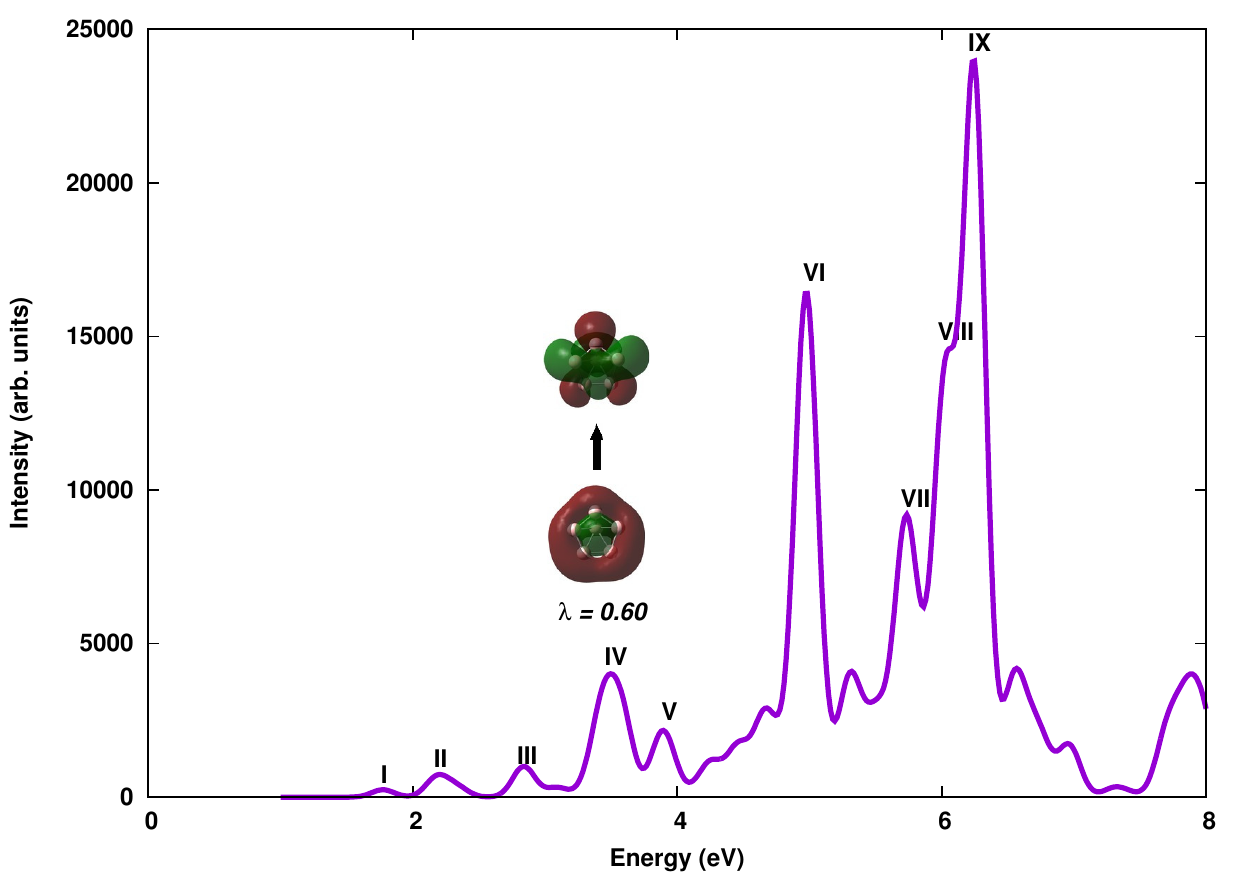}
\caption{\label{fig:cationic-plot-bulged-wheel}  The linear optical absorption spectrum
of bulged wheel B$_{6}^{+}$ isomer, calculated using the \ac{CIS} approach, along with the natural transition orbitals
involved in the excited states corresponding to the peak IV (3.58 eV). 
Parameter $\lambda$ refers to a fraction of the \ac{NTO} pair contribution to a given electronic excitation.}
\end{figure}

Bulged wheel structure is the next low lying isomer of B$_{6}^{+}$ with just 0.023 eV above the global minimum.
However, as compared to the neutral one, this geometry has C$_{1}$ symmetry due to the significant bond length 
reordering. Our computed geometries are consistent with the results of Refs. \citenst{b6-isomerization, b6-dft}. 
The optical absorption spectrum is presented in Fig. \ref{fig:cationic-plot-bulged-wheel}.
The many-particle wavefunctions of excited states contributing to various peaks are presented in Table
\ref{Tab:table-cationic-bulged-wheel}. The spectrum is distinctly different with a large number of smaller 
peaks and a stronger peak at 6.24 eV. The onset of spectrum occurs at 1.76 eV dominated by $H_{\beta} \rightarrow L_{\beta} $ 
 and $H_{\beta} - 1 \rightarrow L_{\beta} $  configurations.

\begin{figure}[h!]
\centering
\includegraphics[width=8.3cm]{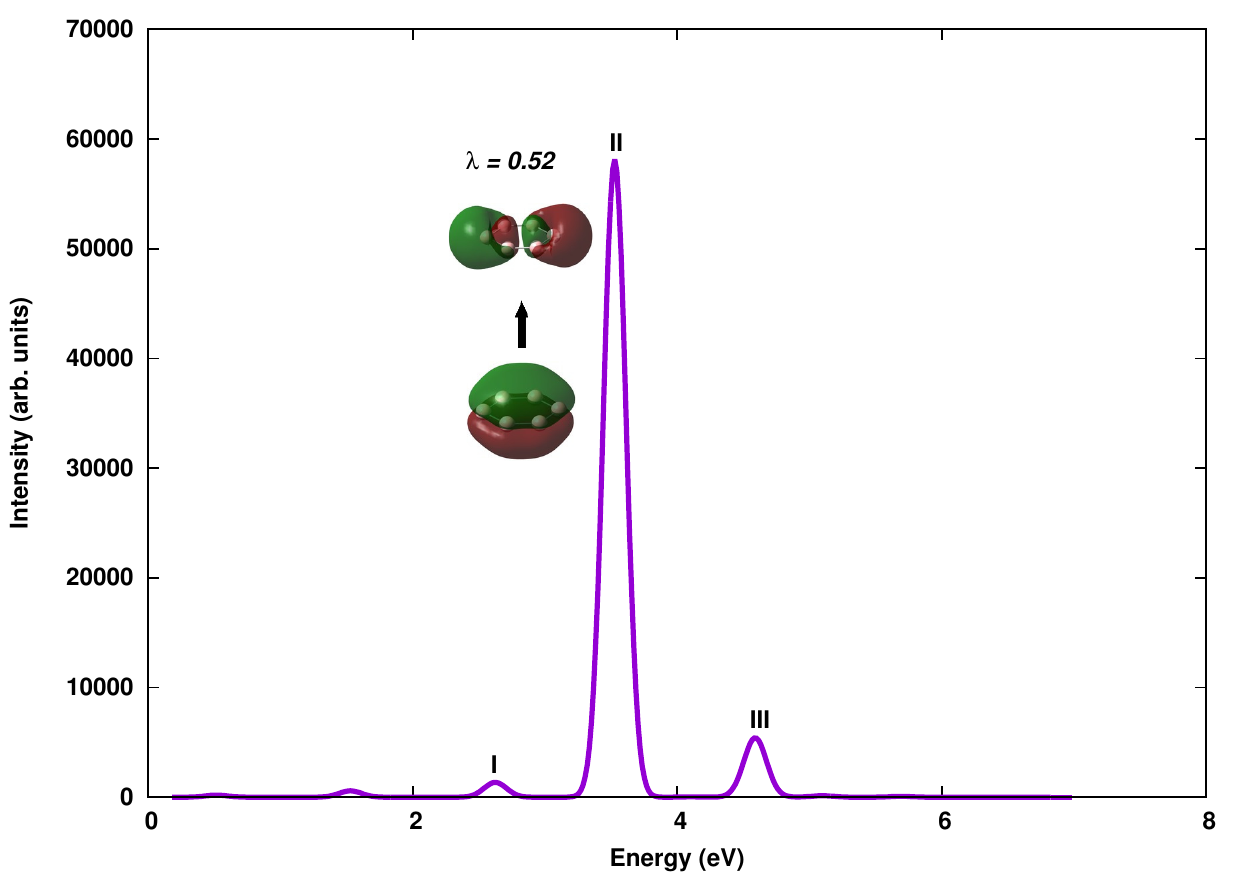}
\caption{\label{fig:cationic-plot-planar-d2h-ring}  The linear optical absorption spectrum
of planar ring  B$_{6}^{+}$ isomer II, calculated using the \ac{CIS} approach, along with the natural transition orbitals
involved in the excited states corresponding to the peak II (3.52 eV). 
Parameter $\lambda$ refers to a fraction of the \ac{NTO} pair contribution to a given electronic excitation.}
\end{figure}

Another ring-like structure with D$_{2h}$ symmetry and doublet multiplicity lies next in the energy order. The hexagonal 
benzene type structure has 1.66 \AA{} and 1.53 \AA{} as unique bond lengths, which are somewhat larger than those 
reported in the literature \cite{b6-isomerization,b6-dft}. The optical absorption spectrum presented in Fig. 
\ref{fig:cationic-plot-planar-d2h-ring}, is fairly simple and has well resolved peaks. The many-particle wavefunctions 
of excited states contributing to various peaks are presented in Table \ref{Tab:table-cationic-planar-d2h-ring}.
All absorption peaks are due to the polarization along the plane of the isomer. The strongest absorption is seen at
3.52 eV with fully delocalized to partially localized nature of transition.

\begin{figure}[h!]
\centering
\includegraphics[width=8.3cm]{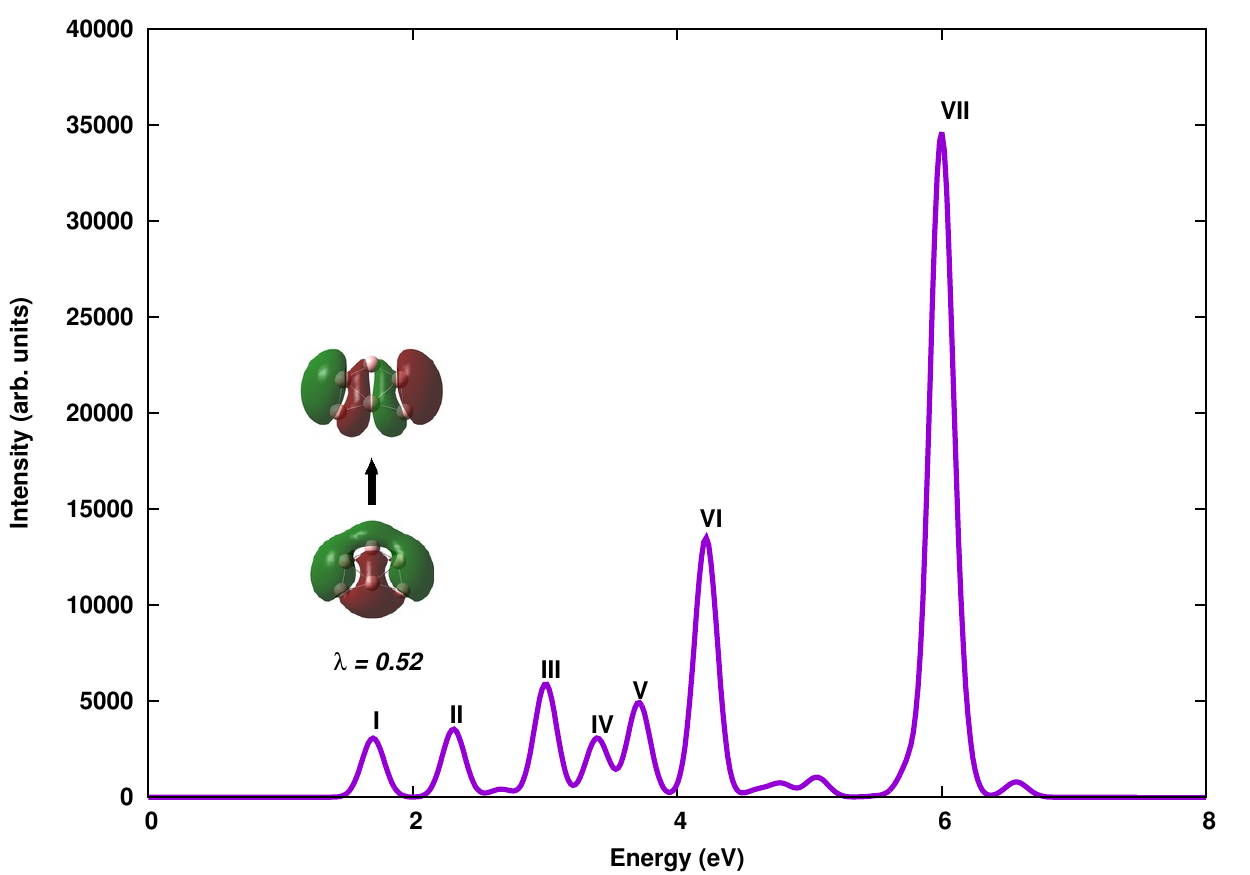}
\caption{\label{fig:cationic-plot-incomplete-wheel}  The linear optical absorption spectrum
of incomplete wheel  B$_{6}^{+}$ isomer, calculated using the \ac{CIS} approach, along with the natural transition orbitals
involved in the excited states corresponding to the peak I (1.69 eV). 
Parameter $\lambda$ refers to a fraction of the \ac{NTO} pair contribution to a given electronic excitation.}
\end{figure}

Next low lying isomer is a planar incomplete wheel structure with C$_{2v}$ point group symmetry and quartet multiplicity.
This multiplicity and computed geometry is consistent with results of Ref. \citenst{b6-dft}. The optical absorption 
(\emph{cf.} Fig. \ref{fig:cationic-plot-incomplete-wheel}) starts at 1.69 eV, with 
polarization transverse to the plane of the isomer. The many particle wave-functions of excited states contributing to various peaks 
are presented in Table \ref{Tab:table-cationic-incomplete-wheel}. The configurations contributing to the first peak are
$H_{\beta} - 2 \rightarrow L_{\beta}$. 

\begin{figure}[h!]
\centering
\includegraphics[width=8.3cm]{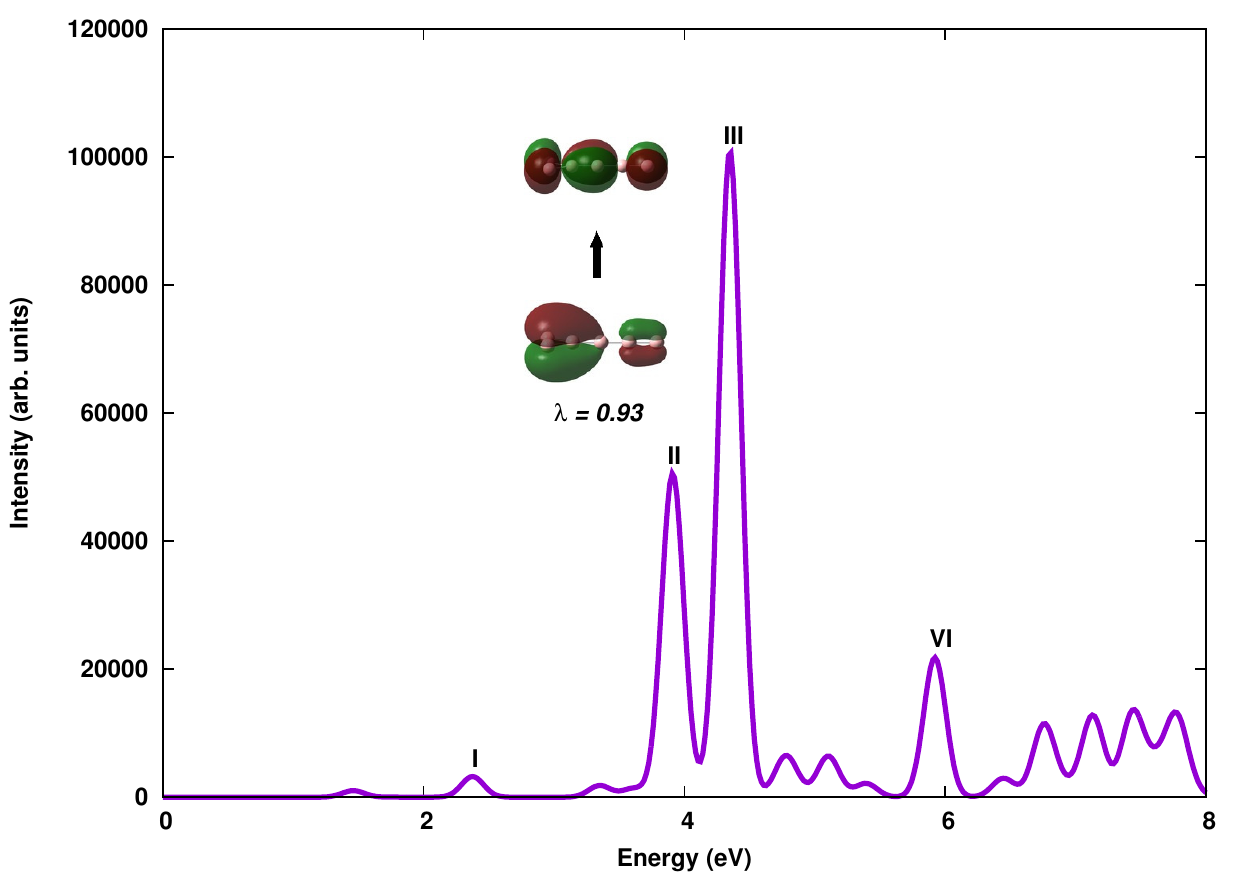}
\caption{\label{fig:cationic-plot-threaded-trimer}  The linear optical absorption spectrum
of threaded trimer  B$_{6}^{+}$ isomer, calculated using the \ac{CIS} approach, along with the natural transition orbitals
involved in the excited states corresponding to the peak II (3.90 eV). 
Parameter $\lambda$ refers to a fraction of the \ac{NTO} pair contribution to a given electronic excitation.}
\end{figure}

Another isomer with the same point group symmetry and multiplicity as that of the previous one, but having a 
geometry of linear chain with an isosceles triangle at the end, is the next low lying isomer of cationic B$_{6}^{+}$.
Our results about geometry are in good agreement with the Ref. \citenst{b6-dft}. 
The optical absorption spectrum (\emph{cf.} Fig. \ref{fig:cationic-plot-threaded-trimer})
has three major peaks, coupled with a number of minor peaks. The first major peak occurs at 3.90 eV, with polarization along the plane 
of the isomer, and has dominant contribution from $H_{3\alpha} \rightarrow L_{\alpha} + 1 $  (\emph{cf.} Table \ref{Tab:table-cationic-threaded-trimer}). 

\begin{figure}[h!]
\centering
\includegraphics[width=8.3cm]{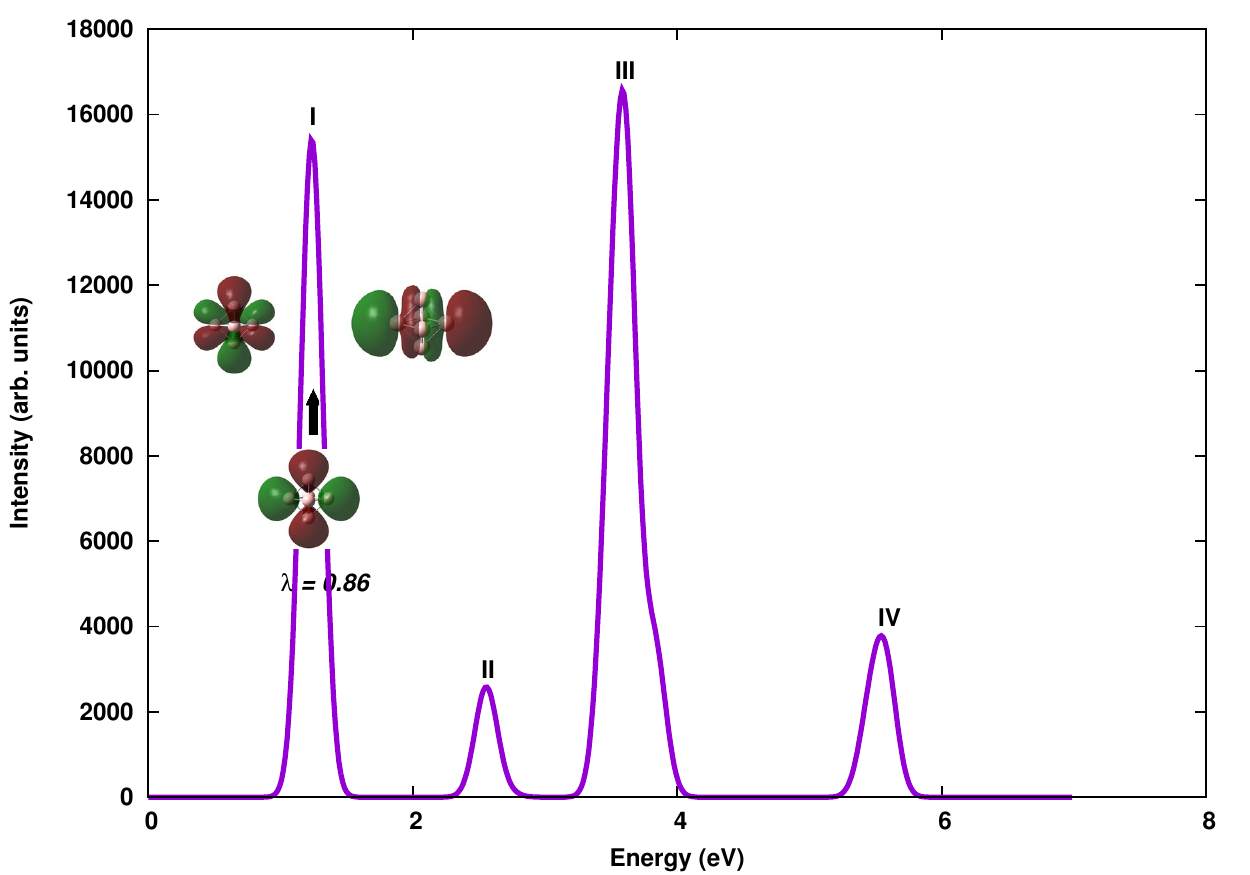}
\caption{\label{fig:cationic-plot-bipyramid}  The linear optical absorption spectrum
of triangular bipyramid B$_{6}^{+}$ isomer , calculated using the \ac{CIS} approach, along with the natural transition orbitals
involved in the excited states corresponding to the peak I (1.23 eV). 
Parameter $\lambda$ refers to a fraction of the \ac{NTO} pair contribution to a given electronic excitation.}
\end{figure}

Tetragonal bipyramid forms the next stable isomer of cationic B$_{6}^{+}$, with D$_{4h}$ point group symmetry 
and doublet multiplicity. This is in good agreement with the geometries reported in Refs. \citenst{b6-isomerization}
and \citenst{b6-dft}. The optical absorption spectrum (\emph{cf.} Fig. \ref{fig:cationic-plot-bipyramid}) has well defined
small number of peaks. $H_{1\alpha} \rightarrow L_{\alpha}$ and
$H_{\alpha} - 4 \rightarrow L_{\alpha}$ contributes dominantly to peaks in the visible range at 1.23 eV and 2.55 eV,
 respectively (\emph{cf.} Table \ref{Tab:table-cationic-triangular-bipyramid}).

\begin{figure}[h!]
\centering
\includegraphics[width=8.3cm]{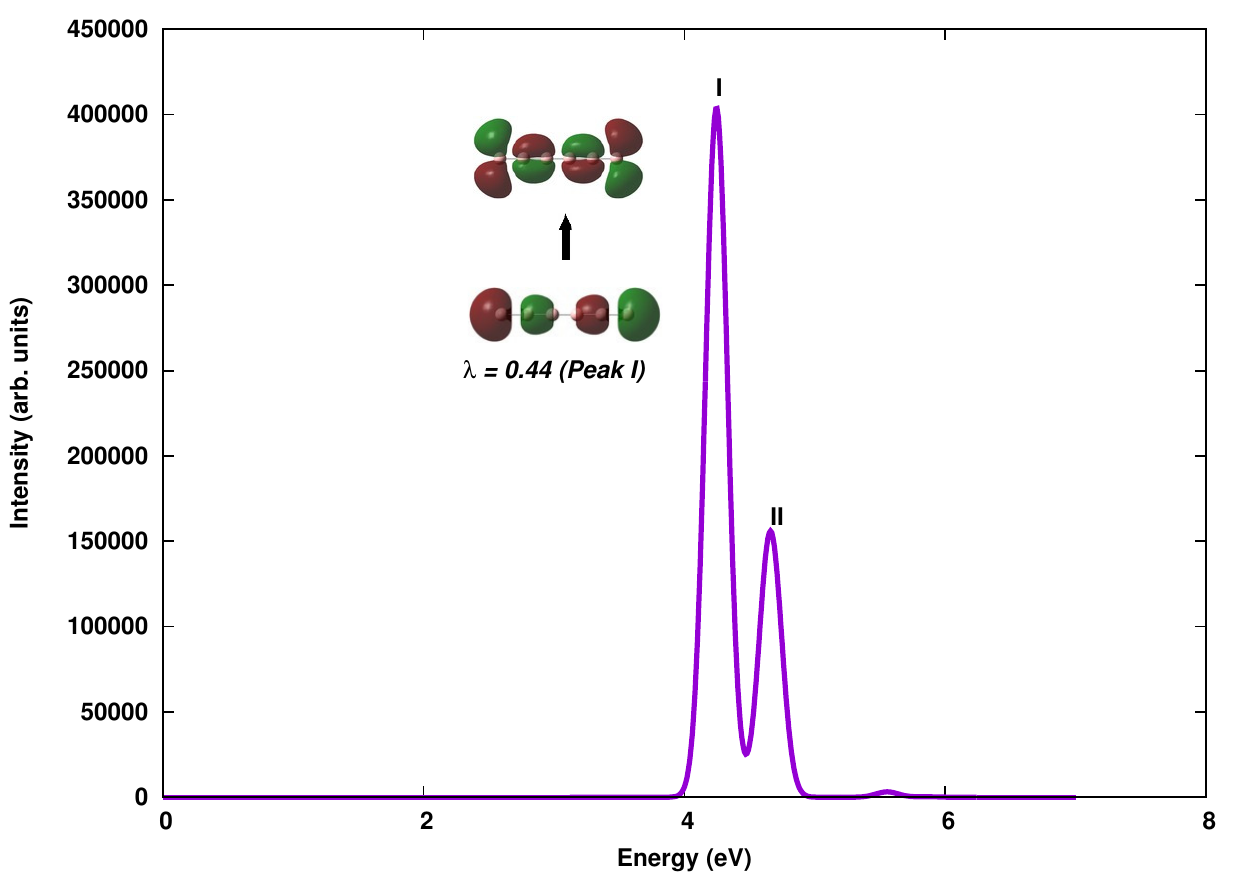}
\caption{\label{fig:cationic-plot-linear}  The linear optical absorption spectrum
of linear B$_{6}^{+}$ isomer , calculated using the \ac{CIS} approach, along with the natural transition orbitals
involved in the excited states corresponding to the peak I (4.25 eV).
Parameter $\lambda$ refers to a fraction of the \ac{NTO} pair contribution to a given electronic excitation. }
\end{figure}

\begin{figure}[h!]
\centering
\includegraphics[width=8.3cm]{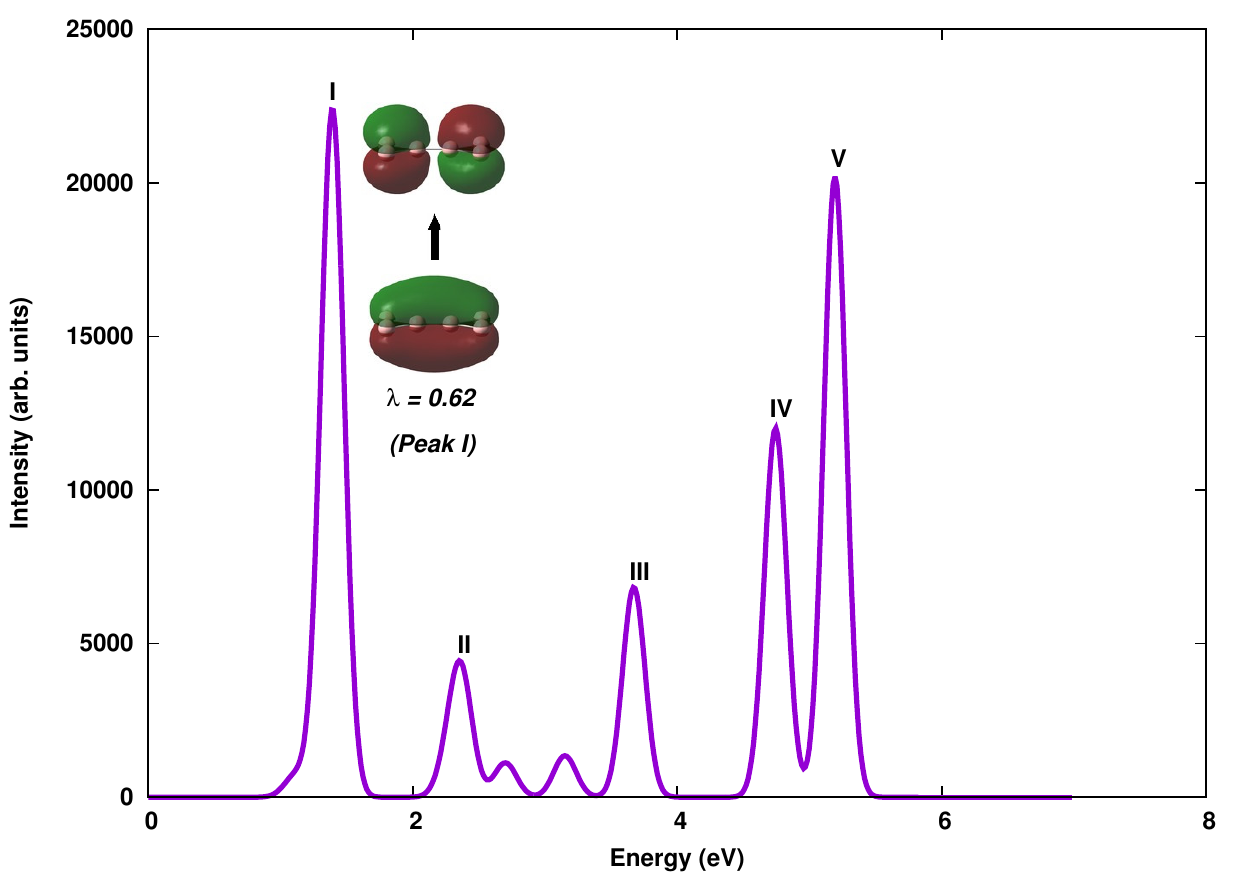}
\caption{\label{fig:cationic-plot-planar-trimers}  The linear optical absorption spectrum
of planar trimers B$_{6}^{+}$ isomer , calculated using the \ac{CIS} approach, along with the natural transition orbitals
involved in the excited states corresponding to the peak I (1.40 eV). 
Parameter $\lambda$ refers to a fraction of the \ac{NTO} pair contribution to a given electronic excitation.}
\end{figure}

Two more structures were found stable \emph{i.e.} (a) a planar structure with two trimers joined together and,
(b) a linear one. These isomers are much above the global minimum energy, it rules out their room temperature existence.
In the linear isomer the absorption spectrum (\emph{cf.} Fig. \ref{fig:cationic-plot-linear})
 is red-shifted as compared to the neutral one, with major peak at 4.25 eV
with dominant contribution from $H_{\beta} - 1 \rightarrow L_{\beta} + 1$ and $H_{\beta} \rightarrow L_{\beta}$ configurations. 
In case of planar trimers structure, the spectrum (\emph{cf.} Fig. \ref{fig:cationic-plot-planar-trimers}) 
also seems to be red shifted as 
 compared to the neutral one. The first peak is found at 1.40 eV with $H_{\beta} \rightarrow L_{\beta}$ and 
 $H_{\beta} - 2 \rightarrow L_{\beta} + 1$, characterized by $\pi \rightarrow \pi^*$ transition, as dominant 
 contribution to the wavefunction of the excited state.

\section{Summary}
\label{sec:conclusion-epjd}
A large number of randomly selected initial structures of neutral B$_{6}$
and cationic B$_{6}^{+}$ clusters are taken into consideration for 
locating the global and local minimas on the potential energy curves. A careful 
geometry optimization is done for all those structures at a correlated level.
The optical absorption spectra of different low-lying isomers of both neutral
and cationic isomers are reported here. A singles configuration interaction approach was 
used to compute excited state energies and the absorption spectra of various clusters.
Spectra of cationic clusters appear slightly red-shifted with respect to the neutral one.
A comparison of spectra with \ac{CIS} as well as more sophisticated \ac{EOM-CCSD} method is presented 
in light of nature of excitations involved in the spectra. In all closed shell systems, a 
complete agreement on the nature of configurations involved is observed in both methods. On the other
hand, for open-shell systems, minor contribution from double excitations are observed. Also,
the spectra computed using \ac{EOM-CCSD} approach is generally red-shifted as compared to the \ac{CIS} ones.
Such comparisons can be used to benchmark the \ac{CIS} results. 

Different isomers exhibit distinct optical response, even though they are isoelectronic 
and many of them are almost degenerate. 
This signals a strong-structure property relationship, which can be exploited 
for experimental identification of these isomers; something which is not possible with the 
conventional mass spectrometry.
A strong mixture of configurations in the many-body wavefunctions of various excited states
are observed, indicating the plasmonic nature of the photoexcited states \cite{plasmon}.


  \lhead{{\chaptername\ \thechapter.}{  Aluminum Clusters Al$_{n}$ ($n$=2 -- 5)}}
   \chapter{\label{chap:main_smallal}Theory of Linear Optical Absorption in Various Isomers of Aluminum Clusters Al$_{n}$ ($n$=2 -- 5)}
\emph{This chapter is based on a submitted manuscript, available on arxiv.org:\textbf{1303.2511} \\ by Ravindra Shinde and Alok Shukla.}
\par

Metal clusters are promising candidates in the era of nanotechnology. The reason behind growing interest in clusters lies in their interesting
properties, applicability of simple theoretical models to describe their properties, and a vast variety of 
potential technological applications \cite{julius_book, clustnano_book, alonso_book, deheer_rmp,na_dehaar_prl}.

Various jellium models have successfully described electronic structures of alkali metal clusters, because alkali metals have free valence
electrons \cite{deheer_rmp}. This beautifully explains the higher abundance of certain
clusters. However, in case of aluminum clusters, the experimental 
results often provide conflicting evidence about the size at which the jellium model would work \cite{wang_sp_hybrid_prl, nonjellium-to-jellium}. 
The theoretical explanation also depends on the valency of aluminum atoms considered. Since \emph{s - p} orbital energy 
separation in aluminum atom is 4.99 eV, and it decreases with the cluster size, the valency should be
changed from one to three \cite{rao_jena_ele_struct_al}. Perturbed jellium model, which takes orbital anisotropy
into account, has successfully explained the mass abundance of aluminum clusters \cite{upton_elec_struct_small_al,free_ele_metal_cluster_clemenger}.

Shell structure and \emph{s - p} hybridization in anionic aluminum clusters
were probed using photoelectron spectroscopy by Gantef\"{o}r and Eberhardt \cite{Gante_shell_struct_al}, and 
Li \emph{et al}. \cite{wang_sp_hybrid_prl}. Evolution of electronic structure and other properties of aluminum clusters has been studied in many 
reports \cite{rao_jena_ele_struct_al, wang_sp_hybrid_prl, upton_elec_struct_small_al, reinhart_big_al_cluster,sachdev_dft_al_small,truhler_theory_validation_jpcb,whetten_struct_bind_prb,cox-al3-experiment,jones_simul_anneal_al,jones_struct_bind_prl,manninen_ionization_al,martinez_all_vs_core,mojtaba_stat_dyna_pol_al}.
Structural properties of aluminum clusters were studied using density functional theory by Rao and Jena \cite{rao_jena_ele_struct_al}.
An all electron and model core potential study of various Al clusters was carried out by Martinez \emph{et al}. \cite{martinez_all_vs_core}. 
Upton performed chemisorption calculations on aluminum clusters and reported that
Al$_{6}$ is the smallest cluster that will absorb H$_{2}$. \ac{DFT} alongwith molecular dynamics were used to study electronic and
structural properties of aluminum clusters \cite{jones_simul_anneal_al}.
 
Although the photoabsorption in alkali metal clusters has been studied by many authors at various levels of 
theory \cite{optical_na_deheer, na_opt_kappes_jcp, na_opt_bethe_salp, li_na_tddft, yanno-frag-absorption-prl89, yanno-optical-response-pra91, yanno-evolution-optical-prb93, yanno-surface-plasmon-prb98},  
very few similar studies exist for aluminum clusters \cite{kanhere_prb_optical_al, smith_tunable_icosa}.
Optical absorption in several aluminum clusters corresponding to the minimum energy configurations
has been studied by Deshpande \emph{et al.} using \ac{TDDFT} \cite{kanhere_prb_optical_al}. Xie \emph{et
al.} presented TDDFT optical absorption spectra of various caged icosahedral
aluminum clusters \cite{smith_tunable_icosa}. However, no theoretical or experimental study has been done on optical absorption in various
low lying isomers of aluminum clusters. The distinction of different isomers of a cluster has to be made using an experimental or theoretical
technique in which the properties are size, as well as shape, dependent. Conventional mass spectrometry only distinguishes clusters according
to the masses. Hence, our theoretical results can be coupled with the experimental measurements of optical absorption, to distinguish 
between different isomers of a cluster. We have recently reported results of such calculation on small boron clusters \cite{smallboron}. 

In this chapter, we present results of systematic calculations of optical absorption in various low lying isomers of small aluminum clusters
using \emph{ab initio} large-scale \ac{MRSDCI} method.
The nature of optical excitations involved in absorption has also been investigated by analyzing the wavefunctions of the excited states.

Remainder of the chapter is organized as follows. Next section discusses theoretical and computational details of the calculations, followed
by section \ref{sec:results-jcp}, in which results are presented and discussed. Conclusions and future directions are presented
in section \ref{sec:conclusions-jpc}. A detailed information about wavefunctions of excited states contributing to various
 photoabsorption peaks is presented in the Appendix \ref{app:aluminum}.

\section{\label{sec:theory-jcp}Theoretical and Computational Details}

\begin{figure*}[!t]
\centering
\subfloat[\textbf{Al$_{\mathbf{2}}$, D$_{\mathbf{\boldsymbol{\infty}h}}$,
$^{\mathbf{3}}\mathbf{\boldsymbol{\Pi_{u}}}$ \label{subfig:subfig-al2}}]{\includegraphics[width=2.8cm]{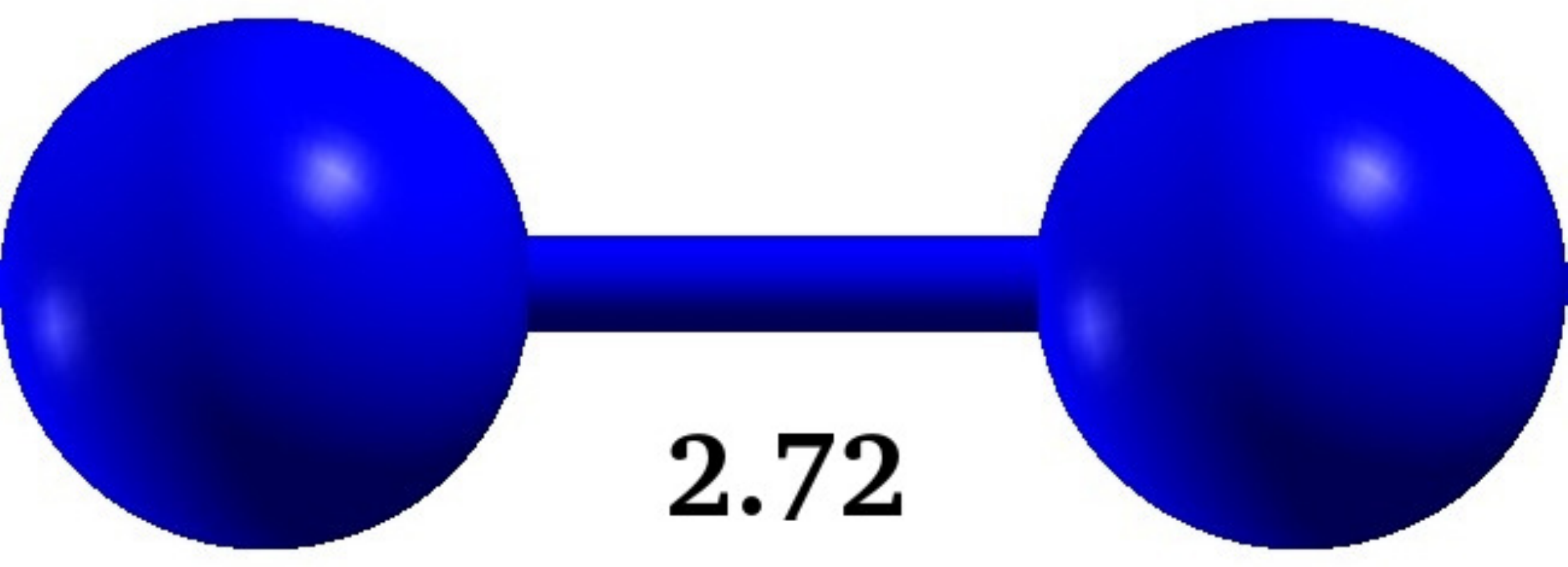}

}\hfill \subfloat[\textbf{Al$_{\boldsymbol{3}}$, D$_{\boldsymbol{3h}}$, $\boldsymbol{^{2}A_{1}^{'}}$}]{\includegraphics[width=0.12\paperwidth]{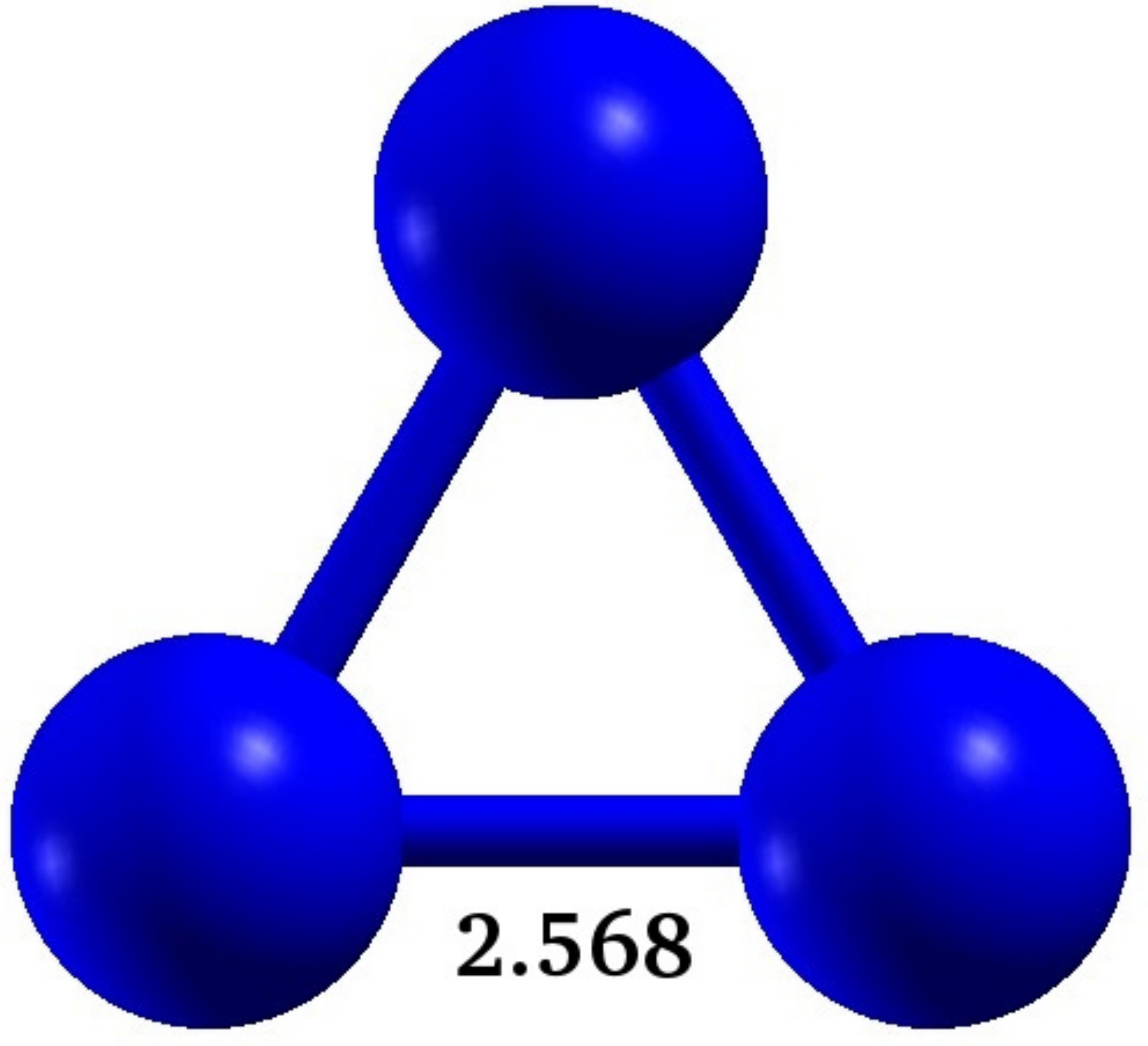}

}\hfill \subfloat[\textbf{Al$_{\boldsymbol{3}}$, C}$_{\boldsymbol{2v}}$, $\boldsymbol{^{4}A_{2}}$]{\includegraphics[width=0.14\paperwidth]{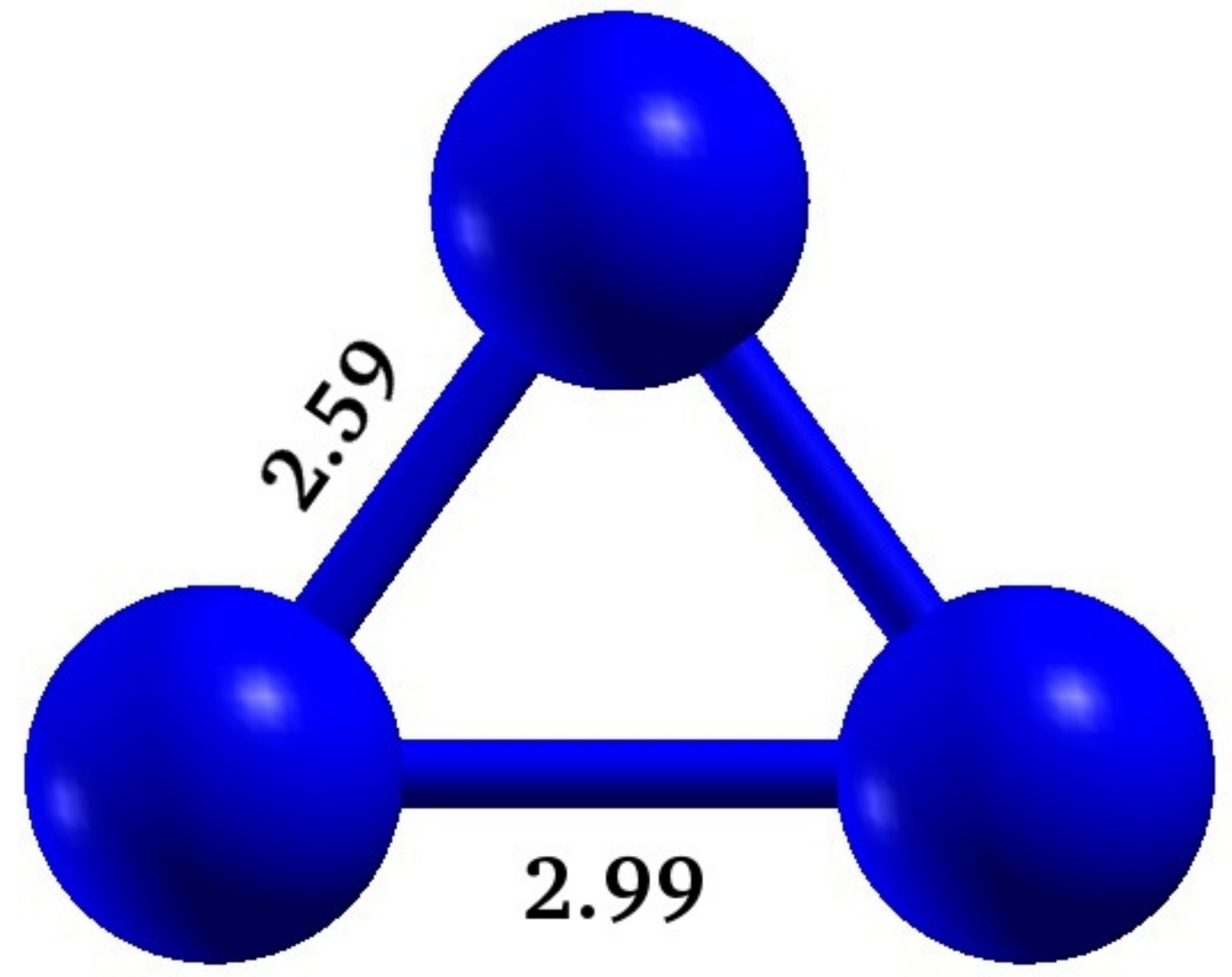}

}\hfill \subfloat[\textbf{Al$_{\boldsymbol{3}}$, D$_{\boldsymbol{\infty h}}$, $\boldsymbol{^{4}\Sigma_{u}}$}]{\includegraphics[width=0.2\paperwidth]{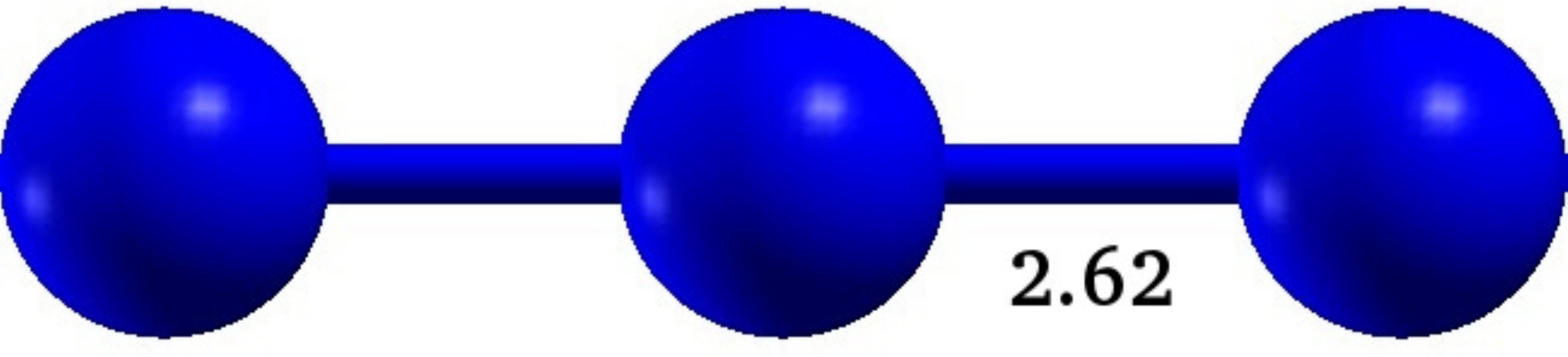}

}

\subfloat[\textbf{Al$_{\boldsymbol{4}}$, D$_{\boldsymbol{2h}}$, $\boldsymbol{^{3}B{}_{2g}}$}]{\includegraphics[width=0.17\paperwidth]{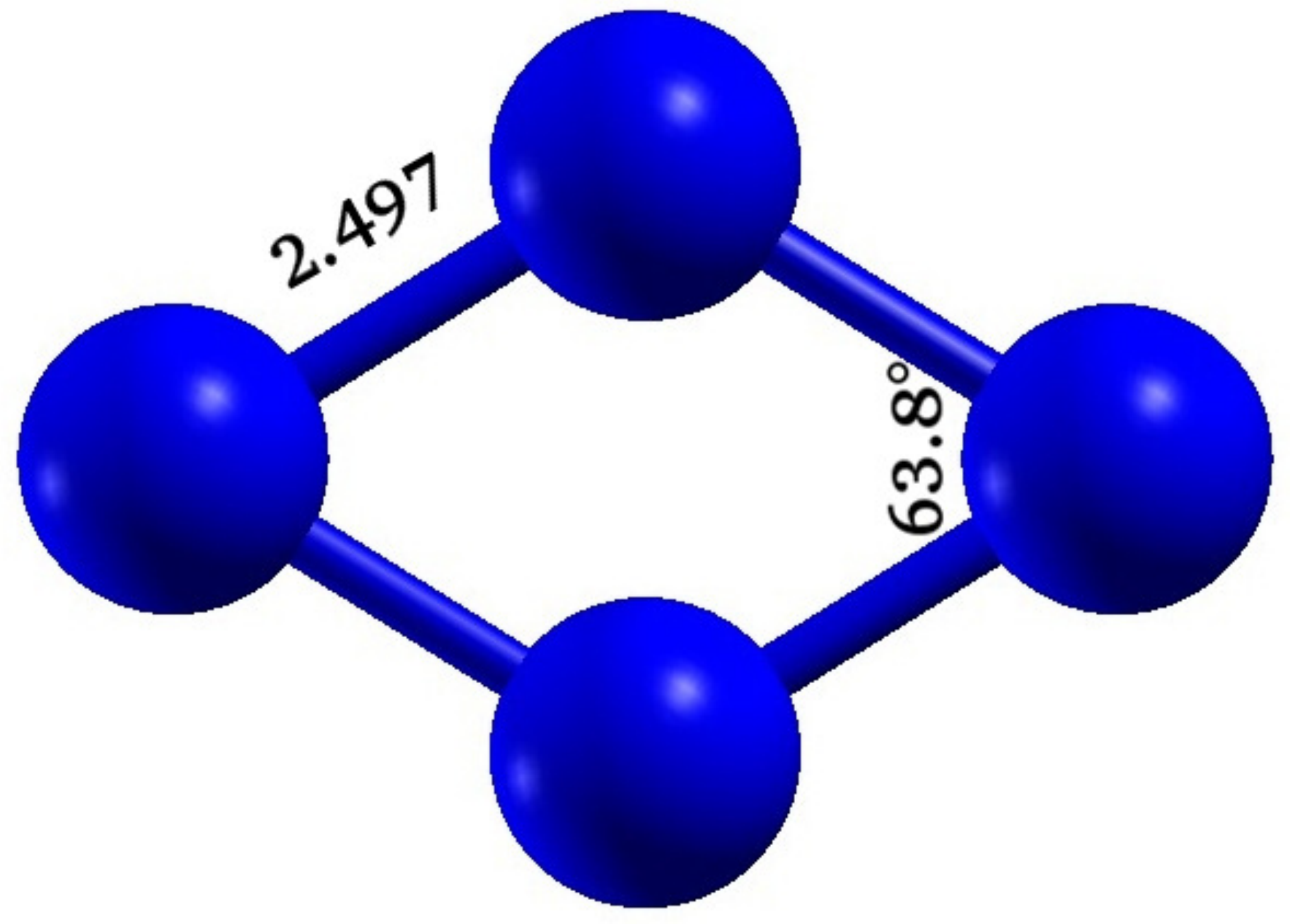}

}\hfill \subfloat[\textbf{Al$_{\boldsymbol{4}}$, D$_{\boldsymbol{4h}}$, $\boldsymbol{^{3}B{}_{3u}}$}]{\includegraphics[width=0.124\paperwidth]{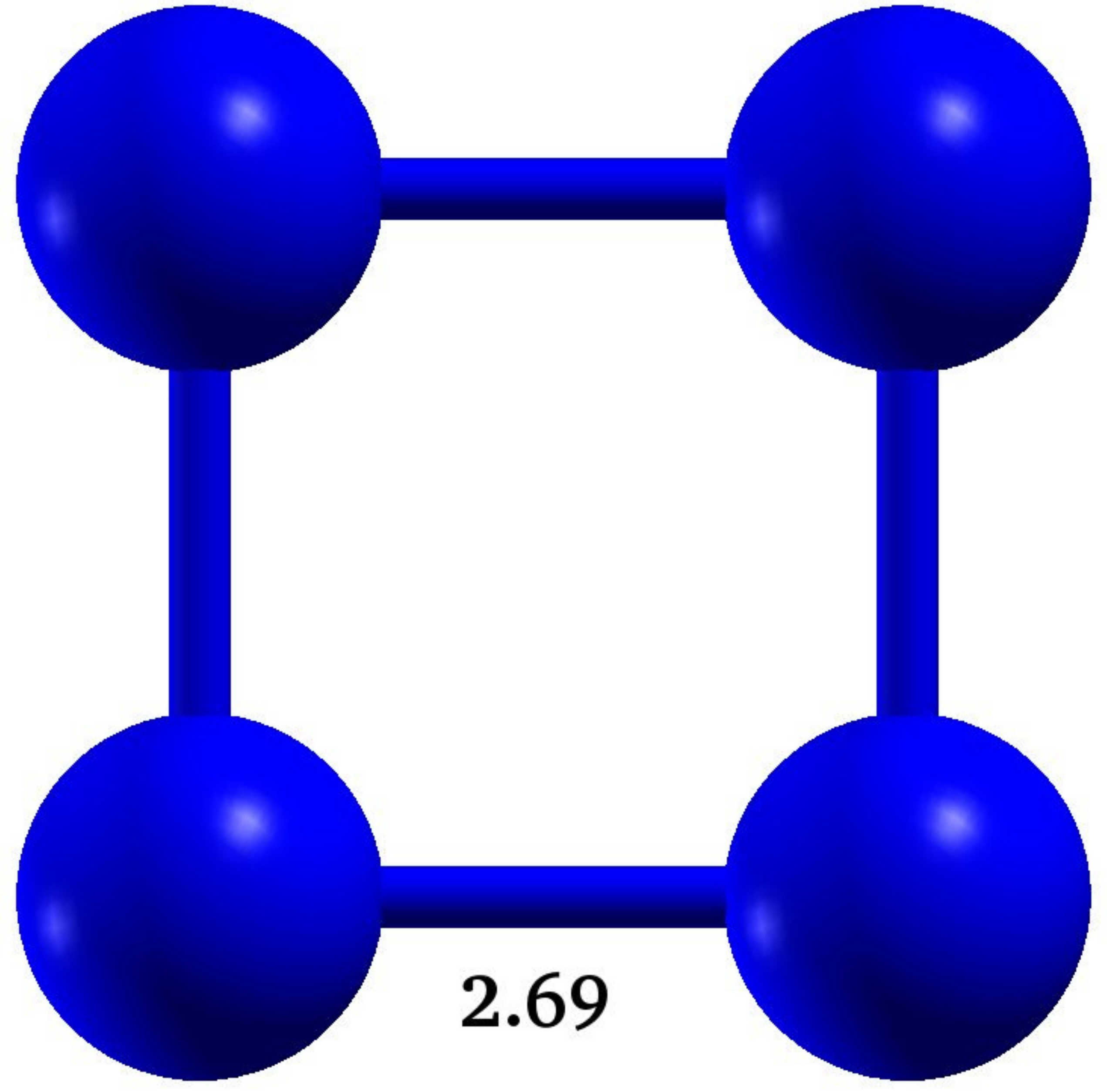}

}\hfill \subfloat[\textbf{Al$_{\boldsymbol{5}}$, C$_{\boldsymbol{2v}}$, $\boldsymbol{^{2}A{}_{1}}$}]{\includegraphics[width=0.2\paperwidth]{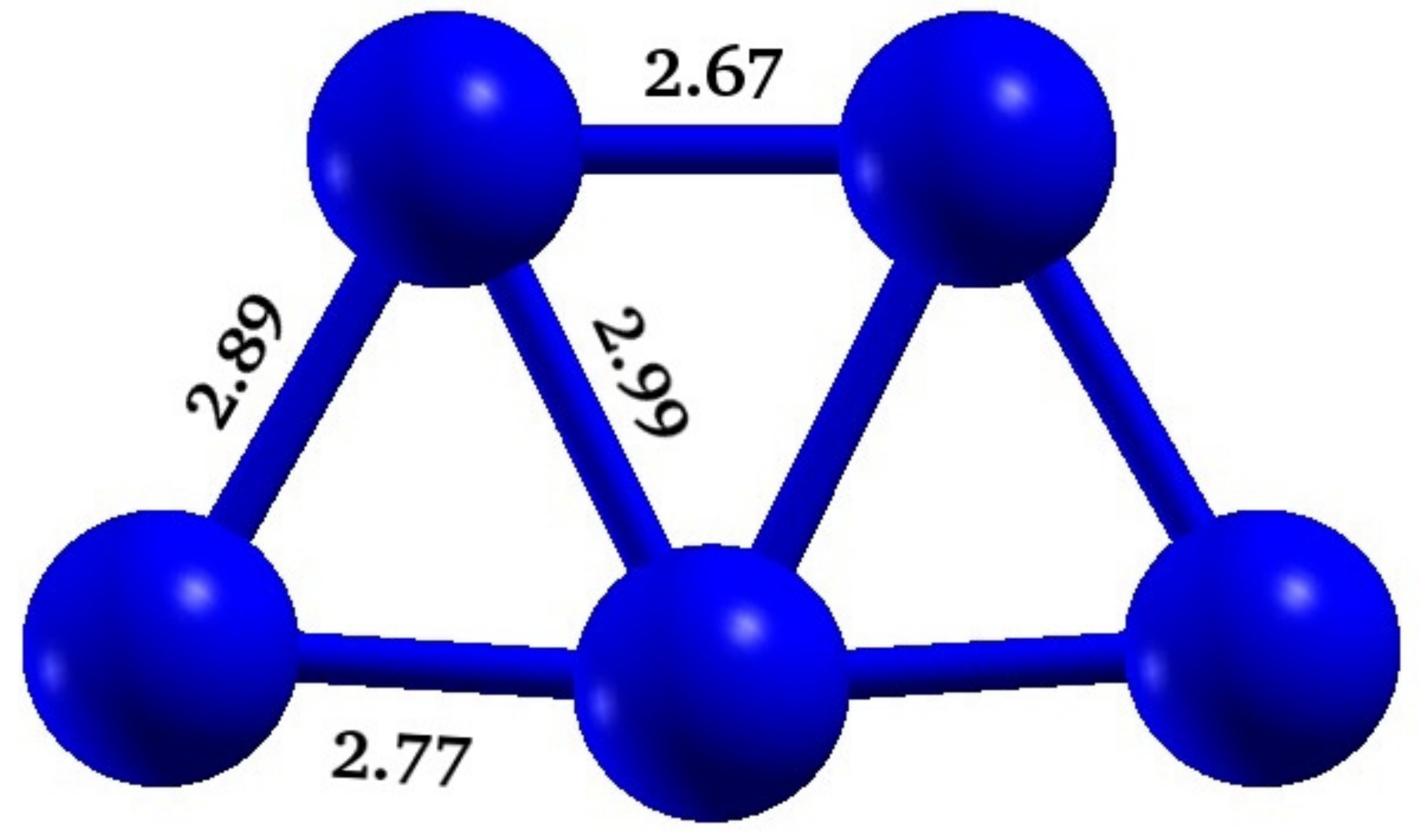}\label{subfig:geom-pentagon}

}\hfill \subfloat[\textbf{Al$_{\boldsymbol{5}}$, C$_{\boldsymbol{4v}}$,} $\boldsymbol{^{2}A{}_{1}}$]{\includegraphics[width=0.12\paperwidth]{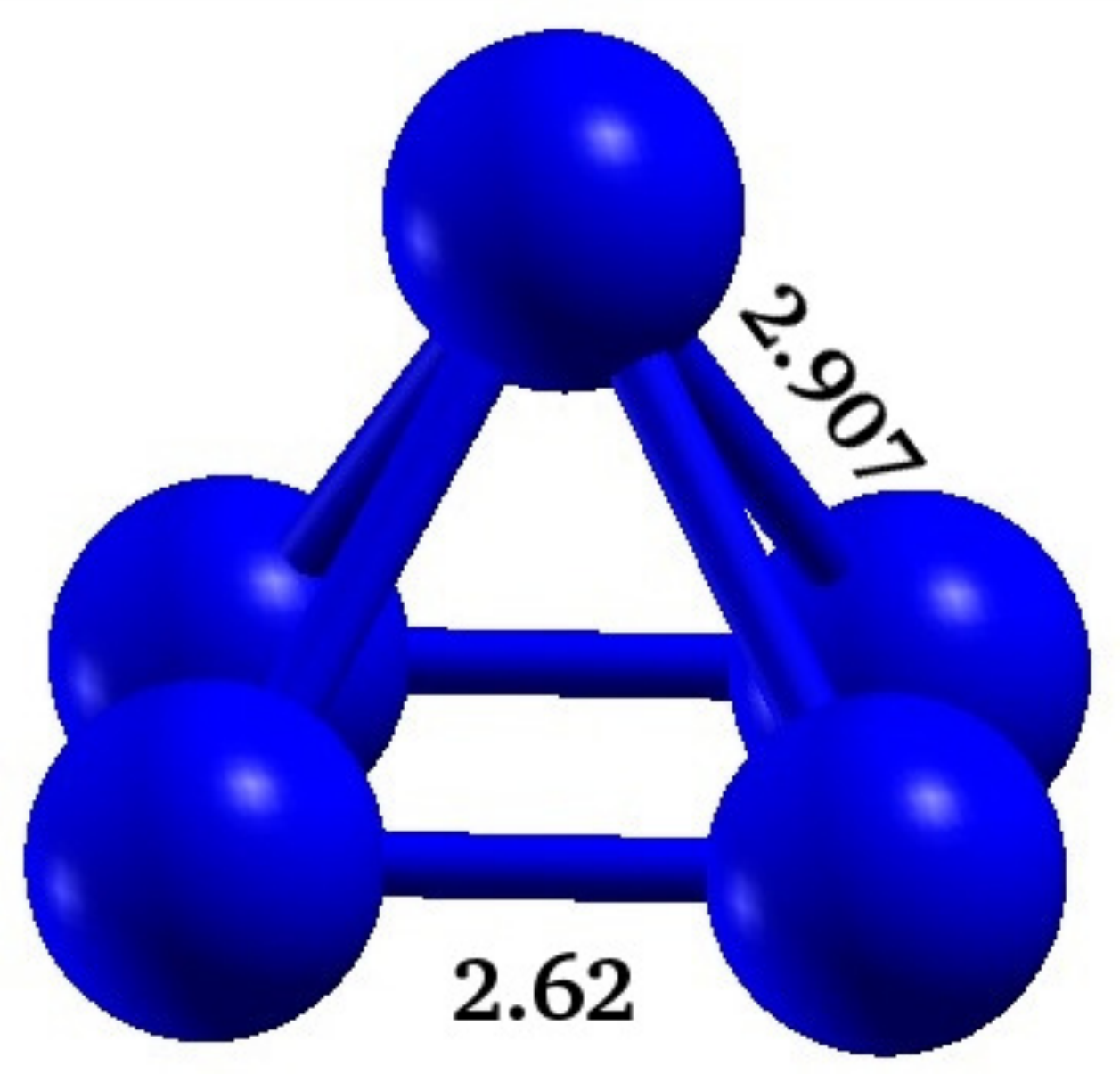}\label{fig:geom-pyra   }

} \vspace{0.2cm}

\caption{\label{fig:geometry}Geometry optimized structures
of aluminum clusters with point group symmetry and the electronic
ground state at the CCSD level. All numbers are in $\textrm{\AA}$
unit. }
\end{figure*}

The geometry of various isomers were optimized using the size-consistent coupled-cluster singles-doubles (CCSD) method, as implemented in the
\textsc{gaussian09} package \cite{gaussian09}. A basis set of 6-311++G(2d,2p)
was used which was included in the \textsc{gaussian 09} package itself. This basis set is optimized for the ground state calculations. Since an
even numbered electron system can have singlet, triplet, or higher spin multiplicity, we repeated the optimization for singlet and triplet systems
 to look for the true ground state geometry. Similarly, for odd numbered electron systems, doublet and quartet multiplicities were considered 
in the geometry optimization. To initiate the optimization, raw geometries, reported by Rao and Jena, based on density functional method 
were used \cite{rao_jena_ele_struct_al}. Figure \ref{fig:geometry} shows the final optimized geometries of the isomers studied in this chapter.

The linear photoabsorption spectra of various isomers of the boron clusters were computed using MRSDCI method, as described in 
subsection \ref{subsection-mrsdci}.

Our group has extensively used such approach in performing large-scale correlated calculations of linear optical absorption spectra of conjugated 
polymers \cite{mrsd_jcp_09,mrsd_prb_07, mrsd_prb_05, mrsd_prb_02}, and atomic clusters \cite{smallboron, borozene-sahu}.
We have also frozen the chemical core of aluminum atom from virtual excitations and have put a cap on total number of virtual orbitals taking 
part in optical absorption based on the energy of the orbital, as done previously with boron clusters.\cite{smallboron}


\section{\label{sec:results-jcp}Results and Discussion}

In this section, first we present a systematic study of the convergence
of our results and various approximations used. In the latter part,
we discuss the results of our calculations on various clusters.

\subsection{Convergence of calculations}

In this section we discuss the convergence of photoabsorption calculations
with respect to the choice of the basis set, and the size of the active
orbital space.

\subsubsection{Choice of basis set}

\begin{figure}
\centering
\includegraphics[width=8cm]{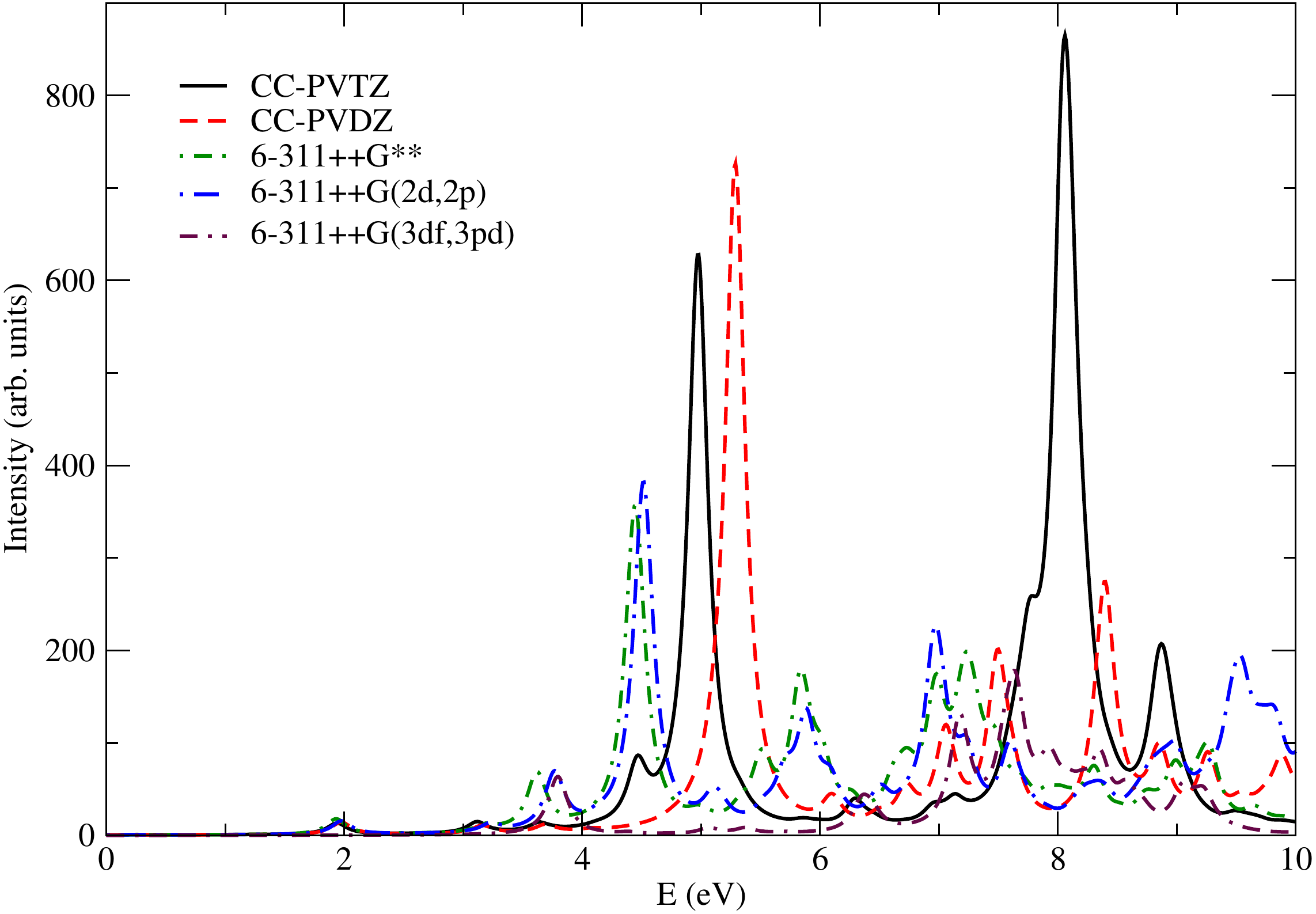} \vspace{0.2cm}
\caption{\label{fig:basis-study}Optical absorption in Al$_{2}$
calculated using various Gaussian contracted basis sets.}
\end{figure}

In the literature several optimized basis sets are available for specific purposes,
such as ground state optimization, excited state calculations etc. 
In an earlier work (see chapter \ref{chap:main_smallboron}), we have reported a systematic basis set dependence of photoabsorption
of boron cluster \cite{smallboron}. Similarly, here we have checked
the dependence of photoabsorption spectrum of aluminum dimer on basis
sets used \cite{emsl_bas1,emsl_bas2} as shown in Fig. \ref{fig:basis-study}.
The 6-311 type Gaussian contracted basis sets are known to be good
for ground state calculations. The correlation consistent (CC) basis
sets, namely, CC-polarized valence double-zeta and CC-polarized valence
triple zeta (cc-pVTZ) give a good description of excited states of various systems.
The latter is found to be more sophisticated in describing the high
energy excitations, which were also confirmed using results of an
independent TDDFT calculation \cite{basis-set-core}. Therefore, in this work,  we have used 
the cc-pVTZ basis set for the optical absorption calculations.

\subsubsection{Orbital truncation scheme}

\begin{figure}
\centering
\includegraphics[width=8cm]{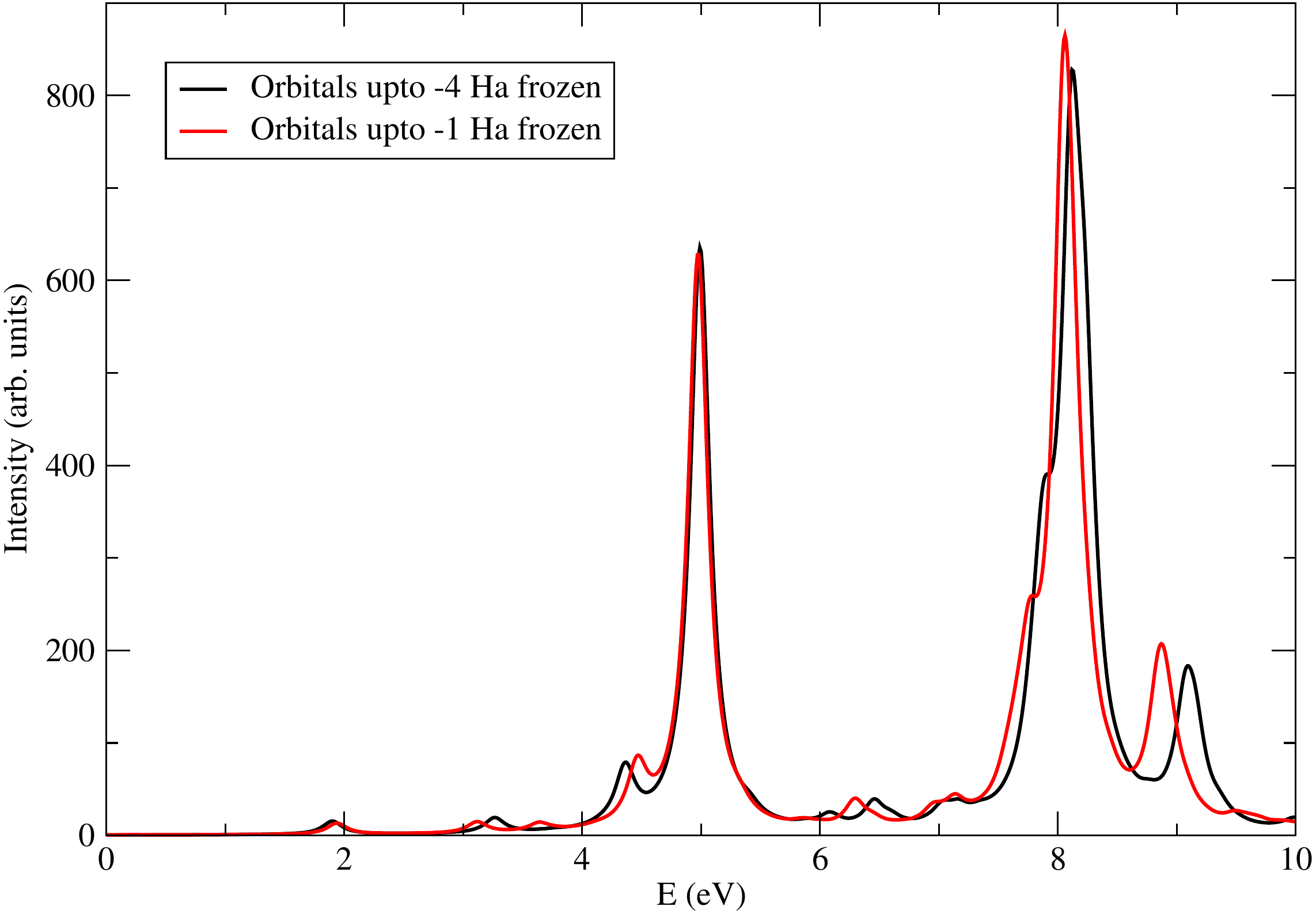} \vspace{0.2cm}
\caption{\label{fig:Core-Study}The effect of freezing the core
orbitals of aluminum atoms on optical absorption spectrum of Al$_{2}$.
It renders little effect on optical absorption spectrum, with significant
reduction in the computational cost.}
\end{figure}

With respect to the total number of orbitals N in the system, the
computational time in configuration interaction calculations scales
as $\thickapprox$ N$^{6}$. Therefore, such calculations become intractable
for moderately sized systems, such as those considered here. So, in
order to ease those calculations, the lowest lying molecular orbitals
are constrained to be doubly occupied in all the configurations, implying
that no optical excitation can occur from those orbitals. It reduces
the size of the CI Hamiltonian matrix drastically. In fact, this approach
is recommended in quantum chemical calculations, because the basis
sets used are not optimized to incorporate the correlations in core
electrons \cite{szabo_book}. The effect of this approximation on the spectrum is as
shown in Fig. \ref{fig:Core-Study}. Since, calculations with all
electrons in active orbitals were unfeasible, we have frozen occupied
orbitals upto -4 Hartree of energy for the purpose of demonstration.
The effect of freezing the core is negligibly small in the low energy
regime, but shows disagreement in the higher energy range. However,
for very high energy excitations, photodissociation may occur, hence
absorption spectra at those energies will cease to have meaning. Thus,
the advantage of freezing the core subdues this issue. Therefore,
in all the calculations, we have frozen the chemical core from optical
excitations.

\begin{figure}
\centering
\includegraphics[width=8cm]{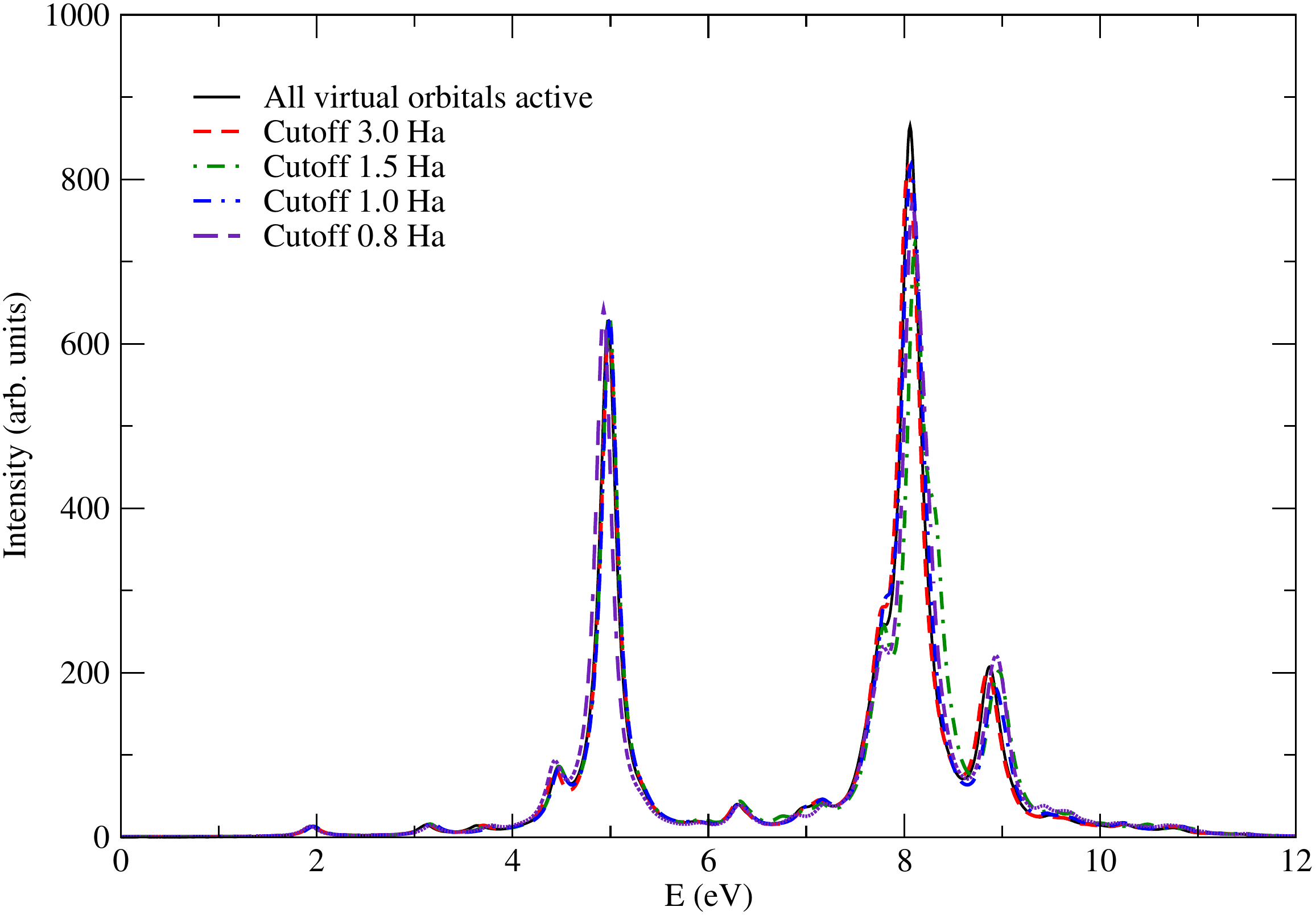} \vspace{0.2cm}
\caption{\label{fig:nref-study}The effect of the number of
active orbitals (N$_{act}$) on the optical absorption spectrum of
Al$_{2}$. Until N$_{act}$=46, the optical spectrum does not exhibit
any significant change. It corresponds to 1.0 Hartree ($\approx27.2$
eV) virtual orbital energy. }
\end{figure}

Not only occupied, but high energy virtual (unoccupied) orbitals can
also be removed from the calculations to make them tractable. In this
case the high lying orbitals are constrained to be unoccupied in all
the configurations. This move is justifiable, because it is unlikely
that electrons would prefer partial filling of high energy orbitals
in an attempt to avoid other electrons. However, this will only be
applicable if the orbitals are sufficiently high in energy. Fig. \ref{fig:nref-study}
shows the effect of removing orbitals having more than the specified
energy. From the figure it is clear that photoabsorption spectra exhibits
no difference at all upto 1 Hartree cutoff on virtual orbitals. Below
0.8 Ha cutoff, the spectra start deviating from each other. Hence,
we have ignored the virtual orbitals having energy more than 1 Ha. 


\subsubsection{Size of the CI expansion}

\begin{table*}
\small
\centering
\begin{threeparttable}[b]
  \caption{The average number of reference configurations (N$_{ref}$), and average
number of total configurations (N$_{total}$) involved in MRSDCI calculations,
ground state (GS) energies (in Hartree) at the MRSDCI level, relative
energies and correlation energies (in eV) of various isomers of aluminum
clusters.\label{tab:energies}}
\par

{\begin{tabular}{c>{\centering\arraybackslash}m{2cm}cccc>{\centering\arraybackslash}m{1.5cm}}
\hline
Cluster  & Isomer  & N$_{ref}$  & N$_{total}$  & GS energy & Relative & \multicolumn{1}{c}{Correlation energy\tnote{2}}
\tabularnewline
 &  &  &  & (Ha) & energy (eV) & \multicolumn{1}{c}{per atom(eV)}\tabularnewline
\hline 
Al$_{2}$  & Linear  & 40 & 445716 & -483.9063281 & 0.00 & 1.69\tabularnewline
 &  &  &  &  &  & \tabularnewline
Al$_{3}$  & Equilateral triangular  & 40 & 1917948 & -725.9053663 & 0.00 & 2.38\tabularnewline
 & Isosceles triangular & 22 & 1786700 & -725.8748996 & 0.83 & 2.36\tabularnewline
 & Linear & 18 & 1627016 & -725.8370397 & 1.85 & 2.16\tabularnewline
 &  &  &  &  &  & \tabularnewline
Al$_{4}$  & Rhombus & 13 & 3460368 & -967.8665897 & 0.00 & 1.82\tabularnewline
 & Square & 21 & 1940116 & -967.8258673 & 1.11 & 1.80\tabularnewline
 &  &  &  &  &  & \tabularnewline
Al$_{5}$  & Pentagonal  & 7 & 3569914 & -1209.8114803 & 0.00 & 1.73\tabularnewline
 & Pyramidal  & 8 & 3825182 & -1209.7836568 & 0.76 & 1.77\tabularnewline
\hline
\end{tabular}}
  \begin{tablenotes}
    \item[2] The difference in Hartree-Fock energy and MRSDCI correlated energy of the ground state.
  \end{tablenotes}
\end{threeparttable}

\end{table*}

In the multi-reference CI method, the size of
the Hamiltonian matrix increases exponentially with the number of molecular
orbitals in the system. Also, accurate correlated results can only
be obtained if sufficient number of reference configurations are included
in the calculations. In our calculations, we have included those configurations
which are dominant in the wavefunctions of excited states for a given
absorption peak. Also, for ground state calculations, we included
configurations until the total energy converges within a pre-defined
tolerance. Table \ref{tab:energies} shows the average number of reference
configurations and average number of total configurations involved
in the CI calculations of various isomers. For a given isomer, the
average is calculated across different irreducible representations
needed in these symmetry adapted calculations of the ground and various
excited states. For the simplest cluster, the total configurations
are about half a million and for the biggest cluster considered here,
it is around four million for each symmetry subspace of Al$_{5}$.
The superiority of our calculations can also be judged from the correlation
energy defined here (\emph{cf.} Table \ref{tab:energies}), which
is the difference in the total energy of a system at the MRSDCI level
and the Hartree-Fock level. The correlation energy per atom seems
to be quite high for all the clusters, making our calculations stand
out among other electronic structure calculations, especially single
reference DFT based calculations.

\subsection{MRSDCI photoabsorption spectra of various clusters}

In this section, we describe the photoabsorption spectra of various
isomers of the aluminum clusters studied. Graphical presentation of
molecular orbitals involved are also given in each subsection below.

\subsubsection{Al$_{2}$}

Aluminum dimer is the most widely studied cluster of aluminum. It
has now been confirmed that the Al$_{2}$ (\emph{cf}. Fig. \ref{fig:geometry}\subref{subfig:subfig-al2})
has $^{3}\Pi_{u}$ ground state. The bond length obtained using geometry
optimization at CCSD level was 2.72 $\textrm{\AA}$ with D$_{\infty h}$
point group symmetry. This is in very good agreement with available
data, such as Martinez \emph{et al}. obtained 2.73 $\textrm{\AA}$
as dimer length using all electron calculations \cite{martinez_all_vs_core},
2.71 $\textrm{\AA}$ \cite{jones_struct_bind_prl} and 2.75 $\textrm{\AA}$ 
\cite{langhoff_3piu} as bond lengths using DFT and configuration
interaction methods, and 2.86 $\textrm{\AA}$ obtained using DFT with
generalized gradient approximation \cite{rao_jena_ele_struct_al}.
The experimental bond length of aluminum dimer is 2.70 $\textrm{\AA}$ \cite{bondybey_expt_3piu_dimer}.
Another metastable state of the dimer exists with $^{3}\Sigma_{g}^{-}$
electronic state, and 2.48 $\textrm{\AA}$ in bond length.


\begin{figure*}
\centering
\subfloat[]{\includegraphics[width=8.3cm]{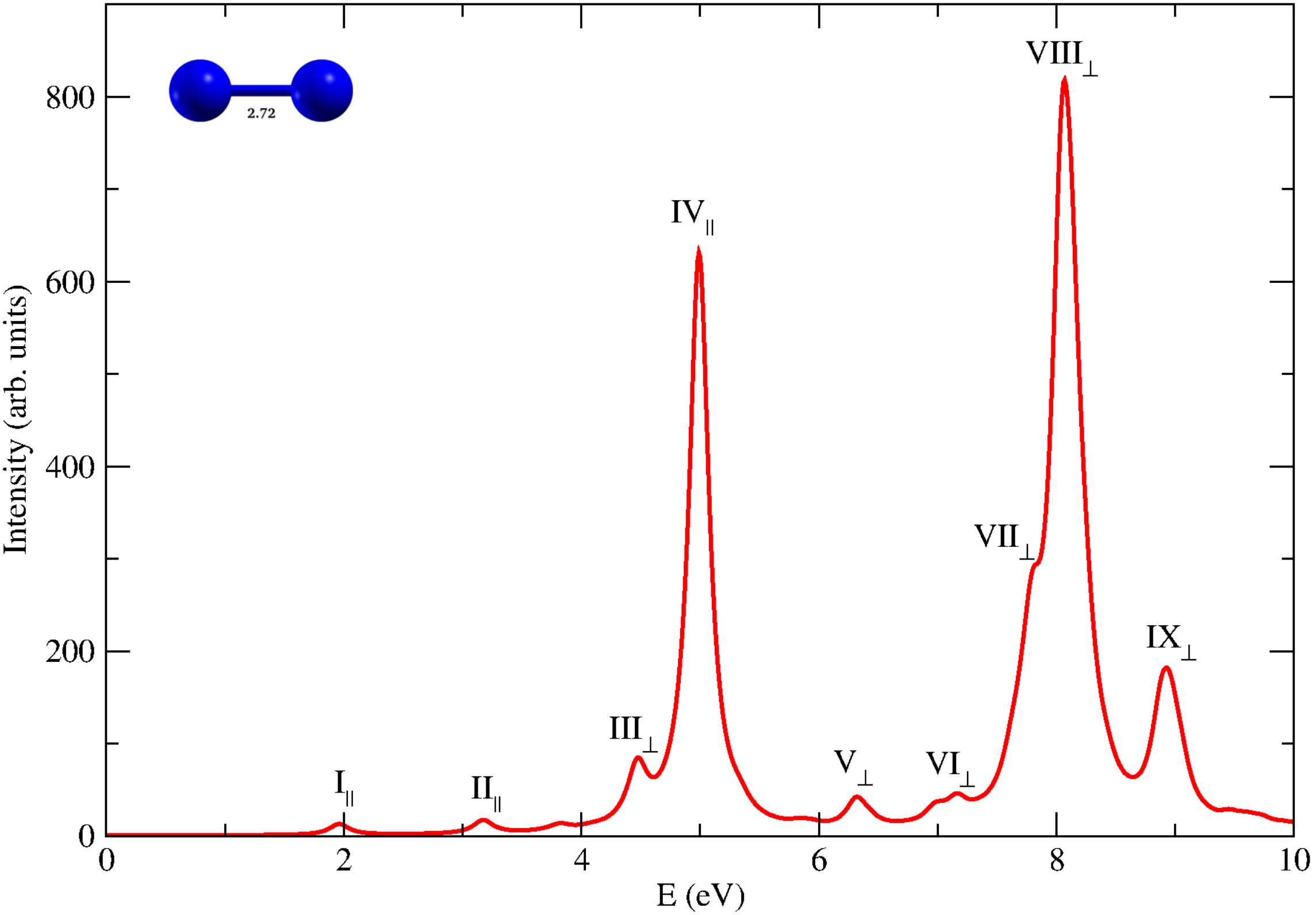}\label{subfig:al2-lin-plot}} \hfill
\subfloat[]{\includegraphics[width=6.cm]{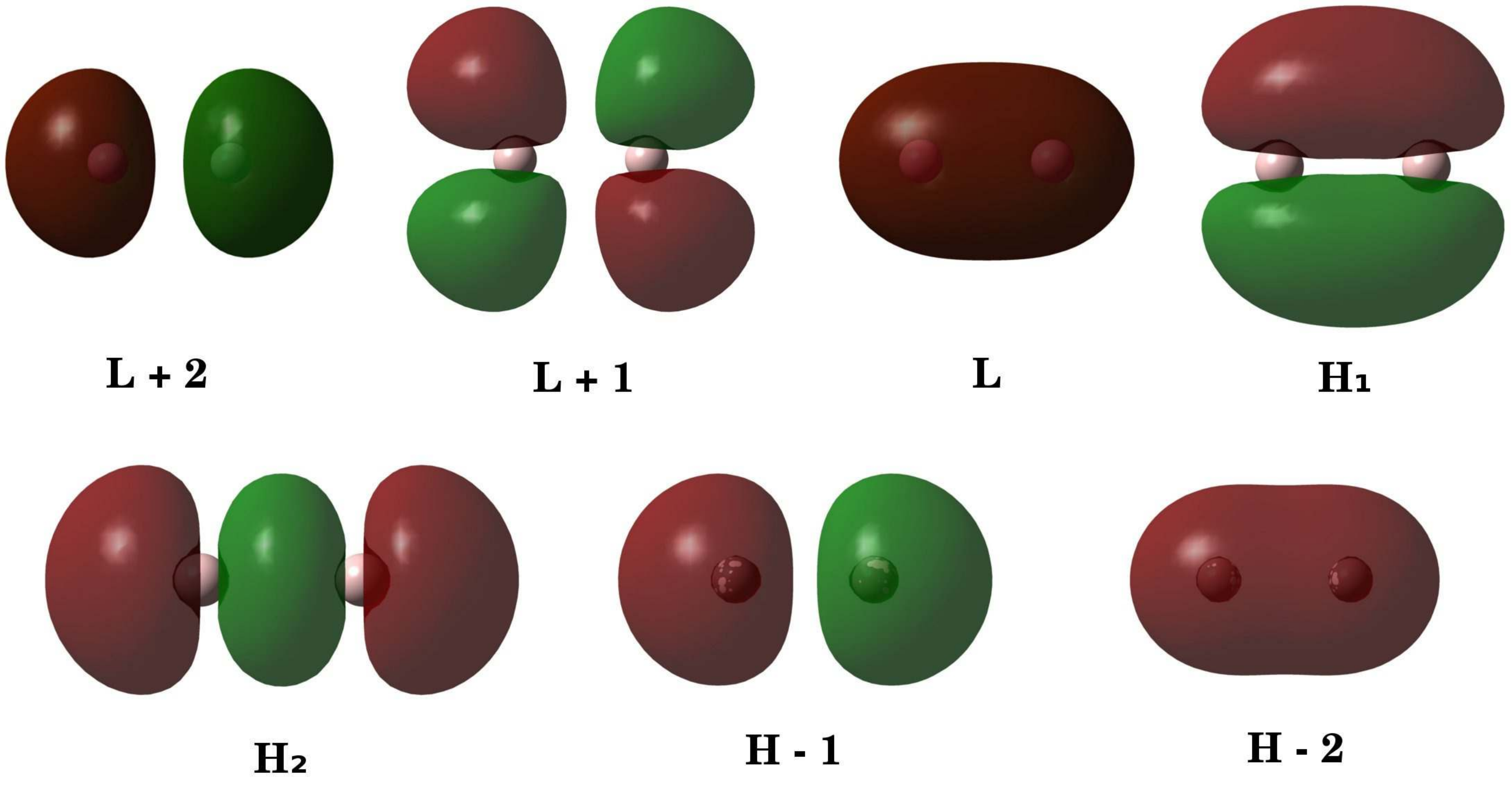}\label{subfig:al2-lin-mo}}
\caption{\label{fig:plot-al2-linear} \protect\subref{subfig:al2-lin-plot} The linear optical absorption
spectrum of Al$_{2}$, calculated using the MRSDCI approach. The peaks
corresponding to the light polarized along the molecular axis are
labeled with the subscript $\parallel$, while those polarized perpendicular
to it are denoted by the subscript $\perp$. For plotting the spectrum,
a uniform linewidth of 0.1 eV was used. \protect\subref{subfig:al2-lin-mo} Molecular orbitals of aluminum dimer. $H$ and $L$
stands for HOMO and LUMO respectively, and $H_{1}$ and $H_{2}$ are singly occupied degenerate molecular orbitals.}
\end{figure*}

The many-particle wavefunction of Al$_{2}$ consists of two degenerate
singly occupied molecular orbitals (to be denoted by $H_{1}$ and
$H_{2}$, henceforth), because it is a spin triplet system. Similarly,
the configurations involving excitations from occupied molecular orbitals
to the unoccupied orbitals, form excited state wavefunctions. The
computed photoabsorption spectrum of Al$_{2}$, as shown in Fig. \ref{fig:plot-al2-linear}\subref{subfig:al2-lin-plot},
is characterized by weaker absorption at lower energies and couple
of intense peaks at higher energies. The many-particle wavefunctions
of excited states contributing to the peaks are presented in Table \ref{Tab:table_al2_lin}.  

The spectrum starts with a small absorption
peak at around 2 eV, characterized by $H_{2}\rightarrow L+1$ and
light polarized along the direction of axis of the dimer. It is followed
by a couple of small intensity peaks, until a dominant absorption
is seen at 5 eV. This is characterized by $H_{1}\rightarrow L+3$.
Another dominant peak is observed at 8 eV having $H-2\rightarrow L$
as dominant configuration, with absorption due to light polarized
perpendicular to the axis of the dimer. 

Our spectrum differs from the one obtained with the time-dependent
local density approximation (TDLDA) method \cite{kanhere_prb_optical_al} 
in both the intensity and the number of peaks. 
However, we agree with TDLDA\cite{kanhere_prb_optical_al} in 
predicting two major peaks at 5 eV and 8 eV. 
Unlike our calculations, the number of peaks is much more in TDLDA results and the spectrum is almost
continuous. Minor peaks at 3.2 eV and 6.3 eV are also observed in the TDLDA spectrum
of dimer. The latter one is, contrary to TDLDA spectrum, found
to be a small peak in our calculations.


\subsubsection{Al$_{3}$}

Among the possible isomers of aluminum cluster Al$_{3}$, the equilateral
triangular isomer is found to be the most stable. We have considered
three isomers of Al$_{3}$, namely, equilateral triangle, isosceles
triangle, and a linear chain. The most stable isomer has $D{}_{3h}$ point
group symmetry, and $^{2}A_{1}^{'}$ electronic state. The optimized
bond length 2.57 $\textrm{\AA}$, is in good agreement with reported
theoretical values 2.61 $\textrm{\AA}$ \cite{rao_jena_ele_struct_al},
2.62 $\textrm{\AA}$ \cite{upton_elec_struct_small_al}, 2.56 $\textrm{\AA}$ \cite{martinez_all_vs_core},
and 2.52 $\textrm{\AA}$ \cite{sachdev_dft_al_small,truhler_theory_validation_jpcb}.
The doublet ground state is also confirmed with the results of magnetic
deflection experiments \cite{cox-al3-experiment}.

The next isomer, which lies 0.83 eV higher in energy, is the isosceles triangular
isomer. The optimized geometry has 2.59 $\textrm{\AA}$, 2.59 $\textrm{\AA}$
and 2.99 $\textrm{\AA}$ as sides of triangle, with a quartet ground state ($^{4}A_{2}$). Our results are in agreement
with other theoretical results \cite{upton_elec_struct_small_al,martinez_all_vs_core,jones_simul_anneal_al}.

Linear Al$_{3}$ isomer again with quartet multiplicity is the next
low-lying isomer. The optimized bond length is 2.62 $\textrm{\AA}$.
This is in good agreement with few available reports \cite{martinez_all_vs_core,jones_simul_anneal_al,sachdev_dft_al_small}.

\begin{figure*}
\centering
\subfloat[]{\includegraphics[width=8.3cm]{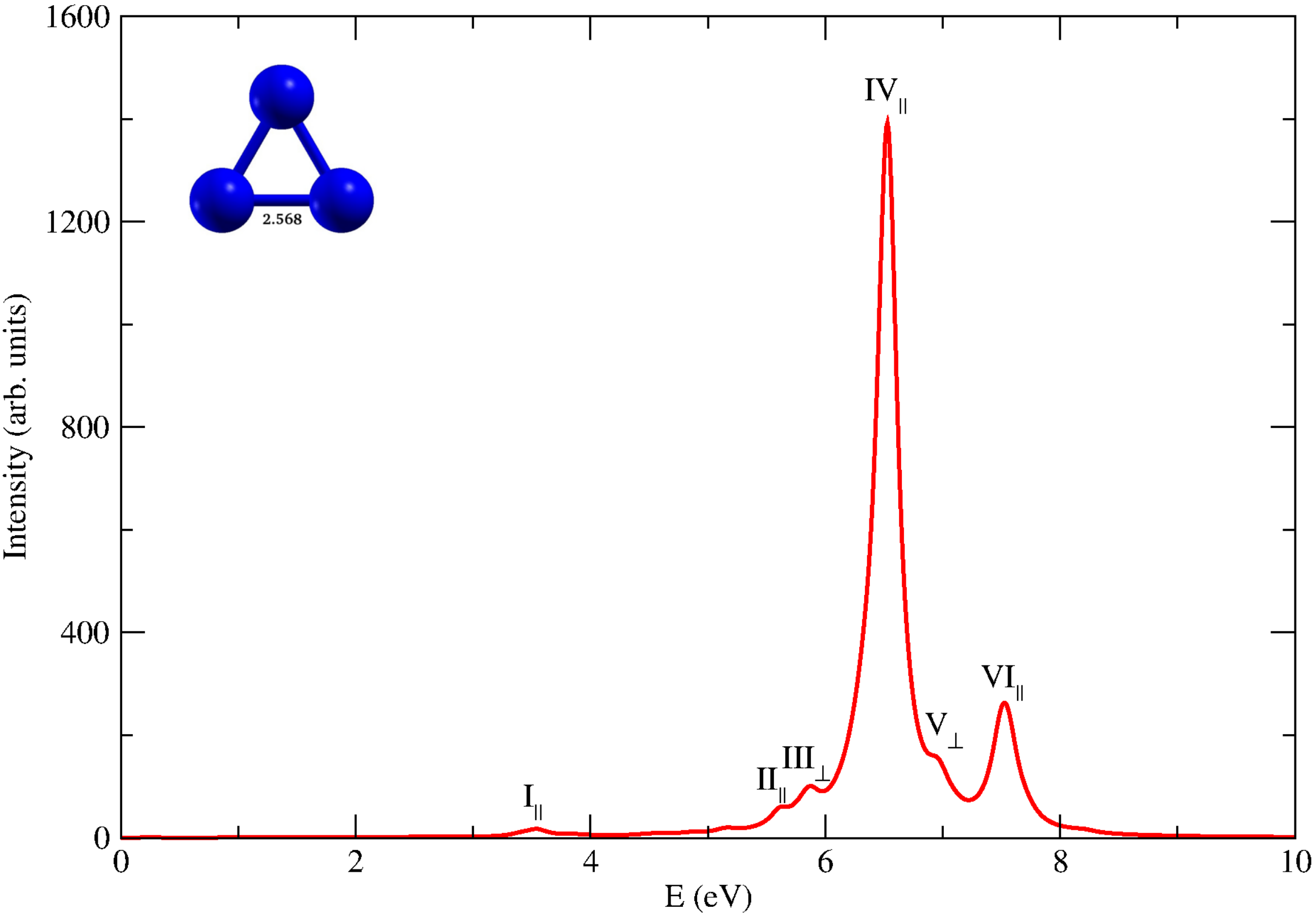} \label{subfig:plot-al3-equil}} \hfill
\subfloat[]{\includegraphics[width=6.5cm]{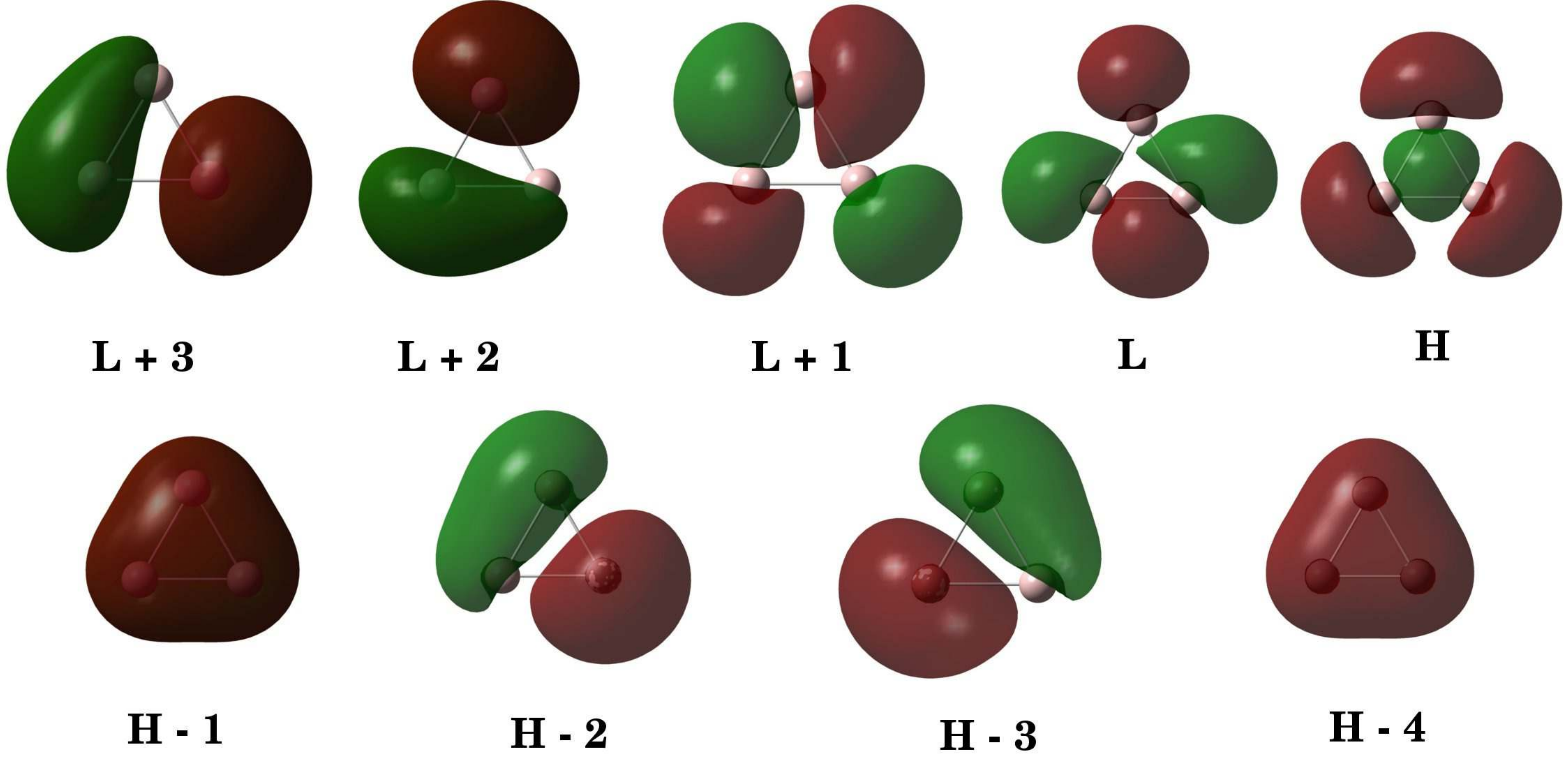} \label{subfig:al3-equil-mo}}
\caption{\label{fig:plot-al3-equil} \protect\subref{subfig:plot-al3-equil} The linear optical absorption 
spectrum of Al$_{3}$ equilateral triangle isomer, calculated using
the MRSDCI approach. The peaks corresponding to the light polarized
along the molecular plane are labeled with the subscript $\parallel$,
while those polarized perpendicular to it are denoted by the subscript
$\perp$. For plotting the spectrum, a uniform linewidth of 0.1 eV
was used. \protect\subref{subfig:al3-equil-mo} Molecular orbitals of equilateral triangular aluminum
trimer. $H$ and $L$ stands for HOMO and LUMO respectively. 
($H-2$, $H-3$), ($L$, $L+1$) and ($L+2$, $L+3$) are degenerate pairs.}
\end{figure*}

\begin{figure*}
\centering
\subfloat[]{\includegraphics[width=8.3cm]{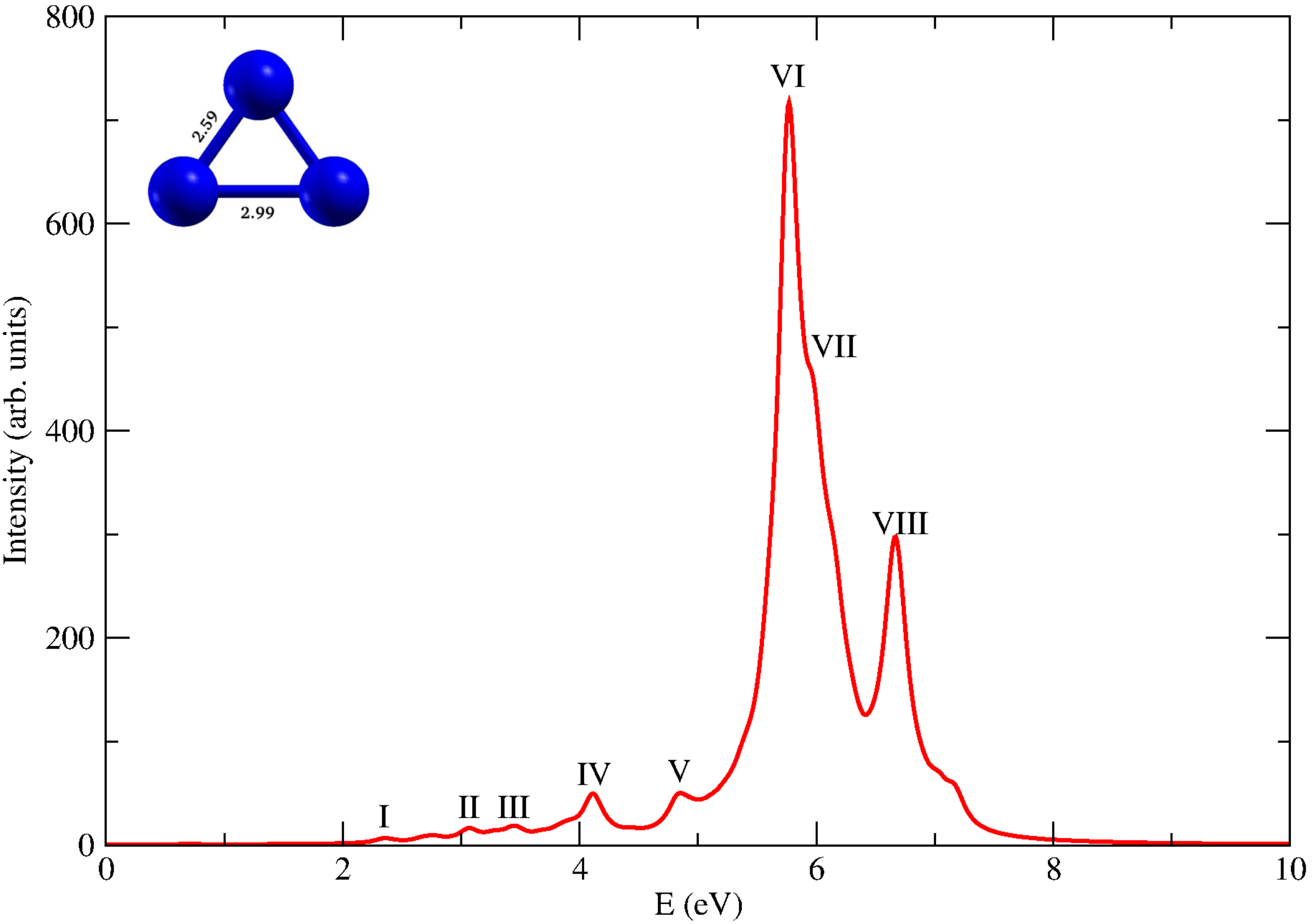} \label{subfig:plot-al3-iso}} \hfill
\subfloat[]{\includegraphics[width=6.5cm]{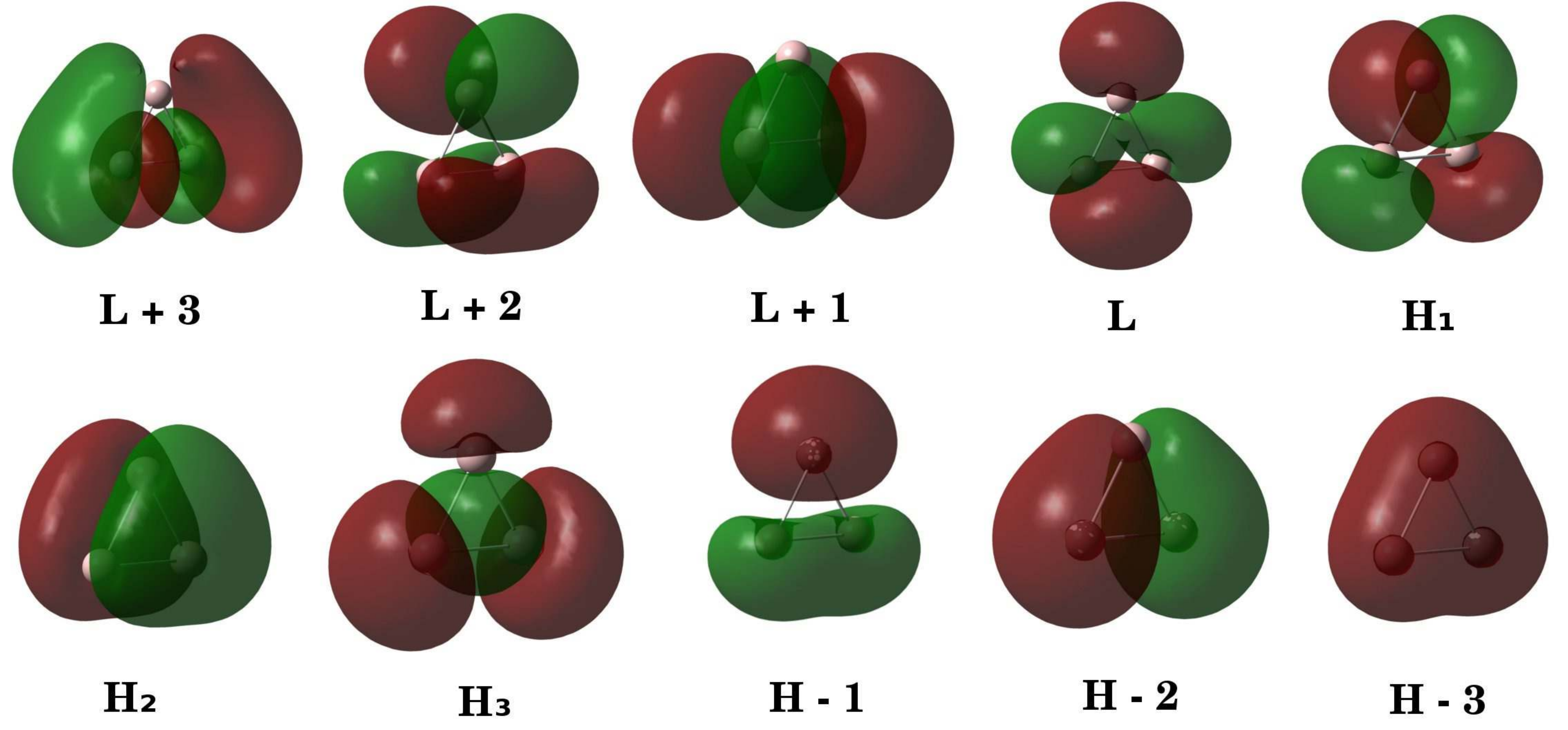} \label{subfig:mo-al3-iso}}
\caption{\label{fig:plot-al3-iso} \protect\subref{subfig:plot-al3-iso} The linear optical absorption
spectrum of Al$_{3}$ isosceles triangle isomer, calculated using
the MRSDCI approach. All peaks labeled above correspond to the light polarized
along the molecular plane. For plotting the spectrum, a uniform linewidth
of 0.1 eV was used. \protect\subref{subfig:mo-al3-iso} Molecular orbitals of isosceles triangular aluminum
trimer. $H$ and $L$ stands for HOMO and LUMO respectively, and $H_{1}$,
$H_{2}$, and $H_{3}$ are singly occupied molecular orbitals.}
\end{figure*}

\begin{figure*}
\centering
\subfloat[]{\includegraphics[width=8.3cm]{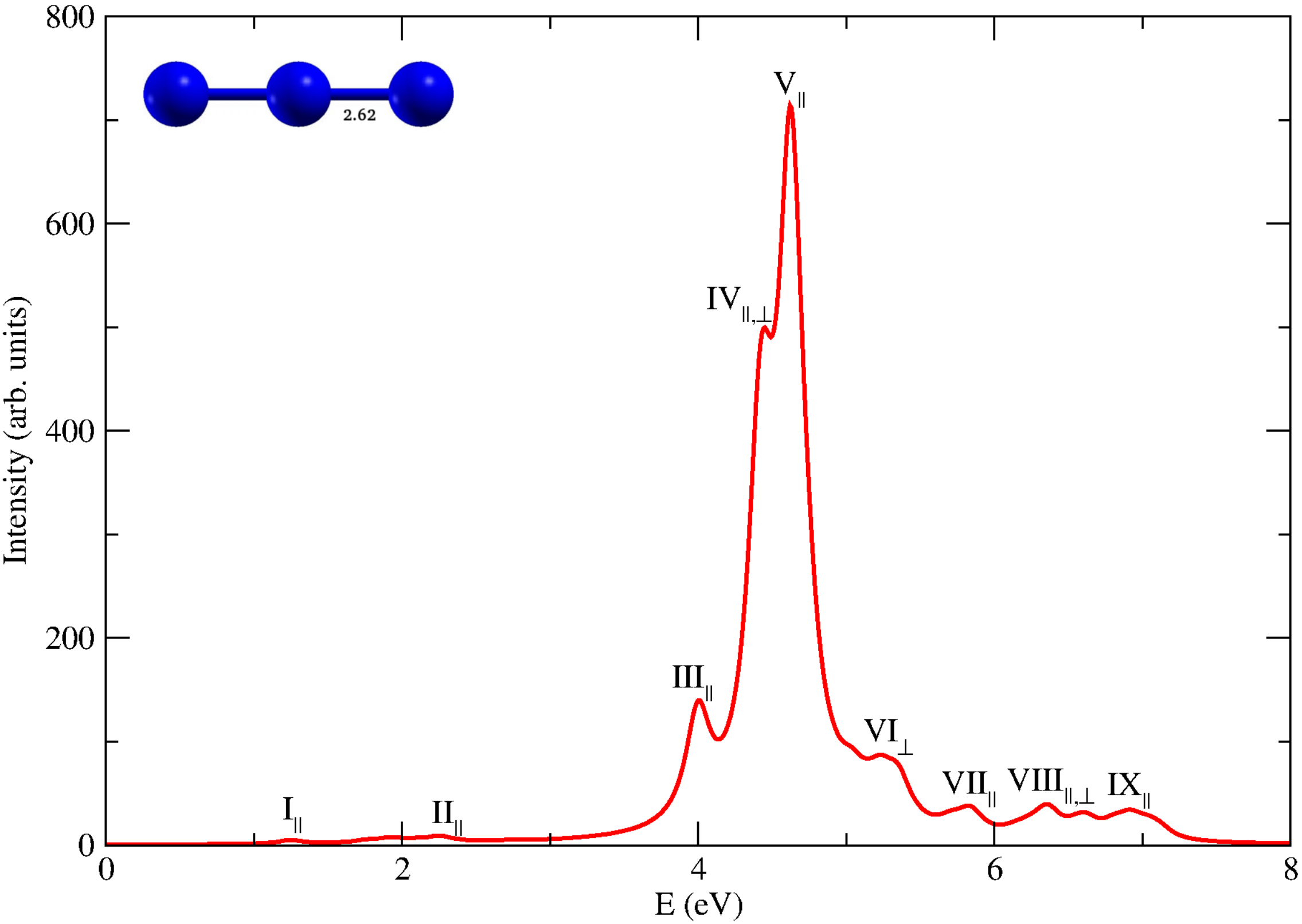} \label{subfig:plot-al3-lin} } \hfill
\subfloat[]{\includegraphics[width=6.5cm]{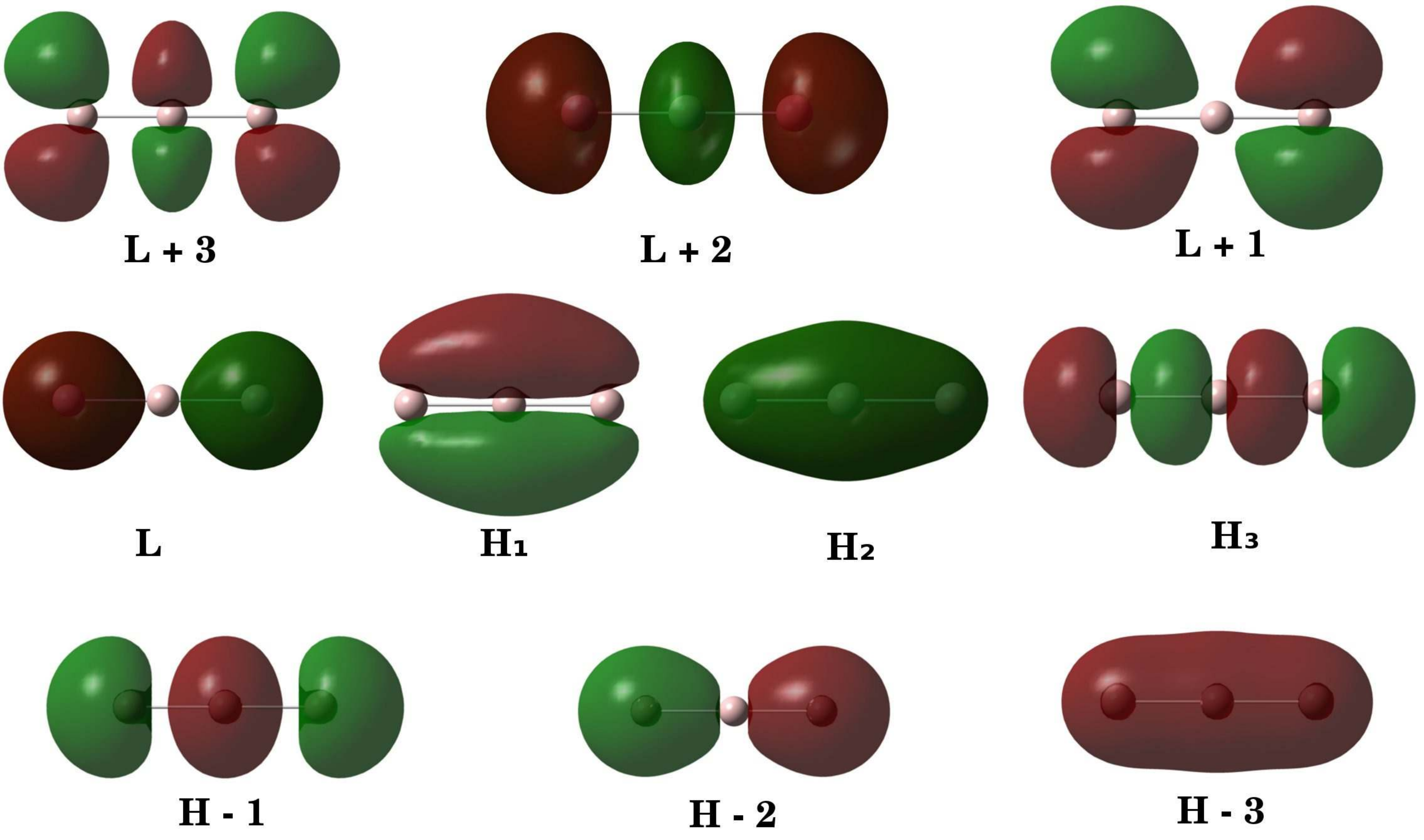} \label{subfig:mo-al3-lin}}
\caption{\label{fig:plot-al3-lin} \protect\subref{subfig:plot-al3-lin} The linear optical absorption
spectrum of Al$_{3}$ linear isomer, calculated using the MRSDCI approach.
The peaks corresponding to the light polarized along the molecular
axis are labeled with the subscript $\parallel$, while those polarized
perpendicular to it are denoted by the subscript $\perp$. For plotting
the spectrum, a uniform linewidth of 0.1 eV was used. \protect\subref{subfig:mo-al3-lin} Molecular 
orbitals of linear aluminum trimer. $H$ and $L$ stands for HOMO and LUMO respectively, and $H_{1}$, $H_{2}$,
and $H_{3}$ are singly occupied molecular orbitals.}
\end{figure*}

The photoabsorption spectra of these isomers are presented in Figs.
\ref{fig:plot-al3-equil}\subref{subfig:plot-al3-equil}, \ref{fig:plot-al3-iso}\subref{subfig:plot-al3-iso} and \ref{fig:plot-al3-lin}\subref{subfig:plot-al3-lin}.
The corresponding many-body wavefunctions of excited states corresponding
to various peaks are presented in Table \ref{Tab:table_al3_equil_tri}, \ref{Tab:table_al3_iso_tri} and \ref{Tab:table_al3_lin_tri} respectively.
In the equilateral triangular isomer, most of the intensity is concentrated
at higher energies. The same is true for the isosceles triangular
isomer. However, the spectrum of isosceles triangular isomer appears
slightly red shifted with respect to the equilateral counterpart.
Along with this shift, there appears a split pair of peaks at 5.8
eV. This splitting of oscillator strengths might be due to distortion
accompanied by symmetry breaking. The absorption spectrum of linear
isomer is altogether different with bulk of the oscillator strength
carried by peaks in the range 4 -- 5 eV, and, due to the polarization
of light absorbed parallel to the axis of the trimer. 

The optical absorption spectrum of equilateral triangular isomer consists
of very feeble low energy peaks at 3.5 eV, 5.6 eV and 5.8 eV characterized
by $H-3\rightarrow L+5$, a double excitation $H-2\rightarrow L+5;H-1\rightarrow L+5$,
and $H-3\rightarrow L+2$ respectively. The latter peak is due to
the light polarized perpendicular to the plane of the isomer. It is
followed by an intense peak at around 6.5 eV with dominant contribution
from $H\rightarrow L+6$ and $H\rightarrow L+4$ configurations. A
semi-major peak is observed at 7.5 eV characterized mainly due to
double excitations.


Two major peaks at 6.5 eV and 7.5 eV in the spectrum of Al$_{3}$ equilateral isomer, obtained in our calculations
are also found in the spectrum of TDLDA calculations, with the difference
that the latter does not have a smaller intensity in TDLDA \cite{kanhere_prb_optical_al}. Other
major peaks obtained by Deshpande \emph{et al.} \cite{kanhere_prb_optical_al}
in the spectrum of aluminum trimer are not observed, or have very small
intensity in our results. 

As compared to the equilateral triangle spectra, the isosceles triangular
isomer exhibit several small intensity peaks (\emph{cf}. Fig. \ref{fig:plot-al3-iso})
in the low energy regime. The majority of contribution to peaks of
this spectrum comes from in-plane polarized transitions, with
negligible contribution from transverse polarized light. The spectrum
starts with a feeble peak at 2.4 eV with contribution from doubly-excited
configuration $H\rightarrow L+1;H-2\rightarrow L+2$. One of the dominant
contribution to the oscillator strength comes from two closely-lying
peaks at 5.8 eV. The wavefunctions of excited states corresponding
to this peak show a strong mixing of doubly-excited configurations,
such as $H-3\rightarrow L+1;H-2\rightarrow L$ and $H-2\rightarrow L+1;H-4\rightarrow L$.
The peak at 6.7 eV shows absorption mainly due to $H\rightarrow L+10$.

Linear trimer of aluminum cluster also shows low activity in the low
energy range. Very feeble peaks are observed at 1.2 eV and 2.3 eV,
both characterized by $H-3\rightarrow H-2$. This configuration also
contributes to the semi-major peak at 4 eV along with $H-4\rightarrow H$.
Two closely lying peaks at 4.3 eV and 4.6 eV carry the bulk of the
oscillator strength. Major contribution to the former comes from $H-1\rightarrow L+2$
along with $H-3\rightarrow H-2$ being dominant in both the peaks.
Again, as expected, the absorption due to light polarized along the
trimer contributes substantially to the spectrum.

It is obvious from the spectra presented above that the location of the most 
intense absorption is quite sensitive to the structure, and thus can be 
used to distinguish between the three isomers.

\subsubsection{Al$_{4}$}

\begin{figure*}
\centering
\subfloat[]{\includegraphics[width=8.3cm]{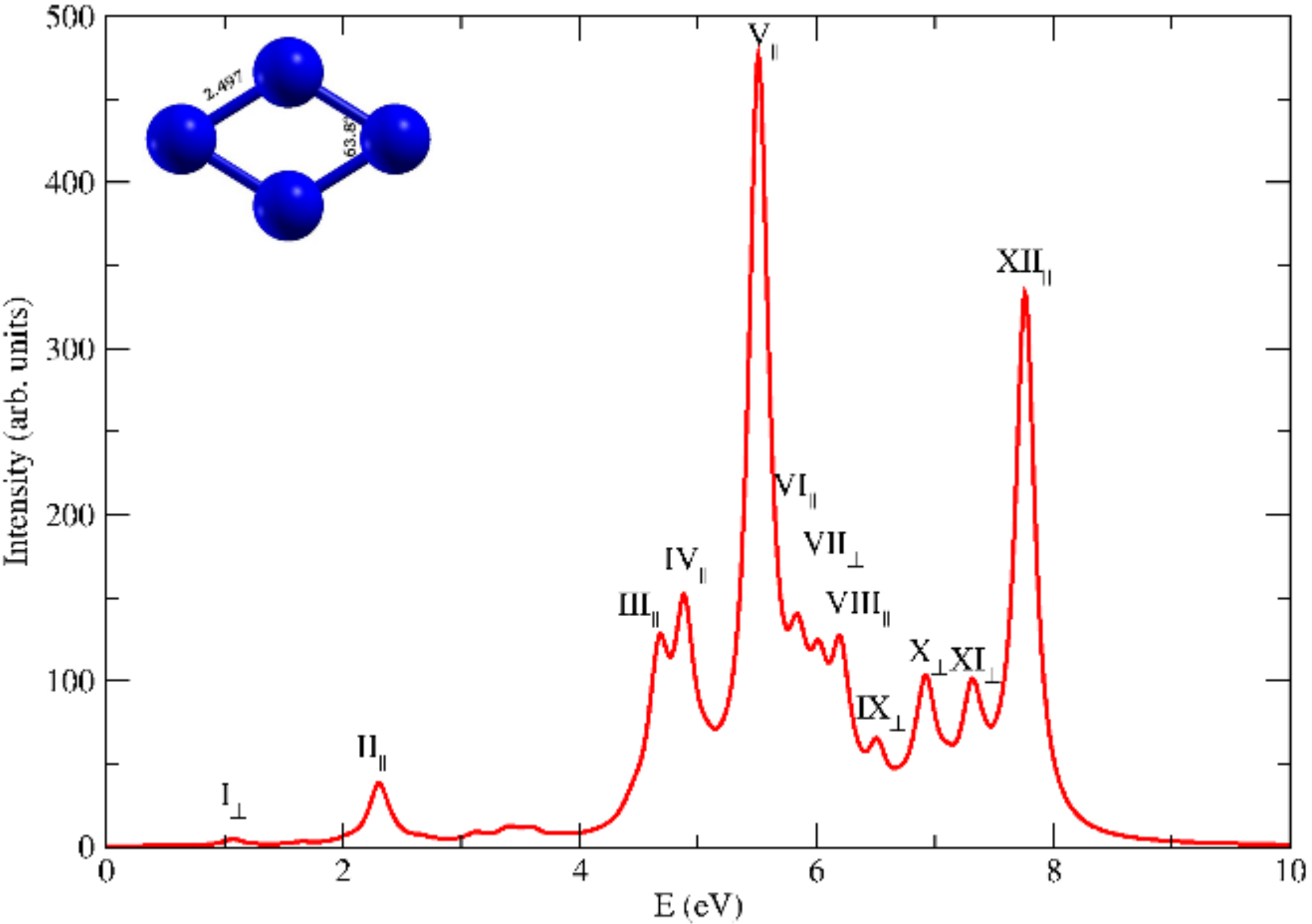} \label{subfig:plot-al4-rho} } \hfill
\subfloat[]{\includegraphics[width=6.5cm]{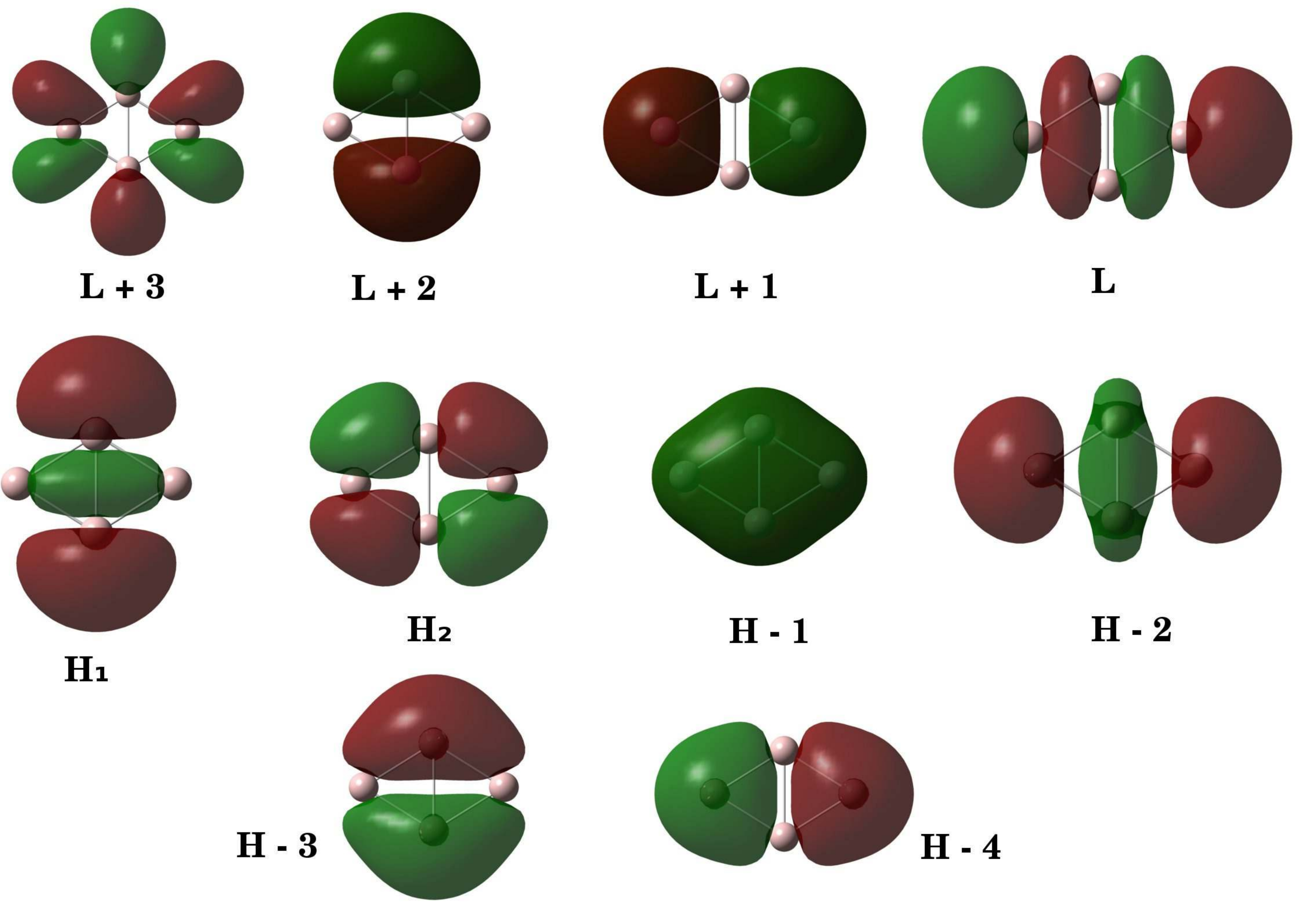} \label{subfig:mo-al4-rho} }
\caption{\label{fig:plot-al4-rho} \protect\subref{subfig:plot-al4-rho} The linear optical absorption
spectrum of rhombus Al$_{4}$, calculated using the MRSDCI approach.
The peaks corresponding to the light polarized along the molecular
axis are labeled with the subscript $\parallel$, while those polarized
perpendicular to it are denoted by the subscript $\perp$. For plotting
the spectrum, a uniform linewidth of 0.1 eV was used.\protect\subref{subfig:mo-al4-rho}  Molecular 
orbitals of rhombus-shaped aluminum tetramer. $H$ and $L$ stands for HOMO and LUMO respectively, and $H_{1}$
and $H_{2}$ are singly occupied molecular orbitals.}
\end{figure*}


Tetramer of aluminum cluster has many low lying isomers due to its
flat potential energy curves. Among them, rhombus structure is the
most stable with $^{3}B_{2g}$ electronic ground state. Our optimized
bond length for rhombus structure is 2.50 $\textrm{\AA}$ and $63.8^\circ$
as the acute angle. This is to be compared with corresponding reported
values of 2.56 $\textrm{\AA}$ and $69.3^\circ$ reported by Martinez
\emph{et al.} \cite{martinez_all_vs_core}, 2.51 $\textrm{\AA}$
and $56.5^\circ$ computed by Jones \cite{jones_struct_bind_prl},
2.55 $\textrm{\AA}$ and $67.6^\circ$ obtained by Schultz \emph{et
al} \cite{truhler_theory_validation_jpcb}. We note that bond lengths
are in good agreement but bond angles appear to vary a bit.

The other isomer studied here is an almost square shaped tetramer
with optimized bond length of 2.69 $\textrm{\AA}$. The electronic
ground state of this $D_{4h}$ symmetric cluster is\textbf{ $^{3}B{}_{3u}$}\emph{.}
This optimized geometry is in accord with 2.69 $\textrm{\AA}$ reported
by Martinez \emph{et al}. \cite{martinez_all_vs_core}, however,
it is somewhat bigger than 2.57 $\textrm{\AA}$ calculated by Yang
\emph{et al.} \cite{sachdev_dft_al_small} and 2.61 $\textrm{\AA}$
obtained by Jones \cite{jones_simul_anneal_al}.

For planar clusters, like rhombus and square shaped Al$_{4}$, two
types of optical absorptions are possible: (a) planar -- those
polarized in the plane of the cluster, and (b) transverse -- the ones
polarized perpendicular to that plane. The many-particle wavefunctions
of excited states contributing to the peaks are presented in Table \ref{Tab:table_al4_rho} and \ref{Tab:table_al4_sqr}.
 The onset of optical absorption in rhombus isomer occurs at around 1 eV with transversely polarized
absorption characterized by $H_{1}\rightarrow L+1$. It is followed
by an in-plane polarized absorption peak at 2.3 eV with dominant contribution
from $H-2\rightarrow H_{1}$. Several closely lying peaks are observed
in a small energy range of 4.5 -- 8 eV. Peaks split from each other
are seen in this range confirming that after shell closure, in perturbed
droplet model, Jahn Teller distortion causes symmetry breaking usually
associated with split absorption peaks. The most intense peak is observed
at 5.5 eV characterized by $H-3\rightarrow L+4$.

\begin{figure*}
\centering
\subfloat[]{\includegraphics[width=8.3cm]{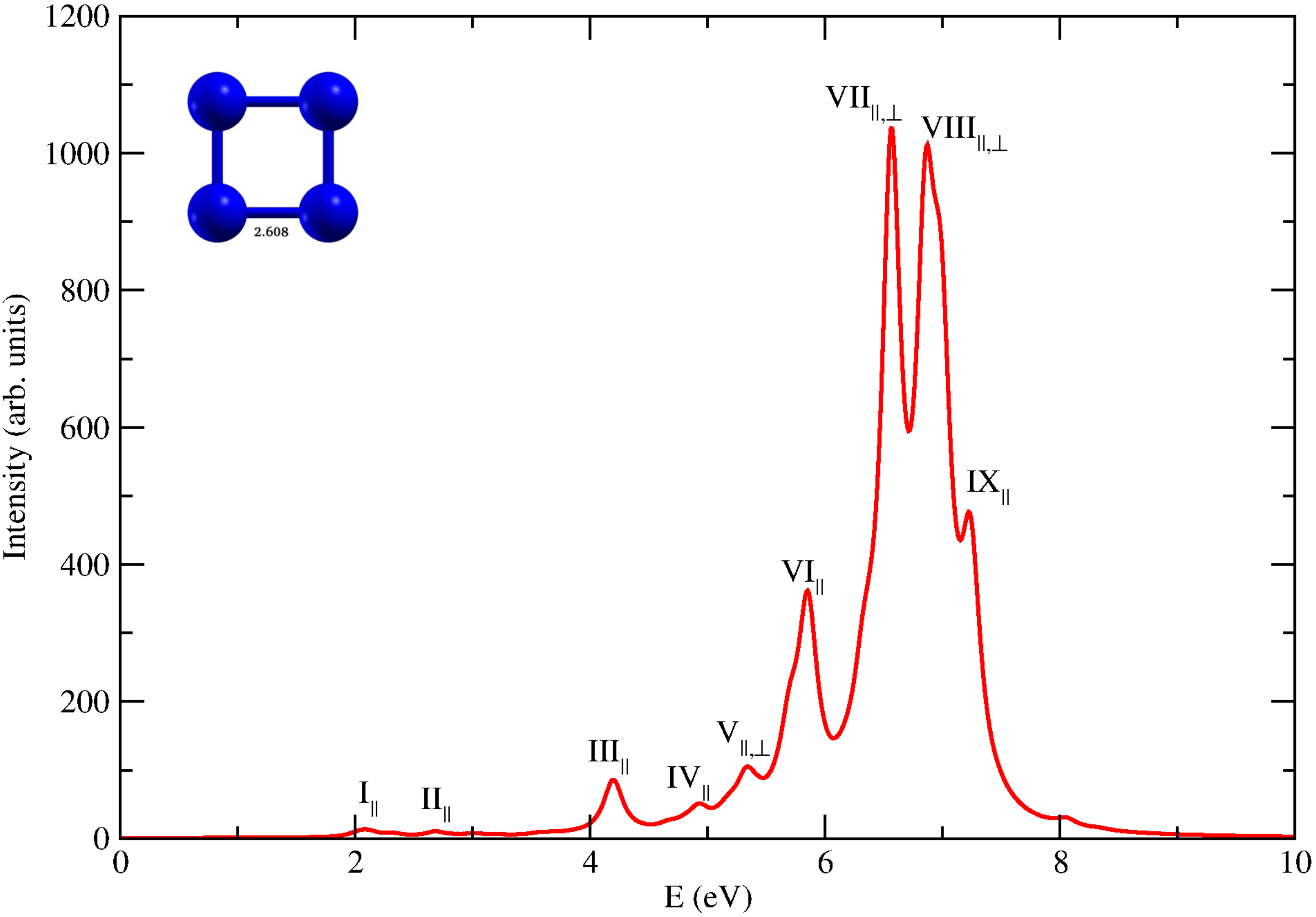} \label{subfig:plot-al4-sqr} } \hfill
\subfloat[]{\includegraphics[width=6.5cm]{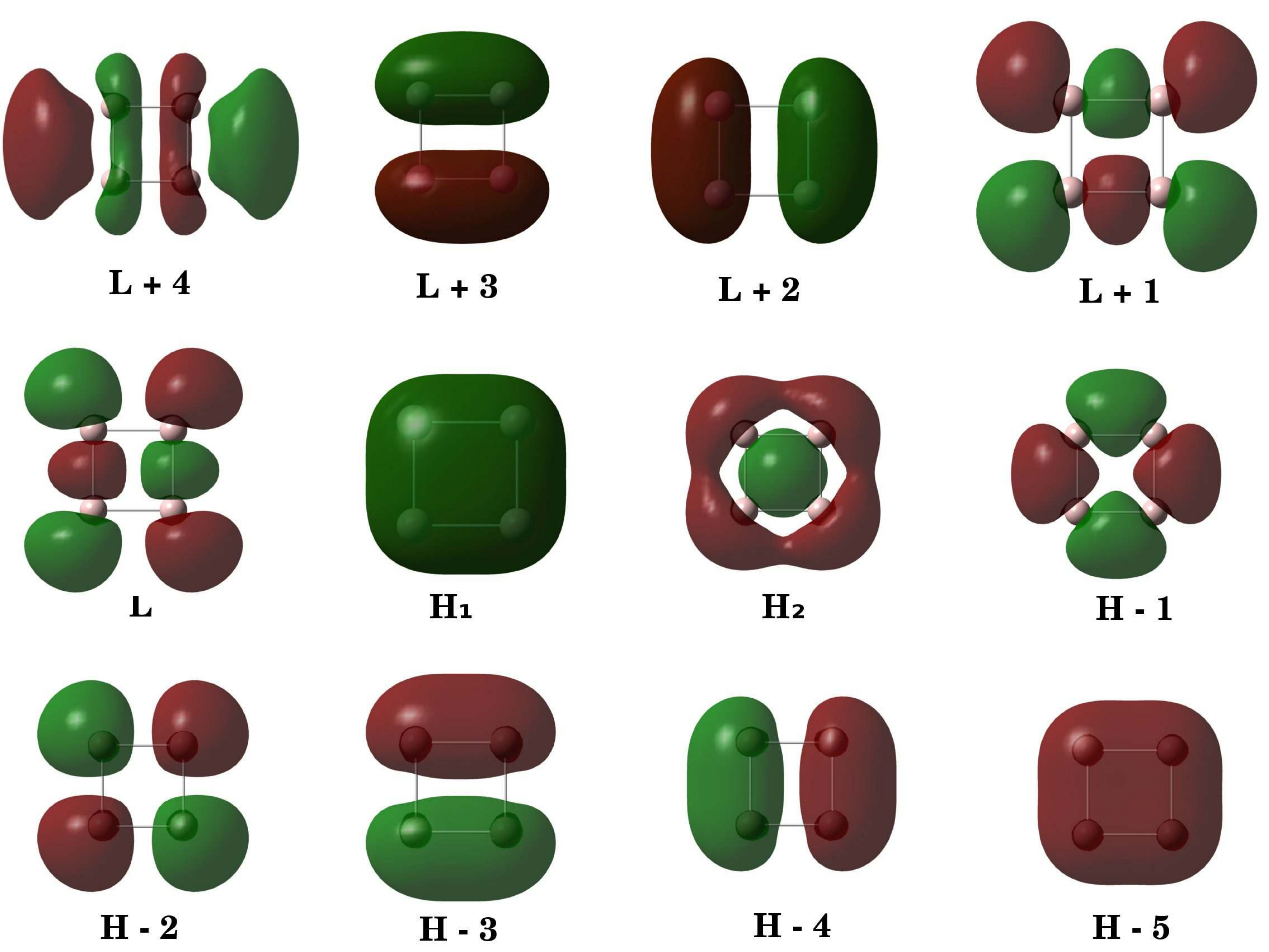} \label{subfig:mo-al4-sqr} } 
\caption{\label{fig:plot-al4-sqr}  \protect\subref{subfig:plot-al4-sqr} The linear optical absorption spectrum of square Al$_{4}$, calculated using the MRSDCI approach.
The peaks corresponding to the light polarized along the molecular
plane are labeled with the subscript $\parallel$, while those polarized
perpendicular to it are denoted by the subscript $\perp$. For plotting
the spectrum, a uniform linewidth of 0.1 eV was used.  \protect\subref{subfig:mo-al4-sqr}  Molecular orbitals of square-shaped 
aluminum tetramer. $H$ and $L$ stands for HOMO and LUMO respectively, and $H_{1}$
and $H_{2}$ are singly occupied molecular orbitals.}
\end{figure*}


The absorption spectrum of square shaped isomer begins with a couple
of low in-plane polarized absorption peaks at 2.1 eV and 2.7
eV characterized by $H-1\rightarrow L$ and $H_{2}\rightarrow L+1$
respectively. The peak at 4.2 and 4.9 eV have $H-2\rightarrow L$
and $H_{1}\rightarrow L+2$ as respective dominant configurations.
A major peak at 5.85 eV is observed with absorption due to in-plane
polarization having $H-2\rightarrow L+2$ and a double excitation
$H_{1}\rightarrow L+2;H-2\rightarrow L+2$ as dominant configurations.
These configurations also make dominant contribution to the peak at
6.5 eV. This peak along with the one at 6.9 eV are two equally and most
intense peaks of the spectrum. The latter has additional contribution
from $H_{1}\rightarrow L+1;H-2\rightarrow L$. A shoulder peak is
observed at 7.2 eV.

The TDLDA spectrum \cite{kanhere_prb_optical_al} of aluminum rhombus tetramer differs from the one presented here.
Peaks labeled III to XII in our calculated spectrum are
also observed in the TDLDA results\cite{kanhere_prb_optical_al}, however, the relative 
intensities tend to disagree.  For example, the strongest absorption
peak of TDLDA calculations is located around 7.9 eV, while in our spectrum we obtain the 
second most intense peak at that location.    The highest absorption
peak in our calculations is at 5.5 eV, while TDLDA does report a strong
peak at the same energy\cite{kanhere_prb_optical_al}, it is not the highest of the spectrum. 

Our calculations also reveal a strong structure-property relationship as far as the location 
of the most intense peak in the absorption spectra of the two isomers is considered, a 
feature which can be utilized in their optical detection.

\subsubsection{Al$_{5}$}

The lowest lying pentagonal isomer of aluminum has $C_{2v}$ symmetry
and has an electronic ground state of $^{2}A_{1}$. The bond lengths
are as shown in Fig. \ref{fig:geometry}\subref{subfig:geom-pentagon}.
These are slightly bigger than those obtained by Rao and Jena \cite{rao_jena_ele_struct_al}
and Yang \emph{et al.} \cite{sachdev_dft_al_small} using the DFT approach.
Many other reports have confirmed that the planar pentagon is the
most stable isomer of $\textrm{Al}_{_{5}}$.

The other optimized structure of pentamer is perfect pyramid with
$C_{4v}$ symmetry and $^{2}A_{1}$ electronic ground state. This
lies 0.76 eV above the global minimum structure. This is the only
three dimensional structure of Al cluster studied for optical absorption.
The optimized geometry is consistent with those reported earlier by
Jones \cite{jones_simul_anneal_al}. However, it should be noted
that there exists many more similar or slightly distorted structure
lying equally close the the global minimum. 

\begin{figure*}
\centering
\subfloat[]{\includegraphics[width=8.3cm]{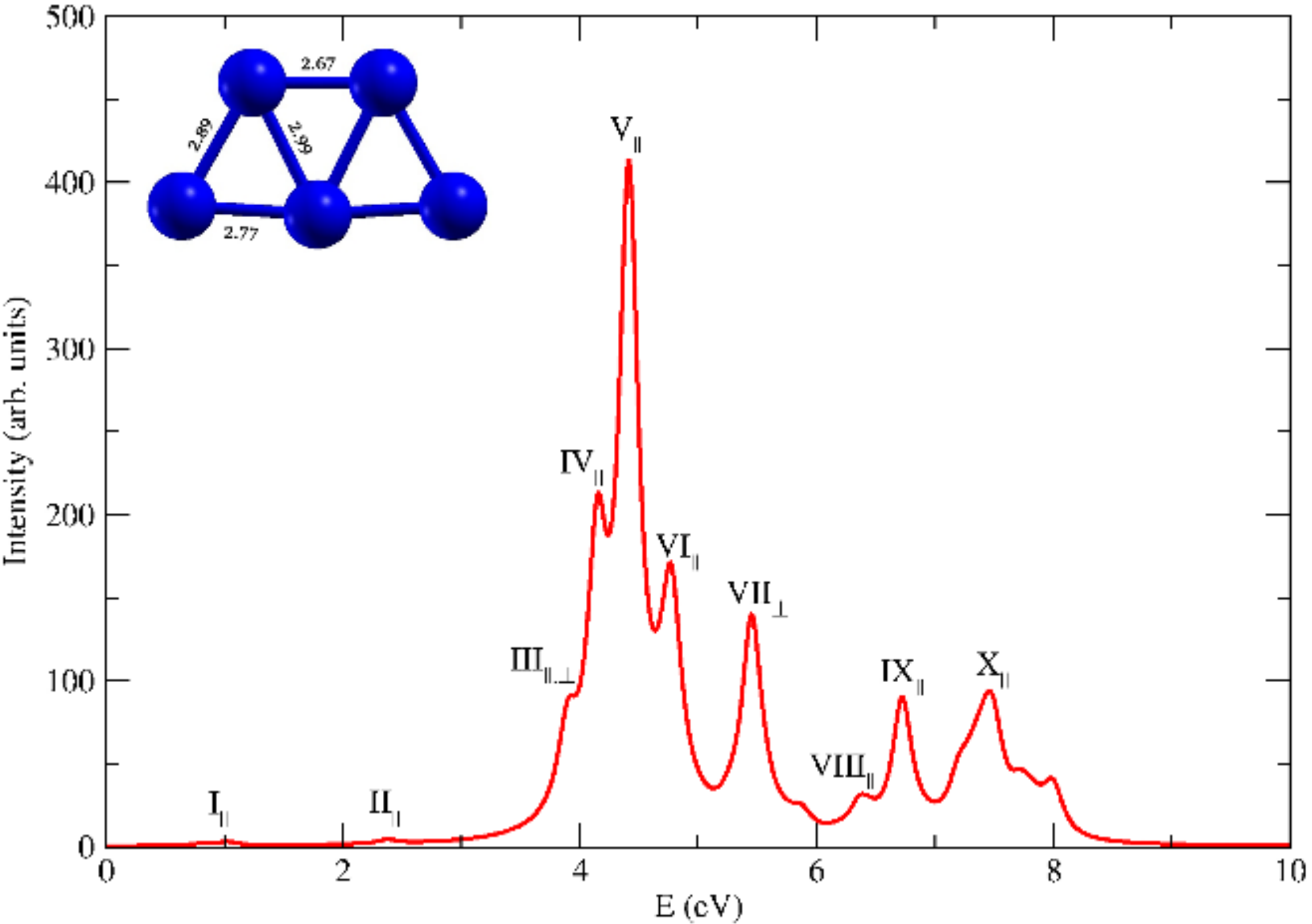} \label{subfig:plot-al5-penta} }\hfill
\subfloat[]{\includegraphics[width=6.5cm]{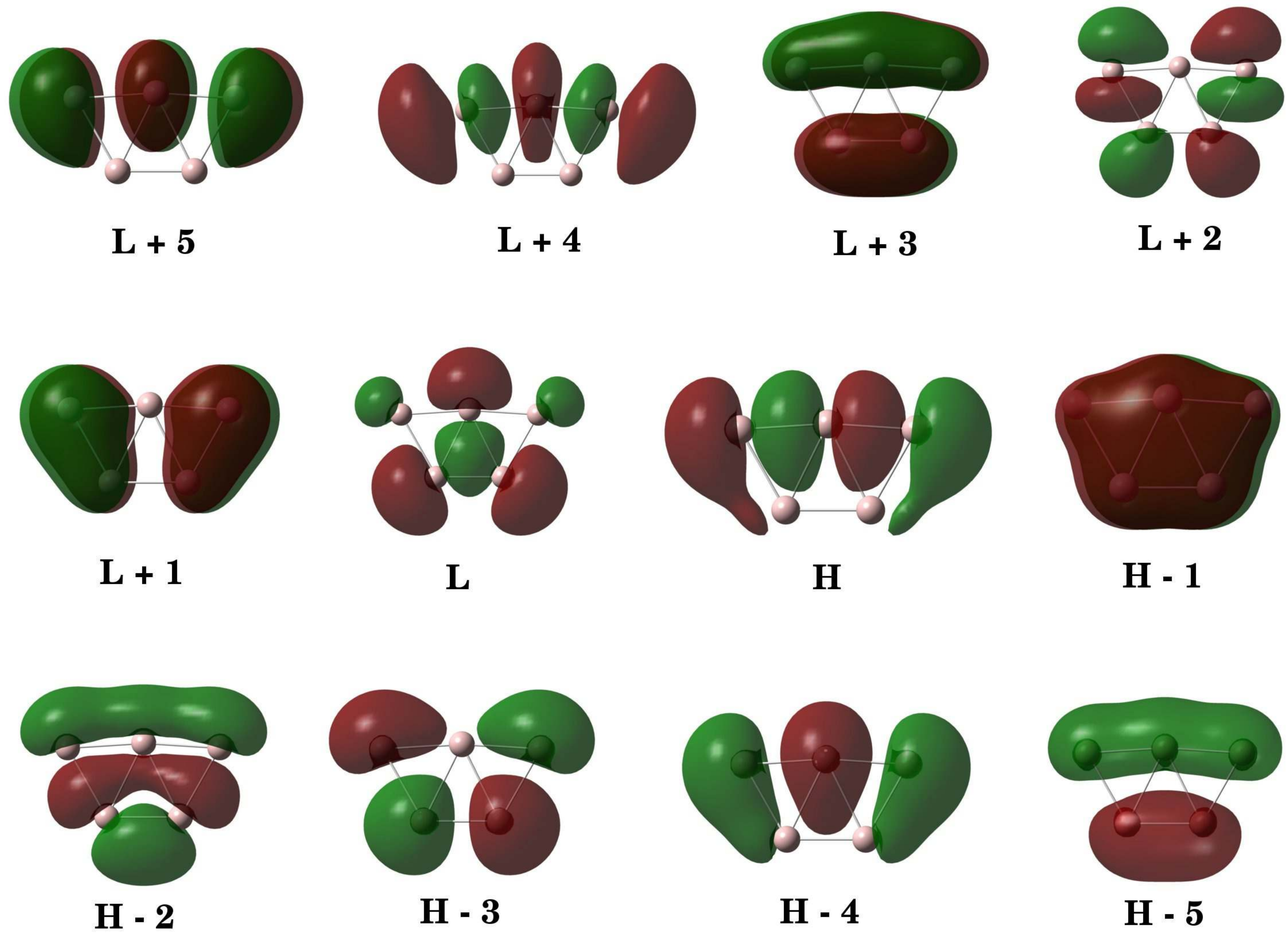} \label{subfig:mo-al5-penta}}
\caption{\label{fig:plot-al5-penta} \protect\subref{subfig:plot-al5-penta} The linear optical absorption
spectrum of pentagonal Al$_{5}$, calculated using the MRSDCI approach.
The peaks corresponding to the light polarized along the molecular
axis are labeled with the subscript $\parallel$, while those polarized
perpendicular to it are denoted by the subscript $\perp$. For plotting
the spectrum, a uniform linewidth of 0.1 eV was used. \protect\subref{subfig:mo-al5-penta} Molecular orbitals 
of pentagonal aluminum pentamer. $H$ and $L$ stands for HOMO and LUMO respectively.}
\end{figure*}


The many-particle wavefunctions of excited states contributing to the 
peaks are presented in Table \ref{Tab:table_al5_pentagon} and \ref{Tab:table_al5_pyramid}.
The optical absorption spectrum of pentagonal Al$_{5}$ has few low
energy peaks followed by major absorption at 4.4 eV. It has dominant
contribution from $H-1\rightarrow L+5$ configuration. Pentagonal
isomer shows more optical absorption in the high energy range, with
peaks within regular intervals of energy.

Few feeble peaks occur in the low energy range in the optical absorption
of pyramidal isomer. The major absorption peak at 4.2 eV is slightly
red-shifted as compared to the pentagonal counterpart. It is characterized
by $H-3\rightarrow L+2$. A peak at 6 eV is seen in this absorption
spectrum having dominant contribution from $H\rightarrow L+13$, which
is missing in the spectrum of pentagon. These differences can lead
to identification of isomers produced experimentally.

\begin{figure*}
\centering
\subfloat[]{\includegraphics[width=8.3cm]{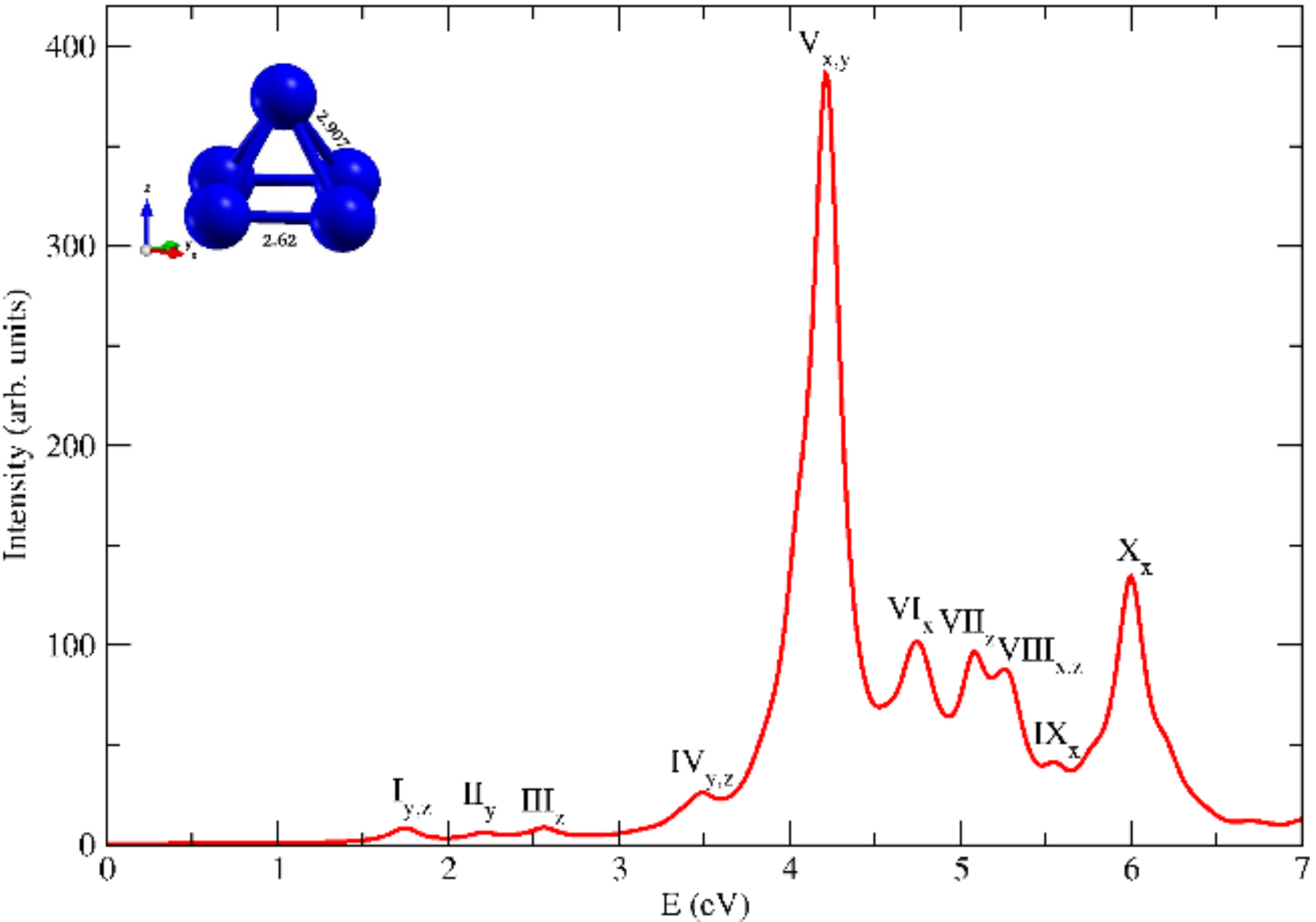} \label{subfig:plot-al5-pyra} } \hfill
\subfloat[]{\includegraphics[width=6.5cm]{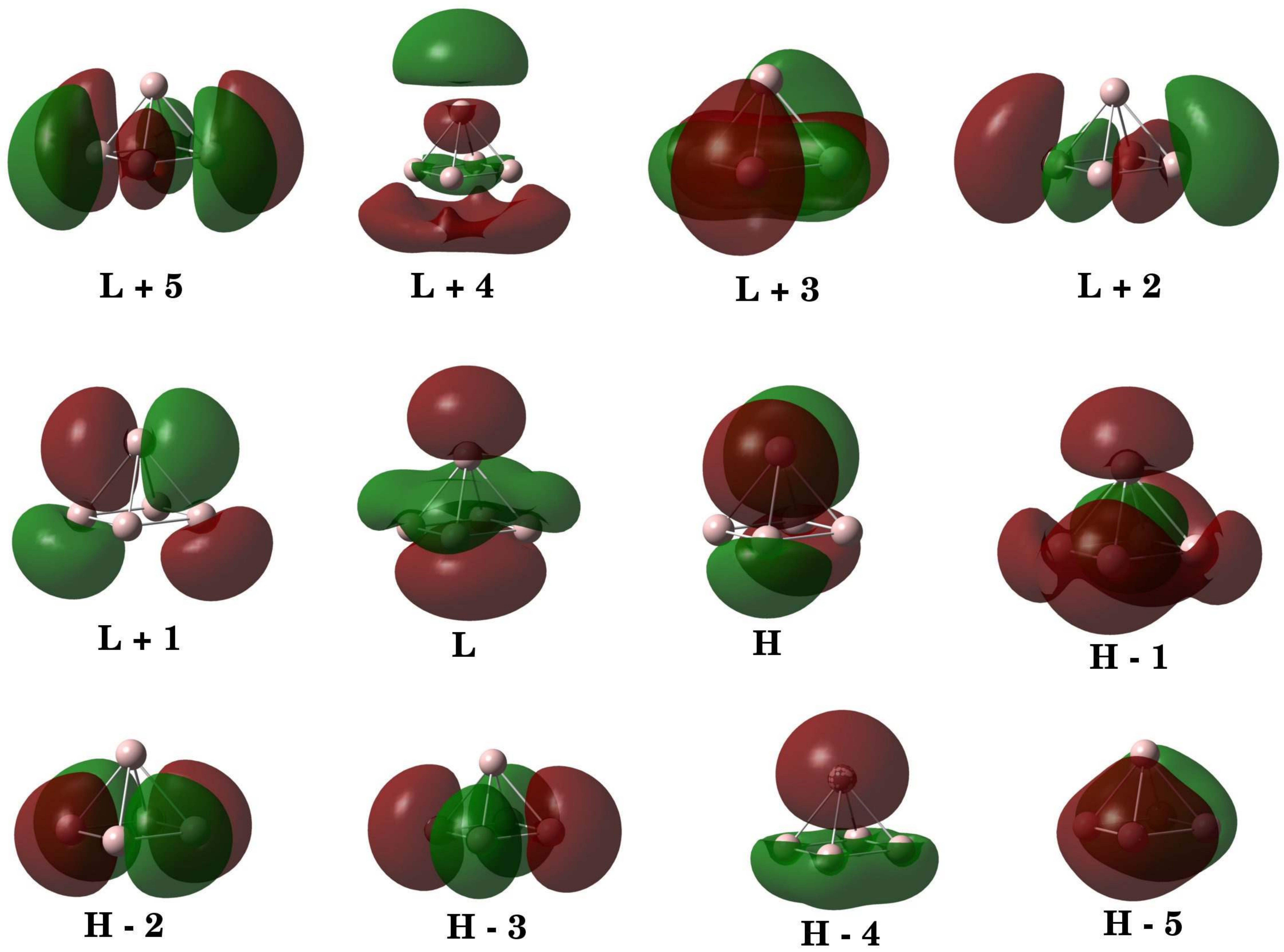} \label{subfig:mo-al5-pyra} } 
\caption{\label{fig:plot-al5-pyra}  \protect\subref{subfig:plot-al5-pyra} The linear optical absorption
spectrum of pyramidal Al$_{5}$, calculated using the MRSDCI approach.
The peaks corresponding to the light polarized along the Cartesian
axes are labeled accordingly. For plotting the spectrum, a uniform
linewidth of 0.1 eV was used. \protect\subref{subfig:plot-al5-pyra} Molecular orbitals of pyramidal aluminum pentamer.
$H$ and $L$ stands for HOMO and LUMO respectively.}
\end{figure*}


In the range of spectrum studied in our calculations, the TDLDA calculated spectrum 
\cite{kanhere_prb_optical_al} of pentagonal isomer is found to be similar to the one presented here as far as the peak locations are concerned, albeit
the intensity profile differs at places. A small peak at 2.4 eV is observed in both the spectra, followed by peaks at 3.9 eV, 4.2 eV and 4.4 eV.
These three peaks are also observed in TDLDA results with a little
bit of broadening. Again, the peak at 5.4 eV matches with each other
calculated from both the approaches. Peak found at 6.7 eV is also
observed in the TDLDA calculation\cite{kanhere_prb_optical_al}. Within the energy 
range studied here, the strongest peak position and intensity of this work is
in good agreement with that of its TDLDA counterpart\cite{kanhere_prb_optical_al}.


\subsection{Nature of Optical Excitations}
If an absorption peak is caused by an interaction among many particle-hole
excitations (i.e. configurations) with comparable weights, it suggests a plasmon-like collective excitation as compared to molecular 
excitation dominated by a single configuration.\cite{plasmon} The wavefunctions of the excited states contributing to 
most of the peaks in the optical absorption spectra of clusters studied here exhibit 
strong configuration mixing, instead of being dominated by single configurations, pointing
to the plasmonic nature of the optical excitations \cite{plasmon}. 

\begin{figure}
\centering
\includegraphics[width=8.3cm]{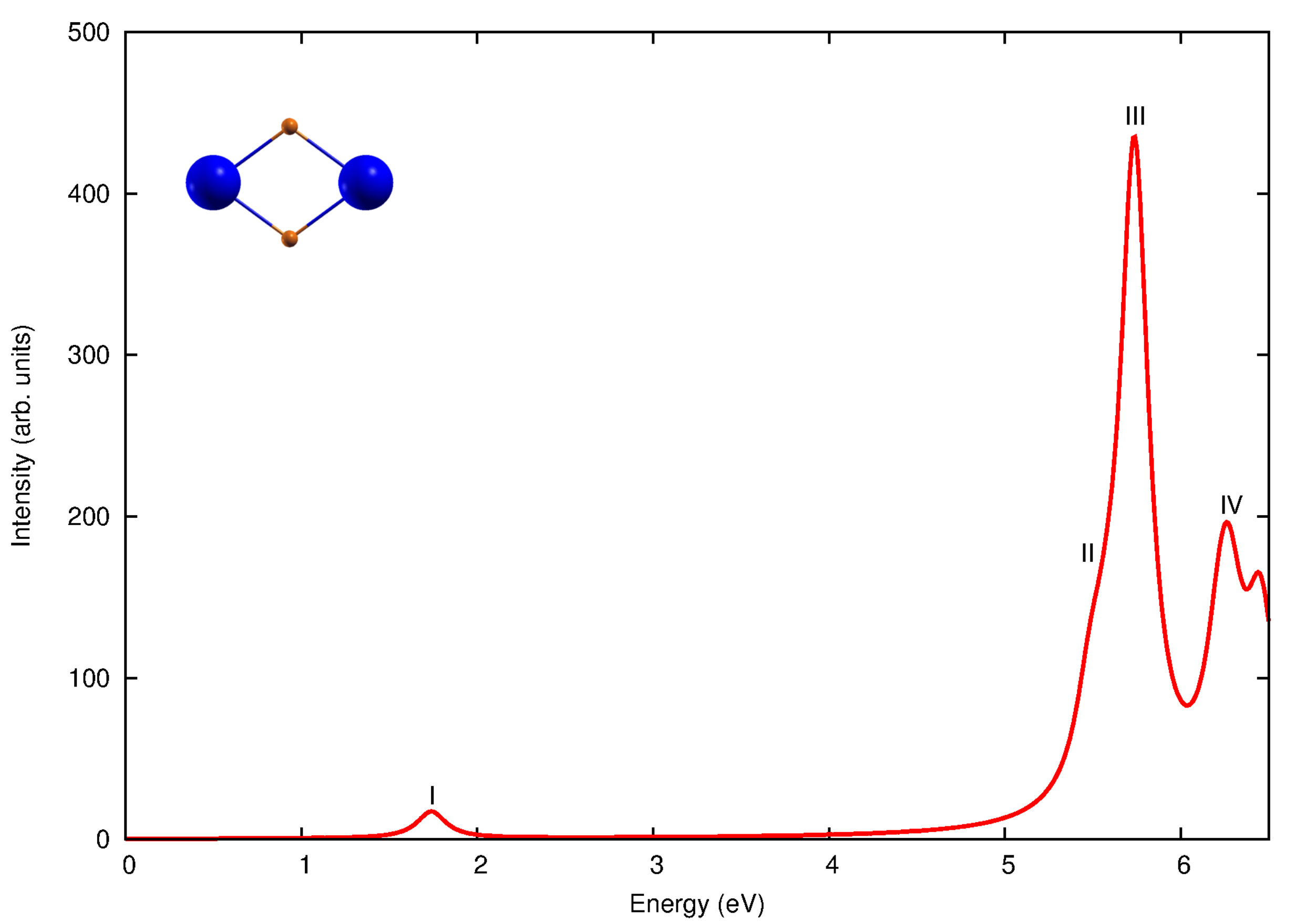} \vspace{0.2cm}
\caption{\label{fig:plot-al2h2}The linear optical absorption
spectrum of Al$_{2}$H$_{2}$, calculated using the MRSDCI approach.
 For plotting the spectrum, a uniform linewidth of 0.1 eV was used.}
\end{figure}

In order to draw a distinction between two cases, we provide the results of 
calculations of Al$_{2}$H$_{2}$ cluster. The MRSDCI calculated optical absorption spectrum is as shown Fig. \ref{fig:plot-al2h2}.

 \begin{footnotesize}	
\begin{table}[h!]
\centering
  \caption{ Excitation energies ($E$) and many-particle wavefunctions of excited
states corresponding to the peaks in the linear absorption spectrum
of Al$_{2}$H$_{2}$
, along with the oscillator strength ($f_{12}$) of the transitions. 
In the wavefunction, the bracketed numbers are the CI coefficients
of a given electronic configuration. Symbols $H$, and $L$, denote HOMO and LUMO
orbitals respectively. $HF$ denotes the Hartree-Fock configuration. }  
\label{Tab:table_al2h2}
\begin{tabular}{cccl}
\hline
Peak &  $E$ (eV) & $f_{12}$  & Wave Function  \tabularnewline \hline 
\tabularnewline
GS
 &  &    & $|HF\rangle$ (0.8961)\tabularnewline
 &  &    & $|H \rightarrow L + 6 \rangle$(0.1234)\tabularnewline
 &  &    & \tabularnewline
I   & 1.73  & 0.1455  & $|H \rightarrow L \rangle$(0.8874) \tabularnewline
 &  &    & $|H \rightarrow L + 8 \rangle$(0.1654) \tabularnewline
 &  &    & \tabularnewline
II  & 5.57  & 0.3465  & $|H \rightarrow L + 3\rangle$(0.8000) \tabularnewline
 &  &    & $|H - 1 \rightarrow L + 1 \rangle$(0.2991) \tabularnewline
  &  &    & \tabularnewline
III  & 5.74 & 3.5591  & $|H \rightarrow L + 4 \rangle$(0.8041) \tabularnewline
 &  &    & $|H \rightarrow L ; H \rightarrow L + 2 \rangle$(0.2745) \tabularnewline
 &  &    & \tabularnewline
IV  & 6.23  & 0.5937  & $|H - 1 \rightarrow L + 2 \rangle$(0.7316) \tabularnewline
 &  &    & $|H \rightarrow L + 4 ; H \rightarrow L + 2  \rangle$(0.3112) \tabularnewline
\hline
\end{tabular} 
\end{table}
 \end{footnotesize}

The many-particle wavefunctions of the excited states contributing to the peaks of Al$_{2}$H$_{2}$  (\emph{cf.} Table \ref{Tab:table_al2h2})
 do not show a strong configuration mixing, rather, one single
hole-particle excitation is seen to be dominant till 6.2 eV.  Therefore, the nature of 
excitation in Al$_{2}$H$_{2}$  is of molecular type. Comparing the excited state wave 
functions of Al$_{2}$H$_{2}$ with those of the Al$_{2}$ cluster, we see a 
clear indication of strong configuration mixing in the latter,  suggesting plasmonic 
nature of optical excitations. In other words, the closed-shell hydrogenated Al dimer has 
molecular type of optical excitations, while the open-shell bare Al dimer exhibits 
plasmonic type excitations.

\section{\label{sec:conclusions-jpc}Summary}

In this study, we have presented large-scale all-electron correlated
calculations of optical absorption spectra of several low-lying isomers
of aluminum clusters Al$_{n}$, (n=2--5). Both ground and excited
state calculations were performed at \ac{MRSDCI} level, which take electron
correlations into account at a sophisticated level. We have analyzed
the nature of low-lying excited states. Isomers of a given cluster
show a distinct signature spectrum, indicating a strong-structure
property relationship. This fact can be used in experiments to distinguish
between different isomers of a cluster. Owing to the sophistication of
our calculations, our results can be used for benchmarking of the
absorption spectra. The optical excitations involved are found to
be collective type, and plasmonic in nature. 

Our results were found to be significantly different as compared to the 
TDLDA results\cite{kanhere_prb_optical_al}, for the clusters studied here.
Given the fact that the \ac{MRSDCI} calculations incorporate electron-correlation effect 
quite well both for the ground and the excited states,
they could be treated as benchmarks, and be used to design superior \ac{TDDFT} approaches.

  \lhead{{\chaptername\ \thechapter.}{  Boron Wheel Clusters B$_{7}$, B$_{8}$ and B$_{9}$}}
   \chapter{\label{chap:main_benchmarking}Linear Optical Absorption in Boron Wheel-Like Clusters B$_{7}$, B$_{8}$ and B$_{9}$: Benchmarking TDDFT against EOM-CCSD}


Planar boron clusters, in particular, have proven to be interesting because of the multiple aromaticities and extreme coordination
environments.\cite{kiran_wang, coord-chem-review,boron-8-ligand, wheel_wang, boron19-natchem} Atoms of boron can adopt 
such an arrangement that they form a miniature wheel, with one atom at the center. 
B$_{13}^{+}$ and B$_{19}^{-}$ perform Wankel motor action when shined by circularly polarized light.\cite{wankel-motor-heine, alina-b13-wankel, boron-19-rotor}
Such clusters have been synthesized and their electronic structure is now well known.\cite{wheel_wang} 
Boron with seven, eight and nine atoms form such truly planar wheel structures with radii of 1.65 \AA{}, 1.80 \AA{} and 2.0 \AA{} respectively.
Although a lot of information about planarity, electronic structure and chemical bonding is now available, optical absorption 
of these clusters remains unexplored. 

In order to describe the ground state of a system, normally, a reference state is used, which is a good approximation to the exact ground state
of the system. Typically, one prefers variational approaches, although that is not the only possible way. This single
reference state may not be the natural choice when it comes to the computation of excited states. Coupled cluster methods offer 
insensitivity to such single reference state for excited state calculations owing to the exponential treatment of single excitation effects.\cite{barlett-eomccsd} 

Equation-of-motion coupled cluster (EOM-CC) is one of the approaches which can effectively and unambiguously describe excited states of 
molecules, polyradicals where ground or excited state is often degenerate. Since it does not make any assumptions about nature of the states,
it is easy to use single reference method. It is most accurate in calculating one electron vertical transition states. Often, the cluster 
expansion is terminated at doubles, for computational feasibility without
serious compromise on quality of results. The results can always be systematically improved by including more excitation levels. However,
the method scales as N$^{6}$, N being the number of basis functions, thereby making it intractable for large molecules. 

We give here an account of other less expensive excited state calculation methods that can approximate the accurate EOM-CCSD results.
In particular, the time-dependent density functional theory with adiabatic approximation continues to remain favorite for study of large 
variety of systems. The exact exchange and correlation functional required in this approach is not known, several approximations are made in
that respect. This adds to the puzzle to choose the right functional for a given type of calculation.
This work will provide qualitative as well as quantitative analysis of benchmarking results of single reference quantum chemical
methods -- TDDFT with different functionals against \ac{EOM-CCSD} method. This study will help in identifying computationally 
the least expensive functional which mimic more accurate \ac{EOM-CCSD} results of optical absorption.

The rest of the chapter is organized as follows. Next section discusses theoretical and computational details of the calculations, followed
by section \ref{sec:results-wheel}, in which results are presented and discussed. Conclusions and future directions are presented
in section \ref{sec:conclusions-wheel}.

\section{\label{sec:theory-wheel}Theoretical and Computational Details}

\subsection{Geometry Optimization}

\begin{figure*}[!t]
\centering
\subfloat[B$_{7}$, D$_{2h}$, $^{2}B_{2g}$]{\label{subfig:subfig-geom-b7}\includegraphics[width=1.2in]{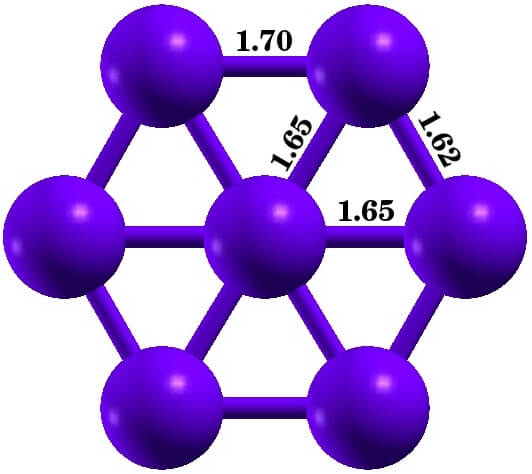}}\hspace{0.02\paperwidth}
\subfloat[B$_{8}$, D$_{7h}$, $^{3}A_{g}$]{\label{subfig:subfig-geom-b8}\includegraphics[width=1.3in]{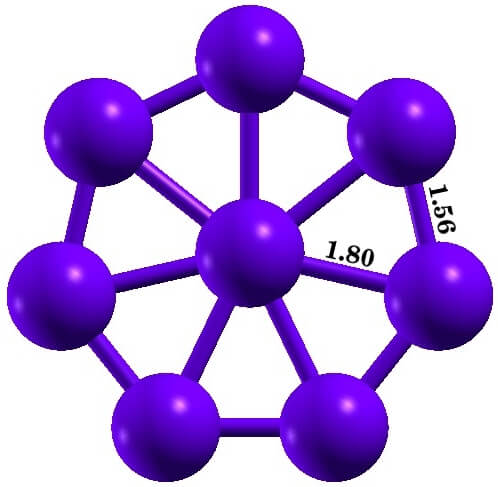}}\hspace{0.02\paperwidth}
\subfloat[B$_{9}$, D$_{2h}$, $^{2}B_{1g}$]{\label{subfig:subfig-geom-b9}\includegraphics[width=1.4in]{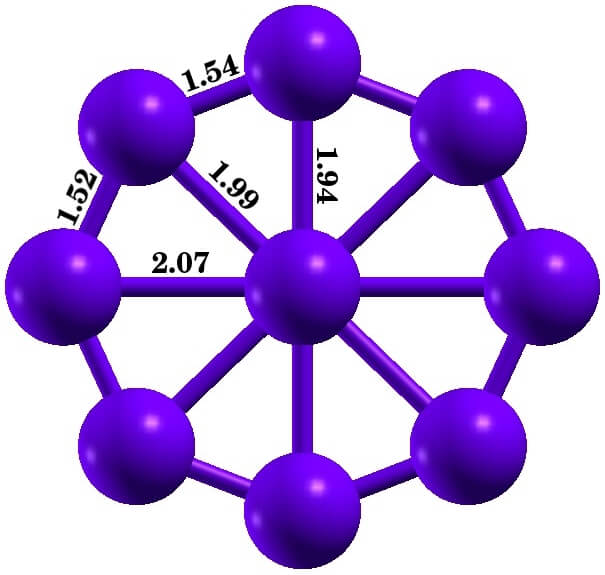}}
\caption{\label{fig:geometries_wheel}Geometry optimized structures of boron wheels with point group symmetry and the electronic ground
state. All numbers are in Angstrom unit. }
\end{figure*}

The geometries of the boron wheel clusters, B$_{7}$, B$_{8}$ and B$_{9}$ were optimized using the computer 
code \textsc{gaussian 09}\cite{gaussian09} employing a 6-311++G(2d,2p) basis set, and  using the size-consistent \ac{CCSD} method. 
To initiate the optimization, raw geometries, reported by Wang \emph {et al}, based on density functional method were used.\cite{wheel_wang} 
 The results of optimization are in accordance with the available report. \cite{wheel_wang} These optimized geometries were further used 
in the calculations of optical absorption spectra. Figure \ref{fig:geometries_wheel} shows the final optimized geometries of the clusters studied 
in this paper. 

The optical absorption spectra of these optimized geometries of the clusters are then calculated using \ac{EOM-CCSD} and \ac{TDDFT}.
 Various exchange and correlation functionals were used to compute the optical absorption spectra using \ac{TDDFT} approach.
An augmented correlation consistent polarized valence double zeta (aug-cc-pVDZ) basis set was used for all methods mentioned above.

\subsection{Excited State Calculation Methods}

Coupled cluster method is known to include electron correlation in a systematic manner. Coupled cluster is an exact formalism if all
possible excitation are taken into account. Often, the excitation level is terminated at doubles, which gives rise to \ac{CCSD} 
method. This method is extended to excited state calculations through what is known as Equation-of-Motion CCSD (EOM-CCSD).\cite{eom-ccsd-jcp, eom-ccsd-ann-rev} 
The EOM-CC approach amounts to diagonalizing the effective Hamiltonian $e^{-T} H e^{-T}$.
The computational time for this approach scales as N$^{6}$, where N is the number of basis functions, thereby limiting its use to smaller molecules.

 Development of Density functional theory has led to enormous progress in the understanding of properties of various systems.
However, the main drawback is, results depend upon the choice of energy functional used to perform the calculation. 
Many different functionals are proposed for various kinds of 
calculations and the number is still increasing. In the adiabatic approximation, the time-dependent counterpart of DFT, also uses the same 
functionals to investigate excited state properties. Here, we have considered various DFT functionals of various types to study the 
excited state properties and optical absorption in boron clusters. The set of functionals includes 
(a) Hybrid Generalized-Gradient Approximation (H-GGA) -- PBE0\cite{adamo-pbe0-jcp}, B3LYP\cite{becke-b3lyp-jcp}, B3PW91\cite{perdew-b3pw91-prb}
(b) Global Hybrid-Meta GGA (HM-GGA) -- M06\cite{truhlar-m06-tca}, M06-2X\cite{truhlar-m06-tca} (c) Long-range 
corrected -- $\omega$B97xD\cite{martin-head-gordon-wb97xd-pccp}, CAM-B3LYP\cite{yanai-cam-b3lyp-cpl}, LC-$\omega$PBE\cite{vydrov-lc-wpbe-jcp}.

\section{\label{sec:results-wheel}{Benchmarking: Photoabsorption Spectra of Wheel Clusters}}
In this section, we present the photoabsorption spectra of boron wheel clusters studied. Graphical presentation of
natural transition orbitals involved in the case of TDDFT calculations are also given below. The many-particle wavefunctions
of excited states contributing to the peaks of EOM-CCSD spectra are presented in
 Table \ref{Tab:table-b7-eomccsd}, \ref{Tab:table-b8-eomccsd} and \ref{Tab:table-b9-eomccsd}.  


\subsection{B$_{7}$}
\begin{figure*}
 \centering
\includegraphics[width=8.3cm]{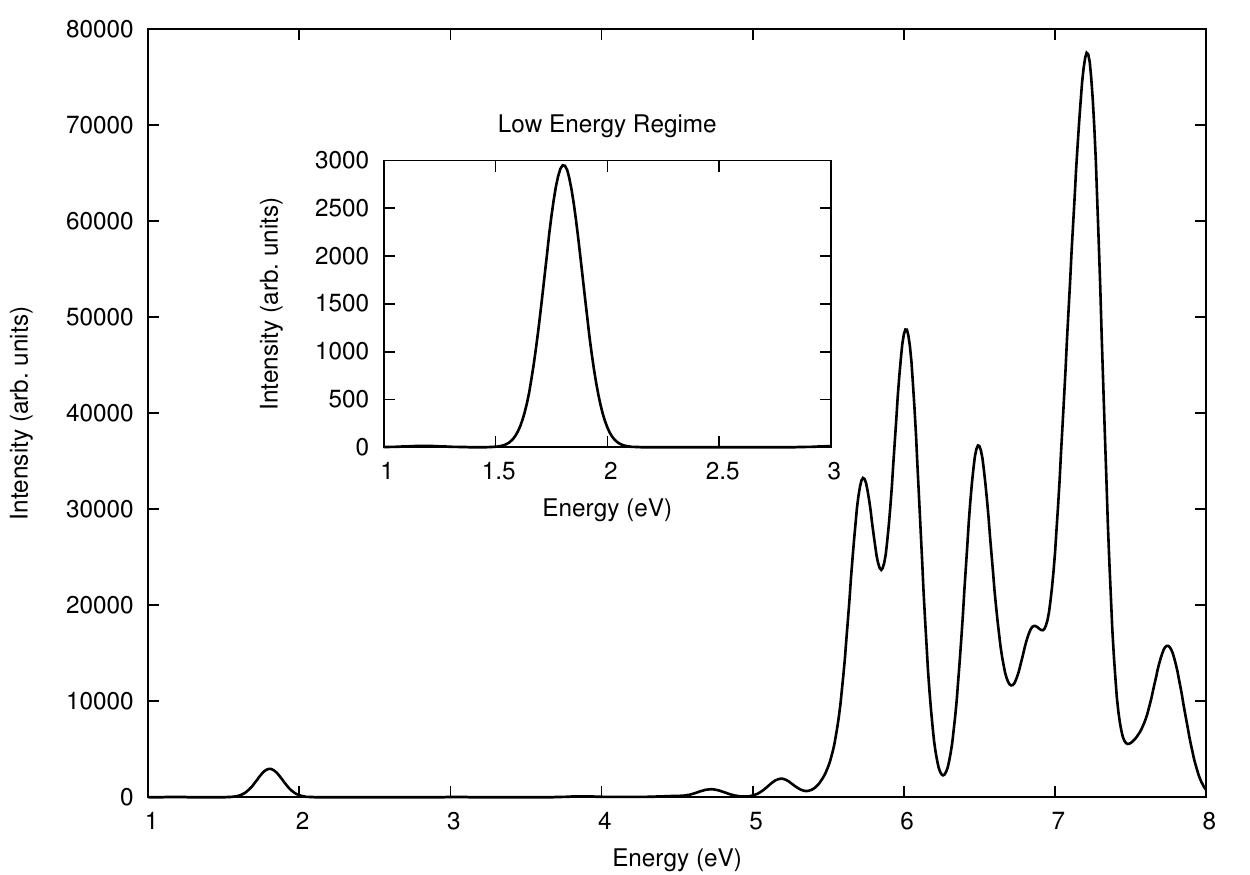}
\caption{\label{fig:b7-eom-cis-rpa-plot}The linear optical absorption spectrum of boron 
wheel B$_{7}$ cluster, calculated using the EOM-CCSD approach. }
\end{figure*}

\begin{figure}[!t]
 \centering
\subfloat[\label{subfig:subfig-b7-eom-tddft-plot} Optical absorption spectrum of B$_{7}$ cluster ] {\includegraphics[width=8.3cm]{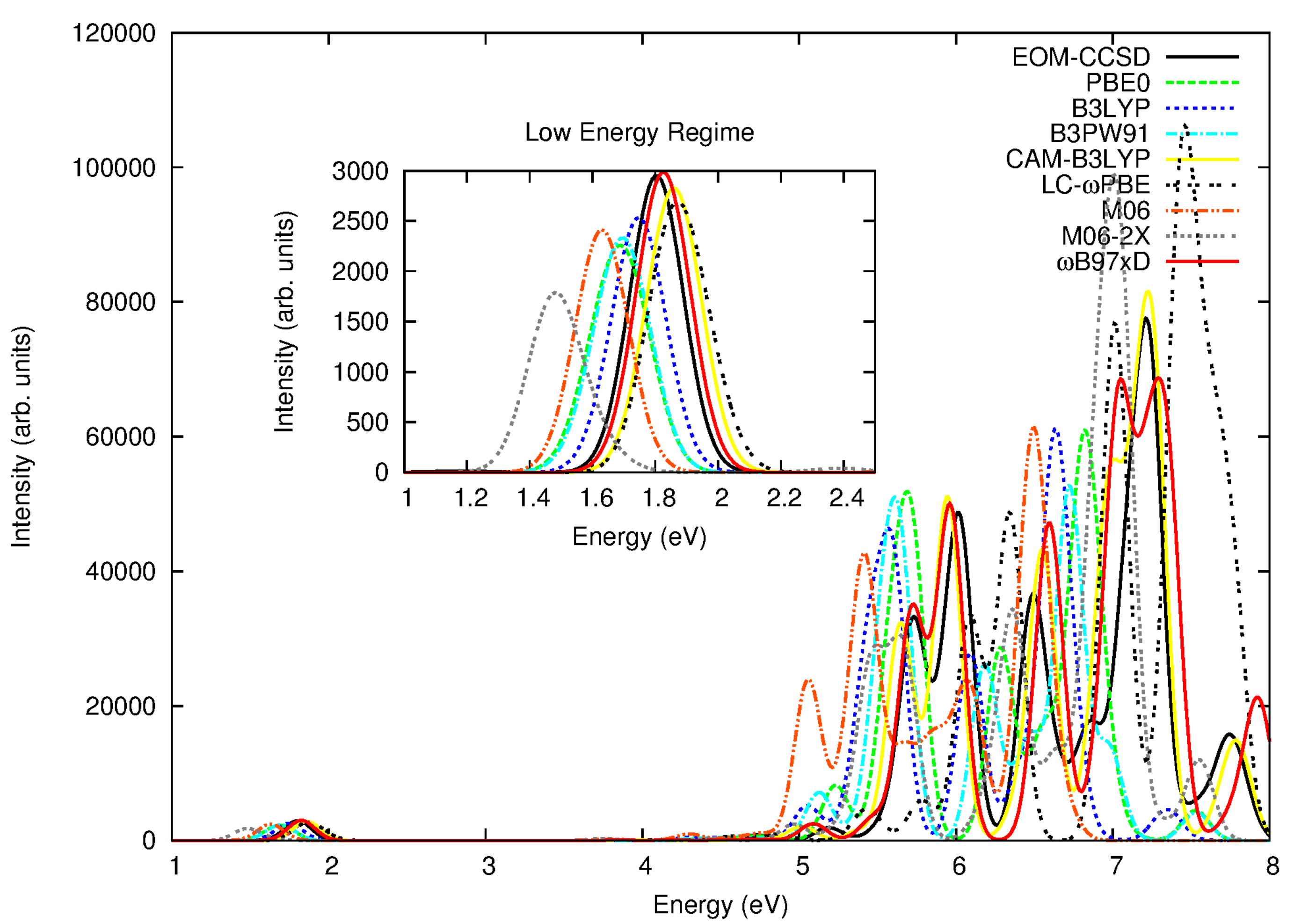}} \hfill
 \raisebox{2cm}{\subfloat[\label{subfig:subfig-b7-nto-1-2} $\lambda$ = 0.56 at 1.63eV]{\includegraphics[width=1.4cm]{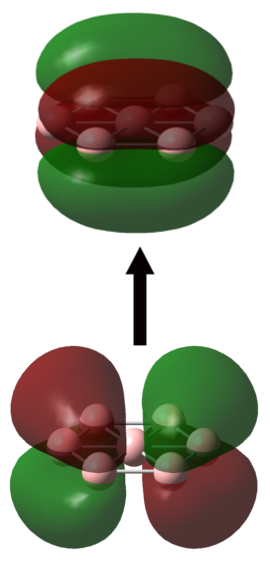} }} \hfill
 \raisebox{2cm}{\subfloat[\label{subfig:subfig-b7-nto-3-4} $\lambda$ = 0.35 at 5.72eV]{\includegraphics[width=1.4cm]{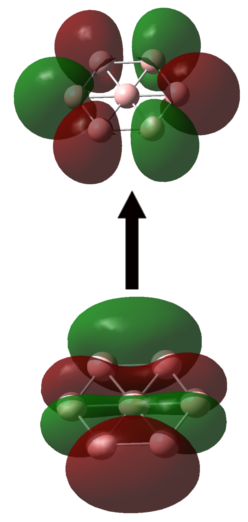} }}
\vfill
\caption{\label{fig:nto-boron7} \protect\subref{subfig:subfig-b7-eom-tddft-plot} The linear optical absorption spectrum of boron 
wheel B$_{7}$ cluster, calculated using the TDDFT approach using various functionals, and compared to the spectrum calculated using EOM-CCSD.
\protect\subref{subfig:subfig-b7-nto-1-2} and \protect\subref{subfig:subfig-b7-nto-3-4}: Natural transition orbitals (NTO) involved in the excited states of B$_{7}$ cluster, calculated using PBE0 method
corresponding to the peak at 1.63 eV with $\lambda$ = 0.56  ($H_{\alpha}-1\rightarrow L_{\alpha}$) and to the peak at 5.72 eV 
with $\lambda$ = 0.35 ($H_{1\alpha}\rightarrow L_{\alpha}+4$). Parameter $\lambda$ refers to a fraction of the NTO pair contribution to a given electronic excitation.}
\end{figure}

An excellent agreement is observed between EOM-CCSD and TDDFT results 
(\emph{cf.} Fig. \ref{fig:nto-boron7}\subref{subfig:subfig-b7-eom-tddft-plot}) with $\omega$B97xD and CAM-B3LYP functionals. This agreement 
holds good both for excitation energies and oscillator strengths. However,
$\omega$B97xD deviates from CAM-B3LYP spectra after 7 eV. M06 and M06-2X functionals show consistently red-shifted absorption
throughout the spectrum. Other functionals, such as, PBE0 and B3PW91 almost overlap to each other in the low energy regime. However,
former blue-shifts from B3PW91 at higher energies. NTO analysis of PBE0 spectrum (\emph{cf.} Fig. \ref{fig:nto-boron7}\subref{subfig:subfig-b7-nto-1-2}, 
\ref{fig:nto-boron7}\subref{subfig:subfig-b7-nto-3-4}) reveals that the nature of excitation
for the peak at 1.63 eV is $\pi \rightarrow \pi^*$, and at 5.72 eV, $\sigma \rightarrow \sigma^*$ dominates the excitation.

\subsection{B$_{8}$}
\begin{figure*}
 \centering
\includegraphics[width=8.3cm]{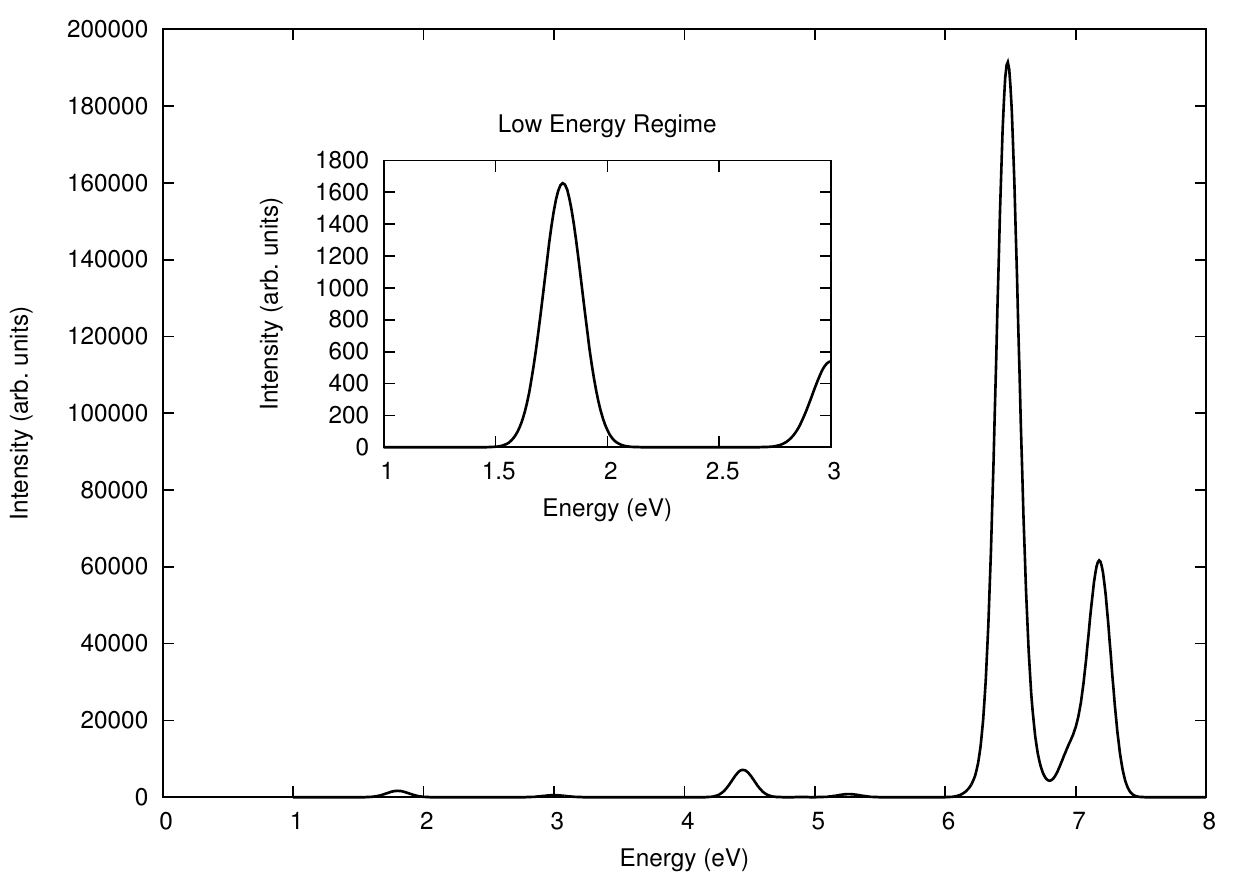}
\caption{\label{fig:b8-eom-cis-rpa-plot}The linear optical absorption spectrum of boron 
wheel B$_{8}$ cluster, calculated using the EOM-CCSD approach. }
\end{figure*}

\begin{figure*}
 \centering
\subfloat[\label{subfig:b8-eom-tddft-plot} ]{\includegraphics[width=8.3cm]{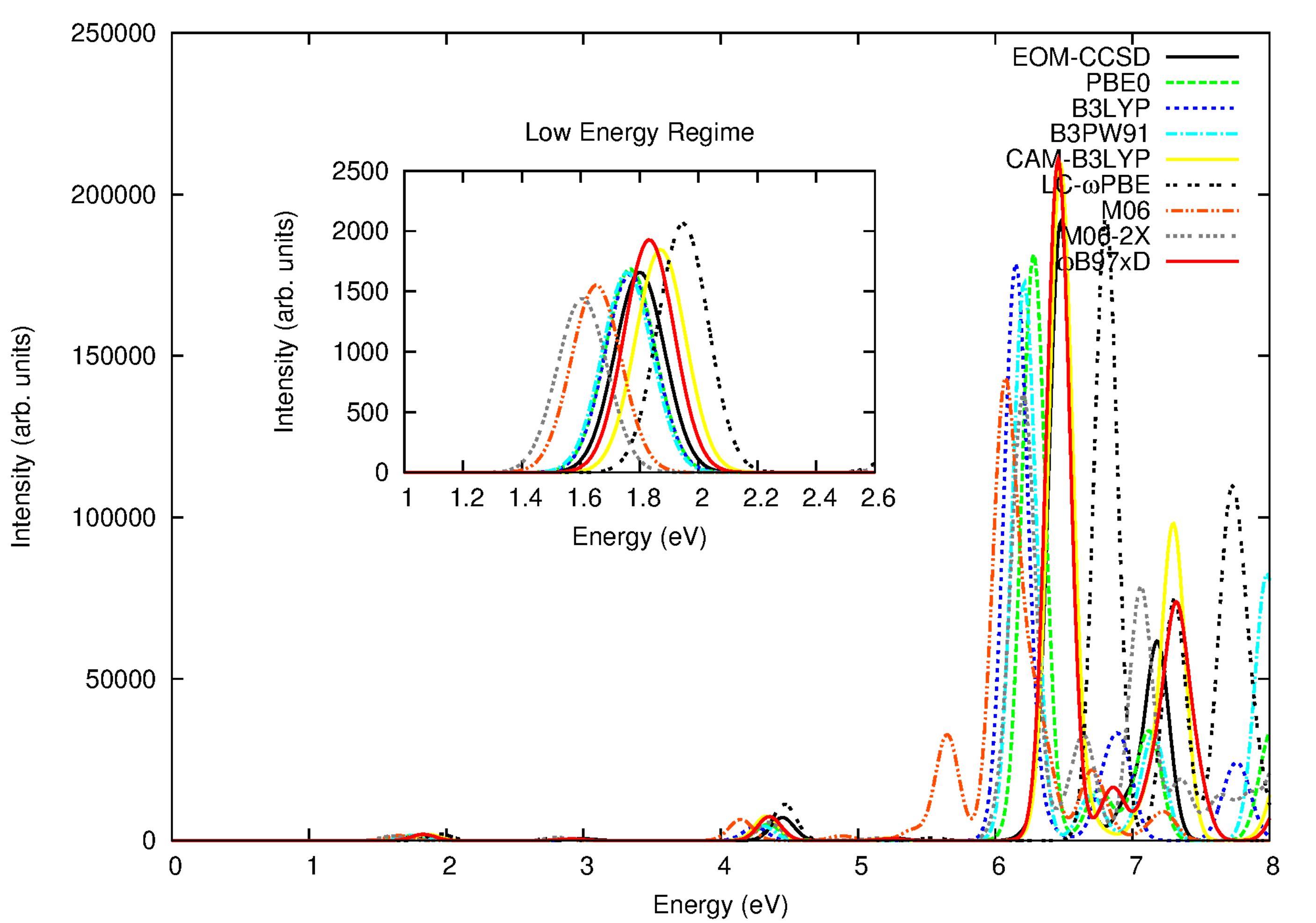}} \\ 
\subfloat[\label{subfig:subfig-b8-nto-1-2}  $\lambda$ = 0.99 at 1.77eV]{\includegraphics[width=1.2cm]{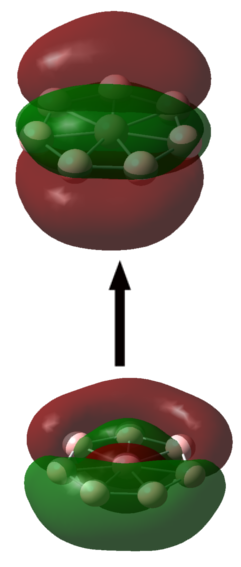} \hfill \includegraphics[width=1.5cm]{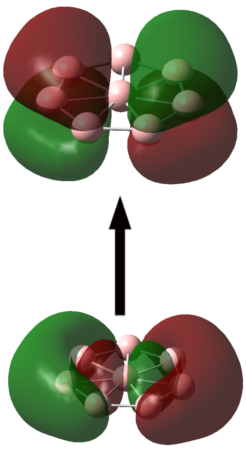}} \hfill
\subfloat[\label{subfig:subfig-b8-nto-5-6} $\lambda$ = 0.11 at 6.28eV]{\includegraphics[width=1.8cm]{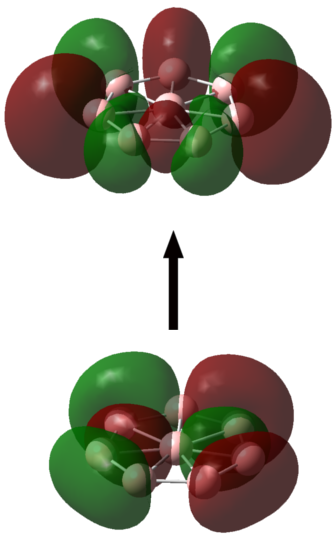}} \hfill
\subfloat[\label{subfig:subfig-b8-nto-7-8} $\lambda$ = 0.25 at 6.28eV]{\includegraphics[width=1.4cm]{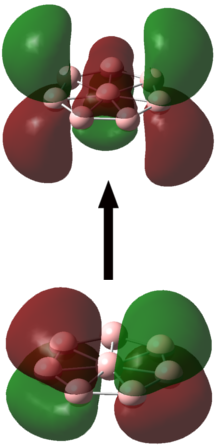}} \hfill
\subfloat[\label{subfig:subfig-b8-nto-9-10} $\lambda$ = 0.29 at 6.28eV]{\includegraphics[width=1.7cm]{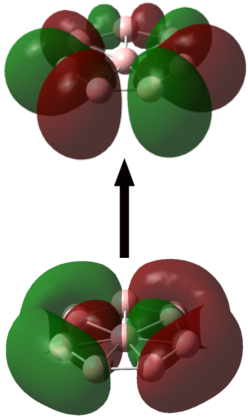}} \hfill
\subfloat[\label{subfig:subfig-b8-nto-11-12} $\lambda$ = 0.29 at 6.28eV]{\includegraphics[width=1.5cm]{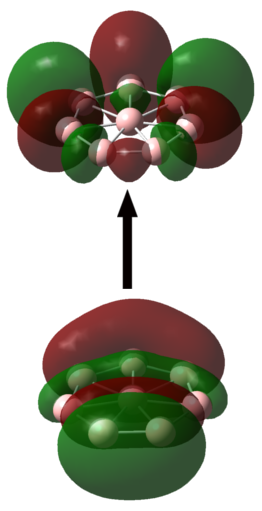}} \hfill
\caption{\label{fig:b8-eom-tddft-plot} \protect\subref{subfig:b8-eom-tddft-plot} The linear optical absorption spectrum of boron 
wheel B$_{8}$ cluster, calculated using the  TDDFT approach using various functionals, and compared to the spectrum calculated using EOM-CCSD. 
 Natural transition orbitals involved in the excited states of B$_{8}$ cluster, calculated using PBE0 
method corresponding to \protect\subref{subfig:subfig-b8-nto-1-2} the peak at 
1.77 eV with $\lambda$ = 0.99 ($H_{\beta}-1\rightarrow H_{1\beta}$ and $H_{\beta}-2\rightarrow H_{2\beta}$) and \protect\subref{subfig:subfig-b8-nto-5-6} to the 
peak at 6.28 eV with $\lambda$ = 0.11 ($H_{\alpha}-2\rightarrow L_{\alpha}+3$), \protect\subref{subfig:subfig-b8-nto-7-8} 0.25 ($H_{\alpha}-1\rightarrow L_{\alpha}+2$)
, \protect\subref{subfig:subfig-b8-nto-9-10} 0.29 ($H_{\beta}-2\rightarrow L_{\beta}+1$), \protect\subref{subfig:subfig-b8-nto-11-12} 0.29 ($H_{\beta}-1\rightarrow L_{\beta}$).
Parameter $\lambda$ refers to a fraction of the NTO pair contribution to a given electronic excitation.}
\end{figure*}

The optical absorption spectrum of B$_8$, Figs. \ref{fig:b8-eom-cis-rpa-plot} and \ref{fig:b8-eom-tddft-plot},
 shows a small number of well-separated peaks. CAM-B3LYP and $\omega$B97xD gives an excellent agreement
on absorption spectrum when compared to EOM-CCSD results. At the strongest absorption peak, both energies and intensity of spectrum 
of these functionals match very well with each other in this case. M06 provide poor performance in this case, with bands are shifted to lower 
energies. Also, there are extra peaks observed at higher energies for this functional. M06-2X seems to correct the latter behavior, while 
retaining the red-shift of bands. PBE0, B3LYP and B3PW91 spectra agree with each other at lower energies, but start deviating afterwards. 
The long-range corrected functional LC-$\omega$PBE has same intensity profile as that of EOM-CCSD, but the peaks are generally blue-shifted 
by 0.2--0.4 eV. A pair of $\sigma \rightarrow \pi^*$ transition takes place at 1.77 eV, as seen in the PBE0 NTO analysis in 
Fig. \ref{fig:b8-eom-tddft-plot}\subref{subfig:subfig-b8-nto-1-2}.

\subsection{B$_{9}$}

\begin{figure*}
 \centering
\includegraphics[width=8.3cm]{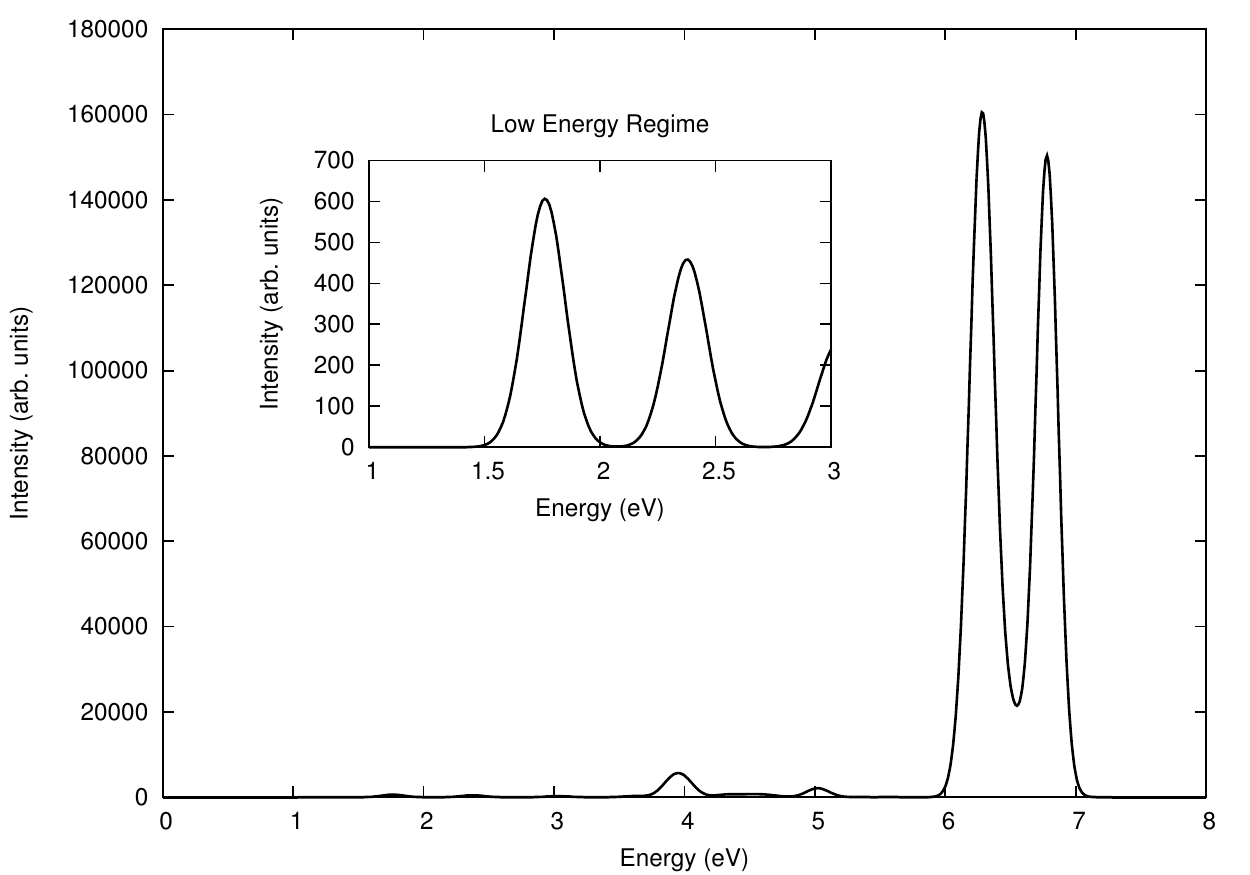}
\caption{\label{fig:b9-cis-eom-pbe-plot}The linear optical absorption spectrum of boron 
wheel B$_{9}$ cluster, calculated using the EOM-CCSD approach.}
\end{figure*}

\begin{figure*}[!t]
 \centering
\subfloat[\label{subfig:subfig-b9-eom-tddft-plot}]{\includegraphics[width=8cm]{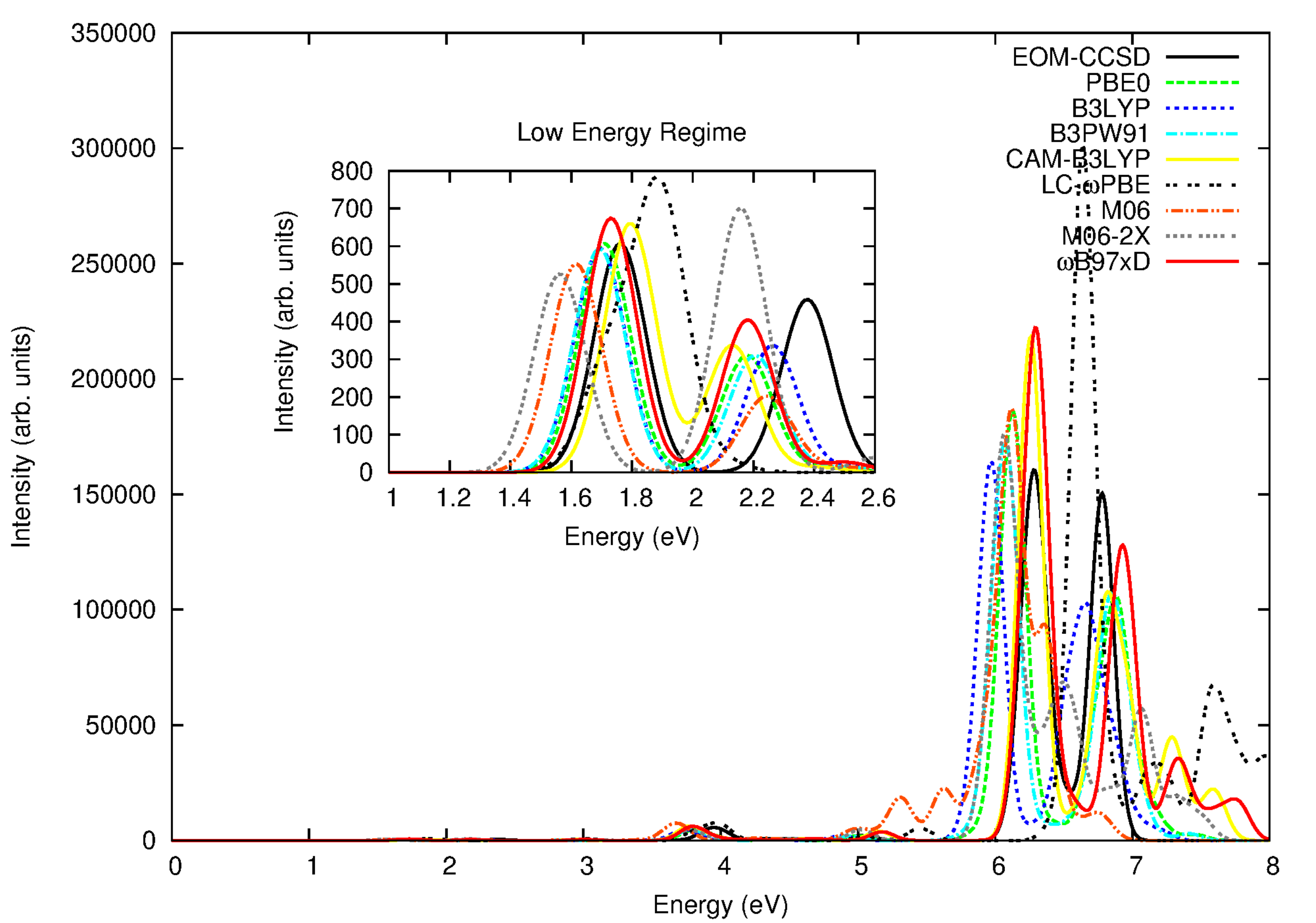}}  \hfill 
\raisebox{2cm}{\subfloat[\label{subfig:subfig-b9-nto-1-2}  $\lambda$ = 0.99 at 1.71eV]{\includegraphics[width=1.5cm]{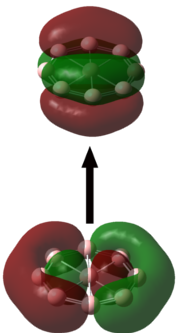}}} \hfill
\raisebox{2cm}{\subfloat[\label{subfig:subfig-b9-nto-3-4} $\lambda$ = 0.25 at 6.15eV]{\includegraphics[width=1.3cm]{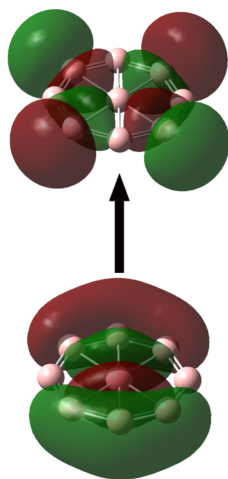}}} \hfill
\raisebox{2cm}{\subfloat[\label{subfig:subfig-b9-nto-5-6} $\lambda$ = 0.32 at 6.15eV]{\includegraphics[width=1.55cm]{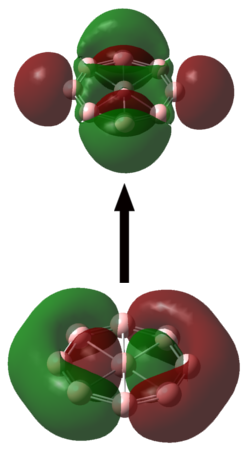}}} \hfill
\caption{\label{fig:plot-nto-boron9}
\protect\subref{subfig:subfig-b9-eom-tddft-plot} The linear optical absorption spectrum of boron wheel B$_{9}$ cluster, calculated using the 
TDDFT approach using various functionals, and compared to the spectrum calculated using EOM-CCSD approach.
\protect\subref{subfig:subfig-b9-nto-1-2}  Natural transition orbitals (NTO) involved in the excited states of B$_{9}$ cluster, calculated using PBE0 method
corresponding to the peak at 1.71 eV with $\lambda$ = 0.99  ($H_{\beta}-3\rightarrow H_{1\beta}$), and \protect\subref{subfig:subfig-b9-nto-3-4} to the peak at 6.15 eV 
with $\lambda$ = 0.25 ($H_{\beta}-3\rightarrow L_{\beta}+1$) and \protect\subref{subfig:subfig-b9-nto-5-6}  $\lambda$ = 0.32 ($H_{\beta}-2\rightarrow L_{\beta}$). 
Parameter $\lambda$ refers to a fraction of the NTO pair contribution to a given electronic excitation.}
\end{figure*}

The optical absorption spectrum of B$_9$ also has a very few intense peaks. Almost negligible absorption is seen at lower energies.
Compared to EOM-CCSD results, the functionals PBE0, B3LYP, B3PW91, M06 and M06-2X underestimate the peak position 
(\emph{cf.} Fig. \ref{fig:plot-nto-boron9}\subref{subfig:subfig-b9-eom-tddft-plot}). 
The spectrum of B$_9$ is extremely overestimated by LC-$\omega$PBE both in position and in the oscillator strengths. CAM-B3LYP 
and $\omega$B97xD again provides an excellent agreement with EOM-CCSD results on intensities as well as position of energy bands.
Peak at 1.71 eV is dominated by $\sigma \rightarrow \pi^*$ and at 6.15 eV, $\sigma \rightarrow \sigma^*$ transitions take place, as 
evident from the NTO analysis shown in Fig. \ref{fig:plot-nto-boron9}.

\section{\label{sec:conclusions-wheel}Summary}
The goal of the present study is to benchmark various exchange-correlation functionals used in TDDFT for calculating optical absorption 
spectra of planar boron wheel clusters. We compared results of TDDFT with eight different functionals to the results of a wavefunction 
based EOM-CCSD approach. 

Hybrid GGA functionals -- PBE0, B3LYP and B3PW91 -- are poor performers as they tend to underestimate the excitations energies.
Meta-GGA functionals M06 and M06-2X --which includes terms that depend on kinetic energy density -- also underestimate the excitation 
energies. Among the long-range corrected functionals CAM-B3LYP provides the best agreement with EOM results on the basis of excitation 
energies as well as spectrum profile. This fact has also been confirmed in other benchmarking studies done previously.\cite{jctc_oscillator,jctc_excitation} 

The contribution of configurations to the many-body wavefunctions of various excited states suggest that the excitations involved are of 
molecular type.\cite{plasmon,vlasta-cis-excited} Since most of the absorption takes place at higher energies, these clusters could potentially be used 
as ultraviolet absorbers.

Although this study neither includes all the functionals available nor does the test cases are comprehensive, it helps in providing 
reasonable comparison between the current gold standard single reference method, namely, EOM-CCSD and TDDFT, by identifying the functionals 
which provide results as good as EOM-CCSD in light of optical absorption calculations. These findings can be tested against more sophisticated 
multi-reference calculations. Such high-level calculations are necessary to design superior yet less time-consuming TDDFT approaches.


  \lhead{{\chaptername\ \thechapter.}{  Magnesium Clusters Mg$_{n}$ ($n$=2 -- 5)}}
   \chapter{\label{chap:main_magnesium}Theory of Linear Optical Absorption in Various Isomers of Magnesium Clusters Mg$_{n}$ ($n$=2 -- 5)}

Clusters of group II elements, such as magnesium, are of special interest because they have two valence electrons, quasi-filled closed shells,
and in bulk they are metals. In the case of small clusters, the bonding between atoms is expected to be of van der Walls type. This is evident in the case of extensively studied 
magnesium dimer. It exhibits large bond length of 3.92 \AA{} and 0.034 eV/atom binding energy.  However it is seen that, for larger clusters
this bonding becomes stronger. Thus, study of divalent metals is appropriate for evolution of various cluster properties and to test 
various theoretical methods. Involvement of metal atoms in the clusters makes theoretical treatment a demanding task, mainly because of 
several nearly degenerate electronic states. In such situations, only multi-reference configuration interaction methods or coupled cluster
singles doubles with perturbative triples (CCSD(T)) is known to provide best qualitative results. \cite{ahlrichs_mg_pccp} Since in this chapter,
we are dealing with small sized clusters of magnesium, treated at a large-scale multi-reference configuration interaction singles doubles level of theory,
the results will be superior to other \emph{ab initio} quantum chemical methods.

There has been an enormous study of equilibrium geometry and electronic structure of small magnesium clusters. 
\cite{ahlrichs_mg_pccp, car_kumar_magnesium_prb, exp_mg_dimer_spectra_jcp, akola_mg_epjd, ground_excited_mg2_jcp, jellinek_mg2-mg5_jpca, 
manninen_mg_evolution_epjd, kaplan_mg3_jcp, ele-struct-magnesium-pra} Andrey \emph{et al.}\cite{ele-struct-magnesium-pra} studied evolution of electronic structure of 
magnesium clusters with cluster size using all-electron density functional theoretical method. An evolution from non-metal to metal was explained
using a gradient-corrected \ac{DFT} calculations by Jellinek and Acioli\cite{jellinek_mg2-mg5_jpca} and by Akola \emph{et al.} \cite{manninen_mg_evolution_epjd}
Larger clusters were studied at \ac{DFT} level by K\"{o}hn \emph{et al.}\cite{ahlrichs_mg_pccp} Kumar and Car performed \emph{ab initio} density 
functional molecular dynamics study of smaller magnesium clusters within local density approximation.\cite{car_kumar_magnesium_prb} 
Stevens and Krauss calculated electronic structure of ground and excited states of Mg dimer using multi-configurational self consistent field 
method. \cite{ground_excited_mg2_jcp} Kaplan, Roszak and Leszczynski investigated the nature of binding in the magnesium trimer 
at \ac{MP4} level.\cite{kaplan_mg3_jcp} 

The optical absorption in dimer was studied experimentally by McCaffrey and Ozin in Ar, Kr and Xe matrices \cite{exp_mg_dimer_spectra_jcp}
and by Balfour and Douglas. \cite{exp_mg_dimer_douglas} Solov'yov \emph{et al.} calculated optical absorption spectra of global minimum 
structures of magnesium clusters using \ac{TDDFT} and compared the spectra with results of classical Mie theory.\cite{optical_mg_jpb}
However, to best of our knowledge, no other experimental or theoretical study exists for optical absorption and excited states calculations of various 
low-lying isomers of magnesium clusters. The distinction of different isomers of a cluster has to be made using an experimental or theoretical
technique in which the properties are size, as well as shape, dependent. Conventional mass spectrometry only distinguishes clusters according
to the masses. We have addressed this issue by performing large-scale correlated calculations of optical absorption spectra of various isomers of 
magnesium clusters Mg$_{n}$ (n=2--5), at \ac{MRSDCI} level of theory. Hence, our theoretical results can be coupled with the 
experimental measurements of optical absorption, to distinguish between different isomers of a cluster. Using this approach, we have reported 
results of such calculation on small boron and aluminum clusters \cite{smallboron, aluminum-ravi}.

In this chapter, we present results of systematic calculations of optical absorption in various low-lying isomers of magnesium clusters
using \emph{ab initio} large-scale \ac{MRSDCI} method. The nature of optical excitations involved in absorption has also been 
investigated by analyzing the wavefunctions of the excited states. Also, wherever possible, the results are compared with available literature.

Remainder of the chapter is organized as follows. Next section discusses theoretical and computational details of the calculations, followed
by section \ref{sec:results-magnesium}, in which results are presented and discussed. Conclusions and future directions are presented
in section \ref{sec:conclusions-magnesium}. A detailed information about wavefunctions of excited states contributing to various
 photoabsorption peaks is presented in the Appendix Table \ref{Tab:table_mg2_lin} -- \ref{Tab:table_mg5_param}.

\section{\label{sec:theory-magnesium}Theoretical and Computational Details}

\begin{figure}
\centering
\subfloat[\textbf{Mg$_{\mathbf{2}}$, D$_{\mathbf{\boldsymbol{\infty}h}}$,
$^{\mathbf{1}}\mathbf{\boldsymbol{\Sigma_{g}}}$ \label{subfig:subfig-mg2}}]{\includegraphics[width=3cm]{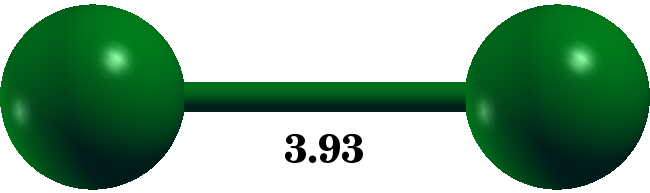} } \hfill 
\subfloat[\textbf{Mg$_{\boldsymbol{3}}$, D$_{\boldsymbol{3h}}$, $\boldsymbol{^{1}A_{1}^{'}}$}]{\includegraphics[width=2.8cm]{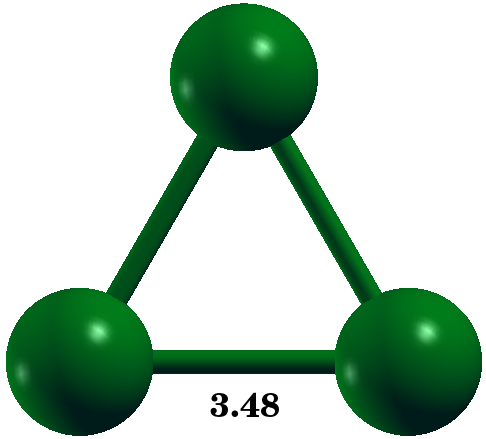} } \hfill 
\subfloat[\textbf{Mg$_{\boldsymbol{3}}$, D$_{\mathbf{\boldsymbol{\infty}h}}$, $\boldsymbol{^{3}\Pi_{u}}$}]{\includegraphics[width=4.3cm]{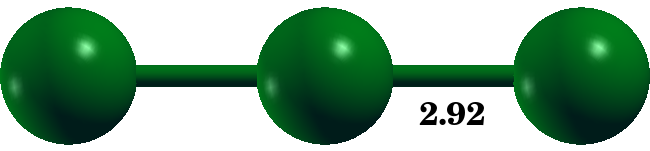} }\hfill 
\subfloat[\textbf{Mg$_{\boldsymbol{3}}$, C}$_{\boldsymbol{2v}}$, $\boldsymbol{^{3}A_{2}}$]{\includegraphics[width=2.5cm]{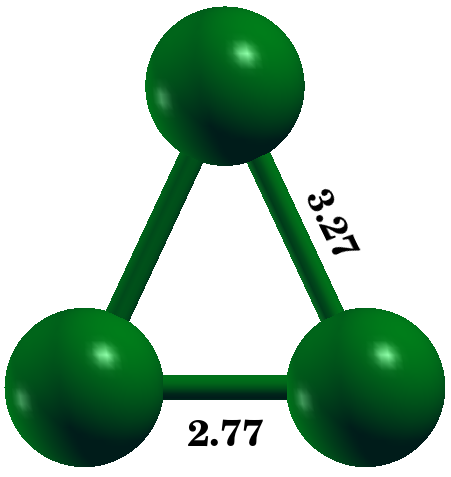} } \hfill 
\subfloat[\textbf{Mg$_{\boldsymbol{3}}$, C$_{\boldsymbol{2v}}$, $\boldsymbol{^{3}B_{1}}$}]{\includegraphics[width=3.cm]{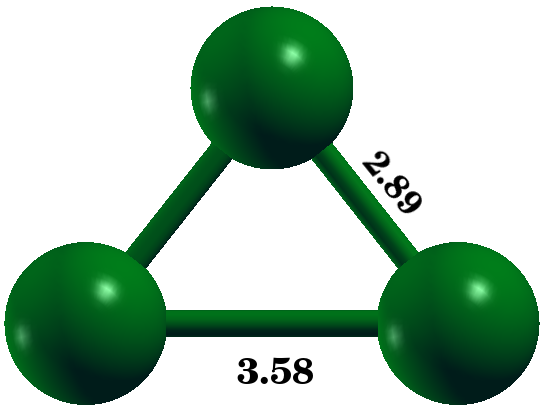} } \hfill 
\subfloat[\textbf{Mg$_{\boldsymbol{4}}$, T$_{\boldsymbol{d}}$, $\boldsymbol{^{1}A{}_{1}}$}]{\includegraphics[width=2.8cm]{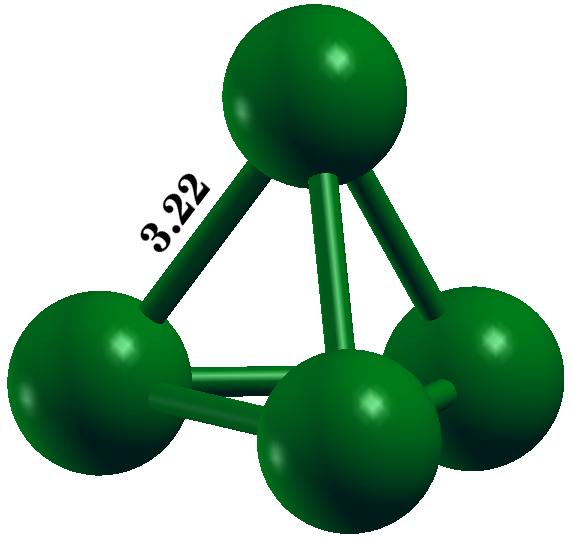} }\hfill 
\subfloat[\textbf{Mg$_{\boldsymbol{4}}$, D$_{\boldsymbol{2h}}$, $\boldsymbol{^{3}B{}_{3u}}$}]{\includegraphics[width=3.9cm]{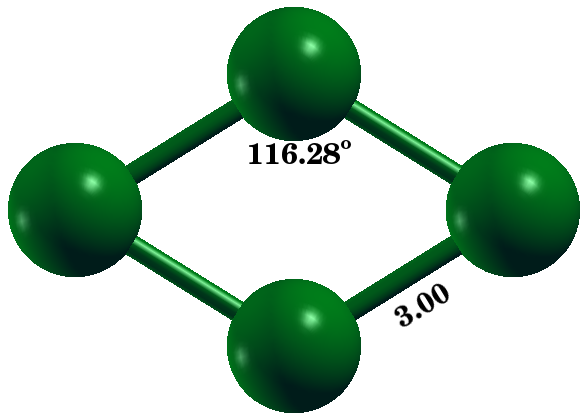} }\hfill 
\subfloat[\textbf{Mg$_{\boldsymbol{4}}$, D$_{\boldsymbol{4h}}$,} $\boldsymbol{^{3}B{}_{2u}}$]{\includegraphics[width=2.8cm]{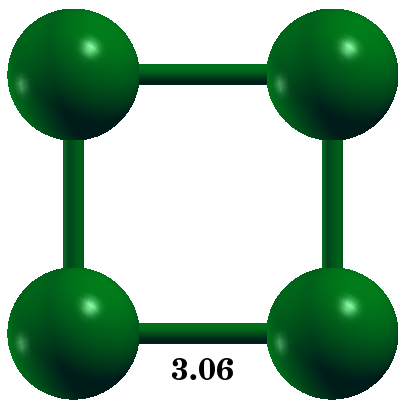} }\hfill 
\subfloat[\textbf{Mg$_{\boldsymbol{5}}$, D$_{\boldsymbol{3h}}$, $\boldsymbol{^{1}A{}_{1}^{'}}$}]{\includegraphics[width=3cm,angle=90]{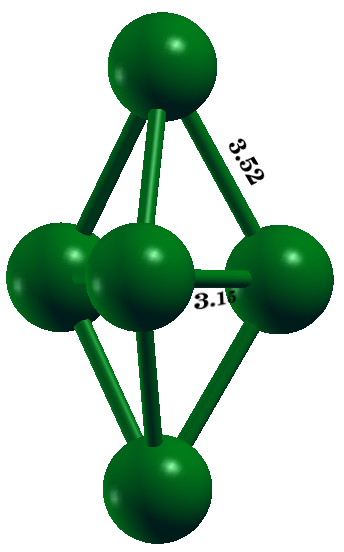} }\hfill 
\subfloat[\textbf{Mg$_{\boldsymbol{5}}$, C$_{\boldsymbol{4v}}$,} $\boldsymbol{^{1}A{}_{1}}$]{\includegraphics[width=4.5cm]{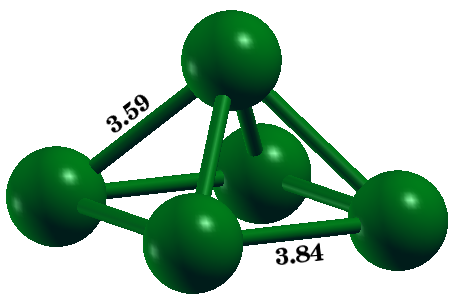}
} \vspace{0.2cm}
\caption{\label{fig:geometry-magnesium}Geometry optimized structures of magnesium clusters with point group symmetry and the electronic
ground state at the CCSD level. All numbers are in $\textrm{\AA}$ unit. }
\end{figure}

The geometry of various isomers were optimized using the size-consistent coupled-cluster singles-doubles (CCSD) method, as implemented in the
\textsc{gaussian09} package \cite{gaussian09}. A basis set of 6-311+G(d) was used which was included in the \textsc{gaussian 09} package 
itself. This basis set is optimized for the ground state calculations. Different spin multiplicities of the isomers were taken into account for 
the optimization to determine the true ground state geometry. The process of optimization was initiated by using the
geometries reported by Lyalin \emph{et al.}\cite{ele-struct-magnesium-pra}, based upon first principles DFT based calculations. 
Figure \ref{fig:geometry-magnesium} shows the final optimized geometries of the isomers studied in this chapter.

If the total number of orbitals used in a \ac{CI} expansion is $N$, the number of configurations in the calculation proliferates as $\approx N^{6}$,
which can become intractable for large values of $N$. To reduce the computation, we employed the so-called ``frozen-core approximation'', in 
which no virtual transitions are allowed from the chemical core orbitals of magnesium.

The linear photoabsorption spectrum of magnesium dimer was computed using full configuration interaction method. The spectra of 
various isomers of the remaining magnesium clusters were computed using MRSDCI method, as described in subsection \ref{subsection-mrsdci}.

\subsection{Choice of Basis Set}
\begin{figure}
\centering
\includegraphics[width=8cm]{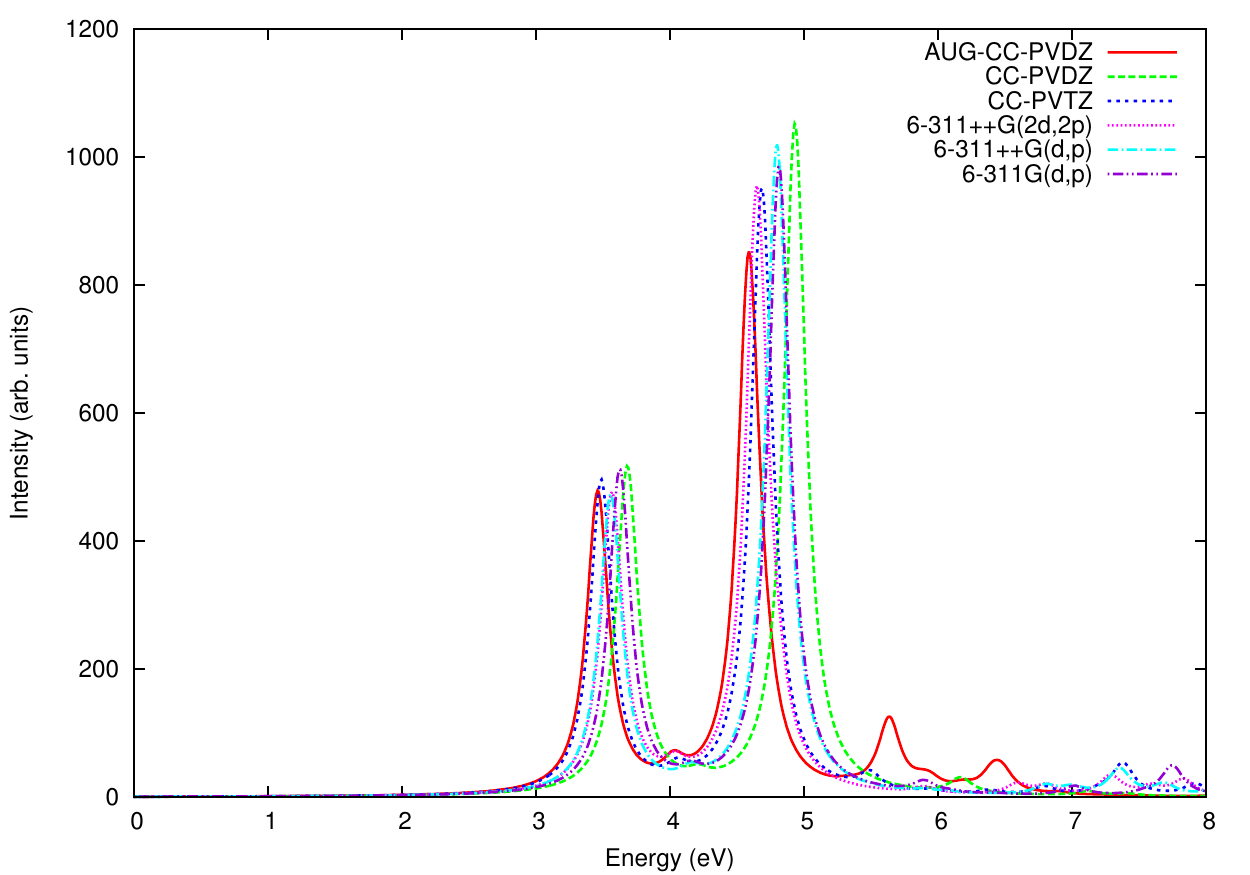} \vspace{0.2cm}
\caption{\label{fig:magnesium-basis-study}Optical absorption in Mg$_{2}$
calculated using various Gaussian contracted basis sets.}
\end{figure}

Electronic structure calculations generally depend upon the size and the quality of basis set used. To explore the basis set dependence
of computed spectra, we used several basis sets\cite{emsl_bas1,emsl_bas2} to compute the optical absorption spectrum of the magnesium dimer. 
For the purpose, we used basis sets named aug-cc-pVDZ, cc-pVDZ, cc-pVTZ, 6-311++G(2d,2p), 6-311++G(d,p) and 6-311G(d,p),
 which consist of polarization functions along with diffuse exponents.\cite{emsl_bas1,emsl_bas2} From the calculated spectra
presented in Fig. \ref{fig:magnesium-basis-study} the following trends emerge: the spectra computed by various correlation consistent  
basis sets (aug-cc-pVDZ, cc-pVDZ, cc-pVTZ) are in good agreement with each other in the energy range up to 5 eV, while those obtained 
using the other basis sets (6-311++G(2d,2p), 6-311++G(d,p) and 6-311G(d,p)) disagree with them substantially, particularly in the higher
energy range. Peaks at 5.6 eV and 6.5 eV are seen only in the spectrum calculated using augmented basis set. Because of the fact that augmented basis 
sets are considered superior for molecular calculations, we decided to perform calculations on the all the clusters using the aug-cc-pVDZ basis set.

\subsection{Size of the CI Expansion}

The electron correlation effects, both in ground state as well as excited states, were taken into account in our calculations by inclusion of 
relevant configurations in the reference space of \ac{MRSDCI} expansion. Larger the reference configuration space, larger will be the 
\ac{CI} expansion, which is prohibitive for bigger systems. A good chemical accuracy is obtained by moderately sized \ac{CI} expansion.  
 In Table \ref{tab:energies-irrep-magnesium} we present
the average number of reference states (N$_{ref}$) included in the \ac{MRSDCI} expansion and average number of configurations (N$_{total}$)
for different isomers. For a given isomer, the average has been calculated across different irreducible representations needed in
the calculations in order to compute the ground and various excited states. The extensiveness of our calculations
can be seen from the number N$_{total}$, which is $\approx$ 45000 for the simplest cluster, and around three million for each symmetry
subspace of Mg$_{5}$. 

Before we discuss the absorption spectrum for each isomer, we present the ground state energies along with the relative energies of each
isomer are given in Table \ref{tab:energies-irrep-magnesium}. 

\begin{table*}
\small
\centering
\begin{threeparttable}
  \caption{The average number of reference configurations (N$_{ref}$), and average
number of total configurations (N$_{total}$) involved in \ac{MRSDCI} calculations,
ground state (GS) energies (in Hartree) at the \ac{MRSDCI} level, relative
energies and correlation energies (in eV) of various isomers of magnesium clusters.\label{tab:energies-irrep-magnesium}}
\par

{\begin{tabular}{cccccc}
\hline
Cluster  	& Isomer  & N$_{ref}$  	& N$_{total}$  & GS energy 	& Relative \tabularnewline
		&  		 &  			&  			& (Ha) 		& energy (eV) \tabularnewline
\hline 
Mg$_{2}$  & Linear  			& 1\tnote{1}	&  44796		& -399.2847413 	& 0.00 \tabularnewline
		&  					&  			&  			&  			&  				\tabularnewline
Mg$_{3}$  & Equilateral Triangular  &    30		& 239465  	& -598.9270344 	& 0.00 \tabularnewline
	      & Linear		 		&  	55		&  460187		& -598.8759291 	& 1.39 \tabularnewline
	      & Isosceles Triangular-1	&   	34		&  516337		& -598.8569875	&  1.91\tabularnewline
	      & Isosceles Triangular-2 	&  	32		&  359780		& -598.8093768	&  3.20\tabularnewline
	      &  					&  			&  			&  			&  \tabularnewline
Mg$_{4}$  & Pyramidal 			&   32		& 2962035 	& -798.5781385 	& 0.00 \tabularnewline
	      & Rhombus 			&   29		& 1278632  	& -798.5405148 	&  1.02 \tabularnewline
	      & Square 				&   35		& 1319301  	& -798.5278160 	&  1.37 \tabularnewline
	      &  					&  			&  			&  				&  \tabularnewline
Mg$_{5}$  & Bipyramidal 		&  	11		& 3242198	&  -998.2044402  	& 0.00 \tabularnewline
	      & Pyramidal  			& 	28 		& 2215749 	& -998.1980062 	& 0.18 \tabularnewline
\hline
\end{tabular}}
  \begin{tablenotes}
    \item[1] Full Configuration Interaction calculation performed for Mg dimer.
  \end{tablenotes}
\end{threeparttable}
\end{table*}

\section{\label{sec:results-magnesium}MRSDCI Photoabsorption Spectra of Magnesium Clusters}

Next we present and discuss the results of our photoabsorption calculations for each isomer. 

\subsection{Mg$_{2}$}

The simplest cluster of magnesium is Mg$_{2}$ with D$_{\infty h}$ point group symmetry. 
We obtained its \ac{CCSD} optimized bond length to be 3.93 \AA{} (\emph{cf.} Fig. 
\ref{fig:geometry-magnesium}\subref{subfig:subfig-mg2}), which is in excellent agreement with the experimental 
value 3.89 \AA{}.\cite{ahlrichs_mg_pccp,exp_mg_dimer_spectra_jcp}
Using a DFT based methodology, several other theoretical values reported are in excellent agreement with our optimized bond length of 
magnesium dimer, i.e., Kumar and Car reported dimer bond length to be 3.88  \AA{} \cite{car_kumar_magnesium_prb} using density functional
molecular dynamics with simulated annealing, Janecek \emph{et al.} computed bond length to be 3.70 \AA{}\cite{wahl_mg_epjd} using \ac{DFT}
 with \ac{LDA} approximation, 3.8 \AA{} bond  length was reported by Stevens and Krauss using multi-configuration self-consistent 
field approach\cite{ground_excited_mg2_jcp}, 3.91 \AA{} bond length of dimer was computed by Jellinek and Acioli using DFT with BP86 exchange-correlation
functional\cite{jellinek_mg2-mg5_jpca} and Lyalin \emph{et al.} reported 3.926 \AA{} bond length using DFT with B3LYP exchange-correlation 
functional\cite{ele-struct-magnesium-pra}.

\begin{figure}
\centering
\includegraphics[width=8.3cm]{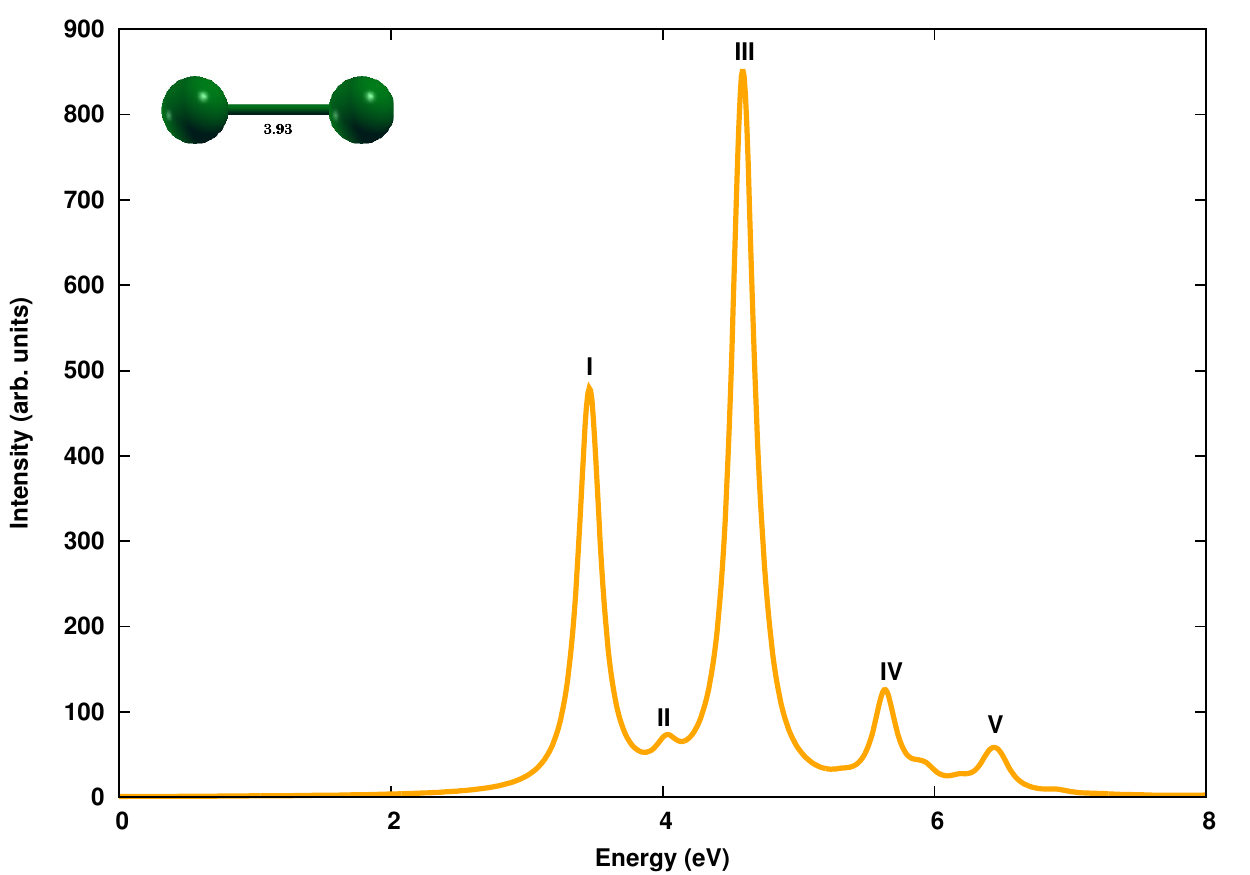}
\caption{\label{fig:plot-mg2-linear} The linear optical absorption
spectrum of Mg$_{2}$, calculated using the MRSDCI approach. For plotting the spectrum, a uniform linewidth of 0.1 eV was used.}
\end{figure}

The computed photoabsorption spectra of Mg$_{2}$, as shown in Fig. \ref{fig:plot-mg2-linear},
is characterized couple of intense peaks in the 3 -- 5 eV range and by weaker absorption at higher energies. The many-particle wavefunctions
of excited states contributing to the peaks are presented in Table \ref{Tab:table_mg2_lin}.  The first peak at 3.46 eV with absorption due to 
longitudinally polarized absorption is characterized by $ H \rightarrow L + 1 $ followed by a weaker absorption at 4 eV 
characterized by a transverse polarized $ H \rightarrow L + 8 $. The most intense peak occurs at 4.6 eV with dominant contribution
 from $ H - 1 \rightarrow L $ and $ H \rightarrow L + 3 $. This peak is also observed in the experimental photoabsorption spectrum at 4.62 eV
\cite{exp_mg_dimer_spectra_jcp}.  All these states exhibit strong mixing of singly-excited configurations. 
The wavefunctions of the excited states contributing to all the peaks exhibit strong configuration mixing,
instead of being dominated by single configurations, pointing to the plasmonic nature of the optical excitations.\cite{plasmon} 

The spectrum calculated using \ac{TDDFT} by Solov'yov \emph{et al.}\cite{optical_mg_jpb} is in excellent agreement with our results. 
In their calculations, first peak is seen at 3.3 eV followed by the most intense peak at 4.6 eV. The overall photoabsorption profile 
is also in accordance with our results.

\subsection{Mg$_{3}$}

We have optimized four low-lying geometries of magnesium trimer. The lowest energy structure at CCSD optimized level has equilateral
triangular shape with D$_{3h}$ symmetry and bond lengths of 3.48 \AA{}. This is in good agreement with other theoretical results reported, 3.51 \AA{} \cite{wahl_mg_epjd},
 3.48 \AA{} \cite{jellinek_mg2-mg5_jpca}, and 3.475 \AA{} \cite{ele-struct-magnesium-pra}. The next low-lying isomer of magnesium trimer 
is linear with D$_{\infty h}$ symmetry. The optimized bond length is found to be 2.92 \AA{}. The remaining two low-lying isomers have
 isosceles triangular shape, with C$_{2v}$ point group symmetry.  Not much has yet been reported on the bond lengths and electronic 
structure of these isomers. 

\begin{figure}
\centering
\includegraphics[width=8.3cm]{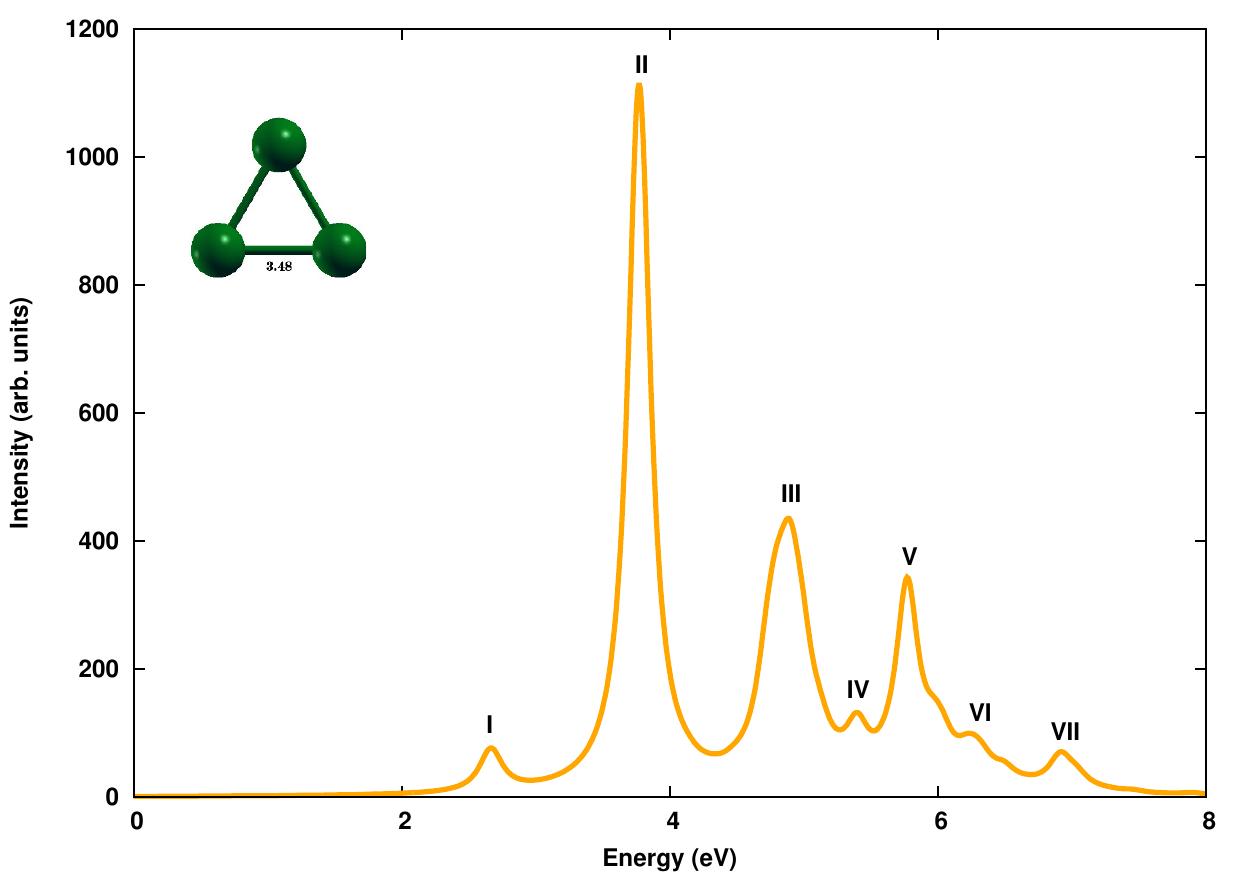}
\caption{\label{fig:plot-mg3-equil} The linear optical absorption spectrum of Mg$_{3}$ equilateral triangle isomer, calculated using
the MRSDCI approach. For plotting the spectrum, a uniform linewidth of 0.1 eV was used.}
\end{figure}

\begin{figure}
\centering
\includegraphics[width=8.3cm]{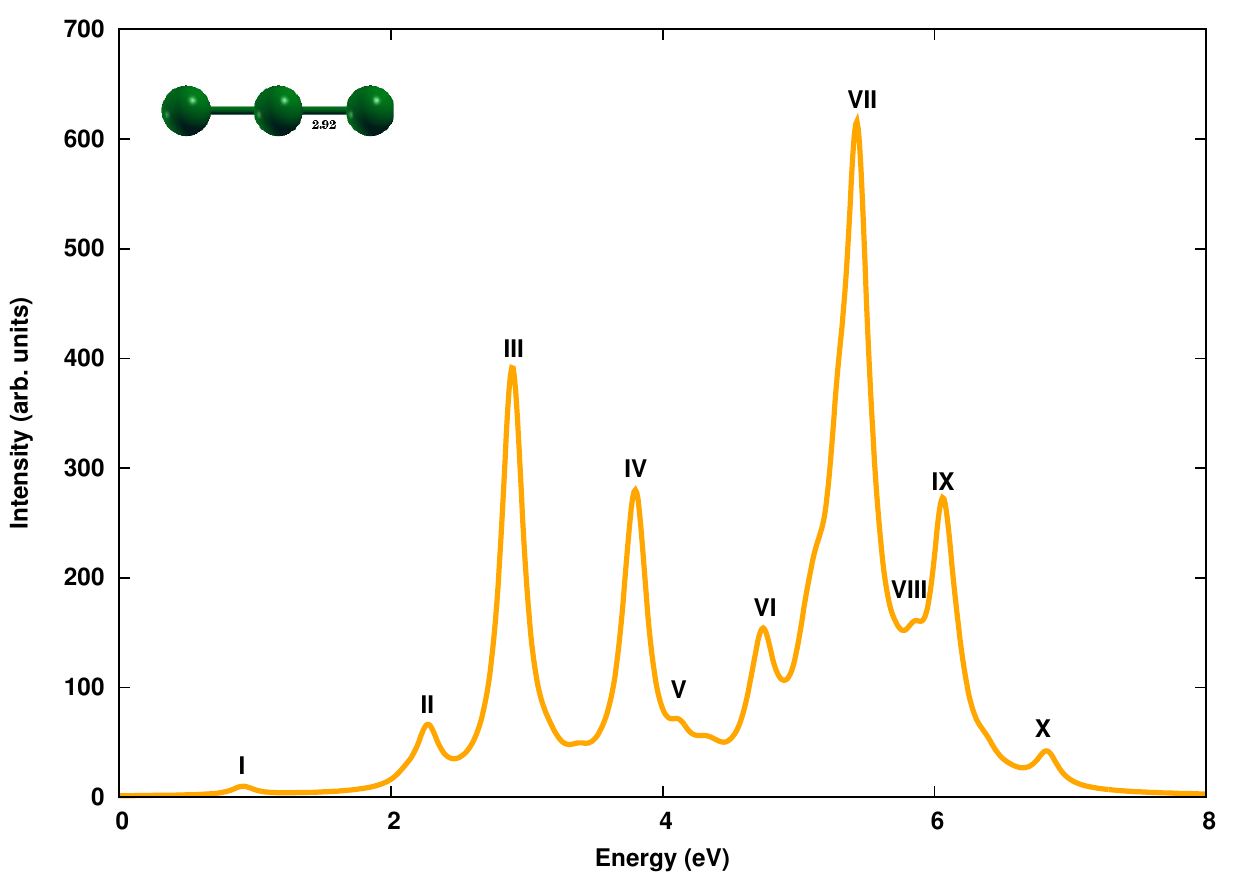}
\caption{\label{fig:plot-mg3-lin}The linear optical absorption spectrum of Mg$_{3}$ linear isomer, calculated using the MRSDCI approach.
For plotting the spectrum, a uniform linewidth of 0.1 eV was used. }
\end{figure}

\begin{figure}
\centering
\includegraphics[width=8.3cm]{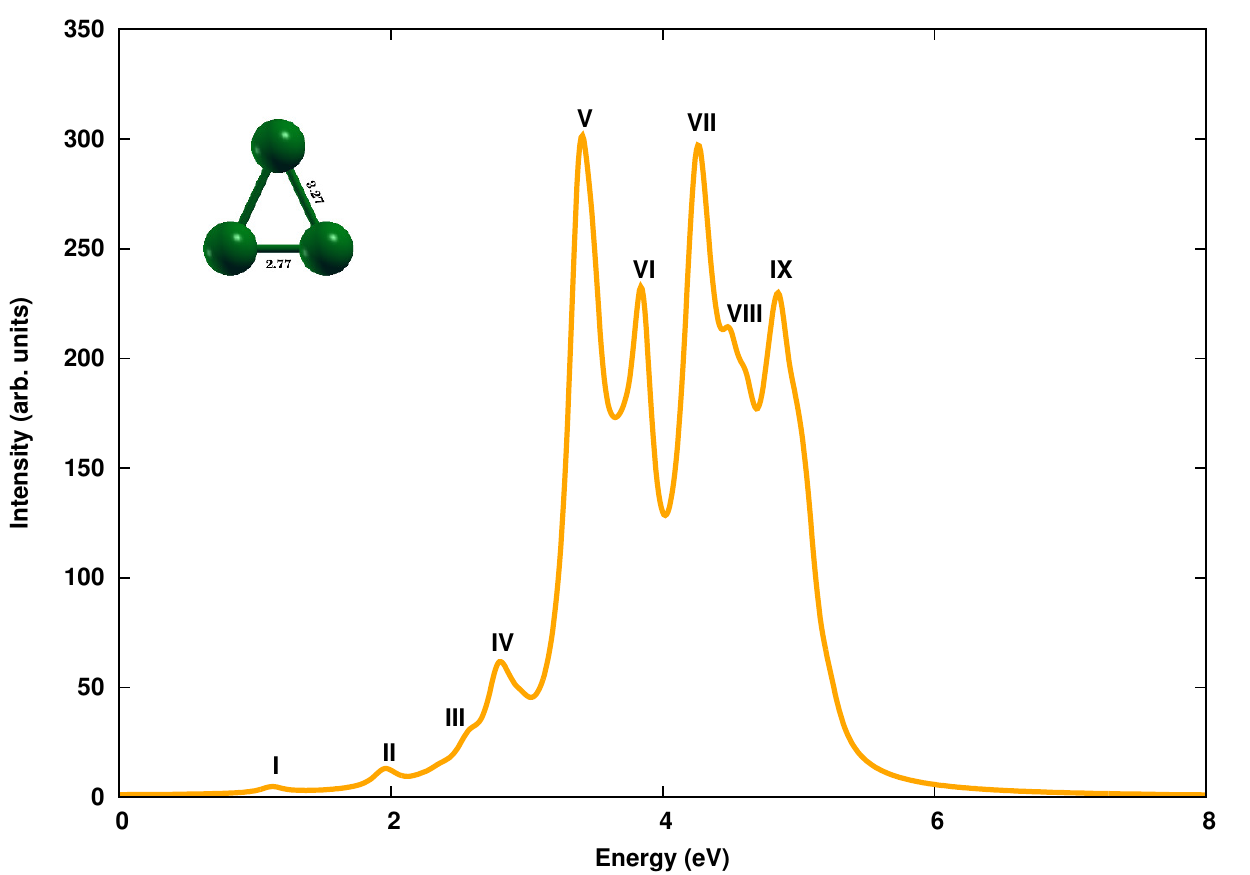}
\caption{\label{fig:plot-mg3-iso1}The linear optical absorption spectrum of Mg$_{3}$ isosceles triangle isomer-I, calculated using
the MRSDCI approach. For plotting the spectrum, a uniform linewidth of 0.1 eV was used.}
\end{figure}

\begin{figure}
\centering
\includegraphics[width=8.3cm]{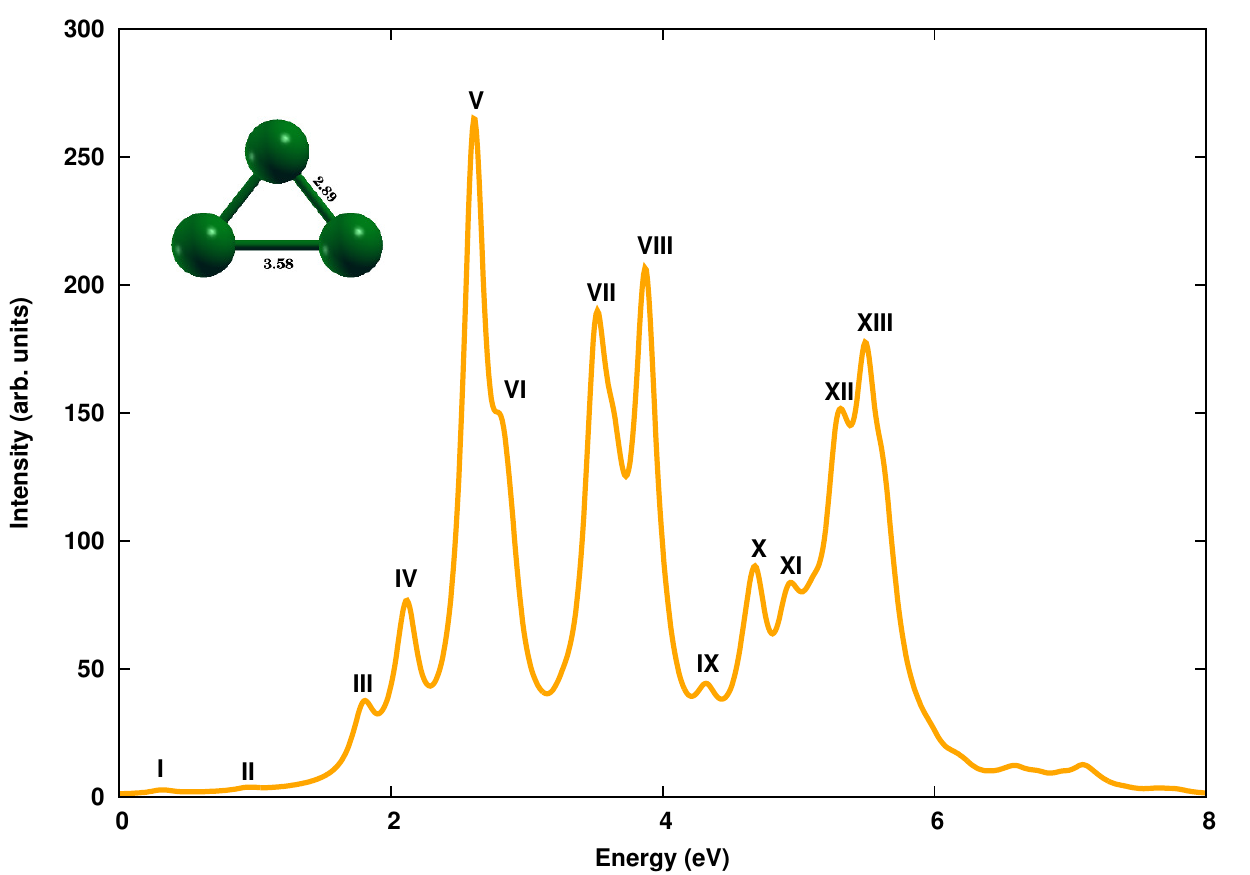}
\caption{\label{fig:plot-mg3-iso2}The linear optical absorption spectrum of Mg$_{3}$ isosceles triangle isomer-II, calculated using
the MRSDCI approach. For plotting the spectrum, a uniform linewidth of 0.1 eV was used.}
\end{figure}

The photoabsorption spectra of these isomers are presented in Figs.
\ref{fig:plot-mg3-equil}, \ref{fig:plot-mg3-lin}, \ref{fig:plot-mg3-iso1}  and \ref{fig:plot-mg3-iso2}.
The corresponding many-body wavefunctions of excited states corresponding
to various peaks are presented in Table \ref{Tab:table_mg3_equil}, \ref{Tab:table_mg3_lin}, \ref{Tab:table_mg3_iso1} and 
\ref{Tab:table_mg3_iso2} respectively. 
In the equilateral triangular isomer, bulk of the oscillator strength carried by peak at 3.7 eV.  
The absorption spectrum of linear isomer is altogether different with a number of peaks spread out in wide energy range, and, due to the 
polarization of light absorbed both parallel and perpendicular the axis of the trimer. 
On contrary, most of the oscillator strength in the absorption spectrum of isosceles triangular isomer-I is carried in the range of 3 -- 5 eV. 
The spectrum of isosceles triangular isomer-II shows a slightly red shifted with respect to the isosceles isomer-I, while peaks are observed 
in the entire ultraviolet range.

The optical absorption spectrum of equilateral triangular isomer consists of a weaker absorption peak at 2.6 eV characterized
by $H\rightarrow L$ and $H\rightarrow L+4$. This is followed by the most intense peak at 3.7 eV due to the light polarized both parallel and 
perpendicular to the plane of the isomer, and with dominant contribution from $H\rightarrow L$, $H\rightarrow L+2$ and $H-1\rightarrow L$.
This is confirmed by an experimental result of photoabsorption of Mg trimer in argon matrix, which shows a peak at 3.64 eV. \cite{exp_mg_dimer_spectra_jcp} 
Semi-major peaks at around 4.7 eV and 5.8 eV get dominant contribution from $H\rightarrow L+7$, $H\rightarrow L+5$ and $H\rightarrow L+9$ 
configurations. The latter being characterized by light polarized perpendicular to the plane of isomer.

Comparing our results with the spectrum obtained by \ac{TDDFT} calculations\cite{optical_mg_jpb}, we see a good agreement on overall profile
 of spectrum and excitation energies. First peak is observed at 2.5 eV followed by most intense one at 3.7 eV in the \ac{TDDFT} spectrum. 
Excitation energies and relative oscillator strengths are also in good agreement with our results. 

Because the ground state of Mg$_{3}$ linear isomer is a spin triplet, its many-particle
wavefunction predominantly consists of a configuration with two degenerate singly occupied molecular orbitals referred to as $H_{1}$
and $H_{2}$ in rest of the discussion. Naturally, the excited state wavefunctions will consist of configurations involving electronic excitations
from the occupied MOs (including singly occupied) to the unoccupied MOs starting from \ac{LUMO}. 
Linear trimer of magnesium cluster records absorption in the entire energy range. Very feeble peaks are observed at 0.9 eV and 2.3 eV,
 characterized by $H_1\rightarrow L+8$ and $H_1\rightarrow L+4$ respectively. The semi-major peak at 2.9 eV get dominant 
contribution from $H_1\rightarrow L+3$. The most intense peak at 5.4 eV has almost equal contribution from $H - 2 \rightarrow L$  and
$H - 1\rightarrow L+2$. The absorption due to light polarized along the trimer contributes to the lower energy part of the spectrum, while light
polarized perpendicular to the trimer contributes to the remaining higher energy part of the spectrum.

Both isosceles triangular isomers have a spin triplet ground state, hence their excited state wavefunctions will consist of configurations 
involving electronic excitations from singly occupied $H_1$ and $H_2$ molecular orbitals, in addition to other doubly occupied orbitals.
In the case of isosceles triangular isomer - I, the spectrum starts with a very feeble peak at 1.1 eV with contribution from $H_1 \rightarrow L + 1$
configuration. However, most of the absorption takes place in the energy range of 3 -- 5 eV, with two equally intense peaks at 3.4 eV and 4.2 eV.
The former is characterized by $H - 2 \rightarrow L$, $H_1\rightarrow L+3$ and $H - 1 \rightarrow L+1$. 
The wavefunctions of excited states corresponding to most of the peaks show a strong mixing of doubly-excited configurations, 
as is the case with the strongest peak at 4.2 eV.

The isosceles triangular isomer -II shows a red-shifted spectrum in comparison to the former isomer, with a distinction of well separated peaks.
The most intense peak at 2.6 eV gets dominant contribution from $H_2 \rightarrow L$ and a doubly-excited 
configuration $H_2 \rightarrow L; H_1\rightarrow L+10$. Two almost equally intense peaks of absorption due to in-plane polarization 
at 3.5 eV and 3.9 eV are characterized by $H - 2 \rightarrow H_1$ configurations, along with $H_1 \rightarrow L + 7$ and 
$H_1 \rightarrow L + 17$ respectively. This isomer also exhibits a strong mixing of doubly-excited configurations in the excited states.

\subsection{Mg$_{4}$}

The most stable isomer of Mg$_{4}$ cluster is a pyramidal / tetrahedron type isomer, with T$_{d}$ point group symmetry and 3.22 \AA{} optimized bond length.
Previously reported bond lengths 3.09 \AA{} \cite{ahlrichs_mg_pccp}, 3.33 \AA{} \cite{wahl_mg_epjd},
3.18 \AA{} \cite{jellinek_mg2-mg5_jpca}, 3.31 \AA{} \cite{manninen_mg_evolution_epjd} and 3.32 \AA{} \cite{kaplan_mg3_jcp} are in very good
agreement with our results. The rhombus isomer with D$_{2h}$ point group symmetry lies 1.02 eV above 
the global minimum structure. The optimized bond length is 3.0 \AA{} with acute angle of 63.5. Square isomer with D$_{4h}$ point group symmetry
and ${}^3B_{2u}$ electronic ground state lies 1.37 eV from the most stable structure. 

\begin{figure}
\centering
\includegraphics[width=8.3cm]{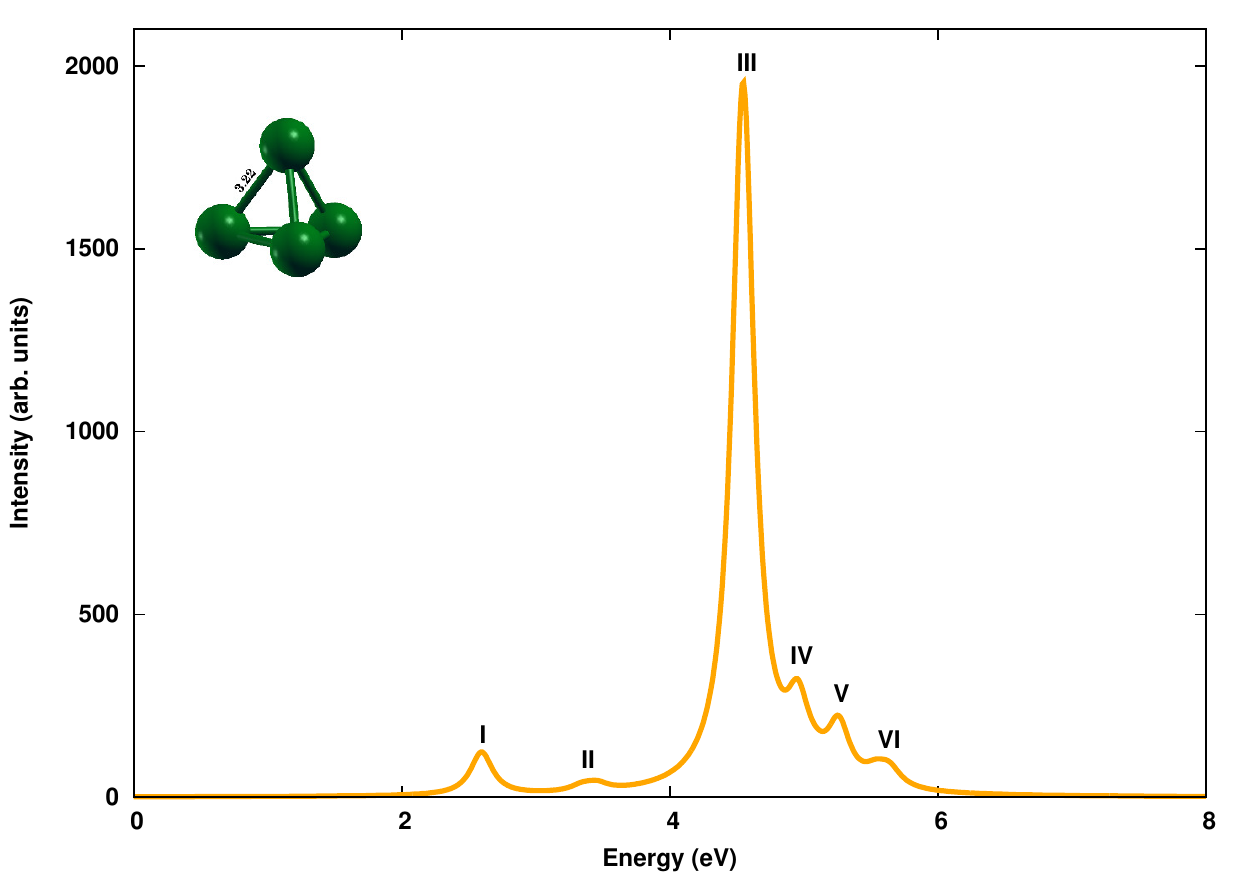}
\caption{\label{fig:plot-mg4-pyra}  The linear optical absorption spectrum of pyramidal Mg$_{4}$ isomer, calculated using the MRSDCI approach.
For plotting the spectrum, a uniform linewidth of 0.1 eV was used.}
\end{figure}

\begin{figure}
\centering
\includegraphics[width=8.3cm]{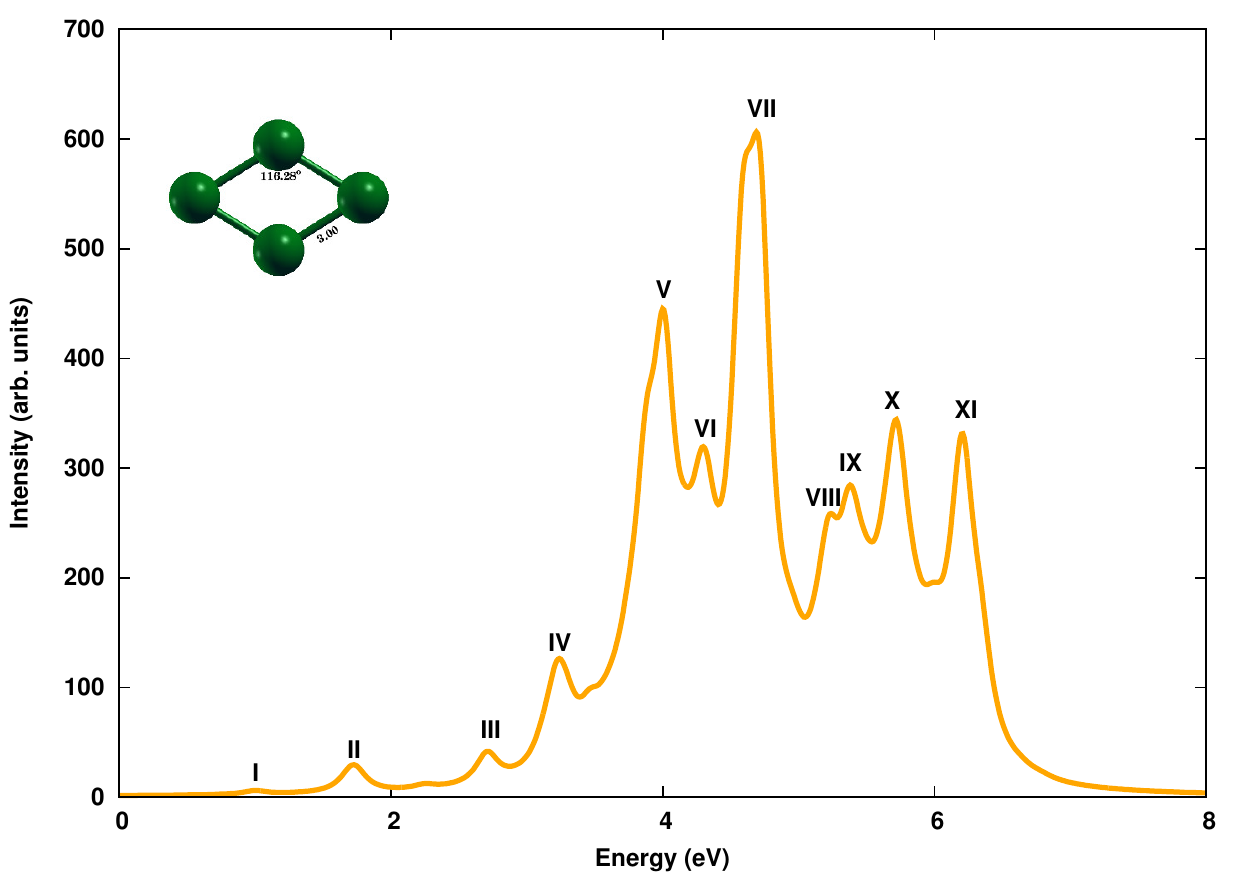}
\caption{\label{fig:plot-mg4-rho} The linear optical absorption spectrum of rhombus Mg$_{4}$, calculated using the MRSDCI approach.
For plotting the spectrum, a uniform linewidth of 0.1 eV was used.}
\end{figure}

\begin{figure}
\centering
\includegraphics[width=8.3cm]{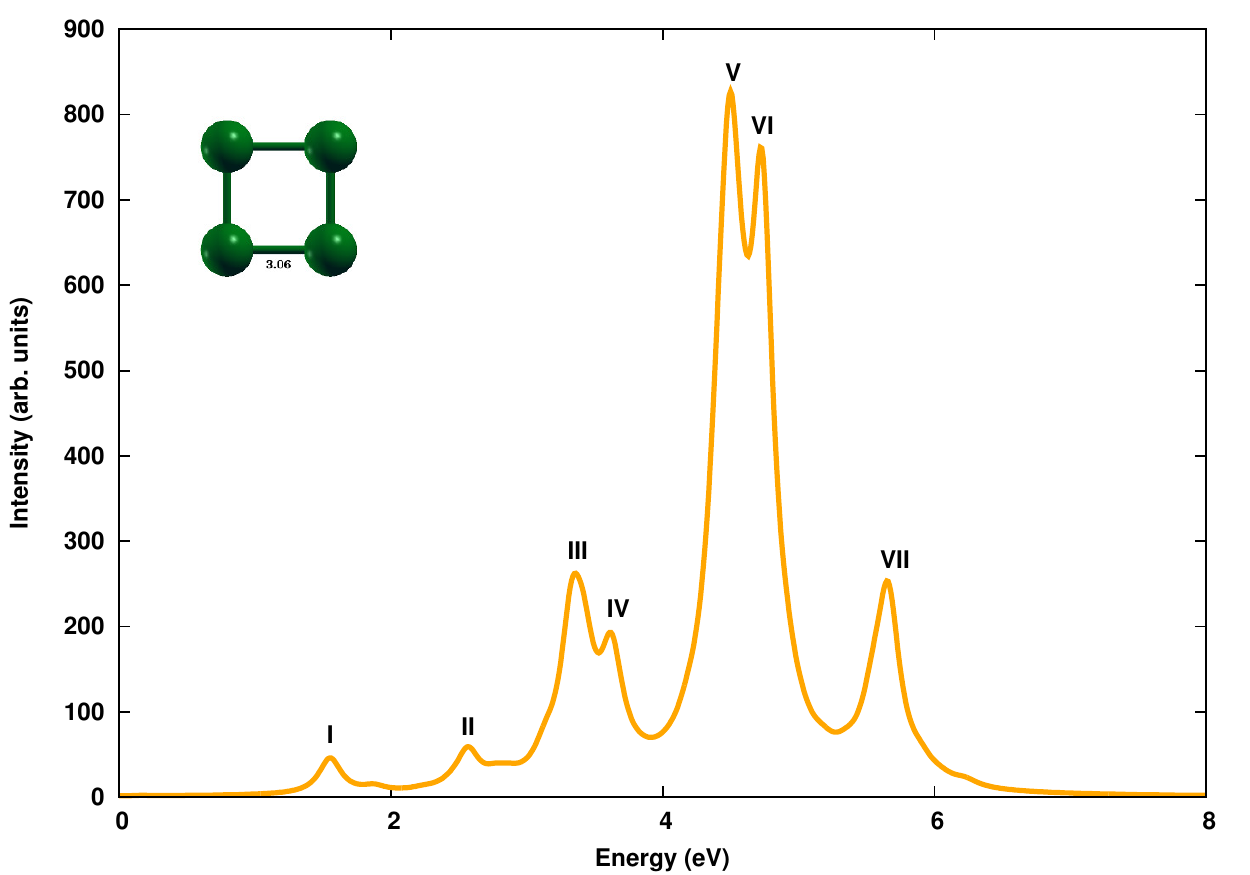}
\caption{\label{fig:plot-mg4-sqr}  The linear optical absorption spectrum of square Mg$_{4}$, calculated using the MRSDCI approach.
For plotting the spectrum, a uniform linewidth of 0.1 eV was used.  }
\end{figure}

The absorption spectra of pyramidal, rhombus and square isomers are presented in Figs. \ref{fig:plot-mg4-pyra}, \ref{fig:plot-mg4-rho},
 and \ref{fig:plot-mg4-sqr} respectively and many-particle wavefunctions of excited states contributing to various 
peaks are presented in Table \ref{Tab:table_mg4_pyra}, \ref{Tab:table_mg4_rho} and \ref{Tab:table_mg4_sqr} respectively. 
The onset of absorption spectrum of pyramidal isomer is seen at 2.6 eV with dominant contribution coming 
from $H - 1 \rightarrow L $, $H\rightarrow L $ and $H - 2 \rightarrow L$ configurations. The absorption spectrum of pyramidal isomer
 shows a very strong absorption at 4.5 eV due to light polarized in all three directions. It is exhibited 
by $H\rightarrow L+ 2$, $H-1\rightarrow L+1$ \emph{etc}. electronic excitations.

The \ac{TDDFT} absorption spectrum reported by Solov'yov \emph{et al.}\cite{optical_mg_jpb}, is slightly red-shifted, however its 
absorption pattern is similar compared to our calculated spectrum. A single most intense peak is seen at 4.2 eV followed by several less intense peaks.

In case of rhombus isomer, the bulk of the oscillator strength is distributed in the energy range 4 -- 6 eV. Several equally intense and closely-lying
peaks are observed in this range. The most intense peak, at 4.7 eV, is characterized by $H-1\rightarrow L+1$ along with its shoulder at 4.6 eV,
which gets dominant contribution from $H_2\rightarrow L+8$. Both are due to light polarized in the plane of the isomer. 

The absorption spectrum of the square structure is slightly blue-shifted as compared to the rhombus
and red-shifted as compared to pyramidal isomer, with the majority of absorption occurring in the energy range 3--6 eV. 
The onset of absorption spectrum occurs at 1.5 eV with peak due to light polarized in the plane of isomer, and characterized 
by $H_1\rightarrow L+15$ and $H_1\rightarrow L+10$. Square isomer also exhibits two very closely spaced most intense peaks, as is observed in 
rhombus counterpart. These peaks at 4.5 eV and 4.7 eV have contribution from singly-excited configurations such as $H -1 \rightarrow L$ and
$H_1\rightarrow L+24$ respectively as well as doubly-excited configurations.

\subsection{Mg$_{5}$}
We optimized geometries of two isomers of Mg$_{5}$: (a) bipyramid with the D$_{3h}$ symmetry, and (b) a pyramidal with the C$_{4v}$ point group 
symmetry. The lowest lying pentagon isomer, has $^{1}$A$_{1}^{'}$ electronic ground state, and is just 0.18 eV lower in energy as
 compared to the pyramid structure. 
Our optimized geometry for the bipyramid corresponds to unique bond lengths of 3.15 \AA{} and 3.52 \AA{}, as against 3.00 \AA{}, 3.33 \AA{}
reported by J. Jellinek and Acioli \cite{jellinek_mg2-mg5_jpca}; and 3.09 \AA{}, 3.44 \AA{} reported by Andrey \emph{et al}. \cite{ele-struct-magnesium-pra}

\begin{figure}
\centering
\includegraphics[width=8.3cm]{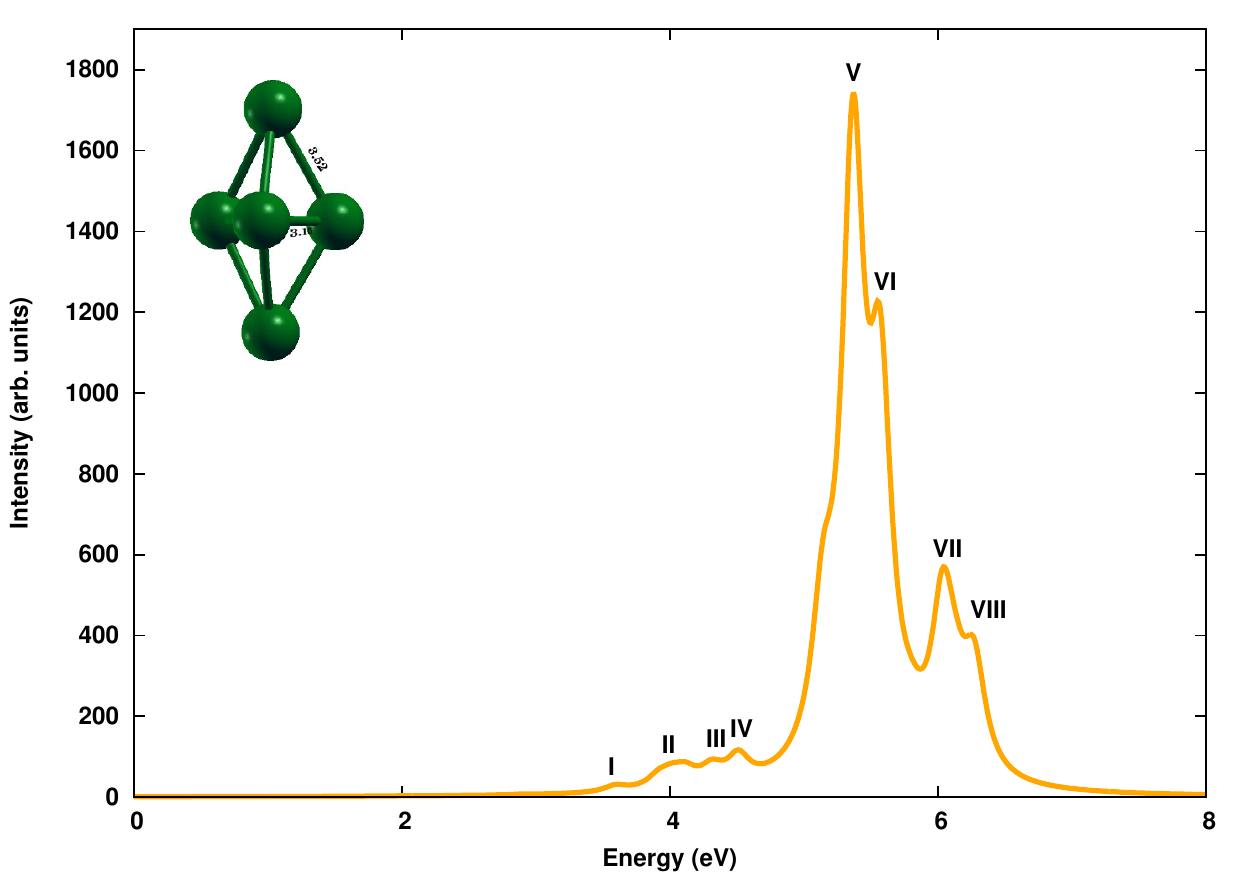}
\caption{\label{fig:plot-mg5-biprism} The linear optical absorption spectrum of bipyramidal Mg$_{5}$ isomer, calculated using the MRSDCI approach.
For plotting the spectrum, a uniform linewidth of 0.1 eV was used.}
\end{figure}

The bipyramidal isomer of Mg$_{5}$ cluster, exhibits an absorption spectrum very different from other isomers, as displayed 
in Fig. \ref{fig:plot-mg5-biprism}. The many-particle wavefunctions of excited states contributing to the 
peaks are presented in Table \ref{Tab:table_mg5_biprism}. The optical absorption spectrum of bipyramidal Mg$_{5}$ has no absorption until 3.5 eV,
while most of the absorption takes place in a narrow energy range 5.3 -- 6.3 eV. The absorption spectrum begins at 3.6 eV, with a very feeble 
peak with contribution from  $H-1\rightarrow L+4$ configuration. This is followed by several such smaller peaks. The most intense peak at 5.4 eV
has dominant contribution from  $H\rightarrow L+1$ and  $H-1\rightarrow L+3$ with absorption polarized along $y-$direction, which is in the 
plane of the triangle of the pyramid. A shoulder peak at 5.6 eV however has absorption due to light polarized along $z-$ direction, which is along
the larger dimension of the isomer.

The \ac{TDDFT} spectrum computed by Solov'yov \emph{et al.}\cite{optical_mg_jpb} shows optical activity in the energy range of 2 --4 eV, which is not observed in our calculated 
spectrum. However, a quasi-continuous spectrum is seen at higher energies in both calculations. 

The entire absorption spectrum of pyramidal isomer is highly red-shifted as compared to the bipyramidal isomer. 
A few feeble peaks occur in the low energy range in the optical absorption of pyramidal isomer. 
The many-particle wavefunctions of excited states contributing to the 
peaks are presented in Table \ref{Tab:table_mg5_param}. The onset of spectrum occurs at 2.2 eV with 
absorption due to polarization both perpendicular as well as in the plane of base of pyramid. It is characterized by  $H - 1 \rightarrow L$ and
 $H\rightarrow L+2$. An intense peak at 3.5 eV separates itself from the most intense one at 4.2 eV. The former gets dominant contribution
from  $H-1\rightarrow L+3$ and  $H-2\rightarrow L$ configurations. While the most intense peak is due to light polarized perpendicular to the basal
plane of pyramid, and contributed from  $H -2 \rightarrow L+1$ and  $H-2\rightarrow L+3$.  
Pyramidal isomer shows more optical absorption in the high energy range, with peaks within regular intervals of energy with declining intensities,
in contrast to single major peak observed in the spectrum of bipyramidal isomer. These differences can lead to identification of isomers 
produced experimentally.

\begin{figure}
\centering
\includegraphics[width=8.3cm]{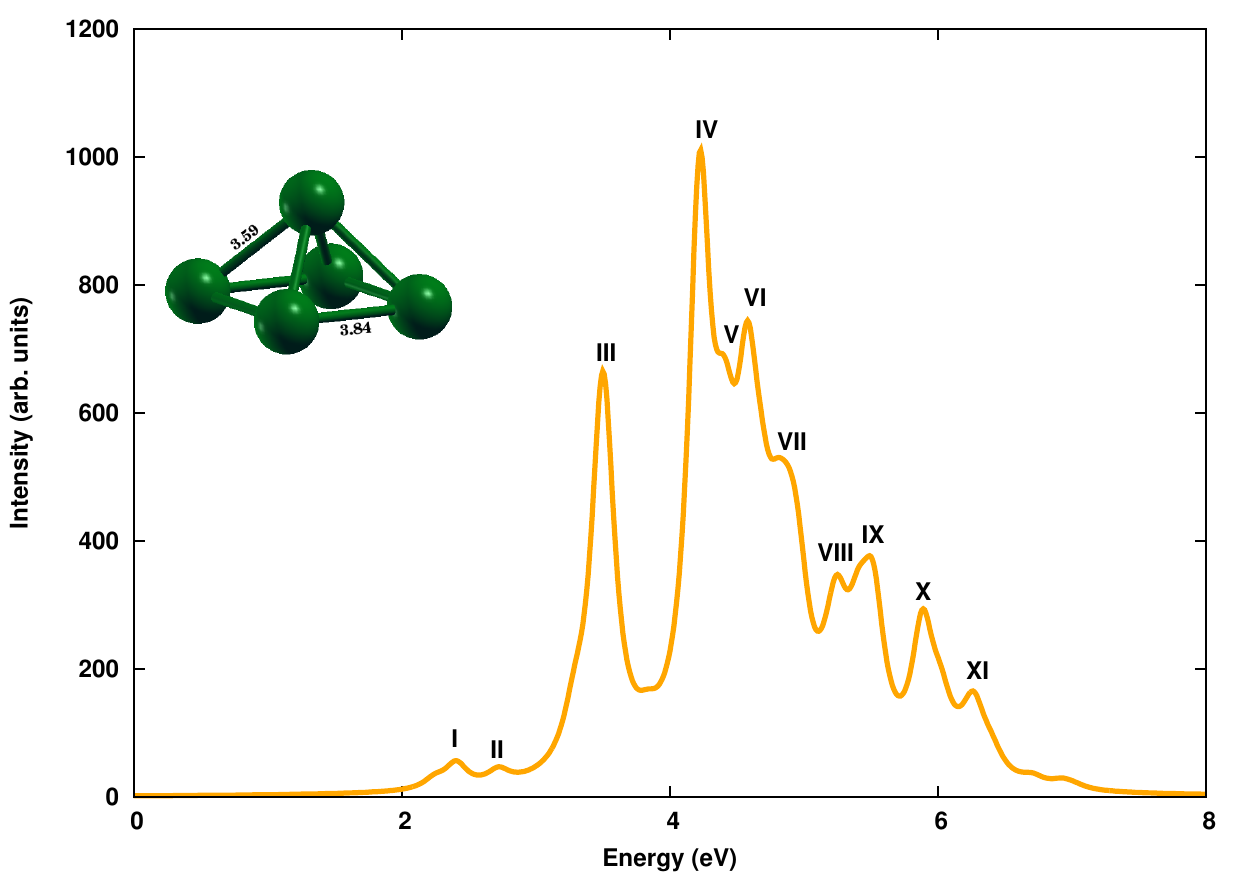}
\caption{\label{fig:plot-mg5-param} The linear optical absorption spectrum of pyramidal Mg$_{5}$, calculated using the MRSDCI approach.
For plotting the spectrum, a uniform linewidth of 0.1 eV was used. }
\end{figure}

\FloatBarrier
 \section{\label{sec:conclusions-magnesium}Summary}

we have presented large-scale all-electron correlated calculations of optical absorption spectra of several low-lying isomers
of magnesium clusters Mg$_{n}$, (n=2--5). We computed the ground state and excited states of magnesium dimer using one of the best possible 
electronic structure methods, namely full configuration interaction with frozen-core approximation. In case of remaining clusters, both ground and excited
state calculations were performed at \ac{MRSDCI} level, which take electron correlations into account at a sophisticated level. We have analyzed
the nature of low-lying excited states. Isomers of a given cluster show a distinct signature spectrum, indicating a strong-structure
property relationship. This fact can be used in experiments to distinguish between different isomers of a cluster. 
Our calculations suggests that the optical excitations involved are found to be collective type because of a strong mixing of configurations, 
and hence are plasmonic in nature. Owing to the sophistication of our calculations, our results can be used for benchmarking of the 
absorption spectra as well as for designing functionals for superior \ac{TDDFT} results.

\FloatBarrier
  \lhead{{\chaptername\ \thechapter.}{  Conclusion and Outlook}}
   \chapter{\label{chap:main_conclusion}Conclusions and Outlook}

 We performed systematic large-scale all-electron correlated calculations on boron B$_{n}$, aluminum Al$_{n}$ and 
magnesium Mg$_{n}$ clusters (n=2--5), to study their linear optical absorption spectra. Several possible isomers of each cluster were 
considered, and their geometries were optimized at the coupled-cluster singles doubles (CCSD) level of theory. Using the optimized 
ground-state geometries, excited states of different clusters were computed using the multi-reference singles-doubles 
configuration-interaction (MRSDCI) approach, which includes electron correlation effects at a sophisticated 
level. These CI wavefunctions were used to compute the transition dipole matrix elements connecting the ground and various excited states
of different clusters, eventually leading to their linear absorption spectra. The convergence of our results with respect to the basis
sets, and the size of the CI expansion was carefully examined. 
Isomers of a given cluster show a distinct signature spectrum, indicating a strong-structure 
property relationship. This fact can be used in experiments to distinguish between different isomers of a cluster. Owing to the sophistication of
our calculations, our results can be used for benchmarking of the absorption spectra and be used to design superior time-dependent density
functional theoretical (TDDFT) approaches. The contribution of configurations to many-body wavefunction of various excited states 
suggests that in most cases optical excitations involved are collective, and plasmonic in nature. \par

In addition, we calculated the optical absorption in various isomers of neutral boron B$_{6}$ and cationic boron B$_{6}^{+}$ clusters using 
computationally less expensive configuration interaction singles (CIS) approach, and benchmarked these results against more sophisticated 
equation-of-motion (EOM) CCSD based approach. In all closed shell systems, a complete agreement on the nature of configurations 
involved is observed in both methods. On the other hand, for open-shell systems, minor contribution from double excitations are observed,
which are not captured in the CIS method. \par

Optical absorption in planar boron clusters in wheel shape, B$_{7}$, B$_{8}$ and B$_{9}$ computed using EOM-CCSD approach, have been 
compared to the results obtained from TDDFT approach with a number of functionals. Hybrid GGA functionals -- PBE0, B3LYP and
 B3PW91 -- are also poor performers as they tend to underestimate the excitations energies.
Meta-GGA functionals M06 and M06-2X --which includes terms that depend on kinetic energy density -- also underestimate the excitation 
energies. Among the long-range corrected functionals CAM-B3LYP provides the best agreement with EOM results on the basis of excitation 
energies as well as spectrum profile. The contribution of configurations to the many-body wavefunctions of various excited states
 suggest that the excitations involved are of molecular type. Since most of the absorption takes place at higher energies, these clusters 
could potentially be used as ultraviolet absorbers. Although this study neither includes all the functionals available nor does the test 
cases are comprehensive, it helps in providing reasonable comparison between the current gold standard single reference method, 
namely, EOM-CCSD and TDDFT, by identifying the functionals which provide results as good as EOM-CCSD in light of optical 
absorption calculations. These findings can be tested against more sophisticated multi-reference calculations. Such high-level calculations 
are necessary to design superior yet less time-consuming TDDFT approaches.

The large-scale configuration interaction based approach can be extended to study non-linear optical phenomenon in clusters, such as excited state
absorption, multi-photon absorption, higher-harmonic generation \emph{etc}. We also look forward to use such a sophisticated approach to study
ab initio photo-emission spectra of atomic and molecular clusters.

For large and extended systems, such as, clusters of ceramic materials (BeO)$_{n}$, (Li$_{2}$O)$_{n}$ and (B$_{2}$O$_{3}$)$_{n}$ (n = 2 -- 20) 
calculations of linear optical absorption using  TDDFT approach are currently under consideration. Results for these systems will be communicated soon. Some of these 
clusters also show reversible hydrogen adsorption properties. For instance, (BeO)$_{n}$ shows remarkable hydrogen adsorption, meeting US Department
of Energy ultimate goal of 7.5 wt/\% hydrogen adsorption. This study was performed in collaboration, whose results are now under revision.

An \emph{ab initio} description of \ac{ARPES} of 2-D surfaces and 3-D bulk, using TDDFT, is also under consideration. This
will help enormously in explaining the experimental \ac{ARPES} spectra, which are usually obtained for bulk or 2-D samples.

 	\pagestyle{fancy}
 	\renewcommand{\sectionmark}[1]{\markright{\thesection\ #1}}
 	\fancyhf{}
 	\lhead{\nouppercase{\leftmark}}
  	\rhead{\href{home.iitb.ac.in/~ravindra.shinde}{ Ravindra Shinde}}
 	\cfoot{Page \thepage\ of \pageref{LastPage}}

\appendix
  \begin{footnotesize}	
  \singlespacing
 \appendixpage
\chapter{Boron Clusters B$_{n}$ (n = 2 -- 5)}

\label{app:wavefunction-boron2-5} In the following tables, we have given the
excitation energies (with respect to the ground state), and the many-body
wavefunctions  of the excited states, corresponding
to the peaks in the MRSDCI photoabsorption spectra of various boron cluster isomers listed
in Fig. \ref{fig:geometries-nanolife}, and discussed in Chapter \ref{chap:main_smallboron}. 
along with the oscillator strength $f_{12}$ of the transitions,
\begin{equation}
f_{12}=\frac{2}{3}\frac{m_{e}}{\hbar^{2}}(E_{2}-E_{1})\sum_{i}|\langle m|d_{i}|G\rangle|^{2}
\end{equation}
where, $|m\rangle$ denotes the excited state in question, $|G\rangle$,
the ground state, and $d_{i}$ is the $i$-th Cartesian component
of the electric dipole operator. The single excitations are with respect to the reference state as given in
respective tables.

\section*{Excited State CI Wavefunctions, Energies and Oscillator Strengths}
\LTcapwidth=\columnwidth
\begin{longtable}{ccccl}
\captionsetup{font=footnotesize}
\caption{\label{Tab:table_b2_lin} Excitation energies, $E$, and many-particle wavefunctions of excited
states corresponding to the peaks in the linear absorption spectrum
of B$_{2}$ (\emph{cf}. Fig. \ref{fig:b2_lin_final}), along with
the oscillator strength of the transitions. Longitudinal and transverse polarization
corresponds to the absorption due to light polarized along and perpendicular
to the molecular axis respectively. In the wavefunction, the bracketed
numbers are the CI coefficients of a given electronic configuration.
Symbols $H_{1}$,$H_{2}$ denote SOMOs discussed earlier, and $H$,
and $L$, denote HOMO and LUMO orbitals respectively. $HF$ denotes
the Hartree-Fock configuration. }  \\

\hline \multicolumn{1}{c}{Peak} & \multicolumn{1}{c}{ $E$ (eV)} & \multicolumn{1}{c}{$f_{12}$ } 
 & \multicolumn{1}{c}{Pol.} &\multicolumn{1}{l}{Wave Function}     \\ \hline 
\endfirsthead

\multicolumn{5}{c}%
{{\scshape \tablename\ \thetable{} -- continued from previous page}} \\ 

\hline \multicolumn{1}{c}{Peak} & \multicolumn{1}{c}{ $E$ (eV)} & \multicolumn{1}{c}{$f_{12}$ } 
 & \multicolumn{1}{c}{Pol.} & \multicolumn{1}{l}{Wave Function}     \\ \hline 
\endhead

\hline \multicolumn{5}{r}{{{\scshape Continued on next page}}} \\ 
\endfoot

\hline \hline
\endlastfoot
GS$^a$
 &  &  &  & $|HF\rangle$ (0.8673)\tabularnewline
 &  &  &  & $|H-1\rightarrow L;H-1\rightarrow L\rangle$(0.2706)\tabularnewline
 &  &  &  & $|H-1\rightarrow L;H-1\rightarrow L+4\rangle$(0.1283)\tabularnewline
 &  &  &  & \tabularnewline
I  & 0.845 & 0.0156 & transverse & $|H_{2}\rightarrow L\rangle$(0.8742) \tabularnewline
 &  &  &  & $|H_{2}\rightarrow L+4\rangle$(0.2194) \tabularnewline
 &  &  &  & \tabularnewline
II  & 4.207 & 0.1463 & longitudinal & $|H-1\rightarrow L\rangle$(0.7599) \tabularnewline
 &  &  &  & $|H_{1}\rightarrow L+7\rangle$(0.2873) \tabularnewline
 &  &  &  & $|H_{2}\rightarrow L+7\rangle$(0.2873) \tabularnewline
 &  &  &  & \tabularnewline
III  & 4.914 & 0.0558 & transverse & $|H-1\rightarrow L;H-1\rightarrow L+2\rangle$(0.7553) \tabularnewline
 &  &  &  & $|H-2\rightarrow L+2\rangle$(0.3186) \tabularnewline
 &  &  &  & $|H-1\rightarrow L+7\rangle$(0.266) \tabularnewline
 &  &  &  & \tabularnewline
IV  & 6.97 & 0.5651 & transverse & $|H_{2}\rightarrow L+4\rangle$(0.5600) \tabularnewline
 &  &  &  & $|H_{2}\rightarrow L+5\rangle$(0.5150) \tabularnewline
 & 7.05 & 2.3743 & longitudinal & $|H-1\rightarrow L\rangle$(0.4462) \tabularnewline
 &  &  &  & $|H_{2}\rightarrow L+3\rangle$(0.3346) \tabularnewline
 &  &  &  & $|H_{1}\rightarrow L+3\rangle$(0.3346) \tabularnewline
 &  &  &  & $|H-1\rightarrow L+4\rangle$(0.2921) \tabularnewline
 &  &  &  & \tabularnewline
V & 7.973 & 0.0906 & longitudinal & $|H-1\rightarrow L;H_{2}\rightarrow L+2\rangle$(0.5613) \tabularnewline
 &  &  &  & $|H-1\rightarrow L;H_{1}\rightarrow L+2\rangle$(0.5613) \tabularnewline
 &  &  &  & $|H-1\rightarrow L+4\rangle$(0.2303) \tabularnewline
\end{longtable}

\FloatBarrier
\LTcapwidth=\columnwidth
\begin{longtable}{ccccl}
\captionsetup{font=footnotesize}
\caption{\label{Tab:table_b3_tri} Excitation energies, $E$, and many-particle wavefunctions of excited
states corresponding to the peaks in the linear absorption spectrum
of triangular B$_{3}$ (\emph{cf}. Fig. \ref{fig:b3_tri_combined}),
along with oscillator strength of transition. The polarization $\Vert$ corresponds
to the absorption due to light polarized in the plane of isomer, while
$\bot$ corresponds to the polarization perpendicular to that plane.
In the wavefunction, the bracketed numbers are the CI coefficients
of a given electronic configuration. Symbol $L$ and $H$ denote LUMO
and SOMO orbitals discussed earlier. $HF$ denotes the Hartree-Fock
configuration. } \\

\hline \multicolumn{1}{c}{Peak} & \multicolumn{1}{c}{ $E$ (eV)} & \multicolumn{1}{c}{$f_{12}$ } 
 & \multicolumn{1}{c}{Pol.} &\multicolumn{1}{l}{Wave Function}     \\ \hline 
\endfirsthead

\multicolumn{5}{c}%
{{\scshape \tablename\ \thetable{} -- continued from previous page}} \\ 

\hline \multicolumn{1}{c}{Peak} & \multicolumn{1}{c}{ $E$ (eV)} & \multicolumn{1}{c}{$f_{12}$ } 
 & \multicolumn{1}{c}{Pol.} & \multicolumn{1}{l}{Wave Function}     \\ \hline 
\endhead

\hline \multicolumn{5}{r}{{{\scshape Continued on next page}}} \\ 
\endfoot

\hline \hline
\endlastfoot
GS$^a$
 &  &  &  & $|HF\rangle$ (0.8229)\tabularnewline
 &  &  &  & $|H-3\rightarrow L\rangle$(0.2162) \tabularnewline
 &  &  &  & $|H\rightarrow L\rangle$(0.2072) \tabularnewline
 &  &  &  & $|H-3\rightarrow H\rangle$(0.1544) \tabularnewline
 &  &  &  & \tabularnewline
I & 0.797 & 0.0119 & $\bot$ & $|H-1\rightarrow H\rangle$(0.7357) \tabularnewline
 &  &  &  & $|H-1\rightarrow L\rangle$(0.4693) \tabularnewline
 &  &  &  & $|H-1\rightarrow L+8\rangle$(0.1142) \tabularnewline
 &  &  &  & \tabularnewline
II & 1.710 & 0.0035 & $\Vert$ & $|H\rightarrow L+1\rangle$(0.4816) \tabularnewline
 &  &  &  & $|H-2\rightarrow H\rangle$(0.4315)\tabularnewline
 &  &  &  & $|H-2\rightarrow L\rangle$(0.3757)\tabularnewline
 &  &  &  & $|H-3\rightarrow L+1\rangle$(0.3709)\tabularnewline
 &  & 0.0037 & $\Vert$ & $|H\rightarrow L\rangle$(0.5041)\tabularnewline
 &  &  &  & $|H-3\rightarrow H\rangle$(0.4703)\tabularnewline
 &  &  &  & \tabularnewline
III & 2.840 & 0.0679 & $\Vert$ & $|H-2\rightarrow H\rangle$(0.5374) \tabularnewline
 &  &  &  & $|H\rightarrow L+1\rangle$(0.5199) \tabularnewline
 &  &  &  & $|H-2\rightarrow L\rangle$(0.3306) \tabularnewline
 &  &  &  & $|H-3\rightarrow L+1\rangle$(0.2216)\tabularnewline
 & 2.872 & 0.0622 & $\Vert$ & $|H\rightarrow L\rangle$(0.4906) \tabularnewline
 &  &  &  & $|H-3\rightarrow L\rangle$(0.4499) \tabularnewline
 &  &  &  & \tabularnewline
IV  & 5.710  & 0.0428 & $\Vert$ & $|H-2\rightarrow L\rangle$(0.4114) \tabularnewline
 &  &  &  & $|H-2\rightarrow H\rangle$(0.3474) \tabularnewline
 & 5.730 & 0.0269 & $\Vert$ & $|H-3\rightarrow L\rangle$(0.4202) \tabularnewline
 &  &  &  & $|H-2\rightarrow H;H-3\rightarrow L\rangle$(0.2050) \tabularnewline
 &  &  &  & \tabularnewline
V  & 5.988 & 0.0663 & $\Vert$ & $|H-3\rightarrow L+1\rangle$(0.3968)\tabularnewline
 &  &  &  & $|H-2\rightarrow L\rangle$(0.3057) \tabularnewline
 &  &  &  & $|H-2\rightarrow H\rangle$(0.1578) \tabularnewline
 & 6.02 & 0.1270 & $\Vert$ & $|H-3\rightarrow L\rangle$(0.2578) \tabularnewline
 &  &  &  & $|H\rightarrow L\rangle$(0.1908)\tabularnewline
 &  &  &  & \tabularnewline
VI  & 7.657 & 0.2151 & $\Vert$ & $|H\rightarrow L+2\rangle$(0.5863) \tabularnewline
 &  &  &  & $|H-3\rightarrow L+2\rangle$(0.2497)\tabularnewline
 &  &  &  & $|H\rightarrow L+14\rangle$(0.2431)\tabularnewline
 &  &  &  & $|H\rightarrow L+4\rangle$(0.2093) \tabularnewline
 & 7.697  & 0.1409 & $\Vert$ & $|H\rightarrow L+3\rangle$(0.2991) \tabularnewline
 &  &  &  & $|H-3\rightarrow L+1\rangle$(0.2648)\tabularnewline
 &  &  &  & $|H-2\rightarrow H\rangle$(0.1794) \tabularnewline
 &  &  &  & $|H-2\rightarrow L\rangle$(0.1521) \tabularnewline
 &  &  &  & \tabularnewline
VII & 7.893 & 0.2797 & $\bot$ & $|H\rightarrow L+5\rangle$(0.7611) \tabularnewline
 &  &  &  & $|H-3\rightarrow L+4\rangle$(0.3477) \tabularnewline
\end{longtable}

\FloatBarrier
\LTcapwidth=\columnwidth
\begin{longtable}{ccccl}
\captionsetup{font=footnotesize}
\caption{\label{Tab:table_b3_lin} Excitation energies, $E$, and many-particle wavefunctions of excited
states corresponding to the peaks in the linear absorption spectrum
of linear B$_{3}$ (\emph{cf}. Fig. \ref{fig:b3_lin_final}), along
with oscillator strength of transition. Longitudinal and transverse polarization
corresponds to the absorption due to light polarized along and perpendicular
to the molecular axis respectively. In the wavefunction, the bracketed
numbers are the CI coefficients of a given electronic configuration.
Symbols $H$ and $L$ denote HOMO and LUMO orbitals respectively,
and $H_{1}$ denotes SOMOs discussed earlier. $HF$ denotes the Hartree-Fock
configuration. } \\

\hline \multicolumn{1}{c}{Peak} & \multicolumn{1}{c}{ $E$ (eV)} & \multicolumn{1}{c}{$f_{12}$ } 
 & \multicolumn{1}{c}{Pol.} &\multicolumn{1}{l}{Wave Function}     \\ \hline 
\endfirsthead

\multicolumn{5}{c}%
{{\scshape \tablename\ \thetable{} -- continued from previous page}} \\ 

\hline \multicolumn{1}{c}{Peak} & \multicolumn{1}{c}{ $E$ (eV)} & \multicolumn{1}{c}{$f_{12}$ } 
 & \multicolumn{1}{c}{Pol.} & \multicolumn{1}{l}{Wave Function}     \\ \hline 
\endhead

\hline \multicolumn{5}{r}{{{\scshape Continued on next page}}} \\ 
\endfoot

\hline \hline
\endlastfoot
GS$^a$
 &  &  &  & $|HF\rangle$ (0.6650)\tabularnewline
 &  &  &  & $|H-1\rightarrow L;H-1\rightarrow L\rangle$(0.3286)\tabularnewline
 &  &  &  & $|H-1\rightarrow L;H-1\rightarrow L\rangle$(0.3277)\tabularnewline
 &  &  &  & $|H\rightarrow L;H-1\rightarrow L+1\rangle$(0.2158)\tabularnewline
 &  &  &  & $|H\rightarrow L;H-1\rightarrow L+1\rangle$(0.2157)\tabularnewline
 &  &  &  & \tabularnewline
I & 0.723 & 0.0199 & longitudinal & $|H\rightarrow L\rangle$(0.8222) \tabularnewline
 &  &  &  & \tabularnewline
II  & 2.707 & 0.0175 & transverse & $|H-1\rightarrow L\rangle$(0.5360) \tabularnewline
 &  &  &  & \tabularnewline
III & 4.338 & 0.5930 & longitudinal & $|H-1\rightarrow L;H-1\rightarrow H\rangle$(0.5826) \tabularnewline
 &  &  &  & $|H-1\rightarrow L;H-1\rightarrow H\rangle$(0.5826) \tabularnewline
 &  &  &  & \tabularnewline
IV  & 5.937 & 0.0301 & longitudinal & $|H-2\rightarrow L\rangle$(0.2497) \tabularnewline
 &  &  &  & \tabularnewline
V  & 7.359 & 1.1053 & longitudinal & $|H-1\rightarrow L+1\rangle$(0.3683) \tabularnewline
 &  &  &  & $|H-1\rightarrow L+1\rangle$(0.3683) \tabularnewline
 &  &  &  & $|H-3\rightarrow L;H\rightarrow L\rangle$(0.3301) \tabularnewline
 &  &  &  & $|H-3\rightarrow H\rangle$(0.2364) \tabularnewline
 &  &  &  & \tabularnewline
VI  & 7.731 & 0.3921 & longitudinal & $|H-2\rightarrow L\rangle$(0.3858) \tabularnewline
 &  &  &  & $|H\rightarrow L;H-1\rightarrow L+4\rangle$(0.2851) \tabularnewline
 &  &  &  & $|H\rightarrow L;H-1\rightarrow L+4\rangle$(0.2851)\tabularnewline
 & 7.786 & 0.3062 & transverse & $|H-1\rightarrow L;H-1\rightarrow L+4\rangle$(0.6438)\tabularnewline
 &  &  &  & $|H-1\rightarrow L;H-1\rightarrow L+4\rangle$(0.3905)\tabularnewline
\end{longtable}

\FloatBarrier

\LTcapwidth=\columnwidth
\begin{longtable}{ccccl}
\captionsetup{font=footnotesize}
\caption{\label{Tab:table_b4_rho} Excitation energies, $E$, and many-particle wavefunctions of excited
states corresponding to the peaks in the linear absorption spectrum
of rhombus B$_{4}$ (\emph{cf}. Fig. \ref{fig:b4_rho_combined}),
along with oscillator strength of transition. The polarization \emph{in-plane}
corresponds to the absorption due to light polarized in the plane
of isomer. In the wavefunction, the bracketed numbers are the CI
coefficients of a given electronic configuration. Symbols $H$/$L$
denote HOMO/LUMO orbitals. $HF$ denotes the Hartree-Fock configuration. }  \\

\hline \multicolumn{1}{c}{Peak} & \multicolumn{1}{c}{ $E$ (eV)} & \multicolumn{1}{c}{$f_{12}$ } 
 & \multicolumn{1}{c}{Pol.} &\multicolumn{1}{l}{Wave Function}     \\ \hline 
\endfirsthead

\multicolumn{5}{c}%
{{\scshape \tablename\ \thetable{} -- continued from previous page}} \\ 

\hline \multicolumn{1}{c}{Peak} & \multicolumn{1}{c}{ $E$ (eV)} & \multicolumn{1}{c}{$f_{12}$ } 
 & \multicolumn{1}{c}{Pol.} & \multicolumn{1}{l}{Wave Function}     \\ \hline 
\endhead

\hline \multicolumn{5}{r}{{{\scshape Continued on next page}}} \\ 
\endfoot

\hline \hline
\endlastfoot
GS$^a$
 &  &  &  & $|HF\rangle$ (0.8787)\tabularnewline
 &  &  &  & $|H\rightarrow L;H\rightarrow L\rangle$(0.1147) \tabularnewline
 &  &  &  & \tabularnewline
I  & 4.159  & 0.4684 & in-plane & $|H\rightarrow L+2\rangle$(0.6566) \tabularnewline
 &  &  &  & $|H\rightarrow L+10\rangle$(0.3210) \tabularnewline
 &  &  &  & $|H-1\rightarrow L+6\rangle$(0.2773) \tabularnewline
 &  &  &  & $|H-1\rightarrow L+17\rangle$(0.1850)\tabularnewline
 &  &  &  & \tabularnewline
II  & 6.118  & 0.3925 & in-plane & $|H-1\rightarrow L+6\rangle$(0.5786) \tabularnewline
 &  &  &  & $|H-1\rightarrow L+17\rangle$(0.3285)\tabularnewline
 &  &  &  & $|H-3\rightarrow L\rangle$(0.2656) \tabularnewline
 &  &  &  & $|H-2\rightarrow L+11\rangle$(0.2544)\tabularnewline
 &  &  &  & $|H-2\rightarrow L+18\rangle$(0.2492)\tabularnewline
 &  &  &  & \tabularnewline
III  & 6.639  & 0.0490 & in-plane & $|H-1\rightarrow L+3\rangle$(0.4496)\tabularnewline
 &  &  &  & $|H-1\rightarrow L+13\rangle$(0.4485)\tabularnewline
 &  &  &  & $|H-4\rightarrow L\rangle$(0.4052) \tabularnewline
 &  &  &  & $|H\rightarrow L+11\rangle$(0.2766) \tabularnewline
 &  &  &  & $|H\rightarrow L+1\rangle$(0.2329) \tabularnewline
 &  &  &  & $|H\rightarrow L+18\rangle$(0.1905) \tabularnewline
 &  &  &  & \tabularnewline
IV  & 7.311  & 0.2528 & in-plane & $|H\rightarrow L+2\rangle$(0.3055) \tabularnewline
 &  &  &  & $|H-2\rightarrow L+11\rangle$(0.2892) \tabularnewline
 &  &  &  & $|H-3\rightarrow L\rangle$(0.2834) \tabularnewline
 &  &  &  & $|H-2\rightarrow L+18\rangle$(0.2487) \tabularnewline
 &  &  &  & $|H-2\rightarrow L+1\rangle$(0.2029) \tabularnewline
 &  &  &  & $|H-2\rightarrow L+5\rangle$(0.1509) \tabularnewline
 &  &  &  & \tabularnewline
V  & 7.842  & 0.3951 & in-plane & $|H\rightarrow L+2\rangle$(0.4233) \tabularnewline
 &  &  &  & $|H\rightarrow L+10\rangle$(0.3270) \tabularnewline
 &  &  &  & $|H-2\rightarrow L+11\rangle$(0.2049) \tabularnewline
 &  &  &  & $|H\rightarrow L+20\rangle$(0.1946) \tabularnewline
 &  &  &  & $|H-2\rightarrow L+1\rangle$(0.1594) \tabularnewline
 &  &  &  & $|H-2\rightarrow L+18\rangle$(0.1582) \tabularnewline
\end{longtable}

\FloatBarrier
\LTcapwidth=\columnwidth
\begin{longtable}{ccccl}
\captionsetup{font=footnotesize}
\caption{\label{Tab:table_b4_sqr} Excitation energies, $E$, and many-particle wavefunctions of excited
states corresponding to the peaks in the linear absorption spectrum
of square B$_{4}$ (\emph{cf}. Fig. \ref{fig:b4_sqr_combined}), along
with oscillator strength of transition. The polarization \emph{in-plane}
corresponds to the absorption due to light polarized in the plane
of isomer. In the wavefunction, the bracketed numbers are the CI
coefficients of a given electronic configuration. Symbols $H$/$L$
denote HOMO/LUMO orbitals. $HF$ denotes the Hartree-Fock configuration. }
 \\

\hline \multicolumn{1}{c}{Peak} & \multicolumn{1}{c}{ $E$ (eV)} & \multicolumn{1}{c}{$f_{12}$ } 
 & \multicolumn{1}{c}{Pol.} &\multicolumn{1}{l}{Wave Function}     \\ \hline 
\endfirsthead

\multicolumn{5}{c}%
{{\scshape \tablename\ \thetable{} -- continued from previous page}} \\ 

\hline \multicolumn{1}{c}{Peak} & \multicolumn{1}{c}{ $E$ (eV)} & \multicolumn{1}{c}{$f_{12}$ } 
 & \multicolumn{1}{c}{Pol.} & \multicolumn{1}{l}{Wave Function}     \\ \hline 
\endhead

\hline \multicolumn{5}{r}{{{\scshape Continued on next page}}} \\ 
\endfoot

\hline \hline
\endlastfoot
GS$^a$
 &  &  &  & $|HF\rangle$ (0.8682)\tabularnewline
 &  &  &  & $|H\rightarrow L;H\rightarrow L\rangle$(0.1765) \tabularnewline
 &  &  &  & $|H-2\rightarrow L;H-2\rightarrow L\rangle$(0.0920) \tabularnewline
 &  &  &  & \tabularnewline
I  & 4.879  & 0.5678 & in-plane & $|H\rightarrow L;H\rightarrow L+1\rangle$(0.2946) \tabularnewline
 &  &  &  & $|H\rightarrow L;H-3\rightarrow L\rangle$(0.1909) \tabularnewline
 &  &  &  & $|H-2\rightarrow L;H-3\rightarrow L\rangle$(0.1651) \tabularnewline
 &  &  &  & \tabularnewline
II  & 5.462 & 0.0235 & in-plane & $|H\rightarrow L;H\rightarrow L+1\rangle$(0.5893) \tabularnewline
 &  &  &  & $|H\rightarrow L;H\rightarrow L+6\rangle$(0.2477) \tabularnewline
 &  &  &  & \tabularnewline
III  & 6.418  & 0.2842 & in-plane & $|H\rightarrow L;H-3\rightarrow L\rangle$(0.5379) \tabularnewline
 &  &  &  & $|H\rightarrow L;H\rightarrow L+1\rangle$(0.2922) \tabularnewline
 &  &  &  & $|H-2\rightarrow L;H-3\rightarrow L\rangle$(0.1744) \tabularnewline
 &  &  &  & \tabularnewline
IV & 7.890 & 1.3888 & in-plane & $|H\rightarrow L+1;H-1\rightarrow L\rangle$(0.2351) \tabularnewline
\end{longtable}

\FloatBarrier
\LTcapwidth=\columnwidth
\begin{longtable}{ccccl}
\captionsetup{font=footnotesize}
\caption{\label{Tab:table_b4_lin} Excitation energies, $E$, and many-particle wavefunctions of excited
states corresponding to the peaks in the linear absorption spectrum
of linear B$_{4}$ (\emph{cf}. Fig. \ref{fig:b4_lin_final}), along
with oscillator strength of transition. Longitudinal and transverse polarization
corresponds to the absorption due to light polarized along and perpendicular
to the molecular axis respectively. In the wavefunction, the bracketed
numbers are the CI coefficients of a given electronic configuration.
Symbols $H$/$L$ denote HOMO/LUMO orbitals. $HF$ denotes the Hartree-Fock
configuration. }  \\

\hline \multicolumn{1}{c}{Peak} & \multicolumn{1}{c}{ $E$ (eV)} & \multicolumn{1}{c}{$f_{12}$ } 
 & \multicolumn{1}{c}{Pol.} &\multicolumn{1}{l}{Wave Function}     \\ \hline 
\endfirsthead

\multicolumn{5}{c}%
{{\scshape \tablename\ \thetable{} -- continued from previous page}} \\ 

\hline \multicolumn{1}{c}{Peak} & \multicolumn{1}{c}{ $E$ (eV)} & \multicolumn{1}{c}{$f_{12}$ } 
 & \multicolumn{1}{c}{Pol.} & \multicolumn{1}{l}{Wave Function}     \\ \hline 
\endhead

\hline \multicolumn{5}{r}{{{\scshape Continued on next page}}} \\ 
\endfoot

\hline \hline
\endlastfoot
GS$^a$
 &  &  &  & $|HF\rangle$ (0.5636)\tabularnewline
 &  &  &  & $|H\rightarrow L\rangle$(0.4737) \tabularnewline
 &  &  &  & $|H-1\rightarrow L;H\rightarrow L+1\rangle$(0.2291) \tabularnewline
 &  &  &  & $|H-1\rightarrow L;H\rightarrow L+1\rangle$(0.2289) \tabularnewline
 &  &  &  & \tabularnewline
I  & 5.363 & 0.0976 & longitudinal & $|H\rightarrow L;H-1\rightarrow L+9\rangle$(0.3005) \tabularnewline
 &  &  &  & $|H\rightarrow L;H-1\rightarrow L+9\rangle$(0.3005) \tabularnewline
 &  &  &  & $|H\rightarrow L;H-1\rightarrow L+8\rangle$(0.2702) \tabularnewline
 &  &  &  & $|H\rightarrow L;H-1\rightarrow L+8\rangle$(0.2702) \tabularnewline
 &  &  &  & $|H\rightarrow L;H-1\rightarrow L+4\rangle$(0.2439) \tabularnewline
 &  &  &  & $|H\rightarrow L;H-1\rightarrow L+4\rangle$(0.2439) \tabularnewline
 &  &  &  & \tabularnewline
II  & 5.947 & 4.1752 & longitudinal & $|H-1\rightarrow L+1\rangle$(0.2957)\tabularnewline
 &  &  &  & $|H-1\rightarrow L+1\rangle$(0.2957)\tabularnewline
 &  &  &  & \tabularnewline
III  & 7.352 & 1.8444 & longitudinal & $|H\rightarrow L;H-2\rightarrow L\rangle$(0.6477)\tabularnewline
 &  &  &  & $|H-3\rightarrow L\rangle$(0.2966)\tabularnewline
 &  &  &  & $|H-1\rightarrow L+1\rangle$(0.2091)\tabularnewline
 &  &  &  & $|H-1\rightarrow L+1\rangle$(0.2091)\tabularnewline
 & 7.380 & 1.0202 & transverse & $|H-1\rightarrow L+13\rangle$(0.5253)\tabularnewline
\end{longtable}

\FloatBarrier
\LTcapwidth=\columnwidth
\begin{longtable}{ccccl}
\captionsetup{font=footnotesize}
\caption{ \label{Tab:table_b4_tetra}  Excitation energies, $E$, and many-particle wavefunctions of excited
states corresponding to the peaks in the linear absorption spectrum
of distorted tetrahedron B$_{4}$ (\emph{cf}. Fig. \ref{fig:b4_tetra_combined}),
along with oscillator strength of transition. The polarization $x$,$y$ and $z$
corresponds to the absorption due to light polarized along $x-$,$y-$
and $z-$ axis respectively. In the wavefunction, the bracketed numbers
are the CI coefficients of a given electronic configuration. Symbols
$H$/$L$ denote HOMO/LUMO orbitals. $HF$ denotes the Hartree-Fock
configuration. }
 \\

\hline \multicolumn{1}{c}{Peak} & \multicolumn{1}{c}{ $E$ (eV)} & \multicolumn{1}{c}{$f_{12}$ } 
 & \multicolumn{1}{c}{Pol.} &\multicolumn{1}{l}{Wave Function}     \\ \hline 
\endfirsthead

\multicolumn{5}{c}%
{{\scshape \tablename\ \thetable{} -- continued from previous page}} \\ 

\hline \multicolumn{1}{c}{Peak} & \multicolumn{1}{c}{ $E$ (eV)} & \multicolumn{1}{c}{$f_{12}$ } 
 & \multicolumn{1}{c}{Pol.} & \multicolumn{1}{l}{Wave Function}     \\ \hline 
\endhead

\hline \multicolumn{5}{r}{{{\scshape Continued on next page}}} \\ 
\endfoot

\hline \hline
\endlastfoot
GS$^a$
 &  &  &  & $|HF\rangle$ (0.6493)\tabularnewline
 &  &  &  & $|H\rightarrow L;H\rightarrow L+2\rangle$(0.4695) \tabularnewline
 &  &  &  & $|H\rightarrow L;H-3\rightarrow L+2\rangle$(0.1547) \tabularnewline
 &  &  &  & \tabularnewline
I  & 1.000 & 0.0068 & x & $|H\rightarrow L\rangle$(0.8127) \tabularnewline
 &  &  &  & $|H\rightarrow L;H-1\rightarrow L+2\rangle$(0.2364)\tabularnewline
 & 1.111 & 0.0094 & y & $|H\rightarrow L+2;H\rightarrow L+1\rangle$(0.5944) \tabularnewline
 &  &  &  & $|H\rightarrow L;H\rightarrow L+1\rangle$(0.5121) \tabularnewline
 &  &  &  & \tabularnewline
II  & 3.609 & 0.0364 & x & $|H\rightarrow L;H-1\rightarrow L+2\rangle$(0.5094) \tabularnewline
 &  &  &  & $|H\rightarrow L;H-1\rightarrow L\rangle$(0.4535)\tabularnewline
 &  &  &  & $|H\rightarrow L+2\rangle$(0.3480) \tabularnewline
 & 3.754 & 0.0519 & y & $|H\rightarrow L+1;H\rightarrow L+2\rangle$(0.2391)\tabularnewline
 &  &  &  & \tabularnewline
III  & 4.48 & 0.0632 & z & $|H-1\rightarrow L\rangle$(0.4418)\tabularnewline
 &  &  &  & \tabularnewline
IV  & 4.96 & 0.0767 & z & $|H-1\rightarrow L\rangle$(0.2678) \tabularnewline
 &  &  &  & $|H\rightarrow L;H-3\rightarrow L\rangle$(0.2227) \tabularnewline
 &  &  &  & \tabularnewline
V & 5.92 & 0.0913 & x & $|H-3\rightarrow L\rangle$(0.2191) \tabularnewline
 &  &  &  & $|H\rightarrow L+2;H-4\rightarrow L\rangle$(0.2191) \tabularnewline
 &  &  &  & \tabularnewline
VI & 6.15 & 0.0895 & y & $|H\rightarrow L;H-3\rightarrow L+1\rangle$(0.2526) \tabularnewline
 &  &  &  & \tabularnewline
VII & 6.508 & 0.0159 & z & $|H-1\rightarrow L+2\rangle$(0.7131) \tabularnewline
 &  &  &  & $|H-1\rightarrow L\rangle$(0.2104) \tabularnewline
 &  &  &  & $|H\rightarrow L+2;H\rightarrow L+2\rangle$(0.2030) \tabularnewline
 &  &  &  & \tabularnewline
VIII & 6.858 & 0.1881 & x & $|H\rightarrow L;H\rightarrow L+10\rangle$(0.6912) \tabularnewline
 &  &  &  & $|H\rightarrow L+2;H\rightarrow L+10\rangle$(0.3415) \tabularnewline
 &  &  &  & $|H\rightarrow L+18;H\rightarrow L\rangle$(0.2759) \tabularnewline
 & 6.99 & 0.1488 & y & $|H\rightarrow L;H\rightarrow L+5\rangle$(0.4169) \tabularnewline
 &  &  &  & $|H\rightarrow L+2;H\rightarrow L+5\rangle$(0.3770) \tabularnewline
 &  &  &  & $|H\rightarrow L+3;H-1\rightarrow L\rangle$(0.2403) \tabularnewline
 &  &  &  & $|H-1\rightarrow L+1\rangle$(0.2156) \tabularnewline
 &  &  &  & \tabularnewline
IX & 7.80 & 0.0686 & z & $|H-1\rightarrow L;H-1\rightarrow L+2\rangle$(0.3386) \tabularnewline
 &  &  &  & $|H\rightarrow L;H-4\rightarrow L+2\rangle$(0.3164) \tabularnewline
 &  &  &  & $|H\rightarrow L;H-4\rightarrow L+2\rangle$(0.2987) \tabularnewline
 &  &  &  & $|H-1\rightarrow L;H-1\rightarrow L\rangle$(0.2722) \tabularnewline
 &  &  &  & $|H\rightarrow L+1;H-1\rightarrow L+3\rangle$(0.2174) \tabularnewline
\end{longtable}
\FloatBarrier

\LTcapwidth=\columnwidth
\begin{longtable}{ccccl}
\captionsetup{font=footnotesize}
\caption{\label{Tab:table_b5_pen} Excitation energies, $E$, and many-particle wavefunctions of excited
states corresponding to the peaks in the linear absorption spectrum
of pentagon B$_{5}$ (\emph{cf}. Fig. \ref{fig:b5_pen_combined}),
along with oscillator strength of transition. The polarization \emph{in-plane}
corresponds to the absorption due to light polarized in the plane
of isomer. In the wavefunction, the bracketed numbers are the CI
coefficients of a given electronic configuration. Symbols $H$ and
$L$ denote HOMO and LUMO orbitals respectively. $HF$ denotes the
Hartree-Fock configuration. }
 \\

\hline \multicolumn{1}{c}{Peak} & \multicolumn{1}{c}{ $E$ (eV)} & \multicolumn{1}{c}{$f_{12}$ } 
 & \multicolumn{1}{c}{Pol.} &\multicolumn{1}{l}{Wave Function}     \\ \hline 
\endfirsthead

\multicolumn{5}{c}%
{{\scshape \tablename\ \thetable{} -- continued from previous page}} \\ 

\hline \multicolumn{1}{c}{Peak} & \multicolumn{1}{c}{ $E$ (eV)} & \multicolumn{1}{c}{$f_{12}$ } 
 & \multicolumn{1}{c}{Pol.} & \multicolumn{1}{l}{Wave Function}     \\ \hline 
\endhead

\hline \multicolumn{5}{r}{{{\scshape Continued on next page}}} \\ 
\endfoot

\hline \hline
\endlastfoot
GS$^a$
 &  &  &  & $|HF\rangle$ (0.8541)\tabularnewline
 &  &  &  & $|H-1\rightarrow L\rangle$(0.1906) \tabularnewline
 &  &  &  & \tabularnewline
I & 1.394 & 0.0237 & $\bot$ to the plane & $|H\rightarrow L+1\rangle$(0.8536) \tabularnewline
 &  &  &  & $|H\rightarrow L+11\rangle$(0.1779) \tabularnewline
 &  &  &  & \tabularnewline
II & 1.818 & 0.0096 & in-plane & $|H\rightarrow L\rangle$(0.6524) \tabularnewline
 &  &  &  & $|H-1\rightarrow H\rangle$(0.4900) \tabularnewline
 &  &  &  & $|H\rightarrow L+3\rangle$(0.2218) \tabularnewline
 &  &  &  & \tabularnewline
III & 3.504  & 0.0848 & in-plane & $|H-2\rightarrow L+1\rangle$(0.5804) \tabularnewline
 &  &  &  & $|H-1\rightarrow H\rangle$(0.5131) \tabularnewline
 &  &  &  & $|H\rightarrow L\rangle$(0.3441) \tabularnewline
 &  &  &  & \tabularnewline
IV & 3.868 & 0.0372 & in-plane & $|H-2\rightarrow L+1\rangle$(0.7968) \tabularnewline
 &  &  &  & $|H-1\rightarrow H\rangle$(0.3416) \tabularnewline
 &  &  &  & \tabularnewline
V & 4.379 & 0.0247 & in-plane & $|H-3\rightarrow H\rangle$(0.8185) \tabularnewline
 &  &  &  & $|H-4\rightarrow L\rangle$(0.1900)\tabularnewline
 &  &  &  & \tabularnewline
VI & 5.378 & 0.1642 & in-plane & $|H-1\rightarrow L\rangle$(0.7309) \tabularnewline
 &  &  &  & $|H-4\rightarrow L\rangle$(0.3583) \tabularnewline
 &  &  &  & $|H-1\rightarrow L+3\rangle$(0.1994) \tabularnewline
 &  &  &  & $|H-3\rightarrow H\rangle$(0.1808) \tabularnewline
 & 5.576  & 0.6640 & in-plane & $|H-2\rightarrow L+1\rangle$(0.4415) \tabularnewline
 &  &  &  & $|H-3\rightarrow L\rangle$(0.4045) \tabularnewline
 &  &  &  & $|H-4\rightarrow H\rangle$(0.4000) \tabularnewline
 &  &  &  & $|H\rightarrow L\rangle$(0.2399) \tabularnewline
 &  &  &  & $|H-1\rightarrow L+4\rangle$(0.2039) \tabularnewline
 &  &  &  & \tabularnewline
VII & 6.305  & 0.6453 & in-plane & $|H-3\rightarrow L\rangle$(0.4073) \tabularnewline
 &  &  &  & $|H-1\rightarrow L+4\rangle$(0.3328) \tabularnewline
 &  &  &  & $|H-1\rightarrow L+9\rangle$(0.2952) \tabularnewline
 &  &  &  & $|H-2\rightarrow L+1\rangle$(0.2870) \tabularnewline
 &  &  &  & $|H-4\rightarrow L+4\rangle$(0.2772) \tabularnewline
 &  &  &  & $|H-4\rightarrow L+9\rangle$(0.2531) \tabularnewline
 & 6.528 & 0.1049 & in-plane & $|H-1\rightarrow L\rangle$(0.5596) \tabularnewline
 &  &  &  & $|H-4\rightarrow L\rangle$(0.3400) \tabularnewline
 &  &  &  & $|H-2\rightarrow L+2\rangle$(0.2761) \tabularnewline
 &  &  &  & $|H-3\rightarrow L+4\rangle$(0.2667) \tabularnewline
 &  &  &  & $|H-3\rightarrow L+9\rangle$(0.2598) \tabularnewline
 &  &  &  & \tabularnewline
VIII & 7.161  & 0.2561 & in-plane & $|H-4\rightarrow L+4\rangle$(0.4305) \tabularnewline
 &  &  &  & $|H-4\rightarrow L+9\rangle$(0.3859) \tabularnewline
 &  &  &  & $|H-3\rightarrow L\rangle$(0.3776) \tabularnewline
 &  &  &  & $|H-1\rightarrow L+4\rangle$(0.2487) \tabularnewline
 & 7.283 & 0.0781 & in-plane & $|H\rightarrow L+6\rangle$(0.7102) \tabularnewline
 &  &  &  & $|H\rightarrow L+9\rangle$(0.3359)\tabularnewline
 &  &  &  & $|H-3\rightarrow L+4\rangle$(0.2091)\tabularnewline
 &  &  &  & $|H-3\rightarrow L+9\rangle$(0.1854)\tabularnewline
 &  &  &  & \tabularnewline
IX & 7.702 & 0.1345 & $\bot$ to the plane & $|H\rightarrow L+11\rangle$(0.7980) \tabularnewline
 &  &  &  & $|H\rightarrow L+14\rangle$(0.2692) \tabularnewline
 & 7.750  & 0.1150 & in-plane & $|H-1\rightarrow L+3\rangle$(0.8332) \tabularnewline
 &  &  &  & $|H-1\rightarrow L+19\rangle$(0.1555) \tabularnewline
\end{longtable}

\FloatBarrier
\LTcapwidth=\columnwidth
\begin{longtable}{ccccl}
\captionsetup{font=footnotesize}
\caption{\label{Tab:table_b5_tet} Excitation energies, $E$, and many-particle wavefunctions of excited
states corresponding to the peaks in the linear absorption spectrum
of distorted triangular-bipyramid B$_{5}$ (\emph{cf}. Fig. \ref{fig:b5_tet_combined}),
along with oscillator strength of transition. The polarization $x$,$y$ and $z$
corresponds to the absorption due to light polarized along $x-$,$y-$
and $z-$ axis respectively. In the wavefunction, the bracketed numbers
are the CI coefficients of a given electronic configuration. Symbols
$H$ and $L$ denote HOMO and LUMO orbitals respectively. $HF$ denotes
the Hartree-Fock configuration. }  \\

\hline \multicolumn{1}{c}{Peak} & \multicolumn{1}{c}{ $E$ (eV)} & \multicolumn{1}{c}{$f_{12}$ } 
 & \multicolumn{1}{c}{Pol.} &\multicolumn{1}{l}{Wave Function}     \\ \hline 
\endfirsthead

\multicolumn{5}{c}%
{{\scshape \tablename\ \thetable{} -- continued from previous page}} \\ 

\hline \multicolumn{1}{c}{Peak} & \multicolumn{1}{c}{ $E$ (eV)} & \multicolumn{1}{c}{$f_{12}$ } 
 & \multicolumn{1}{c}{Pol.} & \multicolumn{1}{l}{Wave Function}     \\ \hline 
\endhead

\hline \multicolumn{5}{r}{{{\scshape Continued on next page}}} \\ 
\endfoot

\hline \hline
\endlastfoot
GS$^a$
 &  &  &  & $|HF\rangle$ (0.8615)\tabularnewline
 &  &  &  & $|H-1\rightarrow H\rangle$(0.2165) \tabularnewline
 &  &  &  & $|H-1\rightarrow H;H-1\rightarrow L\rangle$(0.1371) \tabularnewline
 &  &  &  & $|H-2\rightarrow H;H-1\rightarrow L\rangle$(0.1143) \tabularnewline
 &  &  &  & \tabularnewline
I  & 1.774  & 0.0548 & z & $|H-1\rightarrow L+4\rangle$(0.8689) \tabularnewline
 &  &  &  & \tabularnewline
II  & 3.161  & 0.0763 & y & $|H-2\rightarrow H\rangle$(0.8734) \tabularnewline
 &  &  &  & $|H-1\rightarrow H;H-1\rightarrow L\rangle$(0.2303) \tabularnewline
 &  &  &  & \tabularnewline
III  & 4.274  & 0.0509 & z & $|H-1\rightarrow L+2\rangle$(0.8666) \tabularnewline
 &  &  &  & $|H-1\rightarrow L+6\rangle$(0.1775) \tabularnewline
 &  &  &  & \tabularnewline
IV  & 5.487  & 0.1949 & y & $|H-1\rightarrow H;H-1\rightarrow L\rangle$(0.7812) \tabularnewline
 &  &  &  & $|H-2\rightarrow H\rangle$(0.2336) \tabularnewline
 &  &  &  & $|H-2\rightarrow H;H-1\rightarrow L\rangle$(0.2332) \tabularnewline
 &  &  &  & \tabularnewline
V  & 6.408  & 0.1170 & z & $|H-4\rightarrow H\rangle$(0.7837) \tabularnewline
 &  &  &  & $|H-2\rightarrow L+2\rangle$(0.3357) \tabularnewline
 &  &  &  & $|H-2\rightarrow H;H-1\rightarrow L+1\rangle$(0.2274) \tabularnewline
 &  &  &  & \tabularnewline
VI  & 7.519  & 0.4523 & x & $|H-2\rightarrow H;H-1\rightarrow L+2\rangle$(0.8133) \tabularnewline
 &  &  &  & $|H-2\rightarrow H;H-1\rightarrow L+6\rangle$(0.2669) \tabularnewline
 &  &  &  & \tabularnewline
VII  & 7.744  & 0.0177 & x & $|H-1\rightarrow L;H-1\rightarrow L+1\rangle$(0.6105) \tabularnewline
 &  &  &  & $|H-1\rightarrow L+4\rangle$(0.4305) \tabularnewline
 &  &  &  & $|H-2\rightarrow L+1\rangle$(0.2271) \tabularnewline
\end{longtable}

\chapter{Boron Clusters B$_{6}$ and  B$_{6}^{+}$}
\label{app:wavefunction-epjd}
In the following tables, we have given the
excitation energies (with respect to the ground state), and the many-body
wavefunctions of the excited states, corresponding
to the peaks in the CIS photoabsorption spectra of various isomers listed
in Fig. \ref{fig:geometries-neutral} and Fig. \ref{fig:geometries-cationic}
, along with the oscillator strength $f_{12}$ of the transitions,
\begin{equation}
f_{12}=\frac{2}{3}\frac{m_{e}}{\hbar^{2}}(E_{2}-E_{1})\sum_{i}|\langle m|d_{i}|G\rangle|^{2}
\end{equation}
where, $|m\rangle$ denotes the excited state in question, $|G\rangle$,
the ground state, and $d_{i}$ is the $i$-th Cartesian component
of the electric dipole operator. The single excitations are with respect to the reference state as given in
respective tables.

Similar tables corresponding to the results of select EOM-CCSD calculations are also given below.
\FloatBarrier

\section*{Excited State CIS Wavefunctions, Energies and Oscillator Strengths}
\LTcapwidth=\columnwidth
\begin{longtable}{ccccl}
\captionsetup{font=footnotesize}
\caption{Excitation energies, $E$, and many-particle wavefunctions of excited
states corresponding to the peaks in the CIS linear absorption spectrum
of B$_{6}$ -- planar ring (triplet) isomer (\emph{cf}. Fig. \ref{fig:neutral-plot-planar-ring-triplet}).
The subscript $\Vert$ in the peak number denotes the absorption due to light polarized in the plane of isomer. 
In the wavefunction, the bracketed numbers are the CI coefficients of a given electronic configuration.
Symbols $H_{1\alpha}$,$H_{2\alpha}$ denote SOMOs discussed earlier, and $H$, and $L$, denote HOMO and LUMO orbitals respectively. Note
that, the reference state does not correspond to any peak, instead it represents the reference state from which singles excitations are occurring.}
\\

\hline \multicolumn{1}{c}{Peak} & \multicolumn{1}{c}{ $E$ (eV)} & \multicolumn{1}{c}{$f_{12}$ } 
 & \multicolumn{1}{c}{$\langle S^2 \rangle$} &\multicolumn{1}{l}{Wave Function}     \\ \hline 
\endfirsthead

\multicolumn{5}{c}%
{{\scshape \tablename\ \thetable{} -- continued from previous page}} \\ 

\hline \multicolumn{1}{c}{Peak} & \multicolumn{1}{c}{ $E$ (eV)} & \multicolumn{1}{c}{$f_{12}$ } 
 & \multicolumn{1}{c}{$\langle S^2 \rangle$} & \multicolumn{1}{l}{Wave Function}     \\ \hline 
\endhead

\hline \multicolumn{5}{r}{{{\scshape Continued on next page}}} \\ 
\endfoot

\hline \hline
\endlastfoot
\tabularnewline
Reference 	&	&	& 	& $| H_{1\alpha}^{1}; H_{2\alpha}^1 \rangle $	\\
\tabularnewline
I$_{\Vert}$ 	& 2.85  &0.0099 &3.01	& $| H_{\alpha} - 2 \rightarrow L_{\alpha} \rangle$(0.6589) \\
		&  	&	&	& $| H_{1\alpha} \rightarrow L_{\alpha} \rangle$(0.6102) \\
\tabularnewline
II$_{\Vert}$ 	& 3.42 	&0.0634	&2.90	& $| H_{1\alpha} \rightarrow L_{\alpha} \rangle$(0.5415) \\
		&  	&	&	& $| H_{\beta} \rightarrow L_{\beta} \rangle$(0.4719) \\
\tabularnewline
III$_{\Vert}$ 	& 4.31 	&0.0306	&2.78	& $| H_{\beta} - 1 \rightarrow L_{\beta} + 1 \rangle$(0.7253) \\
		&  	&	&	& $| H_{\beta} - 2 \rightarrow L_{\beta} +  12 \rangle$(0.3011) \\
\tabularnewline
IV$_{\Vert}$ 	& 4.62 	&0.1830	&3.30	& $| H_{\beta} - 3 \rightarrow L_{\beta} +  1 \rangle$(0.6176) \\
		&  	&	&	& $| H_{2\alpha} \rightarrow L_{\alpha} +  10 \rangle$(0.3007) \\
\tabularnewline
V$_{\Vert}$ 	& 4.98 	&0.1673	&2.94	& $| H_{\beta} - 3 \rightarrow L_{\beta} +  1 \rangle$(0.4591) \\
		&  	&	&	& $| H_{\beta} - 2 \rightarrow L_{\beta} +  16 \rangle$(0.3334) \\
\tabularnewline
VI$_{\Vert}$ 	& 5.89 	&0.3505	&2.99	& $| H_{\alpha} - 3 \rightarrow L_{\alpha} +  1 \rangle$(0.4911) \\
		&  	&	&	& $| H_{2\alpha}  \rightarrow L_{\alpha} +  5 \rangle$(0.3587) \\
\tabularnewline
VII$_{\Vert}$ 	& 6.45 	&0.2201	&3.31	& $| H_{\alpha} - 5 \rightarrow L_{\alpha} \rangle$(0.3976) \\
		&  	&	&	& $| H_{\alpha} - 3 \rightarrow L_{\alpha} +  1 \rangle$(0.3848) \\
\tabularnewline
VIII$_{\Vert}$ 	& 6.81 	&0.1848	&2.88	& $| H_{2\alpha} \rightarrow L_{\alpha} +  10 \rangle$(0.4423) \\
		&  	&	&	& $| H_{\beta} - 1 \rightarrow L_{\beta} +  3 \rangle$(0.3970) \\
\label{Tab:table-neutral-planar-ring-triplet}
\end{longtable}

\LTcapwidth=\columnwidth
\begin{longtable}{cccl}
\captionsetup{font=footnotesize}
\caption{Excitation energies, $E$, and many-particle wavefunctions of excited
states corresponding to the peaks in the EOM-CCSD linear absorption spectrum
of B$_{6}$ -- planar ring (triplet) isomer (\emph{cf}. Fig. \ref{fig:neutral-plot-planar-ring-triplet-eomccsd}).
The subscript $\Vert$ in the peak number denotes the absorption due to light polarized in the plane of isomer. 
In the wavefunction, the bracketed numbers are the coupled-cluster amplitudes of a given electronic configuration.
Symbols $H_{1\alpha}$,$H_{2\alpha}$ denote SOMOs discussed earlier, and $H$, and $L$, denote HOMO and LUMO orbitals respectively. Note
that, the reference state does not correspond to any peak, instead it represents the reference state from which singles excitations are occurring.}
\\

\hline \multicolumn{1}{c}{Peak} & \multicolumn{1}{c}{ $E$ (eV)} & \multicolumn{1}{c}{$f_{12}$ } 
  &\multicolumn{1}{l}{Wave Function}     \\ \hline 
\endfirsthead

\multicolumn{4}{c}%
{{\scshape \tablename\ \thetable{} -- continued from previous page}} \\ 

\hline \multicolumn{1}{c}{Peak} & \multicolumn{1}{c}{ $E$ (eV)} & \multicolumn{1}{c}{$f_{12}$ } 
  & \multicolumn{1}{l}{Wave Function}     \\ \hline 
\endhead

\hline \multicolumn{4}{r}{{{\scshape Continued on next page}}} \\ 
\endfoot

\hline \hline
\endlastfoot
\tabularnewline
Reference 	&	&	& $| H_{1\alpha}^{1}; H_{2\alpha}^1 \rangle $	\\
\tabularnewline
I$_{\Vert}$ 	& 1.74  &0.0179 & $| H_{1\alpha} \rightarrow L_{\alpha} \rangle$(0.7968) \\
		&  	&	& $| H_{\alpha} - 2 \rightarrow L_{\alpha} \rangle$(0.2623)\\
		&  	&	& $| H_{\alpha} - 1 \rightarrow L_{\alpha}; H_{\beta} - 1 \rightarrow L_{\beta} + 1 \rangle$(0.1149)\\
\tabularnewline
II$_{\Vert}$ 	& 2.71 	&0.0082	& $| H_{\beta} \rightarrow L_{\beta} \rangle$(0.6442) \\
		&  	&	& $| H_{1\alpha} \rightarrow L_{\alpha} \rangle$(0.2320) \\
\tabularnewline
III$_{\Vert}$ 	& 3.20 	&0.0603	& $| H_{\beta} - 1 \rightarrow L_{\beta} + 1 \rangle$(0.3111) \\
		&  	&	& $| H_{\beta} - 2 \rightarrow L_{\beta} \rangle$(0.4768) \\
\tabularnewline
IV$_{\Vert}$ 	& 3.50 	&0.0371	& $| H_{\beta} - 3 \rightarrow L_{\beta} +  1 \rangle$(0.5933) \\
		&  	&	& $| H_{2\alpha} \rightarrow L_{\alpha} +  10 \rangle$(0.1421) \\
\tabularnewline
V$_{\Vert}$ 	& 4.03	&0.0046	& $| H_{\beta} - 3 \rightarrow L_{\beta} +  1 \rangle$(0.2861) \\
		&  	&	& $| H_{\beta} - 2 \rightarrow L_{\beta} +  16 \rangle$(0.1968) \\
		&  	&	& $| H_{1\alpha} \rightarrow L_{\alpha}; H_{\beta} - 2 \rightarrow L_{\beta} + 1 \rangle$(0.1102)\\
\tabularnewline
VI$_{\Vert}$ 	& 5.05 	&0.4106	& $| H_{2\alpha}  \rightarrow L_{\alpha} +  5 \rangle$(0.3573) \\
		&  	&	& $| H_{\alpha} - 3 \rightarrow L_{\alpha} +  1 \rangle$(0.1820) \\
		&  	&	& $| H_{1\alpha} \rightarrow L_{\alpha}; H_{\beta} - 2 \rightarrow L_{\beta} + 1 \rangle$(0.1612)\\
\tabularnewline
VII$_{\Vert}$ 	& 5.50 	&0.0801	& $| H_{\alpha} - 5 \rightarrow L_{\alpha} \rangle$(0.2023) \\
		&  	&	& $| H_{\alpha} - 4 \rightarrow L_{\alpha} +  7 \rangle$(0.1358) \\
		&  	&	& $| H_{\alpha} - 1 \rightarrow L_{\alpha}; H_{\beta} - 3 \rightarrow L_{\beta} + 1 \rangle$(0.1157)\\
		&  	&	& $| H_{1\alpha} \rightarrow L_{\alpha}; H_{\beta} - 3 \rightarrow L_{\beta} + 16 \rangle$(0.1025)\\
\label{Tab:table-neutral-planar-ring-triplet-eomccsd}
\end{longtable}

\LTcapwidth=\columnwidth
\begin{longtable}{ccccl}
\captionsetup{font=footnotesize}
\caption{Excitation energies, $E$, and many-particle wavefunctions of excited
states corresponding to the peaks in the CIS linear absorption spectrum
of B$_{6}$ -- incomplete wheel isomer (\emph{cf}. Fig. \ref{fig:neutral-plot-incomplete-wheel}).
The subscripts $\Vert$ and $\bot$, in the peak number denote
 the absorption due to light polarized in, and perpendicular to the plane of wheel base, respectively. 
The rest of the information is the same as given in the caption for Table \ref{Tab:table-neutral-planar-ring-triplet}.}
\\

\hline \multicolumn{1}{c}{Peak} & \multicolumn{1}{c}{ $E$ (eV)} & \multicolumn{1}{c}{$f_{12}$ } 
 & \multicolumn{1}{c}{$\langle S^2 \rangle$} &\multicolumn{1}{l}{Wave Function}     \\ \hline 
\endfirsthead

\multicolumn{5}{c}%
{{\scshape \tablename\ \thetable{} -- continued from previous page}} \\ 

\hline \multicolumn{1}{c}{Peak} & \multicolumn{1}{c}{ $E$ (eV)} & \multicolumn{1}{c}{$f_{12}$ } 
 & \multicolumn{1}{c}{$\langle S^2 \rangle$} & \multicolumn{1}{l}{Wave Function}     \\ \hline 
\endhead

\hline \multicolumn{5}{r}{{{\scshape Continued on next page}}} \\ 
\endfoot

\hline \hline
\endlastfoot
\tabularnewline
Reference	&	&	& 	& $| H_{1\alpha}^{1} ; H_{2\alpha}^{1} \rangle$ \\
\tabularnewline
I$_{\Vert}$ 	& 3.09  &0.0447	&2.68 	& $| H_{1\alpha} \rightarrow L_{\alpha} \rangle$(0.6251) \\
		&  	&	&	& $| H_{2\alpha} \rightarrow L_{\alpha} + 1 \rangle$(0.5904) \\
\tabularnewline
II$_{\Vert}$ 	& 3.55 	&0.0346	&2.83	& $| H_{\alpha} - 2 \rightarrow L_{\alpha} \rangle$(0.8721) \\
\tabularnewline
III$_{\bot}$ 	& 3.85 	&0.0623	&2.69	& $| H_{\beta} \rightarrow L_{\beta} \rangle$(0.9095) \\
		&  	&	&	& $| H_{\beta} \rightarrow L_{\beta} + 11 \rangle$(0.2045) \\
\tabularnewline
IV$_{\Vert}$ 	& 4.96 	&0.0651	&3.00	& $| H_{\beta} \rightarrow L_{\beta} + 2 \rangle$(0.3971) \\
		&  	&	&	& $| H_{\alpha} - 3 \rightarrow L_{\alpha} + 1 \rangle$(0.3695) \\
\tabularnewline
V$_{\Vert}$ 	& 5.77 	&0.0320	&2.817	& $| H_{2\alpha}  \rightarrow L_{\alpha} + 25 \rangle$(0.5802) \\
		&  	&	&	& $| H_{2\alpha}  \rightarrow L_{\alpha} + 7 \rangle$(0.3785) \\
\tabularnewline
VI$_{\Vert}$ 	& 6.35 	&0.2942	&3.10	& $| H_{2\alpha}  \rightarrow L_{\alpha} + 7 \rangle$(0.5776) \\
		&  	&	&	& $| H_{\alpha} - 5 \rightarrow L_{\alpha} \rangle$(0.3245) \\
\tabularnewline
VII$_{\Vert}$ 	& 6.69 	&0.0321	&2.82	& $| H_{2\alpha}  \rightarrow L_{\alpha} + 2 \rangle$(0.7301) \\
		&  	&	&	& $| H_{2\alpha} \rightarrow L_{\alpha} + 5 \rangle$(0.4955) \\
\label{Tab:table-neutral-incomplete-wheel}
\end{longtable}

\LTcapwidth=\columnwidth
\begin{longtable}{cccl}
\captionsetup{font=footnotesize}
\caption{Excitation energies, $E$, and many-particle wavefunctions of excited
states corresponding to the peaks in the CIS linear absorption spectrum
of B$_{6}$ -- bulged wheel isomer (\emph{cf}. Fig. \ref{fig:plots-neutral-bulged-wheel}).
The subscripts $\Vert$ and $\bot$, in the peak number denote
 the absorption due to light polarized in, and perpendicular to the plane of 
wheel base, respectively.
In the wavefunction, the bracketed numbers are the CI coefficients of a given electronic configuration.
Symbols $H$ and $L$ denote HOMO and LUMO orbitals respectively. Excitations are 
with respect to the closed-shell Hartree Fock reference state.}
\\

\hline \multicolumn{1}{c}{Peak} & \multicolumn{1}{c}{ $E$ (eV)} & \multicolumn{1}{c}{$f_{12}$ } 
  &\multicolumn{1}{l}{Wave Function}     \\ \hline 
\endfirsthead

\multicolumn{4}{c}%
{{\scshape \tablename\ \thetable{} -- continued from previous page}} \\ 

\hline \multicolumn{1}{c}{Peak} & \multicolumn{1}{c}{ $E$ (eV)} & \multicolumn{1}{c}{$f_{12}$ } 
  & \multicolumn{1}{l}{Wave Function}     \\ \hline 
\endhead

\hline \multicolumn{4}{r}{{{\scshape Continued on next page}}} \\ 
\endfoot

\hline \hline
\endlastfoot
\tabularnewline
I$_{\Vert}$ 	& 3.76  &0.2625	& $ | H - 1 \rightarrow L + 6 \rangle$(0.5622)  \\ 
		&	&	& $ | H \rightarrow L + 6 \rangle$(0.5622) \\
		&  	&	& $| H - 1 \rightarrow L + 2 \rangle$(0.4459)\\
		&	&	& $ | H \rightarrow L + 2 \rangle$(0.4459) \\
\tabularnewline
II$_{\Vert}$ 	& 4.38 	&0.0657	& $| H - 1 \rightarrow L + 3 \rangle $(0.4998)\\
		&	&	& $| H \rightarrow L + 3 \rangle$(0.4998) \\
		&  	&	& $| H - 1 \rightarrow L + 4 \rangle $(0.4981)\\
		&	&	& $| H \rightarrow L + 4 \rangle$(0.4981) \\
\tabularnewline
III$_{\Vert}$ 	& 5.07 	&2.1039	& $| H - 2 \rightarrow L + 1 \rangle $(0.7653)\\
		&	&	& $| H - 2 \rightarrow L \rangle$(0.7653) \\
		&  	&	& $| H - 2 \rightarrow L + 8 \rangle $(0.2857)\\
		&	&	& $| H - 2 \rightarrow L + 7 \rangle$(0.2857) \\
\tabularnewline
IV$_{\bot}$ 	& 5.71 	&0.0810	& $| H - 2 \rightarrow L + 6 \rangle$(0.6266) \\
		&  	&	& $| H - 2 \rightarrow L + 2 \rangle$(0.5564) \\
\tabularnewline
V$_{\Vert}$ 	& 7.05 	&1.0550	& $| H \rightarrow L + 5 \rangle $(0.7645)\\
		&	&	& $| H - 1 \rightarrow L + 5 \rangle$(0.7645) \\
		&  	&	& $| H \rightarrow L + 6 \rangle $(0.4763)\\
		&	&	& $| H - 1 \rightarrow L + 6 \rangle$(0.4763) \\
\label{Tab:table-neutral-bulged-wheel}
\end{longtable}

\LTcapwidth=\columnwidth
\begin{longtable}{cccl}
\captionsetup{font=footnotesize}
\caption{Excitation energies, $E$, and many-particle wavefunctions of excited
states corresponding to the peaks in the EOM-CCSD linear absorption spectrum
of B$_{6}$ -- bulged wheel isomer (\emph{cf}. Fig. \ref{fig:neutral-plot-bulged-wheel-eomccsd}) calculated using EOM-CCSD 
approach. The subscripts $\Vert$ and $\bot$, in the peak number denote
 the absorption due to light polarized in, and perpendicular to the plane of 
wheel base, respectively.
In the wavefunction, the bracketed numbers are the coupled-cluster amplitudes of a given electronic configuration.
Symbols $H$ and $L$ denote HOMO and LUMO orbitals respectively. Excitations are 
with respect to the closed-shell Hartree Fock reference state.}
\\

\hline \multicolumn{1}{c}{Peak} & \multicolumn{1}{c}{ $E$ (eV)} & \multicolumn{1}{c}{$f_{12}$ } 
  &\multicolumn{1}{l}{Wave Function}     \\ \hline 
\endfirsthead

\multicolumn{4}{c}%
{{\scshape \tablename\ \thetable{} -- continued from previous page}} \\ 

\hline \multicolumn{1}{c}{Peak} & \multicolumn{1}{c}{ $E$ (eV)} & \multicolumn{1}{c}{$f_{12}$ } 
  & \multicolumn{1}{l}{Wave Function}     \\ \hline 
\endhead

\hline \multicolumn{4}{r}{{{\scshape Continued on next page}}} \\ 
\endfoot

\hline \hline
\endlastfoot
\tabularnewline
I$_{\Vert}$ 	& 3.21  &0.0088	& $ | H - 1 \rightarrow L + 6 \rangle$(0.3898)  \\ 
		&	&	& $ | H \rightarrow L + 6 \rangle$(0.3898) \\
		&  	&	& $ | H - 1 \rightarrow L + 2 \rangle$(0.3519)\\
		&	&	& $ | H \rightarrow L + 2 \rangle$(0.3519) \\
\tabularnewline
II$_{\Vert}$ 	& 3.43 	&0.0256	& $| H - 1 \rightarrow L + 3 \rangle $(0.3604)\\
		&	&	& $| H \rightarrow L + 3 \rangle$(0.3604) \\
		&  	&	& $| H - 1 \rightarrow L + 4 \rangle $(0.3604)\\
		&	&	& $| H \rightarrow L + 4 \rangle$(0.3604) \\
\tabularnewline
III$_{\Vert}$ 	& 4.61 	&0.2658	& $| H - 2 \rightarrow L + 1 \rangle $(0.5074)\\
		&	&	& $| H - 2 \rightarrow L \rangle$(0.5074) \\
		&  	&	& $| H - 2 \rightarrow L + 8 \rangle $(0.2183)\\
		&	&	& $| H - 2 \rightarrow L + 7 \rangle$(0.2183) \\
\tabularnewline
IV$_{\bot}$ 	& 5.25 	&0.0004	& $| H - 2 \rightarrow L + 6 \rangle$(0.3675) \\
		&  	&	& $| H - 2 \rightarrow L + 2 \rangle$(0.3473) \\
\tabularnewline
V$_{\Vert}$ 	& 6.77 	&0.224	& $| H \rightarrow L + 5 \rangle $(0.5106)\\
		&	&	& $| H - 1 \rightarrow L + 5 \rangle$(0.5106) \\
		&  	&	& $| H \rightarrow L + 6 \rangle $(0.3575)\\
		&	&	& $| H - 1 \rightarrow L + 6 \rangle$(0.3575) \\
\label{Tab:table-neutral-bulged-wheel-eomccsd}
\end{longtable}

\LTcapwidth=\columnwidth
\begin{longtable}{cccl}
\captionsetup{font=footnotesize}
\caption{Excitation energies, $E$, and many-particle wavefunctions of excited
states corresponding to the peaks in the CIS linear absorption spectrum
of B$_{6}$ -- planar ring (singlet) isomer 
(\emph{cf}. Fig. \ref{fig:neutral-plot-planar-ring-singlet}).
The subscripts $\Vert$ and $\bot$, in the peak number denote
 the absorption due to light polarized in, and perpendicular to the plane of 
the isomer, respectively.
The rest of the information is the same as given in the caption for Table \ref{Tab:table-neutral-bulged-wheel}.}
\\

\hline \multicolumn{1}{c}{Peak} & \multicolumn{1}{c}{ $E$ (eV)} & \multicolumn{1}{c}{$f_{12}$ } 
  &\multicolumn{1}{l}{Wave Function}     \\ \hline 
\endfirsthead

\multicolumn{4}{c}%
{{\scshape \tablename\ \thetable{} -- continued from previous page}} \\ 

\hline \multicolumn{1}{c}{Peak} & \multicolumn{1}{c}{ $E$ (eV)} & \multicolumn{1}{c}{$f_{12}$ } 
  & \multicolumn{1}{l}{Wave Function}     \\ \hline 
\endhead

\hline \multicolumn{4}{r}{{{\scshape Continued on next page}}} \\ 
\endfoot

\hline \hline
\endlastfoot
\tabularnewline
I$_{\Vert}$ 	& 3.12  &0.5397	& $| H - 1 \rightarrow L \rangle$(0.9174) \\
		&  	&	& $| H     \rightarrow L + 23 \rangle$(0.2504) \\
\tabularnewline
II$_{\Vert}$ 	& 5.28 	&1.0153	& $| H  \rightarrow L + 7\rangle$(0.4991) \\
		&  	&	& $| H  \rightarrow L + 23\rangle$(0.4093) \\
\tabularnewline
III$_{\Vert}$ 	& 5.61 	&0.1380	& $| H - 1 \rightarrow L + 2 \rangle$(0.5479) \\
		&  	&	& $| H - 1 \rightarrow L + 8 \rangle$(0.5181) \\
\tabularnewline
IV$_{\Vert,\bot}$& 5.88 &0.8199	& $| H \rightarrow L + 7 \rangle$(0.7472) \\
		&  	&	& $| H \rightarrow L + 12 \rangle$(0.3140) \\
\tabularnewline
V$_{\Vert}$ 	& 6.77 	&2.2212	& $| H \rightarrow L + 16 \rangle$(0.4623) \\
		&  	&	& $| H \rightarrow L + 12 \rangle$(0.4600) \\
\tabularnewline
VI$_{\Vert}$ 	& 7.15 	&0.6003	& $| H \rightarrow L + 16 \rangle$(0.6363) \\
		&  	&	& $| H \rightarrow L + 12 \rangle$(0.5687) \\
\label{Tab:table-neutral-planar-ring-singlet}
\end{longtable}

\LTcapwidth=\columnwidth
\begin{longtable}{ccccl}
\captionsetup{font=footnotesize}
\caption{Excitation energies, $E$, and many-particle wavefunctions of excited
states corresponding to the peaks in the CIS linear absorption spectrum
of B$_{6}$ -- octahedron isomer 
(\emph{cf}. Fig. \ref{fig:neutral-plot-octahedron}).
The subscripts $\Vert$ and $\bot$, in the peak number denote the 
absorption due to light polarized in, and perpendicular to the plane of 
pyramidal base, respectively.
The rest of the information is the same as given in the caption for Table \ref{Tab:table-neutral-planar-ring-triplet}.}
\\

\hline \multicolumn{1}{c}{Peak} & \multicolumn{1}{c}{ $E$ (eV)} & \multicolumn{1}{c}{$f_{12}$ } 
 & \multicolumn{1}{c}{$\langle S^2 \rangle$} &\multicolumn{1}{l}{Wave Function}     \\ \hline 
\endfirsthead

\multicolumn{5}{c}%
{{\scshape \tablename\ \thetable{} -- continued from previous page}} \\ 

\hline \multicolumn{1}{c}{Peak} & \multicolumn{1}{c}{ $E$ (eV)} & \multicolumn{1}{c}{$f_{12}$ } 
 & \multicolumn{1}{c}{$\langle S^2 \rangle$} & \multicolumn{1}{l}{Wave Function}     \\ \hline 
\endhead

\hline \multicolumn{5}{r}{{{\scshape Continued on next page}}} \\ 
\endfoot

\hline \hline
\endlastfoot
\tabularnewline
Reference		&	&	&	& $| H_{1\alpha}^{1} ; H_{2\alpha}^{1} \rangle $	\\
\tabularnewline
I$_{\Vert,\bot} $ 	& 0.92  &0.1110	&2.25	& $| H_{1\alpha} \rightarrow L_{\alpha} \rangle$(0.8790) \\
			&	&	&	& $| H_{2\alpha} \rightarrow L_{\alpha} + 2 \rangle$(0.8612) \\
			&  	&	&	& $| H_{2\alpha} \rightarrow L_{\alpha} + 1 \rangle$(0.6610)\\
			&	&	&	& $| H_{1\alpha} \rightarrow L_{\alpha} + 1 \rangle$(0.6128) \\
\tabularnewline
II$_{\Vert,\bot}$ 	& 2.50 	&0.0111	&3.37	& $| H_{\alpha} - 2  \rightarrow L_{\alpha} + 1 \rangle$ (0.6637)\\
			&	&	&	& $| H_{\alpha} - 2  \rightarrow L_{\alpha} + 2 \rangle$ (0.6637)\\
			&  	&	&	& $| H_{\alpha} - 1  \rightarrow L_{\alpha}  \rangle$ (0.6637)\\
			&	&	&	& $| H_{\alpha} - 1  \rightarrow L_{\alpha} + 1 \rangle$ (0.6637)\\
\tabularnewline
III$_{\Vert,\bot}$ 	& 3.69 	&0.0111	&3.65	& $| H_{\alpha} - 4  \rightarrow L_{\alpha}  \rangle$ (0.8186)\\
			&	&	&	& $| H_{\alpha} - 4  \rightarrow L_{\alpha} + 1 \rangle$ (0.8032)\\
			&  	&	&	& $| H_{\alpha} - 4  \rightarrow L_{\alpha} + 2 \rangle$ (0.8032)\\
			&	&	&	& $| H_{\beta}  \rightarrow L_{\beta} + 2 \rangle$ (0.3117)\\
\tabularnewline
IV$_{\Vert,\bot}$ 	& 4.69 	&0.5265	&2.55	& $| H_{\beta}   \rightarrow L_{\beta} + 2 \rangle$ (0.5448)\\
			&	&	&	& $| H_{\beta}   \rightarrow L_{\beta} + 3 \rangle$ (0.5448)\\
			&  	&	&	& $| H_{\beta}   \rightarrow L_{\beta} + 4 \rangle$ (0.5448)\\
			&	&	&	& $| H_{\beta}   \rightarrow L_{\beta} + 10 \rangle$ (0.3673)\\
\label{Tab:table-neutral-octahedron}
\end{longtable}

\LTcapwidth=\columnwidth
\begin{longtable}{ccccl}
\captionsetup{font=footnotesize}
\caption{Excitation energies, $E$, and many-particle wavefunctions of excited
states corresponding to the peaks in the CIS linear absorption spectrum
of B$_{6}$ -- threaded tetramer isomer 
(\emph{cf}. Fig. \ref{fig:neutral-plot-threaded-tetramer}).
The subscript $x$ in the peak number denote the 
absorption due to light polarized along the long axis, 
and, $y,z$ denotes polarization perpendicular to it.
The rest of the information is the same as given in the caption for Table \ref{Tab:table-neutral-planar-ring-triplet}.}
\\

\hline \multicolumn{1}{c}{Peak} & \multicolumn{1}{c}{ $E$ (eV)} & \multicolumn{1}{c}{$f_{12}$ } 
 & \multicolumn{1}{c}{$\langle S^2 \rangle$} &\multicolumn{1}{l}{Wave Function}     \\ \hline 
\endfirsthead

\multicolumn{5}{c}%
{{\scshape \tablename\ \thetable{} -- continued from previous page}} \\ 

\hline \multicolumn{1}{c}{Peak} & \multicolumn{1}{c}{ $E$ (eV)} & \multicolumn{1}{c}{$f_{12}$ } 
 & \multicolumn{1}{c}{$\langle S^2 \rangle$} & \multicolumn{1}{l}{Wave Function}     \\ \hline 
\endhead

\hline \multicolumn{5}{r}{{{\scshape Continued on next page}}} \\ 
\endfoot

\hline \hline
\endlastfoot
\tabularnewline
Reference	&	&	&	& $|H_{1\alpha}^{1} ; H_{2\alpha}^{1} \rangle $	\\
\tabularnewline
I$_{x} $ 	& 3.06  &0.0297	&2.61	& $| H_{2\alpha} \rightarrow L_{\alpha} + 2 \rangle$(0.7532) \\
		&  	&	&	& $| H_{\beta} \rightarrow L_{\beta} + 1 \rangle$(0.2966) \\
\tabularnewline
II$_{x}$ 	& 4.03 	&0.0687	&2.07	& $| H_{1\alpha} \rightarrow L_{\alpha} + 2 \rangle$(0.5156) \\
		&  	&	&	& $| H_{\beta} \rightarrow L_{\beta}  \rangle$(0.3829) \\
		&  	&	&	& $| H_{\beta} - 1 \rightarrow L_{\beta} + 2  \rangle$(0.3347) \\
\tabularnewline
III$_{z}$ 	& 4.48 	&0.3363	&2.05	& $| H_{\beta}  \rightarrow L_{\beta} + 2  \rangle$(0.6089) \\
		&  	&	&	& $| H_{2\alpha}  \rightarrow L_{\alpha} + 1  \rangle$(0.5316) \\
\tabularnewline
IV$_{z}$ 	& 5.67 	&0.0283	&3.13	& $| H_{\beta}  \rightarrow L_{\beta} + 5  \rangle$(0.5808) \\
		&  	&	&	& $| H_{\beta}  \rightarrow L_{\beta} + 20 \rangle$(0.4305) \\
\tabularnewline
V$_{y}$ 	& 6.16 	&0.0384	&2.05	& $| H_{1\alpha} \rightarrow L_{\alpha} + 3 \rangle$(0.5497) \\
		&  	&	&	& $| H_{1\alpha} \rightarrow L_{\alpha} + 5 \rangle$(0.5357) \\
\tabularnewline
VI$_{y}$ 	& 6.36 	&0.2072	&2.32	& $| H_{\beta} - 3  \rightarrow L_{\beta}  \rangle$(0.5076) \\
		&  	&	&	& $| H_{\beta} - 1  \rightarrow L_{\beta} + 2  \rangle$(0.2922) \\
\label{Tab:table-neutral-threaded-tetramer}
\end{longtable}

\LTcapwidth=\columnwidth
\begin{longtable}{ccccl}
\captionsetup{font=footnotesize}
\caption{Excitation energies, $E$, and many-particle wavefunctions of excited
states corresponding to the peaks in the CIS linear absorption spectrum
of B$_{6}$ -- threaded trimer isomer 
(\emph{cf}. Fig. \ref{fig:neutral-plot-threaded-trimer}).
The subscript $\Vert$ , in the peak number denotes the 
absorption due to light polarized along the long 
axis of the isomer.
The rest of the information is the same as given in the caption for Table \ref{Tab:table-neutral-planar-ring-triplet}.}
\\

\hline \multicolumn{1}{c}{Peak} & \multicolumn{1}{c}{ $E$ (eV)} & \multicolumn{1}{c}{$f_{12}$ } 
 & \multicolumn{1}{c}{$\langle S^2 \rangle$} &\multicolumn{1}{l}{Wave Function}     \\ \hline 
\endfirsthead

\multicolumn{5}{c}%
{{\scshape \tablename\ \thetable{} -- continued from previous page}} \\ 

\hline \multicolumn{1}{c}{Peak} & \multicolumn{1}{c}{ $E$ (eV)} & \multicolumn{1}{c}{$f_{12}$ } 
 & \multicolumn{1}{c}{$\langle S^2 \rangle$} & \multicolumn{1}{l}{Wave Function}     \\ \hline 
\endhead

\hline \multicolumn{5}{r}{{{\scshape Continued on next page}}} \\ 
\endfoot

\hline \hline
\endlastfoot
\tabularnewline
Reference	&	&	&	& $| H_{1\alpha}^{1} ; H_{2\alpha}^{1} \rangle $ \\
\tabularnewline
I$_{\Vert} $ 	& 1.16  &0.0063	&2.08	& $| H_{\beta} - 1 \rightarrow L_{\beta} \rangle$(0.6839) \\
		&  	&	& 	& $| H_{\alpha} - 1 \rightarrow L_{\alpha}\rangle$(0.6061) \\
\tabularnewline
II$_{\Vert}$ 	& 2.28 	&0.0313	&2.083	& $| H_{\alpha} - 2 \rightarrow L_{\alpha} \rangle$(0.6226) \\
		&  	&	&	& $| H_{\beta} - 2 \rightarrow L_{\beta} \rangle$(0.6329) \\
\tabularnewline
III$_{\Vert}$ 	& 3.06 	&0.0214	&2.12	& $| H_{2\alpha} \rightarrow L_{\alpha} \rangle$(0.6536) \\
		&  	&	&	& $| H_{\beta} \rightarrow L_{\beta} \rangle$(0.5487) \\
\tabularnewline
IV$_{\Vert}$ 	& 4.03 	&0.6573	&2.08	& $| H_{\beta} \rightarrow L_{\beta} + 1 \rangle$(0.5280) \\
		&  	&	&	& $| H_{2\alpha} \rightarrow L_{\alpha} + 1  \rangle$(0.4969) \\
\tabularnewline
V$_{\Vert}$ 	& 4.73 	&0.596	&2.99	& $| H_{\beta} \rightarrow L_{\beta} + 7 \rangle$(0.3826) \\
		&  	&	&	& $| H_{\alpha} - 3 \rightarrow L_{\alpha} + 1 \rangle$(0.3516) \\
\tabularnewline
VI$_{\Vert}$ 	& 5.37 	&0.1391	&2.07	& $| H_{2\alpha} \rightarrow L_{\alpha} + 7 \rangle$(0.5799) \\
		&  	&	&	& $| H_{\beta} \rightarrow L_{\beta} + 8 \rangle$(0.4231) \\
\tabularnewline
VII$_{\Vert}$ 	& 5.69 	&0.2293	&2.19	& $| H_{\beta} \rightarrow L_{\beta} + 4 \rangle$(0.5125) \\
		&  	&	&	& $| H_{2\alpha} \rightarrow L_{\alpha} + 3  \rangle$(0.4084) \\
\label{Tab:table-neutral-threaded-trimer}
\end{longtable}

\LTcapwidth=\columnwidth
\begin{longtable}{cccl}
\captionsetup{font=footnotesize}
\caption{Excitation energies, $E$, and many-particle wavefunctions of excited
states corresponding to the peaks in the CIS linear absorption spectrum
of B$_{6}$ -- twisted trimers isomer 
(\emph{cf}. Fig. \ref{fig:neutral-plot-twisted-trimers}).
The subscripts $\Vert$ and $\bot$, in the peak number denote the 
absorption due to light polarized along, and perpendicular to 
the long axis of the isomer, respectively.
The rest of the information is the same as given in the caption for Table \ref{Tab:table-neutral-bulged-wheel}.}
\label{Tab:table-neutral-twisted-trimers}
\\

\hline \multicolumn{1}{c}{Peak} & \multicolumn{1}{c}{ $E$ (eV)} & \multicolumn{1}{c}{$f_{12}$ } 
  &\multicolumn{1}{l}{Wave Function}     \\ \hline 
\endfirsthead

\multicolumn{4}{c}%
{{\scshape \tablename\ \thetable{} -- continued from previous page}} \\ 

\hline \multicolumn{1}{c}{Peak} & \multicolumn{1}{c}{ $E$ (eV)} & \multicolumn{1}{c}{$f_{12}$ } 
  & \multicolumn{1}{l}{Wave Function}     \\ \hline 
\endhead

\hline \multicolumn{4}{r}{{{\scshape Continued on next page}}} \\ 
\endfoot

\hline \hline
\endlastfoot
\tabularnewline
I$_{\Vert}$ 	& 1.02  &0.0422	& $| H 	  \rightarrow L      \rangle$(0.6550) \\
		&	&	& $| H -1 \rightarrow L      \rangle$(0.6550) \\
		&  	&	& $| H    \rightarrow L + 1  \rangle$(0.5612) \\
		&	&	& $| H -1 \rightarrow L + 1  \rangle$(0.5612) \\
\tabularnewline
II$_{\bot}$ 	& 2.22 	&0.1279	& $| H - 2 \rightarrow L \rangle$(0.7650) \\
		&  	&	& $| H - 3 \rightarrow L + 1\rangle$(0.5316) \\
\tabularnewline
III$_{\Vert}$ 	& 3.58 	&0.1870	& $| H - 2 \rightarrow L + 2 \rangle$(0.4374) \\
		&	&	& $| H -5 \rightarrow L      \rangle$(0.4374) \\
		&  	&	& $| H - 4 \rightarrow L     \rangle$(0.4371) \\
		&	&	& $| H -2 \rightarrow L + 3  \rangle$(0.4371) \\
\tabularnewline
IV$_{\Vert}$	& 4.72 	&0.4337	& $| H - 2 \rightarrow L + 2 \rangle$(0.4874) \\
		&	&	& $| H - 2 \rightarrow L + 3 \rangle$(0.4874) \\
		&  	&	& $| H     \rightarrow L + 26\rangle$(0.3268) \\
		&	&	& $| H - 1 \rightarrow L + 26  \rangle$(0.3268)\\
\tabularnewline
V$_{\bot}$ 	& 5.26 	&0.3570	& $| H - 5 \rightarrow L + 3 \rangle$(0.5117) \\
		&  	&	& $| H - 4 \rightarrow L + 2 \rangle$(0.5010) \\
\tabularnewline
VI$_{\bot}$ 	& 5.87 	&3.4368	& $| H \rightarrow L + 3 \rangle$(0.5962) \\
		&  	&	& $| H - 1 \rightarrow L + 2 \rangle$(0.5962) \\
\tabularnewline
VII$_{\Vert}$ 	& 6.39  &0.2265 & $| H - 3 \rightarrow L + 2 \rangle$(0.4908) \\
		&	&	& $| H - 3 \rightarrow L + 3 \rangle$ (0.4908) \\
		&  	&	& $| H - 2 \rightarrow L + 3 \rangle$(0.4561) \\
		&	&	& $| H - 2 \rightarrow L + 2 \rangle$ (0.4561) \\
\tabularnewline
VIII$_{\bot}$ 	& 6.98 	&0.6645	& $| H - 3  \rightarrow L + 1\rangle$(0.6479) \\
		&  	&	& $| H - 2  \rightarrow L \rangle$(0.5508) \\
\tabularnewline
IX$_{\Vert}$ 	& 7.25 	&0.3140	& $| H \rightarrow L + 4 \rangle$ (0.5209) \\
		&	&	& $| H - 1 \rightarrow L + 4 \rangle$ (0.5209) \\
		&  	&	& $| H \rightarrow L + 7 \rangle$ (0.4072) \\
		&	&	& $| H - 1 \rightarrow L + 7 \rangle$ (0.4072) \\
\end{longtable}

\LTcapwidth=\columnwidth
\begin{longtable}{cccl}
\captionsetup{font=footnotesize}
\caption{Excitation energies, $E$, and many-particle wavefunctions of excited
states corresponding to the peaks in the CIS linear absorption spectrum
of B$_{6}$ -- planar trimers isomer 
(\emph{cf}. Fig. \ref{fig:neutral-plot-planar-trimers}).
The subscripts $\Vert$ and $\bot$, in the peak number denote the 
absorption due to light polarized in, and perpendicular to 
the plane of the isomer, respectively.
The rest of the information is the same as given in the caption for Table \ref{Tab:table-neutral-bulged-wheel}.}
\\

\hline \multicolumn{1}{c}{Peak} & \multicolumn{1}{c}{ $E$ (eV)} & \multicolumn{1}{c}{$f_{12}$ } 
  &\multicolumn{1}{l}{Wave Function}     \\ \hline 
\endfirsthead

\multicolumn{4}{c}%
{{\scshape \tablename\ \thetable{} -- continued from previous page}} \\ 

\hline \multicolumn{1}{c}{Peak} & \multicolumn{1}{c}{ $E$ (eV)} & \multicolumn{1}{c}{$f_{12}$ } 
  & \multicolumn{1}{l}{Wave Function}     \\ \hline 
\endhead

\hline \multicolumn{4}{r}{{{\scshape Continued on next page}}} \\ 
\endfoot

\hline \hline
\endlastfoot
\tabularnewline
I$_{\bot}$ 	& 0.97  &0.0393	& $| H \rightarrow L  \rangle$ (0.7804) \\
		&  	&	& $| H - 1 \rightarrow L + 1\rangle$(0.5529) \\
\tabularnewline
II$_{\Vert}$	& 2.22 	&0.1261	& $| H - 2 \rightarrow L \rangle$(0.7603) \\
		&  	&	& $| H - 4 \rightarrow L + 1\rangle$(0.5634) \\
\tabularnewline
III$_{\Vert}$ 	& 3.57 	&0.1293	& $| H - 2 \rightarrow L + 2 \rangle$ (0.6070) \\
		&  	&	& $| H - 3 \rightarrow L \rangle$(0.5440) \\
\tabularnewline
IV$_{\Vert}$	& 4.67 	&0.7257	& $| H - 3 \rightarrow L + 2 \rangle$ (0.7242) \\
		&  	&	& $| H \rightarrow L + 12 \rangle$ (0.3824) \\
		&  	&	& $| H - 2 \rightarrow L + 2 \rangle$ (0.5131) \\
		&  	&	& $| H - 3 \rightarrow L \rangle$ (0.4653) \\
\tabularnewline
V$_{\Vert}$ 	& 6.43 	&4.5721	& $| H \rightarrow L + 12 \rangle$(0.5256) \\
		&  	&	& $| H - 3 \rightarrow L + 2 \rangle$(0.4859) \\
\tabularnewline
VI$_{\Vert}$ 	& 6.99 	&1.8466	& $| H - 4 \rightarrow L + 1 \rangle$(0.5794) \\
		&  	&	& $| H \rightarrow L + 7 \rangle$(0.5156) \\
\tabularnewline
VII$_{\Vert}$ 	& 7.34  &0.6386	& $| H \rightarrow L + 12 \rangle$(0.6754) \\
		&  	&	& $| H \rightarrow L + 7 \rangle$(0.4922) \\
\label{Tab:table-neutral-planar-trimers}
\end{longtable}

\LTcapwidth=\columnwidth
\begin{longtable}{cccl}
\captionsetup{font=footnotesize}
\caption{Excitation energies, $E$, and many-particle wavefunctions of excited
states corresponding to the peaks in the CIS linear absorption spectrum
of B$_{6}$ -- convex bowl isomer 
(\emph{cf}. Fig. \ref{fig:neutral-plot-convex-bowl}).
The subscripts $\Vert$ and $\bot$, in the peak number denote the 
absorption due to light polarized in, and perpendicular to 
the plane of the isomer, respectively.
The rest of the information is the same as given in the caption for Table \ref{Tab:table-neutral-bulged-wheel}.}
\\

\hline \multicolumn{1}{c}{Peak} & \multicolumn{1}{c}{ $E$ (eV)} & \multicolumn{1}{c}{$f_{12}$ } 
  &\multicolumn{1}{l}{Wave Function}     \\ \hline 
\endfirsthead

\multicolumn{4}{c}%
{{\scshape \tablename\ \thetable{} -- continued from previous page}} \\ 

\hline \multicolumn{1}{c}{Peak} & \multicolumn{1}{c}{ $E$ (eV)} & \multicolumn{1}{c}{$f_{12}$ } 
  & \multicolumn{1}{l}{Wave Function}     \\ \hline 
\endhead

\hline \multicolumn{4}{r}{{{\scshape Continued on next page}}} \\ 
\endfoot

\hline \hline
\endlastfoot
\tabularnewline
I$_{\Vert}$ 	& 1.58  &0.0486	& $| H - 1 \rightarrow L + 1     \rangle$ (0.9774) \\
\tabularnewline
II$_{\Vert,\bot}$& 1.80 &0.0679	& $| H \rightarrow L + 2 \rangle$(0.9079) \\
		&  	&	& $| H - 1 \rightarrow L + 3\rangle$(0.2748) \\
\tabularnewline
III$_{\Vert}$ 	& 2.43 	&0.6023	& $| H \rightarrow L \rangle$ (0.8644) \\
		&  	&	& $| H - 4 \rightarrow L + 1 \rangle$(0.4364) \\
\tabularnewline
IV$_{\Vert}$	& 2.89 	&0.1811	& $| H - 1 \rightarrow L  \rangle$ (0.8058) \\
		&  	&	& $| H - 3 \rightarrow L + 1\rangle$ (0.5421) \\
\tabularnewline
V$_{\Vert}$ 	& 4.09 	&0.1602	& $| H - 4 \rightarrow L \rangle$(0.5608) \\
		&  	&	& $| H - 3 \rightarrow L + 1 \rangle$(0.5347) \\
\tabularnewline
VI$_{\Vert,\bot}$ & 5.13 &0.0649& $| H - 4 \rightarrow L + 3 \rangle$(0.9070) \\
		&  	&	& $| H - 3 \rightarrow L + 2 \rangle$(0.1753) \\
\tabularnewline
VII$_{\Vert}$ 	& 6.21  &1.4363&  $| H \rightarrow L + 7 \rangle$(0.4118) \\
		&  	&	& $| H \rightarrow L + 4 \rangle$(0.3622) \\
\tabularnewline
VIII$_{\bot}$ 	& 6.39 	&2.2019	& $| H \rightarrow L + 6 \rangle$(0.4335) \\
		&  	&	& $| H - 2  \rightarrow L + 3 \rangle$(0.4081) \\
\tabularnewline
IX$_{\Vert}$ 	& 6.90 	&0.1476	& $| H \rightarrow L + 8 \rangle$(0.4435) \\
		&  	&	& $| H \rightarrow L + 13 \rangle$(0.3534) \\
\label{Tab:table-neutral-convex-bowl}
\end{longtable}

\LTcapwidth=\columnwidth
\begin{longtable}{cccl}
\captionsetup{font=footnotesize}
\caption{Excitation energies, $E$, and many-particle wavefunctions of excited
states corresponding to the peaks in the CIS linear absorption spectrum
of B$_{6}$ -- linear isomer 
 (\emph{cf}. Fig. \ref{fig:neutral-plot-linear}).
The subscripts $\Vert$ and $\bot$, in the peak number denote the 
absorption due to light polarized along, and perpendicular to the axis of 
the isomer, respectively.
The rest of the information is the same as given in the caption for Table \ref{Tab:table-neutral-bulged-wheel}.}
\\

\hline \multicolumn{1}{c}{Peak} & \multicolumn{1}{c}{ $E$ (eV)} & \multicolumn{1}{c}{$f_{12}$ } 
  &\multicolumn{1}{l}{Wave Function}     \\ \hline 
\endfirsthead

\multicolumn{4}{c}%
{{\scshape \tablename\ \thetable{} -- continued from previous page}} \\ 

\hline \multicolumn{1}{c}{Peak} & \multicolumn{1}{c}{ $E$ (eV)} & \multicolumn{1}{c}{$f_{12}$ } 
  & \multicolumn{1}{l}{Wave Function}     \\ \hline 
\endhead

\hline \multicolumn{4}{r}{{{\scshape Continued on next page}}} \\ 
\endfoot

\hline \hline
\endlastfoot
\tabularnewline
I$_{\Vert} $ 	& 5.51	&12.8358& $| H - 1 \rightarrow L + 3 \rangle$(0.6510) \\
		&  	&	& $| H \rightarrow L + 2 \rangle$(0.6489) \\
\tabularnewline
II$_{\bot}$ 	& 6.51 	&1.6532 & $| H \rightarrow L + 4 \rangle $(0.7148)\\
		&	&	& $| H - 1 \rightarrow L + 4 \rangle$(0.7148) \\
		&  	&	& $| H - 3 \rightarrow L + 8 \rangle $(0.3847)\\
		&	&	& $| H - 2 \rightarrow L + 8 \rangle$(0.3847) \\
\label{Tab:table-neutral-linear}
\end{longtable}

\FloatBarrier

\LTcapwidth=\columnwidth
\begin{longtable}{ccccl}
\captionsetup{font=footnotesize}
\caption{Excitation energies, $E$, and many-particle wavefunctions of excited
states corresponding to the peaks in the CIS linear absorption spectrum
of B$_{6}^{+}$ -- planar ring isomer (\emph{cf}. Fig. \ref{fig:cationic-plot-planar-ring}).
The subscripts $\Vert$ and $\bot$, in the peak number denote the 
absorption due to light polarized in, and perpendicular to the plane of 
the isomer, respectively.
 In the wavefunction, the bracketed numbers are the CI coefficients of a given electronic configuration.
Symbols $H$, $H_{1}$ and $L$ denote HOMO, SOMO and LUMO orbitals respectively.}
\\

\hline \multicolumn{1}{c}{Peak} & \multicolumn{1}{c}{ $E$ (eV)} & \multicolumn{1}{c}{$f_{12}$ } 
 & \multicolumn{1}{c}{$\langle S^2 \rangle$} &\multicolumn{1}{l}{Wave Function}     \\ \hline 
\endfirsthead

\multicolumn{5}{c}%
{{\scshape \tablename\ \thetable{} -- continued from previous page}} \\ 

\hline \multicolumn{1}{c}{Peak} & \multicolumn{1}{c}{ $E$ (eV)} & \multicolumn{1}{c}{$f_{12}$ } 
 & \multicolumn{1}{c}{$\langle S^2 \rangle$} & \multicolumn{1}{l}{Wave Function}     \\ \hline 
\endhead

\hline \multicolumn{5}{r}{{{\scshape Continued on next page}}} \\ 
\endfoot

\hline \hline
\endlastfoot
\tabularnewline
Reference	&	&	&	& $|H_{1\alpha}^{1} \rangle $ \\
\tabularnewline
I$_{\bot}$ 	& 4.44  &0.4236	&0.84	& $| H_{\beta} - 1 \rightarrow L_{\beta} \rangle$(0.5530) \\
		&  	&	&	& $| H_{\alpha} - 1 \rightarrow L_{\alpha} \rangle$(0.5077) \\
\tabularnewline
II$_{\Vert}$ 	& 5.32 	&0.1504	&1.02	& $| H_{\beta} - 1 \rightarrow L_{\beta} + 3 \rangle$(0.4780) \\
		&  	&	&	& $| H_{\alpha} - 3 \rightarrow L_{\alpha} + 3 \rangle$(0.4433) \\
\tabularnewline
III$_{\Vert}$ 	& 6.09 	&0.0987	&0.88	& $| H_{\alpha} - 5 \rightarrow L_{\alpha}  \rangle$(0.5713) \\
		&  	&	&	& $| H_{\beta} - 4 \rightarrow L_{\beta} \rangle$(0.5595) \\
\tabularnewline
IV$_{\Vert}$ 	& 6.85 	&0.8066	&0.87	& $| H_{1\alpha}  \rightarrow L_{\alpha} + 6  \rangle$(0.6347) \\
		&  	&	&	& $| H_{\alpha}  \rightarrow L_{\alpha} + 9 \rangle$(0.3584) \\
\label{Tab:table-cationic-planar-ring}
\end{longtable}

\LTcapwidth=\columnwidth
\begin{longtable}{ccccl}
\captionsetup{font=footnotesize}
\caption{Excitation energies, $E$, and many-particle wavefunctions of excited
states corresponding to the peaks in the CIS linear absorption spectrum
of B$_{6}^{+}$ -- bulged wheel isomer (\emph{cf}. Fig. \ref{fig:cationic-plot-bulged-wheel}).
The subscripts $\Vert$ and $\bot$, in the peak number denote the 
absorption due to light polarized in, and perpendicular to the plane 
of the wheel, respectively.
The rest of the information is the same as given in the caption for Table \ref{Tab:table-cationic-planar-ring}.}
\\

\hline \multicolumn{1}{c}{Peak} & \multicolumn{1}{c}{ $E$ (eV)} & \multicolumn{1}{c}{$f_{12}$ } 
 & \multicolumn{1}{c}{$\langle S^2 \rangle$} &\multicolumn{1}{l}{Wave Function}     \\ \hline 
\endfirsthead

\multicolumn{5}{c}%
{{\scshape \tablename\ \thetable{} -- continued from previous page}} \\ 

\hline \multicolumn{1}{c}{Peak} & \multicolumn{1}{c}{ $E$ (eV)} & \multicolumn{1}{c}{$f_{12}$ } 
 & \multicolumn{1}{c}{$\langle S^2 \rangle$} & \multicolumn{1}{l}{Wave Function}     \\ \hline 
\endhead

\hline \multicolumn{5}{r}{{{\scshape Continued on next page}}} \\ 
\endfoot

\hline \hline
\endlastfoot
\tabularnewline
Reference	&	&	& 	& $|H_{1\alpha}^{1} \rangle $ \\
\tabularnewline
I$_{\Vert}$ 	& 1.76  &0.0018	&1.53	& $| H_{\beta} \rightarrow L_{\beta} \rangle$(0.7618) \\
		&  	&	&	& $| H_{\beta} - 1 \rightarrow L_{\beta} \rangle$(0.5218) \\
\tabularnewline
II$_{\Vert}$ 	& 2.19 	&0.0049	&1.02	& $| H_{\beta} - 1 \rightarrow L_{\beta} \rangle$(0.8082) \\
		&  	&	&	& $| H_{\beta} \rightarrow L_{\beta} \rangle$(0.5406) \\
\tabularnewline
III$_{\Vert}$ 	& 2.84 	&0.0074	&2.43	& $| H_{1\alpha} \rightarrow L_{\alpha} \rangle$(0.5381) \\
		&  	&	&	& $| H_{\beta} \rightarrow L_{\beta} + 1 \rangle $(0.4385) \\
\tabularnewline
IV$_{\Vert}$ 	& 3.58 	&0.0183	&1.80	& $| H_{\alpha} - 2 \rightarrow L_{\alpha} + 7 \rangle$(0.4521) \\
		&  	&	&	& $| H_{1\alpha} \rightarrow L_{\alpha} \rangle$(0.4594) \\
\tabularnewline
V$_{\Vert}$ 	& 3.90 	&0.0114	&1.74	& $| H_{1\alpha} \rightarrow L_{\alpha} + 1 \rangle$(0.6233) \\
		&  	&	&	& $| H_{\alpha} - 2 \rightarrow L_{\alpha} + 2 \rangle$(0.5090) \\
\tabularnewline
VI$_{\Vert}$ 	& 4.97  &0.1173	&1.98	& $| H_{\alpha} - 1 \rightarrow L_{\alpha} \rangle$(0.5925) \\
		&  	&	&	& $| H_{\alpha} - 3 \rightarrow L_{\alpha} \rangle$(0.3915) \\
\tabularnewline
VII$_{\Vert}$ 	& 5.74 	&0.0364	&2.04	& $| H_{\alpha} - 4 \rightarrow L_{\alpha} + 1 \rangle$(0.3974) \\
		&  	&	&	& $| H_{\alpha} - 2 \rightarrow L_{\alpha} + 3 \rangle$(0.3930) \\
\tabularnewline
VIII$_{\Vert}$ 	& 6.05 	&0.0054	&2.19	& $| H_{\alpha} - 4 \rightarrow L_{\alpha} + 1 \rangle$(0.4416) \\
		&  	&	&	& $| H_{\alpha} - 3 \rightarrow L_{\alpha} \rangle$(0.3428) \\
\tabularnewline
IX$_{\Vert,\bot}$& 6.24 &0.1731	&1.53	& $| H_{\beta} - 2 \rightarrow L_{\beta} + 2 \rangle$(0.4368) \\
		&  	&	&	& $| H_{\alpha} - 1 \rightarrow L_{\alpha} + 3 \rangle$(0.3700) \\
\label{Tab:table-cationic-bulged-wheel}
\end{longtable}


\LTcapwidth=\columnwidth
\begin{longtable}{ccccl}
\captionsetup{font=footnotesize}
\caption{Excitation energies, $E$, and many-particle wavefunctions of excited
 states corresponding to the peaks in the CIS linear absorption spectrum
 of B$_{6}^{+}$ -- planar ring (II) isomer (\emph{cf}. Fig. \ref{fig:cationic-plot-planar-d2h-ring}).
The subscript $\Vert$ in the peak number denote the 
  absorption due to light polarized in the plane of 
  the isomer.
The rest of the information is the same as given in the caption for Table \ref{Tab:table-cationic-planar-ring}.}
  \\

\hline \multicolumn{1}{c}{Peak} & \multicolumn{1}{c}{ $E$ (eV)} & \multicolumn{1}{c}{$f_{12}$ } 
 & \multicolumn{1}{c}{$\langle S^2 \rangle$} &\multicolumn{1}{l}{Wave Function}     \\ \hline 
\endfirsthead

\multicolumn{5}{c}%
{{\scshape \tablename\ \thetable{} -- continued from previous page}} \\ 

\hline \multicolumn{1}{c}{Peak} & \multicolumn{1}{c}{ $E$ (eV)} & \multicolumn{1}{c}{$f_{12}$ } 
 & \multicolumn{1}{c}{$\langle S^2 \rangle$} & \multicolumn{1}{l}{Wave Function}     \\ \hline 
\endhead

\hline \multicolumn{5}{r}{{{\scshape Continued on next page}}} \\ 
\endfoot

\hline \hline
\endlastfoot
\tabularnewline
Reference	&	&	&	& $| H_{1\alpha}^{1} \rangle $ \\
 \tabularnewline
 I$_{\Vert}$ 	& 2.62  &0.0100	&2.30	& $| H_{\beta} \rightarrow L_{\beta} + 2 \rangle$(0.8930) \\
 		&  	&	&	& $| H_{\alpha} - 2 \rightarrow L_{\alpha} + 1 \rangle$(0.3196) \\
 \tabularnewline
 II$_{\Vert}$ 	& 3.52 	&0.4311	&0.86	& $| H_{\beta} - 1 \rightarrow L_{\beta} \rangle$(0.5075) \\
 		&  	&	&	& $| H_{\beta} \rightarrow L_{\beta} + 1 \rangle$(0.5047) \\
 \tabularnewline
 III$_{\Vert}$ 	& 4.59 	&0.0377	&0.96	& $| H_{\beta} - 3 \rightarrow L_{\beta} \rangle$(0.6407) \\
 		&  	&	&	& $| H_{\alpha} - 4 \rightarrow L_{\alpha} \rangle$(0.4707) \\
\label{Tab:table-cationic-planar-d2h-ring}
\end{longtable}

\LTcapwidth=\columnwidth
\begin{longtable}{ccccl}
\captionsetup{font=footnotesize}
\caption{Excitation energies, $E$, and many-particle wavefunctions of excited
states corresponding to the peaks in the CIS linear absorption spectrum
of B$_{6}^{+}$ -- incomplete wheel (quartet) isomer 
(\emph{cf}. Fig. \ref{fig:cationic-plot-incomplete-wheel}).
The subscripts $\Vert$ and $\bot$, in the peak number denote the 
absorption due to light polarized in, and perpendicular 
to the plane of the isomer, respectively.
In the wavefunction, the bracketed numbers are the CI coefficients of a given electronic configuration.
Symbols $H$, $H_{1\alpha}$, $H_{2\alpha}$, $H_{3\alpha}$ and $L$ denote HOMO, SOMOs and LUMO orbitals respectively.}
\\

\hline \multicolumn{1}{c}{Peak} & \multicolumn{1}{c}{ $E$ (eV)} & \multicolumn{1}{c}{$f_{12}$ } 
 & \multicolumn{1}{c}{$\langle S^2 \rangle$} &\multicolumn{1}{l}{Wave Function}     \\ \hline 
\endfirsthead

\multicolumn{5}{c}%
{{\scshape \tablename\ \thetable{} -- continued from previous page}} \\ 

\hline \multicolumn{1}{c}{Peak} & \multicolumn{1}{c}{ $E$ (eV)} & \multicolumn{1}{c}{$f_{12}$ } 
 & \multicolumn{1}{c}{$\langle S^2 \rangle$} & \multicolumn{1}{l}{Wave Function}     \\ \hline 
\endhead

\hline \multicolumn{5}{r}{{{\scshape Continued on next page}}} \\ 
\endfoot

\hline \hline
\endlastfoot
\tabularnewline
Reference	&	&	&	& $|H_{1\alpha}^{1};H_{2\alpha}^{1};H_{3\alpha}^{1} \rangle $ \\
\tabularnewline
I$_{\bot}$ 	& 1.69  &0.0228	&4.38	& $| H_{\beta} - 2 \rightarrow L_{\beta} \rangle$(0.7070) \\
		&  	&	&	& $| H_{\beta} - 1 \rightarrow L_{\beta} \rangle$(0.5707) \\
\tabularnewline
II$_{\Vert}$	& 2.31 	&0.0263	&4.27	& $| H_{\beta} - 2 \rightarrow L_{\beta} \rangle$(0.6484) \\
		&  	&	&	& $| H_{\beta} - 1 \rightarrow L_{\beta} \rangle$(0.5313) \\
\tabularnewline
III$_{\bot}$ 	& 3.00 	&0.0438	&4.35	& $| H_{\alpha} - 1 \rightarrow L_{\alpha} \rangle$(0.6704) \\
		&  	&	&	& $| H_{2\alpha} \rightarrow L_{\alpha} \rangle$(0.3992) \\
\tabularnewline
IV$_{\Vert}$ 	& 3.39  &0.0227	&4.64	& $| H_{1\alpha} \rightarrow L_{\alpha} \rangle$(0.8546) \\
		&  	&	&	& $| H_{2\alpha} \rightarrow L_{\alpha} + 2  \rangle$(0.2748) \\
\tabularnewline
V$_{\Vert}$	& 3.71	&0.0365	&4.32	& $| H_{3\alpha} \rightarrow L_{\alpha} + 1  \rangle$(0.5299) \\
		&  	&	&	& $| H_{\alpha} - 1 \rightarrow L_{\alpha}  \rangle$(0.5106) \\
\tabularnewline
VI$_{\Vert}$ 	& 4.21 	&0.0987	&4.56	& $| H_{\beta} \rightarrow L_{\beta} + 1 \rangle$(0.8754) \\
		&  	&	&	& $| H_{3\alpha} \rightarrow L_{\alpha} + 9 \rangle$(0.2200) \\
\tabularnewline
VII$_{\bot}$ 	& 5.99  &0.2462	&5.33	& $| H_{\beta} - 1 \rightarrow L_{\beta} + 5 \rangle$(0.5679) \\
		&  	&	&	& $| H_{3\alpha} \rightarrow L_{\alpha} + 4 \rangle$(0.4243) \\
\label{Tab:table-cationic-incomplete-wheel}
\end{longtable}

\LTcapwidth=\columnwidth
\begin{longtable}{ccccl}
\captionsetup{font=footnotesize}
\caption{Excitation energies, $E$, and many-particle wavefunctions of excited
states corresponding to the peaks in the CIS linear absorption spectrum
of B$_{6}^{+}$ -- threaded trimer (quartet) isomer 
(\emph{cf}. Fig. \ref{fig:cationic-plot-threaded-trimer}).
The subscripts $\Vert$ and $\bot$, in the peak number denote the 
absorption due to light polarized along, and perpendicular 
to the long axis of the isomer, respectively.
The rest of the information is the same as given in the caption for Table \ref{Tab:table-cationic-incomplete-wheel}.}
\\

\hline \multicolumn{1}{c}{Peak} & \multicolumn{1}{c}{ $E$ (eV)} & \multicolumn{1}{c}{$f_{12}$ } 
 & \multicolumn{1}{c}{$\langle S^2 \rangle$} &\multicolumn{1}{l}{Wave Function}     \\ \hline 
\endfirsthead

\multicolumn{5}{c}%
{{\scshape \tablename\ \thetable{} -- continued from previous page}} \\ 

\hline \multicolumn{1}{c}{Peak} & \multicolumn{1}{c}{ $E$ (eV)} & \multicolumn{1}{c}{$f_{12}$ } 
 & \multicolumn{1}{c}{$\langle S^2 \rangle$} & \multicolumn{1}{l}{Wave Function}     \\ \hline 
\endhead

\hline \multicolumn{5}{r}{{{\scshape Continued on next page}}} \\ 
\endfoot

\hline \hline
\endlastfoot
\tabularnewline
Reference	&	&	& 	& $|H_{1\alpha}^{1};H_{2\alpha}^{1};H_{3\alpha}^{1} \rangle $ \\
\tabularnewline
I$_{\Vert}$ 	& 2.36  &0.6487	&3.98	& $| H_{\alpha} - 1 \rightarrow L_{\alpha} + 2 \rangle$(0.6487) \\
		&  	&	&	& $| H_{\beta} - 1 \rightarrow L_{\beta} + 2  \rangle$(0.5646) \\
\tabularnewline
II$_{\Vert,\bot}$& 3.90 &0.3729	&4.00	& $| H_{3\alpha} \rightarrow L_{\alpha} + 1 \rangle$(0.6329) \\
		&  	&	& 	& $| H_{\beta} \rightarrow L_{\beta} + 1  \rangle$(0.4181) \\
\tabularnewline
III$_{\Vert}$ 	& 4.34 	&0.7430	&3.79	& $| H_{2\alpha} \rightarrow L_{\alpha} \rangle$(0.6197) \\
		&  	&	&	& $| H_{1\alpha} \rightarrow L_{\alpha} + 1 \rangle$(0.6053) \\
\tabularnewline
III$_{\Vert}$ 	& 5.92 	&0.1573	&3.91	& $| H_{\beta} - 2 \rightarrow L_{\beta}  \rangle$(0.3742) \\
		&  	&	&	& $| H_{\alpha} - 2 \rightarrow L_{\alpha} \rangle$(0.3477) \\
\label{Tab:table-cationic-threaded-trimer}
\end{longtable}

\LTcapwidth=\columnwidth
\begin{longtable}{ccccl}
\captionsetup{font=footnotesize}
\caption{Excitation energies, $E$, and many-particle wavefunctions of excited
states corresponding to the peaks in the CIS linear absorption spectrum
of B$_{6}^{+}$ -- tetragonal bipyramid isomer 
(\emph{cf}. Fig. \ref{fig:cationic-plot-bipyramid}).
The subscripts $\Vert$ and $\bot$, in the peak number denote the 
absorption due to light polarized in, and perpendicular to the square plane of 
bipyramid, respectively.
The rest of the information is the same as given in the caption for Table \ref{Tab:table-cationic-planar-ring}.}
\\

\hline \multicolumn{1}{c}{Peak} & \multicolumn{1}{c}{ $E$ (eV)} & \multicolumn{1}{c}{$f_{12}$ } 
 & \multicolumn{1}{c}{$\langle S^2 \rangle$} &\multicolumn{1}{l}{Wave Function}     \\ \hline 
\endfirsthead

\multicolumn{5}{c}%
{{\scshape \tablename\ \thetable{} -- continued from previous page}} \\ 

\hline \multicolumn{1}{c}{Peak} & \multicolumn{1}{c}{ $E$ (eV)} & \multicolumn{1}{c}{$f_{12}$ } 
 & \multicolumn{1}{c}{$\langle S^2 \rangle$} & \multicolumn{1}{l}{Wave Function}     \\ \hline 
\endhead

\hline \multicolumn{5}{r}{{{\scshape Continued on next page}}} \\ 
\endfoot

\hline \hline
\endlastfoot
\tabularnewline
Reference	&	&	& 	& $| H_{1\alpha}^{1} \rangle $ \\
\tabularnewline
I$_{\bot}$ 	& 1.23  &0.1144	&1.29 	& $| H_{1\alpha} \rightarrow L_{\alpha} \rangle$(0.9530) \\
		&  	&	& 	& $| H_{1\alpha} \rightarrow L_{\alpha} + 1 \rangle$(0.9530) \\
		&  	&	& 	& $| H_{\beta} \rightarrow L_{\beta} + 3 \rangle$(0.1202) \\
\tabularnewline
II$_{\Vert}$ 	& 2.55 	&0.0191	&2.40 	& $| H_{\alpha} - 4 \rightarrow L_{\alpha} \rangle$(0.6498) \\
		&  	&	& 	& $| H_{\alpha} - 3 \rightarrow L_{\alpha} + 1 \rangle$(0.6498) \\
		&  	&	& 	& $| H_{\beta} \rightarrow L_{\beta} + 1 \rangle$(0.2971) \\
\tabularnewline
III$_{\Vert,\bot}$ & 3.61 &0.1002&2.10	& $| H_{\beta} \rightarrow L_{\beta} + 1 \rangle$(0.8844) \\
		&  	&	& 	& $| H_{\alpha} - 4 \rightarrow L_{\alpha} \rangle$(0.2133) \\
		&  	&	& 	& $| H_{\alpha} - 3 \rightarrow L_{\alpha} + 1 \rangle$(0.2133) \\
\tabularnewline
IV$_{\Vert}$ 	& 5.54 	&0.0223	&1.36	& $| H_{1\alpha} \rightarrow L_{\alpha} + 5 \rangle$(0.8524) \\
		&  	&	&	& $| H_{\beta} - 3 \rightarrow L_{\beta} + 3 \rangle$(0.2325) \\
		&  	&	&	& $| H_{\beta} - 2 \rightarrow L_{\beta} + 4 \rangle$(0.2325) \\
\label{Tab:table-cationic-triangular-bipyramid}
\end{longtable}

\LTcapwidth=\columnwidth
\begin{longtable}{ccccl}
\captionsetup{font=footnotesize}
\caption{Excitation energies, $E$, and many-particle wavefunctions of excited
states corresponding to the peaks in the CIS linear absorption spectrum
of B$_{6}^{+}$ -- linear (quartet) isomer 
(\emph{cf}. Fig. \ref{fig:cationic-plot-linear}).
The subscripts $\Vert$ and $\bot$, in the peak number denote the 
absorption due to light polarized along, and perpendicular 
to the long axis of the isomer, respectively.
The rest of the information is the same as given in the caption for Table \ref{Tab:table-cationic-incomplete-wheel}.}
\\

\hline \multicolumn{1}{c}{Peak} & \multicolumn{1}{c}{ $E$ (eV)} & \multicolumn{1}{c}{$f_{12}$ } 
 & \multicolumn{1}{c}{$\langle S^2 \rangle$} &\multicolumn{1}{l}{Wave Function}     \\ \hline 
\endfirsthead

\multicolumn{5}{c}%
{{\scshape \tablename\ \thetable{} -- continued from previous page}} \\ 

\hline \multicolumn{1}{c}{Peak} & \multicolumn{1}{c}{ $E$ (eV)} & \multicolumn{1}{c}{$f_{12}$ } 
 & \multicolumn{1}{c}{$\langle S^2 \rangle$} & \multicolumn{1}{l}{Wave Function}     \\ \hline 
\endhead

\hline \multicolumn{5}{r}{{{\scshape Continued on next page}}} \\ 
\endfoot

\hline \hline
\endlastfoot
\tabularnewline
Reference	&	&	& 	& $| H_{1\alpha}^{1};H_{2\alpha}^{1};H_{3\alpha}^{1} \rangle $ \\
\tabularnewline
I$_{\Vert}$ 	& 4.25  &2.9960	& 5.43	& $| H_{\beta} - 1  \rightarrow L_{\beta} + 1 \rangle$(0.5588) \\
		&  	&	&	& $| H_{\beta}  \rightarrow L_{\beta} \rangle$(0.5588) \\
		&  	&	&	& $| H_{\alpha} - 4  \rightarrow L_{\alpha} \rangle$(0.3565) \\
\tabularnewline
II$_{\bot}$	& 4.66 	&1.1384	&5.62	& $| H_{\alpha} - 4  \rightarrow L_{\alpha} \rangle$(0.5669) \\
		&  	&	&	& $| H_{1\alpha}   \rightarrow L_{\alpha} + 2 \rangle$(0.4295) \\
		&  	&	&	& $| H_{2\alpha}  \rightarrow L_{\alpha} + 1 \rangle$(0.4295) \\
\label{Tab:table-cationic-linear}
\end{longtable}

\LTcapwidth=\columnwidth
\begin{longtable}{ccccl}
\captionsetup{font=footnotesize}
\caption{Excitation energies, $E$, and many-particle wavefunctions of excited
states corresponding to the peaks in the CIS linear absorption spectrum
of B$_{6}^{+}$ -- planar trimers isomer 
(\emph{cf}. Fig. \ref{fig:cationic-plot-planar-trimers}).
The subscripts $\Vert$ and $\bot$, in the peak number denote the 
absorption due to light polarized in, and perpendicular to the plane of 
the isomer, respectively.
The rest of the information is the same as given in the caption for Table \ref{Tab:table-cationic-planar-ring}.}
\\

\hline \multicolumn{1}{c}{Peak} & \multicolumn{1}{c}{ $E$ (eV)} & \multicolumn{1}{c}{$f_{12}$ } 
 & \multicolumn{1}{c}{$\langle S^2 \rangle$} &\multicolumn{1}{l}{Wave Function}     \\ \hline 
\endfirsthead

\multicolumn{5}{c}%
{{\scshape \tablename\ \thetable{} -- continued from previous page}} \\ 

\hline \multicolumn{1}{c}{Peak} & \multicolumn{1}{c}{ $E$ (eV)} & \multicolumn{1}{c}{$f_{12}$ } 
 & \multicolumn{1}{c}{$\langle S^2 \rangle$} & \multicolumn{1}{l}{Wave Function}     \\ \hline 
\endhead

\hline \multicolumn{5}{r}{{{\scshape Continued on next page}}} \\ 
\endfoot

\hline \hline
\endlastfoot
\tabularnewline
Reference	&	&	& 	& $| H_{1\alpha}^{1} \rangle $ \\
\tabularnewline
I$_{\Vert}$ 	& 1.40  &0.1363	&1.17	& $| H_{\beta} \rightarrow L_{\beta} \rangle$(0.8954) \\
		&  	&	&	& $| H_{\beta} - 2 \rightarrow L_{\beta} + 1 \rangle$(0.2557) \\
\tabularnewline
II$_{\bot}$ 	& 2.36 	&0.0293	&1.03	& $| H_{\alpha} - 2 \rightarrow L_{\alpha} \rangle$(0.6214) \\
		&  	&	&	& $| H_{\alpha} - 4 \rightarrow L_{\alpha} + 1 \rangle$(0.4845) \\
\tabularnewline
III$_{\bot}$ 	& 3.66 	&0.0507	&1.03	& $| H_{\alpha} - 2 \rightarrow L_{\alpha} + 2 \rangle$(0.5121) \\
		&  	&	&	& $| H_{\beta} - 1 \rightarrow L_{\beta} + 1 \rangle$(0.3618) \\
\tabularnewline
IV$_{\bot}$ 	& 4.74 	&0.0891	&1.26 	& $| H_{1\alpha} \rightarrow L_{\alpha} + 6 \rangle$(0.4977) \\
		&  	&	&  	& $| H_{\alpha} - 1 \rightarrow L_{\alpha} + 5 \rangle$(0.4270) \\
\tabularnewline
V$_{\Vert}$ 	& 5.19	&0.1498	&1.09	& $| H_{\alpha} - 3 \rightarrow L_{\alpha} + 2 \rangle$(0.6496) \\
		&  	&	&	& $| H_{1\alpha} \rightarrow L_{\alpha} + 4 \rangle$(0.4993) \\
\label{Tab:table-cationic-planar-trimers}
\end{longtable}

\chapter{\label{app:wavefunction_eom_wheel}Boron Clusters B$_{7}$, B$_{8}$ and B$_{9}$}
\section*{Excited State Wavefunctions, Energies and Oscillator Strengths}

\LTcapwidth=\columnwidth
\begin{longtable}{cccl}
 \caption{Excitation energies, $E$, and many-particle wavefunctions of excited
 states corresponding to the peaks in the EOM-CCSD linear absorption spectrum
  of B$_{7}$ boron wheel cluster (\emph{cf}. Fig. \ref{fig:geometries_wheel}\protect\subref{subfig:subfig-geom-b7}).
 The subscripts $\Vert$ and $\bot$, in the peak number denote the absorption due to light polarized in, and perpendicular to the plane of 
wheel base, respectively. In the wavefunction, the bracketed numbers are the CI coefficients of a given electronic configuration.
 Symbols $H_{1\alpha}$ denotes SOMO discussed earlier, and $H$, and $L$, denote HOMO and LUMO orbitals respectively. Note
 that, the reference state does not correspond to any peak, instead it represents the reference state from which singles excitations are occurring.}  \\

\hline \multicolumn{1}{c}{Peak} & \multicolumn{1}{c}{ $E$ (eV)} & \multicolumn{1}{c}{$f_{12}$ } 
  &\multicolumn{1}{l}{Wave Function}     \\ \hline 
\endfirsthead

\multicolumn{4}{c}%
{{\scshape \tablename\ \thetable{} -- continued from previous page}} \\ 

\hline \multicolumn{1}{c}{Peak} & \multicolumn{1}{c}{ $E$ (eV)} & \multicolumn{1}{c}{$f_{12}$ } 
  & \multicolumn{1}{l}{Wave Function}     \\ \hline 
\endhead

\hline \multicolumn{4}{r}{{{\scshape Continued on next page}}} \\ 
\endfoot

\hline \hline
\endlastfoot
Reference 			&		&		& $| H_{1\alpha}^{1} \rangle $	\\
\tabularnewline
I$_{\bot}$ 	& 1.80  	& 0.0219 	& $| H_{\alpha} - 1 \rightarrow L_{\alpha} \rangle$(0.4824) \\
			&  		&		& $| H_{\beta} - 2  \rightarrow H_{1\beta} \rangle$(0.6132) \\
			&  		&		& $| H_{\beta} - 1  \rightarrow L_{\beta} \rangle$(0.5092) \\
\tabularnewline
II$_{\Vert}$ 	& 4.72 	&0.0054	& $| H_{\alpha} - 2 \rightarrow L_{\alpha} + 3 \rangle$(0.3384) \\
			&  		&		& $| H_{\alpha}  - 1 \rightarrow L_{\alpha} + 2 \rangle$(0.3169) \\
			&  		&		& $| H_{1\alpha} \rightarrow L_{\alpha} + 26 \rangle$(0.3076) \\
			&  		&		& $| H_{\beta} - 2  \rightarrow L_{\beta} + 14 \rangle$(0.3325) \\
			&  		&		& $| H_{\beta} - 2  \rightarrow L_{\beta} + 6 \rangle$(0.2516) \\
\tabularnewline
III$_{\Vert}$ 	& 5.18 	&0.0142	& $| H_{\alpha}  - 1 \rightarrow L_{\alpha} + 2 \rangle$(0.3310) \\
			&  		&		& $| H_{\alpha} - 2 \rightarrow L_{\alpha} + 3 \rangle$(0.2395) \\
			&  		&		& $| H_{\beta} - 2  \rightarrow L_{\beta} + 14 \rangle$(0.4377) \\
			&  		&		& $| H_{\beta} - 2  \rightarrow L_{\beta} + 6 \rangle$(0.3537) \\
\tabularnewline
IV$_{\Vert}$ 	& 5.72 	&0.2362	& $| H_{\alpha} - 1 \rightarrow L_{\alpha} +  3 \rangle$(0.4398) \\
			&  		&		& $| H_{1\alpha} \rightarrow L_{\alpha} +  7 \rangle$(0.3412) \\
			&  		&		& $| H_{\beta} - 2 \rightarrow L_{\beta} +  1 \rangle$(0.3830) \\
\tabularnewline
V$_{\Vert}$ 	& 5.90 	&0.0573	& $| H_{\alpha} - 3 \rightarrow L_{\alpha} +  2 \rangle$(0.4912) \\
			&  		&		& $| H_{\alpha} - 3 \rightarrow L_{\alpha} +  9 \rangle$(0.3730) \\
			&  		&		& $| H_{\beta} - 4 \rightarrow L_{\beta} +  14 \rangle$(0.2564) \\
\tabularnewline
VI$_{\Vert}$ 	& 5.98 	&0.0415	& $| H_{\alpha} - 3 \rightarrow L_{\alpha} +  3 \rangle$(0.3226) \\
			&  		&		& $| H_{\alpha} - 2 \rightarrow L_{\alpha} +  2 \rangle$(0.3100) \\
			&  		&		& $| H_{\beta} - 1 \rightarrow L_{\beta} +  14 \rangle$(0.3539) \\
			&  		&		& $| H_{\beta} - 1 \rightarrow L_{\beta} +  6 \rangle$(0.3390) \\
\tabularnewline
VII$_{\Vert}$ 	& 6.02 	&0.3001	& $| H_{\beta} - 4 \rightarrow L_{\beta} +  14 \rangle$(0.4346) \\
			&  		&		& $| H_{\beta} - 2 \rightarrow L_{\beta} +  14 \rangle$(0.3512) \\
\tabularnewline
VIII$_{\Vert}$ 	& 6.49 	&0.2356	& $| H_{1\alpha} \rightarrow L_{\alpha} +  7 \rangle$(0.3764) \\
			&  		&		& $| H_{1\alpha} \rightarrow L_{\alpha} +  3 \rangle$(0.3727) \\
			&  		&		& $| H_{\beta} - 4 \rightarrow L_{\beta} +  1 \rangle$(0.2666) \\
\tabularnewline
IX$_{\Vert}$ 	& 6.66 	&0.0443	& $| H_{\alpha} - 1 \rightarrow L_{\alpha} +  8 \rangle$(0.5698) \\
			&  		&		& $| H_{\alpha} - 1 \rightarrow L_{\alpha} +  9 \rangle$(0.4799) \\
\tabularnewline
X$_{\Vert}$ 	& 6.84 	&0.1057	& $| H_{\beta} - 2 \rightarrow L_{\beta} +  9 \rangle$(0.7094) \\
			&  		&		& $| H_{\beta} - 2 \rightarrow L_{\beta} +  3 \rangle$(0.5419) \\
\tabularnewline
XI$_{\Vert}$ 	& 6.91 	&0.0227	& $| H_{\alpha} - 2 \rightarrow L_{\alpha} +  8 \rangle$(0.5593) \\
			&  		&		& $| H_{\alpha} - 2 \rightarrow L_{\alpha} +  9 \rangle$(0.4252) \\
\tabularnewline
XII$_{\Vert}$ 	& 7.09 	&0.2532	& $| H_{1\alpha} \rightarrow L_{\alpha} + 14 \rangle$(0.2937) \\
			&  		&		& $| H_{1\alpha} \rightarrow L_{\alpha} + 31 \rangle$(0.2827) \\
			&  		&		& $| H_{\beta} - 6 \rightarrow L_{\beta} +  2 \rangle$(0.2682) \\
			&  		&		& $| H_{\beta} - 4 \rightarrow L_{\beta} +  1 \rangle$(0.2429) \\
\tabularnewline
XIII$_{\Vert}$ 	& 7.23 	&0.4993	& $| H_{\alpha} - 3 \rightarrow L_{\alpha} + 2 \rangle$(0.3017) \\
			&  		&		& $| H_{1\alpha} \rightarrow L_{\alpha} + 15 \rangle$(0.2026) \\
			&  		&		& $| H_{\beta} - 4 \rightarrow L_{\beta} +  14 \rangle$(0.3452) \\
			&  		&		& $| H_{\beta} - 4 \rightarrow L_{\beta} +  6 \rangle$(0.3046) \\
\label{Tab:table-b7-eomccsd}
\end{longtable}

\LTcapwidth=\columnwidth
\begin{longtable}{cccl}
 \caption{Excitation energies, $E$, and many-particle wavefunctions of excited
 states corresponding to the peaks in the EOM-CCSD linear absorption spectrum
 of B$_{8}$ boron wheel cluster (\textit{cf}. Fig. \ref{fig:geometries_wheel}\protect\subref{subfig:subfig-geom-b8}).
 The subscripts $\Vert$ and $\bot$, in the peak number denote the absorption due to light polarized in, and perpendicular to the plane of 
wheel base, respectively. In the wavefunction, the bracketed numbers are the CI coefficients of a given electronic configuration.
 Symbols $H_{1\alpha}$ and $H_{2\alpha}$ denotes SOMOs discussed earlier, and $H$, and $L$, denote HOMO and LUMO orbitals respectively. Note
 that, the reference state does not correspond to any peak, instead it represents the reference state from which singles excitations are occurring.} \\

\hline \multicolumn{1}{c}{Peak} & \multicolumn{1}{c}{ $E$ (eV)} & \multicolumn{1}{c}{$f_{12}$ } 
  &\multicolumn{1}{l}{Wave Function}     \\ \hline 
\endfirsthead

\multicolumn{4}{c}%
{{\scshape \tablename\ \thetable{} -- continued from previous page}} \\ 

\hline \multicolumn{1}{c}{Peak} & \multicolumn{1}{c}{ $E$ (eV)} & \multicolumn{1}{c}{$f_{12}$ } 
  & \multicolumn{1}{l}{Wave Function}     \\ \hline 
\endhead

\hline \multicolumn{4}{r}{{{\scshape Continued on next page}}} \\ 
\endfoot

\hline \hline
\endlastfoot
Reference 			&		&		& $| H_{1\alpha}^{1};H_{2\alpha}^{1} \rangle $	\\
\tabularnewline
I$_{\bot}$ 	& 1.80  	& 0.0123 	& $| H_{\beta} - 2  \rightarrow H_{1\beta} \rangle$(0.6609) \\
			&  		&		& $| H_{\beta} - 1  \rightarrow H_{2\beta} \rangle$(0.6609) \\
\tabularnewline
II$_{\Vert}$ 	& 4.45 	&0.0264	& $| H_{2\alpha}  \rightarrow L_{\alpha} + 5 \rangle$(0.5112) \\
			&  		&		& $| H_{1\alpha} \rightarrow L_{\alpha} + 6 \rangle$(0.5112) \\
			&  		&		& $| H_{\beta}  - 3 \rightarrow H_{2\beta}  \rangle$(0.3395) \\
\tabularnewline
III$_{\Vert}$ 	& 4.45 	&0.0264	& $| H_{1\alpha}  \rightarrow L_{\alpha} + 5 \rangle$(0.5112) \\
			&  		&		& $| H_{2\alpha}   \rightarrow L_{\alpha} + 6 \rangle$(0.5112) \\
			&  		&		& $| H_{\beta}  - 3 \rightarrow H_{1\beta}  \rangle$(0.3395) \\
\tabularnewline
IV$_{\Vert}$ 	& 6.27 	&0.0131	& $| H_{\alpha} - 5 \rightarrow L_{\alpha} +  1 \rangle$(0.2415) \\
			&  		&		& $| H_{\alpha} - 6 \rightarrow L_{\alpha}  \rangle$(0.2415) \\
			&  		&		& $| H_{\beta} - 5 \rightarrow L_{\beta} +  2 \rangle$(0.3207) \\
			&  		&		& $| H_{\beta} - 6 \rightarrow L_{\beta} +  1 \rangle$(0.3207) \\
\tabularnewline
V$_{\Vert}$ 	& 6.27 	&0.0131	& $| H_{\alpha} - 5 \rightarrow L_{\alpha}  \rangle$(0.2415) \\
			&  		&		& $| H_{\alpha} - 6 \rightarrow L_{\alpha} + 1  \rangle$(0.2415) \\
			&  		&		& $| H_{\beta} - 5 \rightarrow L_{\beta} +  1 \rangle$(0.3207) \\
			&  		&		& $| H_{\beta} - 6 \rightarrow L_{\beta} +  2 \rangle$(0.3207) \\
\tabularnewline
VI$_{\Vert}$ 	& 6.47 	&0.7062	& $| H_{2\alpha}  \rightarrow L_{\alpha} +  7 \rangle$(0.3368) \\
			&  		&		& $| H_{\alpha} - 1 \rightarrow L_{\alpha} +  1 \rangle$(0.2666) \\
			&  		&		& $| H_{\alpha} - 2 \rightarrow L_{\alpha}  \rangle$(0.2666) \\
			&  		&		& $| H_{\beta} - 2 \rightarrow L_{\beta} +  2 \rangle$(0.2435) \\
			&  		&		& $| H_{\beta} - 1 \rightarrow L_{\beta} +  1 \rangle$(0.2435) \\
\tabularnewline
VII$_{\Vert}$ 	& 6.47 	&0.7032	& $| H_{1\alpha}  \rightarrow L_{\alpha} +  7 \rangle$(0.3368) \\
			&  		&		& $| H_{\alpha} - 1 \rightarrow L_{\alpha}  \rangle$(0.2666) \\
			&  		&		& $| H_{\alpha} - 2 \rightarrow L_{\alpha} + 1 \rangle$(0.2666) \\
			&  		&		& $| H_{\beta} - 2 \rightarrow L_{\beta} +  1 \rangle$(0.2435) \\
			&  		&		& $| H_{\beta} - 1 \rightarrow L_{\beta} +  2 \rangle$(0.2435) \\
\tabularnewline
VIII$_{\bot}$ 	& 6.65 	&0.0586	& $| H_{2\alpha} \rightarrow L_{\alpha} +  4 \rangle$(0.6256) \\
			&  		&		& $| H_{1\alpha} \rightarrow L_{\alpha} +  3 \rangle$(0.6256) \\
\tabularnewline
IX$_{\Vert}$ 	& 6.95 	&0.0423	& $| H_{\beta} - 2 \rightarrow L_{\beta} +  8 \rangle$(0.6389) \\
			&  		&		& $| H_{\beta} - 2 \rightarrow L_{\beta}  \rangle$(0.6067) \\
\tabularnewline
X$_{\Vert}$ 	& 6.95 	&0.0423	& $| H_{\beta} - 1 \rightarrow L_{\beta} +  8 \rangle$(0.6389) \\
			&  		&		& $| H_{\beta} - 1 \rightarrow L_{\beta}  \rangle$(0.6067) \\
\tabularnewline
XI$_{\Vert}$ 	& 7.08 	&0.0186	& $| H_{\alpha} - 2 \rightarrow L_{\alpha} +  10 \rangle$(0.6548) \\
			&  		&		& $| H_{\alpha} - 2 \rightarrow L_{\alpha} +  2 \rangle$(0.5472) \\
\tabularnewline
XII$_{\Vert}$ 	& 7.08 	&0.0186	& $| H_{\alpha} - 1 \rightarrow L_{\alpha} +  10 \rangle$(0.6548) \\
			&  		&		& $| H_{\alpha} - 1 \rightarrow L_{\alpha} +  2 \rangle$(0.5472) \\
\tabularnewline
XIII$_{\Vert}$ 	& 7.18 	&0.2174	& $| H_{\alpha} - 5 \rightarrow L_{\alpha} + 1 \rangle$(0.2584) \\
			&  		&		& $| H_{\alpha} - 6 \rightarrow L_{\alpha} \rangle$(0.2584) \\
			&  		&		& $| H_{2\alpha}  \rightarrow L_{\beta} +  7 \rangle$(0.2437) \\
			&  		&		& $| H_{\beta} - 5 \rightarrow L_{\beta} +  2 \rangle$(0.2437) \\
			&  		&		& $| H_{\beta} - 6 \rightarrow L_{\beta} +  1 \rangle$(0.2521) \\
\tabularnewline
XIV$_{\Vert}$ 	& 7.18 	&0.2174	& $| H_{\alpha} - 5 \rightarrow L_{\alpha} \rangle$(0.2584) \\
			&  		&		& $| H_{\alpha} - 6 \rightarrow L_{\alpha} + 1 \rangle$(0.2584) \\
			&  		&		& $| H_{1\alpha}  \rightarrow L_{\beta} +  7 \rangle$(0.2437) \\
			&  		&		& $| H_{\beta} - 5 \rightarrow L_{\beta} +  1 \rangle$(0.2437) \\
			&  		&		& $| H_{\beta} - 6 \rightarrow L_{\beta} +  2 \rangle$(0.2521) \\
\label{Tab:table-b8-eomccsd}
\end{longtable}

\LTcapwidth=\columnwidth
\begin{longtable}{cccl}
 \caption{Excitation energies, $E$, and many-particle wavefunctions of excited
 states corresponding to the peaks in the EOM-CCSD linear absorption spectrum
 of B$_{9}$ boron wheel cluster (\emph{cf}. Fig. \ref{fig:geometries_wheel}\protect\subref{subfig:subfig-geom-b9}).
Rest of the information is same as given in the caption for Table \ref{Tab:table-b7-eomccsd} }. \\

\hline \multicolumn{1}{c}{Peak} & \multicolumn{1}{c}{ $E$ (eV)} & \multicolumn{1}{c}{$f_{12}$ } 
 &\multicolumn{1}{l}{Wave Function}     \\ \hline 
\endfirsthead

\multicolumn{4}{c}%
{{\scshape \tablename\ \thetable{} -- continued from previous page}} \\ 

\hline \multicolumn{1}{c}{Peak} & \multicolumn{1}{c}{ $E$ (eV)} & \multicolumn{1}{c}{$f_{12}$ } 
  & \multicolumn{1}{l}{Wave Function}     \\ \hline 
\endhead

\hline \multicolumn{4}{r}{{{\scshape Continued on next page}}} \\ 
\endfoot

\hline \hline
\endlastfoot
Reference 			&		&		& $| H_{1\alpha}^{1} \rangle $	\\
\tabularnewline
I$_{\Vert}$ 	& 3.90  	& 0.0247 	& $| H_{\alpha}  - 1 \rightarrow L_{\alpha}  \rangle$(0.3911) \\
			&  		&		& $| H_{1\alpha}   \rightarrow L_{\alpha} + 1 \rangle$(0.4562) \\
			&  		&		& $| H_{\alpha}  - 2 \rightarrow L_{\alpha} + 2 \rangle$(0.3107) \\
			&  		&		& $| H_{\beta} - 1  \rightarrow L_{\beta} + 5 \rangle$(0.4079) \\
			&  		&		& $| H_{\beta} - 1  \rightarrow L_{\beta} + 6 \rangle$(0.2668) \\
\tabularnewline
II$_{\Vert}$ 	& 4.00 	&0.0243	& $| H_{1\alpha}  \rightarrow L_{\alpha}  \rangle$(0.4492) \\
			&  		&		& $| H_{\alpha}  - 1 \rightarrow L_{\alpha} + 1 \rangle$(0.3673) \\
			&  		&		& $| H_{\beta}  - 1 \rightarrow L_{\beta}  + 7 \rangle$(0.4330) \\
\tabularnewline
III$_{\Vert}$ 	& 5.02 	&0.0148	& $| H_{\alpha} - 3 \rightarrow L_{\alpha} + 3 \rangle$(0.4830) \\
			&  		&		& $| H_{\alpha}  - 2 \rightarrow L_{\alpha} + 2 \rangle$(0.3530) \\
			&  		&		& $| H_{\beta}  - 2 \rightarrow L_{\beta}  \rangle$(0.3695) \\
			&  		&		& $| H_{\beta} - 2  \rightarrow L_{\beta} + 1 \rangle$(0.3441) \\
\tabularnewline
IV$_{\Vert}$ 	& 6.17 	&0.0231	& $| H_{\alpha} - 2 \rightarrow L_{\alpha} +  26 \rangle$(0.4842) \\
			&  		&		& $| H_{\beta} - 2 \rightarrow L_{\beta} +  26 \rangle$(0.4492) \\
\tabularnewline
V$_{\bot}$ 	& 6.20 	&0.0113	& $| H_{1\alpha} \rightarrow L_{\alpha}  + 6 \rangle$(0.2252) \\
			&  		&		& $| H_{\beta} - 1 \rightarrow L_{\beta} +  4 \rangle$(0.8200) \\
$_{\Vert}$	& 6.20	&0.2518	& $| H_{\alpha} - 2 \rightarrow L_{\alpha} + 3  \rangle$(0.2805) \\
			& 		&		& $| H_{\alpha} - 3 \rightarrow L_{\alpha} + 26  \rangle$(0.2702) \\
			&  		&		& $| H_{\beta} - 3 \rightarrow L_{\beta} +  26 \rangle$(0.3121) \\
			&  		&		& $| H_{\beta} - 3 \rightarrow L_{\beta}  \rangle$(0.2965) \\
\tabularnewline
VI$_{\Vert}$ 	& 6.29 	&0.2542	& $| H_{\alpha} - 3 \rightarrow L_{\alpha} +  2 \rangle$(0.2638) \\
			&  		&		& $| H_{\alpha} - 3 \rightarrow L_{\alpha} +  26 \rangle$(0.2496) \\
			&  		&		& $| H_{\beta} - 7 \rightarrow L_{\beta}  \rangle$(0.3701) \\
			&  		&		& $| H_{\beta} - 2 \rightarrow L_{\beta} +  1 \rangle$(0.2570) \\
			& 6.29 	&0.6647	& $| H_{1\alpha}  \rightarrow L_{\alpha} +  7 \rangle$(0.2724) \\
			&  		&		& $| H_{\alpha} - 3 \rightarrow L_{\alpha} + 3 \rangle$(0.2535) \\
			&  		&		& $| H_{\beta} - 1 \rightarrow L_{\beta} +  5 \rangle$(0.4263) \\
			&  		&		& $| H_{\beta} - 1 \rightarrow L_{\beta} +  6 \rangle$(0.3587) \\
\tabularnewline
VII$_{\bot}$ 	& 6.33 	&0.0504	& $| H_{1\alpha} \rightarrow L_{\alpha} +  6 \rangle$(0.8079) \\
			&  		&		& $| H_{\beta} - 1 \rightarrow L_{\beta} +  4 \rangle$(0.2667) \\
\tabularnewline
VIII$_{\Vert}$ 	& 6.37 	&0.0500	& $| H_{1\alpha}  \rightarrow L_{\alpha} +  7 \rangle$(0.3945) \\
			&  		&		& $| H_{\alpha} - 2 \rightarrow L_{\alpha} +  4 \rangle$(0.3056) \\
			&  		&		& $| H_{\beta} - 2 \rightarrow L_{\beta}  + 2 \rangle$(0.3094) \\
\tabularnewline
IX$_{\Vert}$ 	& 6.46 	&0.0152	& $| H_{1\alpha}  \rightarrow L_{\alpha} +  7 \rangle$(0.4715) \\
			&  		&		& $| H_{\alpha} - 2 \rightarrow L_{\alpha} +  4 \rangle$(0.3076) \\
			&  		&		& $| H_{\beta} - 1 \rightarrow L_{\beta}  + 6 \rangle$(0.3902) \\
\tabularnewline
X$_{\Vert}$ 	& 6.51 	&0.0877	& $| H_{\alpha} - 3 \rightarrow L_{\alpha} +  4 \rangle$(0.3536) \\
			&  		&		& $| H_{\alpha} - 3 \rightarrow L_{\alpha} +  12 \rangle$(0.2843) \\
			&  		&		& $| H_{\beta} - 3 \rightarrow L_{\beta} +  2 \rangle$(0.3392) \\
			&  		&		& $| H_{\beta} - 3 \rightarrow L_{\beta} +  12 \rangle$(0.2590) \\
\tabularnewline
XI$_{\bot}$ 	& 6.63 	&0.0222	& $| H_{\alpha} - 1 \rightarrow L_{\alpha} +  5 \rangle$(0.8459) \\
			&  		&		& $| H_{\alpha} - 1 \rightarrow L_{\alpha} +  30 \rangle$(0.2594) \\
\tabularnewline
XII$_{\Vert}$ 	& 6.77 	&0.8520	& $| H_{1\alpha}  \rightarrow L_{\alpha} + 7 \rangle$(0.2633) \\
			&  		&		& $| H_{1\alpha}  \rightarrow L_{\alpha} + 20 \rangle$(0.2568) \\
			&  		&		& $| H_{\beta} - 1 \rightarrow L_{\beta} +  6 \rangle$(0.2881) \\
			&  		&		& $| H_{\beta} - 2 \rightarrow L_{\beta}   \rangle$(0.2459) \\
\tabularnewline
XIII$_{\Vert}$ 	& 6.79 	&0.2611	& $| H_{\alpha} - 1 \rightarrow L_{\alpha} + 7 \rangle$(0.6406) \\
			&  		&		& $| H_{\alpha} - 1 \rightarrow L_{\alpha} + 20 \rangle$(0.4094) \\
			&  		&		& $| H_{\beta} - 3 \rightarrow L_{\beta} +  2 \rangle$(0.2385) \\
			&  		&		& $| H_{\beta} - 3 \rightarrow L_{\beta} +  12 \rangle$(0.2018) \\
\label{Tab:table-b9-eomccsd}
\end{longtable}

\chapter{\label{app:aluminum}Aluminum Clusters Al$_{n}$ (n = 2 -- 5)}	

\section*{Excited State CI Wavefunctions, Energies and Oscillator Strengths}
\LTcapwidth=\columnwidth
\begin{longtable}{ccccl}
\captionsetup{font=footnotesize}
\caption{Excitation energies ($E$) and many-particle wavefunctions of excited
states corresponding to the peaks in the linear absorption spectrum
of Al$_{2}$ (\emph{cf}. Fig. \ref{fig:plot-al2-linear}\protect\subref{subfig:al2-lin-plot})
, along with the oscillator strength ($f_{12}$) of the transitions. Longitudinal
and transverse polarization corresponds to the absorption due to light
polarized along and perpendicular to the molecular axis respectively.
In the wavefunction, the bracketed numbers are the CI coefficients
of a given electronic configuration. Symbols $H_{1}$,$H_{2}$ denote
SOMOs discussed earlier, and $H$, and $L$, denote HOMO and LUMO
orbitals respectively. $HF$ denotes the Hartree-Fock configuration. } \\

\hline \multicolumn{1}{c}{Peak} & \multicolumn{1}{c}{ $E$ (eV)} & \multicolumn{1}{c}{$f_{12}$ } 
 & \multicolumn{1}{c}{Pol.} &\multicolumn{1}{l}{Wave Function}     \\ \hline 
\endfirsthead

\multicolumn{5}{c}%
{{\scshape \tablename\ \thetable{} -- continued from previous page}} \\ 

\hline \multicolumn{1}{c}{Peak} & \multicolumn{1}{c}{ $E$ (eV)} & \multicolumn{1}{c}{$f_{12}$ } 
 & \multicolumn{1}{c}{Pol.} & \multicolumn{1}{l}{Wave Function}     \\ \hline 
\endhead

\hline \multicolumn{5}{r}{{{\scshape Continued on next page}}} \\ 
\endfoot

\hline \hline
\endlastfoot
GS\footnotemark[1]
 &  &  &  & $|H_{1}^{1},\, H_{2}^{1}\rangle$ (0.9096)\tabularnewline
 &  &  &  & $|H-1\rightarrow H_{1};H_{2}\rightarrow L\rangle$(0.1139)\tabularnewline
 &  &  &  & $|H-2\rightarrow L;H-1\rightarrow L+2\rangle$(0.0889)\tabularnewline
 &  &  &  & \tabularnewline
I  & 1.96 & 0.1027 & longitudinal & $|H_{2}\rightarrow L+1\rangle$(0.8120) \tabularnewline
 &  &  &  & $|H-1\rightarrow H_{1}\rangle$(0.3685) \tabularnewline
 &  &  &  & \tabularnewline
II  & 3.17 & 0.1249 & longitudinal & $|H-1\rightarrow H_{1}\rangle$(0.6172) \tabularnewline
 &  &  &  & $|H_{1}\rightarrow L+3\rangle$(0.4068) \tabularnewline
 &  &  &  & $|H_{1}\rightarrow L;H-1\rightarrow L\rangle$(0.3190) \tabularnewline
 &  &  &  & \tabularnewline
III  & 4.47 & 0.5149 & transverse & $|H_{2}\rightarrow L+4\rangle$(0.8313) \tabularnewline
 &  &  &  & $|H_{2}\rightarrow L+6\rangle$(0.2024) \tabularnewline
 &  &  &  & \tabularnewline
IV  & 4.99 & 5.4531 & longitudinal & $|H_{1}\rightarrow L+3\rangle$(0.7353) \tabularnewline
 &  &  &  & $|H-1\rightarrow H_{1}\rangle$(0.4104) \tabularnewline
 &  &  &  & \tabularnewline
V & 6.31 & 0.2554 & transverse & $|H_{2}\rightarrow L+6\rangle$(0.4683) \tabularnewline
 &  &  &  & $|H-1\rightarrow L+1\rangle$(0.3894) \tabularnewline
 &  &  &  & $|H-1\rightarrow L;H_{2}\rightarrow L+2\rangle$(0.3886) \tabularnewline
 &  &  &  & \tabularnewline
VI & 7.17 & 0.1549 & transverse & $|H_{2}\rightarrow L+2;H-1\rightarrow L\rangle$(0.4782) \tabularnewline
 &  &  &  & $|H-1\rightarrow L+1\rangle$(0.4327) \tabularnewline
 &  &  &  & $|H_{1}\rightarrow L;H_{2}\rightarrow L+8\rangle$(0.3867) \tabularnewline
 &  &  &  & \tabularnewline
VII & 7.79 & 1.2530 & transverse & $|H-1\rightarrow H_{1};H_{2}\rightarrow L+3\rangle$(0.4833) \tabularnewline
 &  &  &  & $|H_{1}\rightarrow L+7\rangle$(0.3917) \tabularnewline
 &  &  &  & $|H_{1}\rightarrow L;H_{2}\rightarrow L+8\rangle$(0.3791) \tabularnewline
 &  &  &  & \tabularnewline
VIII & 8.05 & 3.5391 & transverse & $|H-2\rightarrow L\rangle$(0.5316) \tabularnewline
 &  &  &  & $|H-1\rightarrow L+2\rangle$(0.3756) \tabularnewline
 &  &  &  & $|H_{1}\rightarrow L+8\rangle$(0.3531) \tabularnewline
 & 8.10 & 1.1418 & transverse & $|H-1\rightarrow H_{1};H_{2}\rightarrow L+3\rangle$(0.4788) \tabularnewline
 &  &  &  & $|H_{2}\rightarrow L+6\rangle$(0.4095) \tabularnewline
 &  &  &  & \tabularnewline
IX & 8.87 & 0.7044 & transverse & $|H_{1}\rightarrow L+11\rangle$(0.5061) \tabularnewline
 &  &  &  & $|H_{1}\rightarrow L;H_{2}\rightarrow L+7\rangle$(0.4162) \tabularnewline
 & 8.95 & 0.6872 & transverse & $|H_{1}\rightarrow L+7\rangle$(0.4932) \tabularnewline
 &  &  &  & $|H_{2}\rightarrow L;H_{1}\rightarrow L+8\rangle$(0.4414) \tabularnewline
 &  &  &  & $|H_{1}\rightarrow L+4;H-1\rightarrow L+1\rangle$(0.3262) \tabularnewline
\label{Tab:table_al2_lin}
\footnotetext[1]{$GS$ does not correspond to any peak, rather it corresponds to the
ground state wavefunction of Al$_{2}$ isomer.}
\end{longtable}

\LTcapwidth=\columnwidth
\begin{longtable}{ccccl}
\captionsetup{font=footnotesize}
 \caption{Excitation energies ($E$) and many-particle wavefunctions of excited
states corresponding to the peaks in the linear absorption spectrum
of Al$_{3}$ equilateral triangle isomer (\emph{cf}. Fig. \ref{fig:plot-al3-equil}\protect\subref{subfig:plot-al3-equil})
, along with the oscillator strength ($f_{12}$) of the transitions.
In-plane and transverse polarization corresponds to the absorption
due to light polarized in and perpendicular to the plane of the triangular
isomer respectively. In the wavefunction, the bracketed numbers are
the CI coefficients of a given electronic configuration. Symbols $H$
and $L$, denote HOMO (singly occupied, in this case) and LUMO orbitals
respectively. $HF$ denotes the Hartree-Fock configuration. } \label{Tab:table_al3_equil_tri} \\

\hline \multicolumn{1}{c}{Peak} & \multicolumn{1}{c}{ $E$ (eV)} & \multicolumn{1}{c}{$f_{12}$ } 
 & \multicolumn{1}{c}{Pol.} &\multicolumn{1}{l}{Wave Function}     \\ \hline 
\endfirsthead

\multicolumn{5}{c}%
{{\scshape \tablename\ \thetable{} -- continued from previous page}} \\ 

\hline \multicolumn{1}{c}{Peak} & \multicolumn{1}{c}{ $E$ (eV)} & \multicolumn{1}{c}{$f_{12}$ } 
 & \multicolumn{1}{c}{Pol.} & \multicolumn{1}{l}{Wave Function}     \\ \hline 
\endhead

\hline \multicolumn{5}{r}{{{\scshape Continued on next page}}} \\ 
\endfoot

\hline \hline
\endlastfoot
GS\footnotemark[2]
 &  &  &  & $|HF\rangle$ (0.8373)\tabularnewline
 &  &  &  & $|H-2\rightarrow L+5\rangle$(0.1329)\tabularnewline
 &  &  &  & \tabularnewline
I  & 3.42 & 0.0376 & in-plane & $|H-3\rightarrow L+5\rangle$(0.2908) \tabularnewline
 &  &  &  & $|H-2\rightarrow L+1\rangle$(0.2439) \tabularnewline
 & 3.54 & 0.1080 & in-plane & $|H-2\rightarrow L+5\rangle$(0.3686) \tabularnewline
 &  &  &  & $|H-2\rightarrow H\rangle$(0.3403) \tabularnewline
 &  &  &  & \tabularnewline
II  & 5.61 & 0.2565 & in-plane & $|H-2\rightarrow L+5;H-1\rightarrow L+5\rangle$(0.4854) \tabularnewline
 &  &  &  & $|H\rightarrow L+1;H-1\rightarrow L+1\rangle$(0.4476) \tabularnewline
 &  &  &  & \tabularnewline
III  & 5.87 & 0.3413 & transverse & $|H-3\rightarrow L+2\rangle$(0.2915) \tabularnewline
 &  &  &  & $|H-2\rightarrow L\rangle$(0.2842) \tabularnewline
 &  &  &  & \tabularnewline
IV & 6.53 & 6.3289 & in-plane & $|H\rightarrow L+6\rangle$(0.4044) \tabularnewline
 &  &  &  & $|H-3\rightarrow L+1\rangle$(0.3965) \tabularnewline
 &  &  &  & $|H-2\rightarrow L+5\rangle$(0.3158) \tabularnewline
 & 6.53 & 5.7925 & in-plane & $|H\rightarrow L+4\rangle$(0.3842) \tabularnewline
 &  &  &  & $|H-3\rightarrow L+5\rangle$(0.2834) \tabularnewline
 &  &  &  & $|H-4\rightarrow L+1\rangle$(0.2256) \tabularnewline
 &  &  &  & \tabularnewline
V & 6.96 & 0.4145 & transverse & $|H-2\rightarrow L\rangle$(0.3140) \tabularnewline
 &  &  &  & $|H-3\rightarrow L+2\rangle$(0.2626) \tabularnewline
 &  &  &  & \tabularnewline
VI & 7.50 & 0.9430 & in-plane & $|H-2\rightarrow L+1;H\rightarrow L+5\rangle$(0.3136) \tabularnewline
 &  &  &  & $|H-3\rightarrow L+5\rangle$(0.2864) \tabularnewline
 & 7.57 & 0.8630 & in-plane & $|H\rightarrow L+5;H-3\rightarrow L+1\rangle$(0.3838) \tabularnewline
 &  &  &  & $|H-3\rightarrow L+1\rangle$(0.2651) \tabularnewline
 &  &  &  & $|H-2\rightarrow L+5\rangle$(0.2590) \tabularnewline
\footnotetext[2]{$GS$ does not correspond to any peak, rather it corresponds to the
ground state wavefunction of Al$_{3}$ equilateral triangle isomer.}
\end{longtable}

\begin{longtable}{ccccl}
\captionsetup{font=footnotesize}
 \caption{Excitation energies ($E$) and many-particle wavefunctions of excited
states corresponding to the peaks in the linear absorption spectrum
of Al$_{3}$ isosceles triangle isomer (\emph{cf}. Fig. \ref{fig:plot-al3-iso}\protect\subref{subfig:plot-al3-iso}), 
along with the oscillator strength ($f_{12}$) of the transitions.
In-plane and transverse polarization corresponds to the absorption
due to light polarized in and perpendicular to the plane of the triangular
isomer respectively. In the wavefunction, the bracketed numbers are
the CI coefficients of a given electronic configuration. Symbols $H_{1}$,
$H_{2}$ and $H_{3}$ denote SOMOs discussed earlier, $H$ and $L$,
denote HOMO and LUMO orbitals respectively. } \label{Tab:table_al3_iso_tri}  

\\

\hline \multicolumn{1}{c}{Peak} & \multicolumn{1}{c}{ $E$ (eV)} & \multicolumn{1}{c}{$f_{12}$ } 
 & \multicolumn{1}{c}{Pol.} &\multicolumn{1}{l}{Wave Function}     \\ \hline 
\endfirsthead

\multicolumn{5}{c}%
{{\scshape \tablename\ \thetable{} -- continued from previous page}} \\ 

\hline \multicolumn{1}{c}{Peak} & \multicolumn{1}{c}{ $E$ (eV)} & \multicolumn{1}{c}{$f_{12}$ } 
 & \multicolumn{1}{c}{Pol.} & \multicolumn{1}{l}{Wave Function}     \\ \hline 
\endhead

\hline \multicolumn{5}{r}{{{\scshape Continued on next page}}} \\ 
\endfoot

\hline \hline
\endlastfoot
GS\footnotemark[3]
 &  &  &  & $|H_{1}^{1},H_{2}^{1},H_{3}^{1}\rangle$ (0.8670)\tabularnewline
 &  &  &  & $|H-1\rightarrow L+10\rangle$(0.1213)\tabularnewline
 &  &  &  & \tabularnewline
I & 2.37 & 0.0358 & in-plane & $|H_{1}\rightarrow L+1;H_{3}\rightarrow L+2\rangle$(0.7066) \tabularnewline
 &  &  &  & $|H-1\rightarrow L+1;H_{1}\rightarrow L\rangle$(0.4052) \tabularnewline
 &  &  &  & \tabularnewline
II & 3.06 & 0.0992 & in-plane & $|H_{3}\rightarrow H_{2};H-2\rightarrow L\rangle$(0.4691) \tabularnewline
 &  &  &  & $|H-1\rightarrow L+1;H_{3}\rightarrow H_{2}\rangle$(0.4070) \tabularnewline
 &  &  &  & \tabularnewline
III  & 3.45 & 0.0967 & in-plane & $|H_{1}\rightarrow L+3\rangle$(0.5566) \tabularnewline
 &  &  &  & $|H-1\rightarrow L+1;H_{1}\rightarrow L\rangle$(0.5209) \tabularnewline
 &  &  &  & \tabularnewline
IV  & 4.11 & 0.3208 & in-plane & $|H_{1}\rightarrow L+4\rangle$(0.6038) \tabularnewline
 &  &  &  & $|H_{3}\rightarrow L+1;H-2\rightarrow L\rangle$(0.5272) \tabularnewline
 &  &  &  & \tabularnewline
V & 4.83 & 0.2242 & in-plane & $|H_{1}\rightarrow L+1;H-2\rightarrow L+1\rangle$(0.5321) \tabularnewline
 &  &  &  & $|H_{1}\rightarrow L+5\rangle$(0.2611) \tabularnewline
 &  &  &  & \tabularnewline
VI & 5.76 & 5.0792 & in-plane & $|H-1\rightarrow L+1;H_{3}\rightarrow L\rangle$(0.3479) \tabularnewline
 &  &  &  & $|H-3\rightarrow L+1;H_{1}\rightarrow L\rangle$(0.2875) \tabularnewline
 &  &  &  & $|H_{2}\rightarrow L+1;H_{1}\rightarrow L+3\rangle$(0.2800) \tabularnewline
 & 5.85 & 0.8553 & in-plane & $|H_{3}\rightarrow L+1;H-2\rightarrow L\rangle$(0.4081) \tabularnewline
 &  &  &  & $|H-1\rightarrow L;H_{3}\rightarrow L\rangle$(0.2400) \tabularnewline
 &  &  &  & \tabularnewline
VII & 5.95 & 1.7094 & in-plane & $|H-1\rightarrow L+2\rangle$(0.3296) \tabularnewline
 &  &  &  & $|H-1\rightarrow L+1;H_{3}\rightarrow L\rangle$(0.3138) \tabularnewline
 & 6.15 & 0.7827 & in-plane & $|H_{1}\rightarrow L+7\rangle$(0.7827) \tabularnewline
 &  &  &  & \tabularnewline
VIII & 6.68 & 1.7774 & in-plane & $|H_{1}\rightarrow L+10\rangle$(0.4548) \tabularnewline
 &  &  &  & $|H_{2}\rightarrow L+1;H_{1}\rightarrow L+6\rangle$(0.2705) \tabularnewline
 &  &  &  & $|H_{1}\rightarrow L+6\rangle$(0.2447) \tabularnewline
\footnotetext[3]{$GS$ does not correspond to any peak, rather it corresponds to the
ground state wavefunction of Al$_{3}$ isosceles triangle isomer.}
\end{longtable}

\begin{longtable}{ccccl}
\captionsetup{font=footnotesize}
 \caption{Excitation energies ($E$) and many-particle wavefunctions of excited
states corresponding to the peaks in the linear absorption spectrum
of Al$_{3}$ linear isomer,  (\emph{cf}. Fig. \ref{fig:plot-al3-lin}\protect\subref{subfig:plot-al3-lin}),
along with the oscillator strength ($f_{12}$) of the transitions.
Longitudinal and transverse polarization corresponds to the absorption
due to light polarized along and perpendicular to the axis of the
linear isomer respectively. In the wavefunction, the bracketed numbers
are the CI coefficients of a given electronic configuration. Symbols
$H_{1}$, $H_{2}$ and $H_{3}$ denote SOMOs discussed earlier, $H$
and $L$, denote HOMO and LUMO orbitals respectively. $HF$ denotes
the Hartree-Fock configuration. } \label{Tab:table_al3_lin_tri} 

\\

\hline \multicolumn{1}{c}{Peak} & \multicolumn{1}{c}{ $E$ (eV)} & \multicolumn{1}{c}{$f_{12}$ } 
 & \multicolumn{1}{c}{Pol.} &\multicolumn{1}{l}{Wave Function}     \\ \hline 
\endfirsthead

\multicolumn{5}{c}%
{{\scshape \tablename\ \thetable{} -- continued from previous page}} \\ 

\hline \multicolumn{1}{c}{Peak} & \multicolumn{1}{c}{ $E$ (eV)} & \multicolumn{1}{c}{$f_{12}$ } 
 & \multicolumn{1}{c}{Pol.} & \multicolumn{1}{l}{Wave Function}     \\ \hline 
\endhead

\hline \multicolumn{5}{r}{{{\scshape Continued on next page}}} \\ 
\endfoot

\hline \hline
\endlastfoot
GS\footnotemark[4]
 &  &  &  & $|H_{1}^{1},H_{2}^{1},H_{3}^{1}\rangle$ (0.8010)\tabularnewline
 &  &  &  & $|H-3\rightarrow H-1;H-2\rightarrow L\rangle$(0.1913)\tabularnewline
 &  &  &  & \tabularnewline
I & 1.24 & 0.0317 & longitudinal & $|H_{2}\rightarrow L+1\rangle$(0.6602) \tabularnewline
 &  &  &  & $|H-1\rightarrow H_{3}\rangle$(0.3636) \tabularnewline
 &  &  &  & \tabularnewline
II & 2.25 & 0.0489 & longitudinal & $|H-1\rightarrow H_{3}\rangle$(0.6856) \tabularnewline
 &  &  &  & $|H-2\rightarrow H_{1}\rangle$(0.3230) \tabularnewline
 &  &  &  & \tabularnewline
III  & 4.01 & 0.9019 & longitudinal & $|H-2\rightarrow H_{1}\rangle$(0.5249) \tabularnewline
 &  &  &  & $|H-1\rightarrow H_{3}\rangle$(0.3471) \tabularnewline
 &  &  &  & \tabularnewline
IV  & 4.43 & 2.8593 & longitudinal & $|H-1\rightarrow H_{3}\rangle$(0.4070) \tabularnewline
 &  &  &  & $|H-1\rightarrow L+4;H_{2}\rightarrow L+6\rangle$(0.2409) \tabularnewline
 & 4.47 & 0.0960 & transverse & $|H_{2}\rightarrow L+2\rangle$(0.5402) \tabularnewline
 &  &  &  & $|H-1\rightarrow H_{3};H_{2}\rightarrow L+6\rangle$(0.3068) \tabularnewline
 &  &  &  & \tabularnewline
V & 4.62 & 5.1747 & longitudinal & $|H-1\rightarrow H_{3}\rangle$(0.4600) \tabularnewline
 &  &  &  & $|H-1\rightarrow L+4;H_{2}\rightarrow L+6\rangle$(0.2862) \tabularnewline
 &  &  &  & \tabularnewline
VI & 5.29 & 0.1070 & transverse & $|H_{2}\rightarrow L+5\rangle$(0.4951) \tabularnewline
 &  &  &  & $|H-1\rightarrow H_{3};H-1\rightarrow L+1\rangle$(0.3284) \tabularnewline
 &  &  &  & $|H-1\rightarrow L+3\rangle$(0.3091) \tabularnewline
 &  &  &  & \tabularnewline
VII & 5.83 & 0.1412 & longitudinal & $|H-1\rightarrow L+2;H_{1}\rightarrow L\rangle$(0.6637) \tabularnewline
 &  &  &  & $|H-2\rightarrow H_{1}\rangle$(0.2225) \tabularnewline
 &  &  &  & $|H-1\rightarrow H_{3}\rangle$(0.2073) \tabularnewline
 &  &  &  & \tabularnewline
VIII & 6.31 & 0.0459 & longitudinal & $|H_{1}\rightarrow L+6;H_{3}\rightarrow L\rangle$(0.5099) \tabularnewline
 &  &  &  & $|H_{1}\rightarrow L;H_{3}\rightarrow L+6\rangle$(0.2706) \tabularnewline
 & 6.37 & 0.0740 & transverse & $|H-1\rightarrow L+3\rangle$(0.3989) \tabularnewline
 &  &  &  & $|H-1\rightarrow H_{2};H_{3}\rightarrow L+6\rangle$(0.2266) \tabularnewline
 &  &  &  & \tabularnewline
IX & 6.89 & 0.1311 & longitudinal & $|H-5\rightarrow L+6\rangle$(0.3920) \tabularnewline
 &  &  &  & $|H_{1}\rightarrow L+4;H_{3}\rightarrow L+6\rangle$(0.3086)
\footnotetext[4]{$GS$ does not correspond to any peak, rather it corresponds to the
ground state wavefunction of Al$_{3}$ linear triangle isomer.}
\end{longtable}

\begin{longtable}{ccccl}
\captionsetup{font=footnotesize}
 \caption{Excitation energies ($E$) and many-particle wavefunctions of excited
states corresponding to the peaks in the linear absorption spectrum
of Al$_{4}$ rhombus isomer, (\emph{cf}. Fig. \ref{fig:plot-al4-rho}\protect\subref{subfig:plot-al4-rho}),
along with the oscillator strength ($f_{12}$) of the transitions.
In-plane and transverse polarization corresponds to the absorption
due to light polarized in and perpendicular to the plane of the rhombus
isomer respectively. In the wavefunction, the bracketed numbers are
the CI coefficients of a given electronic configuration. Symbols $H_{1}$,$H_{2}$
denote SOMOs discussed earlier, and $H$, and $L$, denote HOMO and
LUMO orbitals respectively.} \label{Tab:table_al4_rho} \\

\hline \multicolumn{1}{c}{Peak} & \multicolumn{1}{c}{ $E$ (eV)} & \multicolumn{1}{c}{$f_{12}$ } 
 & \multicolumn{1}{c}{Pol.} &\multicolumn{1}{l}{Wave Function}     \\ \hline 
\endfirsthead

\multicolumn{5}{c}%
{{\scshape \tablename\ \thetable{} -- continued from previous page}} \\ 

\hline \multicolumn{1}{c}{Peak} & \multicolumn{1}{c}{ $E$ (eV)} & \multicolumn{1}{c}{$f_{12}$ } 
 & \multicolumn{1}{c}{Pol.} & \multicolumn{1}{l}{Wave Function}     \\ \hline 
\endhead

\hline \multicolumn{5}{r}{{{\scshape Continued on next page}}} \\ 
\endfoot

\hline \hline
\endlastfoot
GS\footnotemark[5]
 &  &  &  & $|H_{1}^{1},H_{2}^{1}\rangle$ (0.8724)\tabularnewline
 &  &  &  & $|H-3\rightarrow L;H-3\rightarrow L\rangle$(0.1050)\tabularnewline
 &  &  &  & \tabularnewline
I  & 1.07 & 0.0247 & transverse & $|H_{1}\rightarrow L+1\rangle$(0.8489) \tabularnewline
 &  &  &  & $|H-2\rightarrow L+5\rangle$(0.1601) \tabularnewline
 &  &  &  & \tabularnewline
II  & 2.31 & 0.3087 & in-plane & $|H-2\rightarrow H_{1}\rangle$(0.7645) \tabularnewline
 &  &  &  & $|H_{2}\rightarrow L+1\rangle$(0.3113) \tabularnewline
 &  &  &  & \tabularnewline
III  & 4.67 & 0.5709 & in-plane & $|H-2\rightarrow L;H-1\rightarrow L+3\rangle$(0.6036) \tabularnewline
 &  &  &  & $|H-1\rightarrow L+3\rangle$(0.4213) \tabularnewline
 &  &  &  & $|H_{1}\rightarrow L+7\rangle$(0.3113) \tabularnewline
 &  &  &  & \tabularnewline
IV  & 4.88 & 0.9622 & in-plane & $|H-1\rightarrow L;H-1\rightarrow L+3\rangle$(0.6036) \tabularnewline
 &  &  &  & $|H-3\rightarrow L\rangle$(0.4699) \tabularnewline
 &  &  &  & \tabularnewline
V & 5.51 & 3.8316 & in-plane & $|H-3\rightarrow L+4\rangle$(0.7378) \tabularnewline
 &  &  &  & $|H-2\rightarrow H_{1}\rangle$(0.2161) \tabularnewline
 &  &  &  & \tabularnewline
VI & 5.84 & 0.4900 & in-plane & $|H-2\rightarrow L+3\rangle$(0.3889) \tabularnewline
 &  &  &  & $|H-2\rightarrow L;H-3\rightarrow L\rangle$(0.3758) \tabularnewline
 &  &  &  & $|H-3\rightarrow L\rangle$(0.3594) \tabularnewline
 &  &  &  & $|H-1\rightarrow L;H-1\rightarrow L+3\rangle$(0.3591) \tabularnewline
 &  &  &  & \tabularnewline
VII & 6.01 & 0.5332 & transverse & $|H_{2}\rightarrow L+7\rangle$(0.7268) \tabularnewline
 &  &  &  & $|H-3\rightarrow L+2\rangle$(0.3050) \tabularnewline
 &  &  &  & \tabularnewline
VIII & 6.20 & 0.7477 & in-plane & $|H-2\rightarrow L+3\rangle$(0.5195) \tabularnewline
 &  &  &  & $|H-2\rightarrow L;H-3\rightarrow L\rangle$(0.4189) \tabularnewline
 &  &  &  & \tabularnewline
IX & 6.51 & 0.2928 & transverse & $|H-3\rightarrow L+2\rangle$(0.7001) \tabularnewline
 &  &  &  & $|H-2\rightarrow H_{1};H-1\rightarrow L+2\rangle$(0.2232) \tabularnewline
 &  &  &  & $|H-2\rightarrow L;H-3\rightarrow L+2\rangle$(0.2070) \tabularnewline
 &  &  &  & \tabularnewline
X & 6.92 & 0.6053 & transverse & $|H-3\rightarrow L+2\rangle$(0.5144) \tabularnewline
 &  &  &  & $|H-2\rightarrow L;H-3\rightarrow L+2\rangle$(0.3549) \tabularnewline
 &  &  &  & $|H-2\rightarrow L+5\rangle$(0.2676) \tabularnewline
 &  &  &  & \tabularnewline
XI & 7.31 & 0.4328 & transverse & $|H-2\rightarrow L+5\rangle$(0.4033) \tabularnewline
 &  &  &  & $|H-3\rightarrow L;H-1\rightarrow L+1\rangle$(0.3787) \tabularnewline
 &  &  &  & \tabularnewline
XII & 7.76 & 2.7450 & in-plane & $|H_{1}\rightarrow L+8\rangle$(0.4387) \tabularnewline
 &  &  &  & $|H_{1}\rightarrow L+1;H-1\rightarrow L+2\rangle$(0.3435) \tabularnewline
\footnotetext[5]{$GS$ does not correspond to any peak, rather it corresponds to the
ground state wavefunction of Al$_{4}$ rhombus isomer.}
\end{longtable}

\begin{longtable}{ccccl}
\captionsetup{font=footnotesize}
 \caption{Excitation energies ($E$) and many-particle wavefunctions of excited
states corresponding to the peaks in the linear absorption spectrum
of Al$_{4}$ square isomer, (\emph{cf}. Fig. \ref{fig:plot-al4-sqr}\protect\subref{subfig:plot-al4-sqr}),
along with the oscillator strength ($f_{12}$) of the transitions.
In-plane and transverse polarization corresponds to the absorption
due to light polarized in and perpendicular to the plane of the rhombus
isomer respectively. In the wavefunction, the bracketed numbers are
the CI coefficients of a given electronic configuration. Symbols $H_{1}$,$H_{2}$
denote SOMOs discussed earlier, and $H$, and $L$, denote HOMO and
LUMO orbitals respectively.} \label{Tab:table_al4_sqr}  

\\

\hline \multicolumn{1}{c}{Peak} & \multicolumn{1}{c}{ $E$ (eV)} & \multicolumn{1}{c}{$f_{12}$ } 
 & \multicolumn{1}{c}{Pol.} &\multicolumn{1}{l}{Wave Function}     \\ \hline 
\endfirsthead

\multicolumn{5}{c}%
{{\scshape \tablename\ \thetable{} -- continued from previous page}} \\ 

\hline \multicolumn{1}{c}{Peak} & \multicolumn{1}{c}{ $E$ (eV)} & \multicolumn{1}{c}{$f_{12}$ } 
 & \multicolumn{1}{c}{Pol.} & \multicolumn{1}{l}{Wave Function}     \\ \hline 
\endhead

\hline \multicolumn{5}{r}{{{\scshape Continued on next page}}} \\ 
\endfoot

\hline \hline
\endlastfoot
GS\footnotemark[6]
 &  &  &  & $|H_{1}^{1},H_{2}^{1}\rangle$(0.8525)\tabularnewline
 &  &  &  & $|H_{1}\rightarrow L;H-2\rightarrow L\rangle$(0.0972)\tabularnewline
 &  &  &  & \tabularnewline
I  & 2.08 & 0.0278 & in-plane & $|H-1\rightarrow L\rangle$(0.7191) \tabularnewline
 &  &  &  & $|H-1\rightarrow H_{1};H_{2}\rightarrow L+1\rangle$(0.2645) \tabularnewline
 &  &  &  & $|H_{2}\rightarrow L+1\rangle$(0.2536) \tabularnewline
 &  &  &  & $|H_{1}\rightarrow L\rangle$(0.2443) \tabularnewline
 &  &  &  & \tabularnewline
II  & 2.68 & 0.0301 & in-plane & $|H_{2}\rightarrow L+1\rangle$(0.4757) \tabularnewline
 &  &  &  & $|H-1\rightarrow L\rangle$(0.4358) \tabularnewline
 &  &  &  & $|H-1\rightarrow H_{1};H_{2}\rightarrow L+1\rangle$(0.3608)\tabularnewline
 &  &  &  & \tabularnewline
III  & 4.19 & 0.3420 & in-plane & $|H-2\rightarrow L\rangle$(0.5889) \tabularnewline
 &  &  &  & $|H-1\rightarrow L+2\rangle$(0.4283) \tabularnewline
 &  &  &  & $|H_{1}\rightarrow L\rangle$(0.2329) \tabularnewline
 &  &  &  & \tabularnewline
IV  & 4.92 & 0.1131 & in-plane & $|H_{1}\rightarrow L+2\rangle$(0.5780) \tabularnewline
 &  &  &  & $|H-1\rightarrow L+2\rangle$(0.4083) \tabularnewline
 &  &  &  & $|H-2\rightarrow L\rangle$(0.3198) \tabularnewline
 &  &  &  & \tabularnewline
V & 5.17 & 0.1238 & transverse & $|H-2\rightarrow L;H_{1}\rightarrow L+1\rangle$(0.3693)\tabularnewline
 &  &  &  & $|H-2\rightarrow L;H_{1}\rightarrow L+1\rangle$(0.3692)\tabularnewline
 & 5.33 & 0.2470 & in-plane & $|H-2\rightarrow H_{1};H-2\rightarrow L\rangle$(0.5193)\tabularnewline
 &  &  &  & $|H-1\rightarrow L+2\rangle$(0.3915)\tabularnewline
 &  &  &  & $|H-2\rightarrow L+2\rangle$(0.3335)\tabularnewline
 &  &  &  & \tabularnewline
VI & 5.85 & 1.2446 & in-plane & $|H-2\rightarrow L+2\rangle$(0.7184) \tabularnewline
 &  &  &  & $|H-1\rightarrow H_{1};H-2\rightarrow L+2\rangle$(0.2587) \tabularnewline
 &  &  &  & $|H-1\rightarrow L+2\rangle$(0.2579) \tabularnewline
 &  &  &  & \tabularnewline
VII & 6.55 & 3.7894 & in-plane & $|H-2\rightarrow L+2\rangle$(0.5706) \tabularnewline
 &  &  &  & $|H-1\rightarrow H_{1};H-2\rightarrow L+2\rangle$(0.4089) \tabularnewline
 &  &  &  & $|H-1\rightarrow L+2\rangle$(0.3325) \tabularnewline
 & 6.58 & 0.2634 & transverse & $|H_{1}\rightarrow L+1;H-2\rightarrow L\rangle$(0.4375) \tabularnewline
 &  &  &  & $|H_{1}\rightarrow L+1;H-2\rightarrow L\rangle$(0.4375) \tabularnewline
 &  &  &  & $|H-2\rightarrow L+3\rangle$(0.4183) \tabularnewline
 &  &  &  & \tabularnewline
VIII & 6.87 & 2.9702 & in-plane & $|H-2\rightarrow L+2\rangle$(0.5100) \tabularnewline
 &  &  &  & $|H-1\rightarrow L+2\rangle$(0.3495) \tabularnewline
 & 6.93 & 0.2483 & transverse & $|H_{1}\rightarrow L+1;H-2\rightarrow L\rangle$(0.3558) \tabularnewline
 &  &  &  & $|H_{1}\rightarrow L+1;H-2\rightarrow L\rangle$(0.3558) \tabularnewline
 &  &  &  & $|H-2\rightarrow L+3\rangle$(0.2929) \tabularnewline
 &  &  &  & \tabularnewline
IX & 7.22 & 1.4267 & in-plane & $|H-1\rightarrow H_{1};H-2\rightarrow L+2\rangle$(0.4039) \tabularnewline
 &  &  &  & $|H-3\rightarrow H_{1}\rangle$(0.2900) \tabularnewline
 &  &  &  & $|H-1\rightarrow H_{1};H-1\rightarrow L+2\rangle$(0.2848)\tabularnewline
\footnotetext[6]{$GS$ does not correspond to any peak, rather it corresponds to the
ground state wavefunction of Al$_{4}$ square isomer.}
\end{longtable}

\begin{longtable}{ccccl}
\captionsetup{font=footnotesize}
 \caption{Excitation energies ($E$) and many-particle wavefunctions of excited
states corresponding to the peaks in the linear absorption spectrum
of Al$_{5}$ pentagonal isomer, (\emph{cf}. Fig. \ref{fig:plot-al5-penta}\protect\subref{subfig:plot-al5-penta}),
along with the oscillator strength ($f_{12}$) of the transitions.
In-plane and transverse polarization corresponds to the absorption
due to light polarized in and perpendicular to the plane of the pentagonal
isomer respectively. In the wavefunction, the bracketed numbers are
the CI coefficients of a given electronic configuration. Symbols $H$
and $L$, denote HOMO and LUMO orbitals respectively.} \label{Tab:table_al5_pentagon}

\\

\hline \multicolumn{1}{c}{Peak} & \multicolumn{1}{c}{ $E$ (eV)} & \multicolumn{1}{c}{$f_{12}$ } 
 & \multicolumn{1}{c}{Pol.} &\multicolumn{1}{l}{Wave Function}     \\ \hline 
\endfirsthead

\multicolumn{5}{c}%
{{\scshape \tablename\ \thetable{} -- continued from previous page}} \\ 

\hline \multicolumn{1}{c}{Peak} & \multicolumn{1}{c}{ $E$ (eV)} & \multicolumn{1}{c}{$f_{12}$ } 
 & \multicolumn{1}{c}{Pol.} & \multicolumn{1}{l}{Wave Function}     \\ \hline 
\endhead

\hline \multicolumn{5}{r}{{{\scshape Continued on next page}}} \\ 
\endfoot

\hline \hline
\endlastfoot
GS\footnotemark[7]
 &  &  &  & $|(H-2)^{1}\rangle$ (0.8679)\tabularnewline
 &  &  &  & $|H-2\rightarrow L+1;H\rightarrow L+2\rangle$(0.1045)\tabularnewline
 &  &  &  & \tabularnewline
I  & 1.03 & 0.0195 & in-plane & $|H-1\rightarrow L\rangle$(0.8635) \tabularnewline
 &  &  &  & $|H-1\rightarrow L;H\rightarrow L+3\rangle$(0.0880) \tabularnewline
 &  &  &  & \tabularnewline
II  & 2.38 & 0.0219 & in-plane & $|H-3\rightarrow H-2\rangle$(0.8560) \tabularnewline
 &  &  &  & $|H-1\rightarrow L+4\rangle$(0.1387) \tabularnewline
 &  &  &  & \tabularnewline
III  & 3.90 & 0.1042 & transverse & $|H\rightarrow L+4\rangle$(0.8387) \tabularnewline
 &  &  &  & $|H\rightarrow L;H-1\rightarrow L+2\rangle$(0.1944) \tabularnewline
 &  & 0.3362 & in-plane & $|H-4\rightarrow L\rangle$(0.8140) \tabularnewline
 &  &  &  & $|H-2\rightarrow L+9\rangle$(0.1841) \tabularnewline
 &  &  &  & \tabularnewline
IV  & 4.16 & 1.3144 & in-plane & $|H-1\rightarrow L+4\rangle$(0.7276) \tabularnewline
 &  &  &  & $|H-1\rightarrow L+5\rangle$(0.4478) \tabularnewline
 &  &  &  & \tabularnewline
V & 4.42 & 3.3339 & in-plane & $|H-1\rightarrow L+5\rangle$(0.7096) \tabularnewline
 &  &  &  & $|H-1\rightarrow L+4\rangle$(0.4490) \tabularnewline
 &  &  &  & $|H-1\rightarrow L+9\rangle$(0.1535) \tabularnewline
 &  &  &  & \tabularnewline
VI & 4.78 & 1.0471 & in-plane & $|H-2\rightarrow L+9\rangle$(0.7992) \tabularnewline
 &  &  &  & $|H-2\rightarrow L;H\rightarrow L+6\rangle$(0.2058) \tabularnewline
 &  &  &  & \tabularnewline
VII & 5.46 & 1.1014 & transverse & $|H\rightarrow L+13\rangle$(0.8156) \tabularnewline
 &  &  &  & $|H\rightarrow L;H-2\rightarrow L\rangle$(0.1708) \tabularnewline
 &  &  &  & \tabularnewline
VIII & 6.37 & 0.1270 & in-plane & $|H-3\rightarrow L\rangle$(0.7632) \tabularnewline
 &  &  &  & \tabularnewline
IX & 6.73 & 0.7104 & in-plane & $|H-3\rightarrow L\rangle$(0.7370) \tabularnewline
 &  &  &  & $|H\rightarrow L+1\rangle$(0.3698) \tabularnewline
 &  &  &  & $|H-1\rightarrow L;H\rightarrow L+3\rangle$(0.1225) \tabularnewline
 &  &  &  & \tabularnewline
X & 7.49 & 0.3989 & in-plane & $|H\rightarrow L+3\rangle$(0.5087) \tabularnewline
 &  &  &  & $|H-2\rightarrow L+16\rangle$(0.3508) \tabularnewline
 &  &  &  & $|H\rightarrow L;H-1\rightarrow L+1\rangle$(0.2937) \tabularnewline
\footnotetext[7]{$GS$ does not correspond to any peak, rather it corresponds to the
ground state wavefunction of Al$_{5}$ pentagonal isomer.}
\end{longtable} 

\begin{longtable}{ccccl}
\captionsetup{font=footnotesize}
 \caption{Excitation energies ($E$) and many-particle wavefunctions of excited
states corresponding to the peaks in the linear absorption spectrum
of Al$_{5}$ pyramid isomer, (\emph{cf}. Fig. \ref{fig:plot-al5-pyra}\protect\subref{subfig:plot-al5-pyra}),
along with the oscillator strength ($f_{12}$) of the transitions.
In the wavefunction, the bracketed numbers are the CI coefficients
of a given electronic configuration. Symbols $H$ and $L$, denote
HOMO and LUMO orbitals respectively.} \label{Tab:table_al5_pyramid} 

\\

\hline \multicolumn{1}{c}{Peak} & \multicolumn{1}{c}{ $E$ (eV)} & \multicolumn{1}{c}{$f_{12}$ } 
 & \multicolumn{1}{c}{Pol.} &\multicolumn{1}{l}{Wave Function}     \\ \hline 
\endfirsthead

\multicolumn{5}{c}%
{{\scshape \tablename\ \thetable{} -- continued from previous page}} \\ 

\hline \multicolumn{1}{c}{Peak} & \multicolumn{1}{c}{ $E$ (eV)} & \multicolumn{1}{c}{$f_{12}$ } 
 & \multicolumn{1}{c}{Pol.} & \multicolumn{1}{l}{Wave Function}     \\ \hline 
\endhead

\hline \multicolumn{5}{r}{{{\scshape Continued on next page}}} \\ 
\endfoot

\hline \hline
\endlastfoot
GS\footnotemark[8]%
 &  &  &  & $|(H-2)^{1}\rangle$ (0.8591)\tabularnewline
 &  &  &  & $|H-3\rightarrow L+1;H-3\rightarrow L+1\rangle$(0.1138)\tabularnewline
 &  &  &  & \tabularnewline
I  & 1.72 & 0.0046 & y & $|H-3\rightarrow L+1\rangle$(0.6849) \tabularnewline
 &  &  &  & $|H-2\rightarrow L+1\rangle$(0.2887) \tabularnewline
 & 1.75 & 0.0521 & z & $|H\rightarrow L+3\rangle$(0.2887) \tabularnewline
 &  &  &  & \tabularnewline
II  & 2.21 & 0.0296 & y & $|H-3\rightarrow L+1\rangle$(0.7170) \tabularnewline
 &  &  &  & $|H-2\rightarrow L+2\rangle$(0.3402) \tabularnewline
 &  &  &  & $|H-3\rightarrow L+2\rangle$(0.2290) \tabularnewline
 &  &  &  & \tabularnewline
III  & 2.55 & 0.0477 & z & $|H\rightarrow L+3\rangle$(0.5390) \tabularnewline
 &  &  &  & $|H-4\rightarrow H-2\rangle$(0.1296) \tabularnewline
 &  &  &  & \tabularnewline
IV & 3.46 & 0.0399 & y & $|H-3\rightarrow L;H-2\rightarrow L+1\rangle$(0.6131) \tabularnewline
 &  &  &  & $|H-3\rightarrow L+2\rangle$(0.4975) \tabularnewline
  & 3.48 & 0.0769 & z & $|H-4\rightarrow H-2\rangle$(0.7340) \tabularnewline
 &  &  &  & $|H-4\rightarrow L\rangle$(0.3735) \tabularnewline
 &  &  &  & \tabularnewline
V & 4.04 & 0.6432 & x & $|H\rightarrow L+7\rangle$(0.5929) \tabularnewline
 &  &  &  & $|H\rightarrow L+4\rangle$(0.4432) \tabularnewline
 & 4.22 & 3.0735 & y & $|H-3\rightarrow L+2\rangle$(0.8272) \tabularnewline
 &  &  &  & $|H-3\rightarrow L+1\rangle$(0.1580) \tabularnewline
 &  &  &  & \tabularnewline
VI & 4.74 & 0.3474 & x & $|H\rightarrow L\rangle$(0.7617) \tabularnewline
 &  &  &  & $|H\rightarrow L+7\rangle$(0.2542) \tabularnewline
 &  &  &  & \tabularnewline
VII & 5.08 & 0.5494 & z & $|H-2\rightarrow L;H\rightarrow L+2\rangle$(0.5540) \tabularnewline
 &  &  &  & $|H-4\rightarrow L\rangle$(0.4833) \tabularnewline
 &  &  &  & \tabularnewline
VIII & 5.26 & 0.3175 & z & $|H-2\rightarrow L;H\rightarrow L+5\rangle$(0.6251) \tabularnewline
 &  &  &  & $|H-4\rightarrow L\rangle$(0.3902) \tabularnewline
 & 5.27 & 0.1267 & x & $|H-6\rightarrow H-2;H\rightarrow L+1\rangle$(0.6056) \tabularnewline
 &  &  &  & $|H\rightarrow L\rangle$(0.3242) \tabularnewline
 &  &  &  & \tabularnewline
IX & 5.56 & 0.1384 & x & $|H\rightarrow L+11\rangle$(0.7819) \tabularnewline
 &  &  &  & $|H\rightarrow L+13\rangle$(0.3051) \tabularnewline
 &  &  &  & \tabularnewline
X & 6.00 & 1.0052 & x & $|H\rightarrow L+13\rangle$(0.8132) \tabularnewline
 &  &  &  & $|H\rightarrow L+11\rangle$(0.1852) \tabularnewline
\footnotetext[8]{$GS$ does not correspond to any peak, rather it corresponds to the
ground state wavefunction of Al$_{5}$ pyramid isomer. }
\end{longtable}



\chapter{Magnesium Clusters Mg$_{n}$ (n = 2 -- 5)}
\section*{Excited State CI Wavefunctions, Energies and Oscillator Strengths}
\LTcapwidth=\columnwidth
\begin{longtable}{ccccl}
\captionsetup{font=footnotesize}
\caption{Excitation energies ($E$) and many-particle wavefunctions of excited
states corresponding to the peaks in the linear absorption spectrum
of Mg$_{2}$ (\emph{cf}. Fig. \ref{fig:plot-mg2-linear})
, along with the oscillator strength ($f_{12}$) of the transitions. Longitudinal
and transverse polarization corresponds to the absorption due to light
polarized along and perpendicular to the molecular axis respectively.
In the wavefunction, the bracketed numbers are the CI coefficients
of a given electronic configuration. Symbols $H$, and $L$, denote HOMO and LUMO
orbitals respectively. $HF$ denotes the Hartree-Fock configuration. 
$GS$ does not correspond to any peak, rather it corresponds to the
 ground state wavefunction of the isomer.} \\

\hline \multicolumn{1}{c}{Peak} & \multicolumn{1}{c}{ $E$ (eV)} & \multicolumn{1}{c}{$f_{12}$ } 
 & \multicolumn{1}{c}{Polarization} &\multicolumn{1}{l}{Wave Function}     \\ \hline 
\endfirsthead

\multicolumn{5}{c}%
{{\scshape \tablename\ \thetable{} -- continued from previous page}} \\ 

\hline \multicolumn{1}{c}{Peak} & \multicolumn{1}{c}{ $E$ (eV)} & \multicolumn{1}{c}{$f_{12}$ } 
 & \multicolumn{1}{c}{Polarization} & \multicolumn{1}{l}{Wave Function}     \\ \hline 
\endhead

\hline \multicolumn{5}{r}{{{\scshape Continued on next page}}} \\ 
\endfoot

\hline \hline
\endlastfoot
GS
 &  &  &  & $|HF \rangle$ (0.9272)\tabularnewline
 &  &  &  & $|H-1\rightarrow L+6; H \rightarrow L+8\rangle$(0.0821)\tabularnewline
 &  &  &  & \tabularnewline
I  & 3.46 & 4.1237 & longitudinal & $|H \rightarrow L+1\rangle$(0.7556) \tabularnewline
 &  &  &  & $|H \rightarrow L + 11 \rangle$(0.4578) \tabularnewline
 &  &  &  & \tabularnewline
II  & 4.02 & 0.1395 & transverse & $|H\rightarrow L  +8 \rangle$(0.4705) \tabularnewline
 &  &  &  & $|H\rightarrow L+3\rangle$(0.3968) \tabularnewline
 &  &  &  & $|H-1\rightarrow L \rangle$(0.2963) \tabularnewline
 &  &  &  & \tabularnewline
III  & 4.59 & 3.6556 & transverse & $|H -1 \rightarrow L\rangle$(0.5891) \tabularnewline
 &  &  &  & $|H \rightarrow L+3 \rangle$(0.4821) \tabularnewline
 &  &  &  & $|H \rightarrow L+8 \rangle$(0.3344) \tabularnewline
 &  &  &  & \tabularnewline
IV  & 5.64 & 0.4934 & transverse & $|H\rightarrow L+3\rangle$(0.6531) \tabularnewline
 &  &  &  & $|H\rightarrow L+8\rangle$(0.4114) \tabularnewline
 &  &  &  & \tabularnewline
V & 6.46 & 0.1223 & transverse & $|H - 1 \rightarrow L+11\rangle$(0.4476) \tabularnewline
 &  &  &  & $|H-1\rightarrow L+6\rangle$(0.2648) 
\label{Tab:table_mg2_lin}
\end{longtable}

\LTcapwidth=\columnwidth
\begin{longtable}{ccccl}
\captionsetup{font=footnotesize}
 \caption{Excitation energies ($E$) and many-particle wavefunctions of excited
states corresponding to the peaks in the linear absorption spectrum
of Mg$_{3}$ equilateral triangle isomer (\emph{cf}. Fig. \ref{fig:plot-mg3-equil})
, along with the oscillator strength ($f_{12}$) of the transitions.
In-plane and transverse polarization corresponds to the absorption
due to light polarized in and perpendicular to the plane of the triangular
isomer respectively. The rest of the information is the same as given in the caption for Table \ref{Tab:table_mg2_lin}.
} \label{Tab:table_mg3_equil} \\

\hline \multicolumn{1}{c}{Peak} & \multicolumn{1}{c}{ $E$ (eV)} & \multicolumn{1}{c}{$f_{12}$ } 
 & \multicolumn{1}{c}{Polarization} &\multicolumn{1}{l}{Wave Function}     \\ \hline 
\endfirsthead

\multicolumn{5}{c}%
{{\scshape \tablename\ \thetable{} -- continued from previous page}} \\ 

\hline \multicolumn{1}{c}{Peak} & \multicolumn{1}{c}{ $E$ (eV)} & \multicolumn{1}{c}{$f_{12}$ } 
 & \multicolumn{1}{c}{Polarization} & \multicolumn{1}{l}{Wave Function}     \\ \hline 
\endhead

\hline \multicolumn{5}{r}{{{\scshape Continued on next page}}} \\ 
\endfoot

\hline \hline
\endlastfoot
GS
 &  &  &  & $|HF\rangle$ (0.8963)\tabularnewline
 &  &  &  & $|H-1\rightarrow L;H \rightarrow L + 9 \rangle$(0.0573)\tabularnewline
 &  &  &  & \tabularnewline
I  & 2.65 & 0.2867 & in-plane & $|H\rightarrow L\rangle$(0.6687) \tabularnewline
   &  &  &  & $|H\rightarrow L+4\rangle$(0.2850) \tabularnewline
   & 2.67 & 0.3023 & in-plane & $|H\rightarrow L\rangle$(0.6765) \tabularnewline
   &  &  &  & $|H\rightarrow L + 4\rangle$(0.2840) \tabularnewline
   &  &  &  & \tabularnewline
II & 3.76 & 4.7767 & in-plane & $|H\rightarrow L\rangle$(0.4324) \tabularnewline
    &  &  &  & $|H\rightarrow L+2\rangle$(0.3967) \tabularnewline
    & 3.78 & 4.7859 & in-plane & $|H\rightarrow L\rangle$(0.4297) \tabularnewline
    &  &  &  & $|H\rightarrow L + 2\rangle$(0.4051) \tabularnewline
    & 3.78 & 0.2401 & transverse & $|H-1\rightarrow L\rangle$(0.3863) \tabularnewline
    &  &  &  & $|H\rightarrow L + 9\rangle$(0.3334) \tabularnewline
    &  &  &  & \tabularnewline
III  & 4.73 & 0.8483 & in-plane & $|H\rightarrow L+7\rangle$(0.7134) \tabularnewline
      &  &  &  & $|H\rightarrow L+4\rangle$(0.2517) \tabularnewline
      & 4.78 & 0.8654 & in-plane & $|H\rightarrow L+7\rangle$(0.7140) \tabularnewline
      &  &  &  & $|H\rightarrow L + 4\rangle$(0.2327) \tabularnewline
      & 4.89 & 1.6579 & transverse & $|H\rightarrow L+5\rangle$(0.4874) \tabularnewline
      &  &  &  & $|H\rightarrow L + 3\rangle$(0.3940) \tabularnewline
      &  &  &  & \tabularnewline
IV & 5.39 & 0.3035 & in-plane & $|H\rightarrow L+6\rangle$(0.5376) \tabularnewline
     &  &  &  & $|H\rightarrow L+2\rangle$(0.3842) \tabularnewline
     & 5.40 & 0.3192 & in-plane & $|H\rightarrow L+6\rangle$(0.5362) \tabularnewline
     &  &  &  & $|H\rightarrow L+2\rangle$(0.3223) \tabularnewline
     &  &  &  & \tabularnewline
V  & 5.77 & 2.5002 & transverse & $|H\rightarrow L+9\rangle$(0.5042) \tabularnewline
    &  &  &  & $|H\rightarrow L+9\rangle$(0.4931) \tabularnewline
    &  &  &  & \tabularnewline
VI & 6.92 & 0.2915 & transverse & $|H-1\rightarrow L+5\rangle$(0.4447) \tabularnewline
     &  &  &  & $|H\rightarrow L;H\rightarrow L+1\rangle$(0.2492) 
\end{longtable}

\begin{longtable}{ccccl}
\captionsetup{font=footnotesize}
 \caption{Excitation energies ($E$) and many-particle wavefunctions of excited
states corresponding to the peaks in the linear absorption spectrum
of Mg$_{3}$ linear isomer (\emph{cf}. Fig. \ref{fig:plot-mg3-lin}), 
along with the oscillator strength ($f_{12}$) of the transitions.
Symbols $H_{1}$ and  $H_{2}$ denote SOMOs discussed earlier.
The rest of the information is the same as given in the caption for Table \ref{Tab:table_mg2_lin}.
}\label{Tab:table_mg3_lin}  
\\

\hline \multicolumn{1}{c}{Peak} & \multicolumn{1}{c}{ $E$ (eV)} & \multicolumn{1}{c}{$f_{12}$ } 
 & \multicolumn{1}{c}{Polarization} &\multicolumn{1}{l}{Wave Function}     \\ \hline 
\endfirsthead

\multicolumn{5}{c}%
{{\scshape \tablename\ \thetable{} -- continued from previous page}} \\ 

\hline \multicolumn{1}{c}{Peak} & \multicolumn{1}{c}{ $E$ (eV)} & \multicolumn{1}{c}{$f_{12}$ } 
 & \multicolumn{1}{c}{Polarization} & \multicolumn{1}{l}{Wave Function}     \\ \hline 
\endhead

\hline \multicolumn{5}{r}{{{\scshape Continued on next page}}} \\ 
\endfoot

\hline \hline
\endlastfoot
GS
 &  &  &  & $|H_{1}^{1},H_{2}^{1}\rangle$ (0.9143)\tabularnewline
 &  &  &  & $|H-1\rightarrow H_{1}\rangle$(0.0776)\tabularnewline
 &  &  &  & \tabularnewline
I & 0.91 & 0.0662 & transverse & $|H_{1}\rightarrow L+8\rangle$(0.5898) \tabularnewline
 &  &  &  & $|H_{1}\rightarrow L+2\rangle$(0.5022) \tabularnewline
 &  &  &  & \tabularnewline
II & 2.27 & 0.4539 & longitudinal & $|H_{1}\rightarrow L+4\rangle$(0.7559) \tabularnewline
   &  &  &  & $|H-1\rightarrow L + 4\rangle$(0.2878) \tabularnewline
   &  &  &  & \tabularnewline
III  & 2.89 & 3.3232 & longitudinal & $|H_{1} \rightarrow L + 3  \rangle$(0.6158) \tabularnewline
 &  &  &  & $|H_{1}\rightarrow L + 9\rangle$(0.3922) \tabularnewline
 &  &  &  & \tabularnewline
IV  & 3.80 & 1.8806 & longitudinal & $|H_{1}\rightarrow L + 12 \rangle$(0.5049) \tabularnewline
 &  &  &  & $|H_{1}\rightarrow L+3\rangle$(0.3629) \tabularnewline
 &  &  &  & \tabularnewline
V & 4.13 & 0.1273 & longitudinal & $|H_{1} \rightarrow L + 9 \rangle$(0.5919) \tabularnewline
 &  &  &  & $|H_{1} \rightarrow L + 12 \rangle$(0.3121) \tabularnewline
 &  &  &  & \tabularnewline
VI & 4.74 & 0.3048 & transverse & $|H_{2} \rightarrow L + 4; H_{1} \rightarrow L + 8  \rangle$(0.4072) \tabularnewline
 &  &  &  & $|H_{2} \rightarrow L + 4 ; H_{1} \rightarrow L + 2 \rangle$(0.3208) \tabularnewline
 &  & 0.5016 & transverse & $| H - 2 \rightarrow L \rangle$(0.3005) \tabularnewline
 &  &  &  & $|H-  1  \rightarrow  L + 8 \rangle$(0.2711) \tabularnewline
 &  &  &  & \tabularnewline
VII & 5.43 & 4.2854 & transverse & $|H - 2 \rightarrow L \rangle$(0.4131) \tabularnewline
 &  &  &  & $|H-1 \rightarrow L +2 \rangle$(0.4130) \tabularnewline
 &  &  &  & \tabularnewline
VIII & 5.83 & 0.1316 & transverse & $|H_{1} \rightarrow L  + 7 ; H - 1 \rightarrow H_{2}  \rangle$(0.2611) \tabularnewline
 &  &  &  & $|H_{2} \rightarrow L \rangle$(0.2524) \tabularnewline
 &  &  &  & \tabularnewline
IX & 6.05 & 0.9742 & transverse & $|H - 1 \rightarrow L + 8 \rangle$(0.4630) \tabularnewline
 &  &  &  & $|H - 1 \rightarrow L + 2 \rangle$(0.4365) \tabularnewline
 &  &  &  & \tabularnewline
X & 6.82 & 0.2656 & transverse & $|H - 2 \rightarrow L + 11 \rangle$(0.3064) \tabularnewline
&  &  &  & $|H - 2 \rightarrow L + 5 \rangle$(0.2136) 
\end{longtable}

\begin{longtable}{ccccl}
\captionsetup{font=footnotesize}
 \caption{Excitation energies ($E$) and many-particle wavefunctions of excited
states corresponding to the peaks in the linear absorption spectrum
of Mg$_{3}$ isosceles triangle isomer -- I (\emph{cf}. Fig. \ref{fig:plot-mg3-iso1}), 
along with the oscillator strength ($f_{12}$) of the transitions.
Symbols $H_{1}$ and  $H_{2}$ denote SOMOs discussed earlier.
The rest of the information is the same as given in the caption for Table \ref{Tab:table_mg3_equil}.
}\label{Tab:table_mg3_iso1}  
\\

\hline \multicolumn{1}{c}{Peak} & \multicolumn{1}{c}{ $E$ (eV)} & \multicolumn{1}{c}{$f_{12}$ } 
 & \multicolumn{1}{c}{Polarization} &\multicolumn{1}{l}{Wave Function}     \\ \hline 
\endfirsthead

\multicolumn{5}{c}%
{{\scshape \tablename\ \thetable{} -- continued from previous page}} \\ 

\hline \multicolumn{1}{c}{Peak} & \multicolumn{1}{c}{ $E$ (eV)} & \multicolumn{1}{c}{$f_{12}$ } 
 & \multicolumn{1}{c}{Polarization} & \multicolumn{1}{l}{Wave Function}     \\ \hline 
\endhead

\hline \multicolumn{5}{r}{{{\scshape Continued on next page}}} \\ 
\endfoot

\hline \hline
\endlastfoot
GS
 &  &  &  & $|H_{1}^{1},H_{2}^{1}\rangle$ (0.8533)\tabularnewline
 &  &  &  & $|H-1\rightarrow H_{2}; H-1 \rightarrow L+4\rangle$(0.1728)\tabularnewline
 &  &  &  & $|H-1\rightarrow H_{2}; H-1 \rightarrow L+11\rangle$(0.1539)\tabularnewline
 &  &  &  & \tabularnewline
I & 1.13 & 0.0243 & in-plane & $|H_{1}\rightarrow L+1\rangle$(0.6636) \tabularnewline
 &  &  &  & $|H_{1}\rightarrow L+12\rangle$(0.3636) \tabularnewline
 &  &  &  & \tabularnewline
II & 1.94 & 0.0415 & in-plane & $|H-1\rightarrow L+1\rangle$(0.5335) \tabularnewline
 &  &  &  & $|H-1\rightarrow H_{1}\rangle$(0.4266) \tabularnewline
 & 1.96 & 0.0178 & in-plane & $|H-2\rightarrow H_{2}\rangle$(0.5213) \tabularnewline
 &  &  &  & $|H-1\rightarrow L + 2\rangle$(0.3225) \tabularnewline
 & 1.99 & 0.0132 & transverse & $|H-1\rightarrow L; H -1 \rightarrow  H_{2} \rangle$(0.5661) \tabularnewline
 &  &  &  & $|H_{2} \rightarrow L \rangle$(0.3167) \tabularnewline
 &  &  &  & \tabularnewline
III  & 2.59 & 0.0455 & in-plane & $|H - 1 \rightarrow H_{1} \rangle$(0.3585) \tabularnewline
 &  &  &  & $|H-1\rightarrow H_{2}; H_{1}\rightarrow L + 3\rangle$(0.3553) \tabularnewline
 &  &  &  & \tabularnewline
IV  & 2.78 & 0.3038 & in-plane & $|H_{1}\rightarrow L+3\rangle$(0.4121) \tabularnewline
 &  &  &  & $|H_{1}\rightarrow L+13\rangle$(0.3549) \tabularnewline
 & 2.95 & 0.0913 & in-plane & $|H-1\rightarrow H_{2}; H_{1} \rightarrow  L + 6 \rangle$(0.6290) \tabularnewline
 &  &  &  & $|H-1\rightarrow H_{2}; H_{1} \rightarrow  L + 16 \rangle$(0.3040) \tabularnewline
 &  &  &  & \tabularnewline
V & 3.36 & 0.1739 & transverse & $|H-2\rightarrow L \rangle$(0.3532) \tabularnewline
 &  &  &  & $|H - 1 \rightarrow L + 1; H - 1 \rightarrow L \rangle$(0.3185) \tabularnewline
 & 3.40 & 1.5229 & in-plane & $| H_{1} \rightarrow  L + 3 \rangle$(0.4774) \tabularnewline
 &  &  &  & $|H-1 \rightarrow  L + 4 \rangle$(0.3656) \tabularnewline
 & 3.49 & 0.9137 & in-plane & $| H - 1  \rightarrow  L + 1 \rangle$(0.4090) \tabularnewline
 &  &  &  & $|H_{1} \rightarrow  L + 2 \rangle$(0.3670) \tabularnewline
 &  &  &  & \tabularnewline
VI & 3.83 & 1.3497 & in-plane & $|H-1\rightarrow L + 1 \rangle$(0.3930) \tabularnewline
 &  &  &  & $|H - 1 \rightarrow H_{2}; H_{1} \rightarrow L + 13 \rangle$(0.2977) \tabularnewline
 & 3.85 & 0.0181 & transverse & $| H_{1} \rightarrow  L ; H - 1 \rightarrow L + 4 \rangle$(0.3416) \tabularnewline
 &  &  &  & $|H_{1} \rightarrow  L + 5 \rangle$(0.3208) \tabularnewline
 &  &  &  & \tabularnewline
VII & 4.19 & 0.2361 & in-plane & $|H - 1 \rightarrow L + 12 \rangle$(0.3806) \tabularnewline
 &  &  &  & $|H-1\rightarrow H_{2}; H_{1} \rightarrow L +18 \rangle$(0.3183) \tabularnewline
 & 4.28 & 0.8469 & in-plane & $|H_{1} \rightarrow L; H - 1 \rightarrow L + 16 \rangle$(0.2572) \tabularnewline
 &  &  &  & $|H-1\rightarrow L+4 \rangle$(0.2451) \tabularnewline
 &  &  &  & \tabularnewline
VIII & 4.47 & 0.0581 & transverse & $|H_{1} \rightarrow L ; H - 1 \rightarrow L + 2 \rangle$(0.3326) \tabularnewline
 &  &  &  & $|H_{1} \rightarrow L ; H - 1 \rightarrow L + 4 \rangle$(0.2381) \tabularnewline
 & 4.48 & 0.4692 & in-plane & $| H - 1  \rightarrow  L + 10 \rangle$(0.2582) \tabularnewline
 &  &  &  & $|H-1 \rightarrow  L + 3 \rangle$(0.2328) \tabularnewline
 &  &  &  & \tabularnewline
IX & 4.83 & 0.0606 & transverse & $|H_{1} \rightarrow L + 12 ; H - 1 \rightarrow H_{2} \rangle$(0.3662) \tabularnewline
 &  &  &  & $|H - 1 \rightarrow L + 6 \rangle$(0.2672) \tabularnewline
 & 4.86 & 0.4879 & in-plane & $| H - 1  \rightarrow  L + 7 \rangle$(0.3323) \tabularnewline
 &  &  &  & $|H_{1} \rightarrow L + 21 ; H - 1 \rightarrow H_{2} \rangle$(0.2667) 
\end{longtable}

\begin{longtable}{ccccl}
\captionsetup{font=footnotesize}
 \caption{Excitation energies ($E$) and many-particle wavefunctions of excited
states corresponding to the peaks in the linear absorption spectrum
of Mg$_{3}$ isosceles triangle isomer -- II (\emph{cf}. Fig. \ref{fig:plot-mg3-iso2}), 
along with the oscillator strength ($f_{12}$) of the transitions.
The rest of the information is the same as given in the caption for Table \ref{Tab:table_mg3_iso1}.
}\label{Tab:table_mg3_iso2}  
\\

\hline \multicolumn{1}{c}{Peak} & \multicolumn{1}{c}{ $E$ (eV)} & \multicolumn{1}{c}{$f_{12}$ } 
 & \multicolumn{1}{c}{Polarization} &\multicolumn{1}{l}{Wave Function}     \\ \hline 
\endfirsthead

\multicolumn{5}{c}%
{{\scshape \tablename\ \thetable{} -- continued from previous page}} \\ 

\hline \multicolumn{1}{c}{Peak} & \multicolumn{1}{c}{ $E$ (eV)} & \multicolumn{1}{c}{$f_{12}$ } 
 & \multicolumn{1}{c}{Polarization} & \multicolumn{1}{l}{Wave Function}     \\ \hline 
\endhead

\hline \multicolumn{5}{r}{{{\scshape Continued on next page}}} \\ 
\endfoot

\hline \hline
\endlastfoot
GS
 &  &  &  & $|H_{1}^{1},H_{2}^{1}\rangle$ (0.8442)\tabularnewline
 &  &  &  & $|H-1\rightarrow  L+10\rangle$(0.4155)\tabularnewline
 &  &  &  & \tabularnewline
I & 0.30 & 0.0113 & transverse & $|H_{1}\rightarrow L+2\rangle$(0.8197) \tabularnewline
 &  &  &  & $|H_{1}\rightarrow L+8\rangle$(0.3015) \tabularnewline
 &  &  &  & \tabularnewline
II & 0.94 & 0.0098 & transverse & $|H_{1} \rightarrow L+5\rangle$(0.6053) \tabularnewline
 &  &  &  & $|H_{1} \rightarrow L + 13 \rangle$(0.5629) \tabularnewline
 &  &  &  & \tabularnewline
III  & 1.80 & 0.2202 & in-plane & $|H_{1} \rightarrow L  + 15  \rangle$(0.6807) \tabularnewline
 &  &  &  & $|H_{1}\rightarrow L + 6 \rangle$(0.5032) \tabularnewline
 &  &  &  & \tabularnewline
IV  & 2.11 & 0.5248 & in-plane & $|H_{2}\rightarrow L+2\rangle$(0.5894) \tabularnewline
 &  &  &  & $|H_{1}\rightarrow L+17 \rangle$(0.3347) \tabularnewline
 &  &  &  & \tabularnewline
V & 2.61 & 1.9694 & in-plane & $|H_{2} \rightarrow L \rangle$(0.7286) \tabularnewline
 &  &  &  & $|H_{2} \rightarrow L; H_{1} \rightarrow L + 10 \rangle$(0.3752) \tabularnewline
 &  &  &  & \tabularnewline
VI & 2.82 & 0.4311 & transverse & $|H_{1} \rightarrow L + 12 \rangle$(0.5004) \tabularnewline
 &  &  &  & $|H_{1} \rightarrow L + 4 \rangle$(0.4888) \tabularnewline
 &  &  &  & \tabularnewline
VII & 3.51 & 1.2606 & in-plane & $|H_{1} \rightarrow L + 7 \rangle$(0.7050) \tabularnewline
 &  &  &  & $|H-2\rightarrow H_{1}\rangle$(0.3603) \tabularnewline
 &  &  &  & \tabularnewline
VIII & 3.86 & 1.5312 & in-plane & $|H_{1} \rightarrow L + 17 \rangle$(0.4732) \tabularnewline
 &  &  &  & $|H - 2 \rightarrow H_{1} \rangle$(0.4363) \tabularnewline
 &  &  &  & \tabularnewline
IX & 4.33 & 0.1162 & in-plane & $|H - 2 \rightarrow L \rangle$(0.4200) \tabularnewline
 &  &  &  & $|H - 2 \rightarrow L ; H_{1} \rightarrow L  + 17 \rangle$(0.3966) \tabularnewline
 &  &  &  & \tabularnewline
X & 4.67 & 0.2938 & transverse & $|H_{1} \rightarrow L + 21\rangle$(0.4191) \tabularnewline
 &  &  &  & $|H - 2 \rightarrow H_{1} \rangle$(0.3579) \tabularnewline
 & 4.68 & 0.2347 & in-plane & $| H_{2}  \rightarrow  L + 12 \rangle$(0.4609) \tabularnewline
 &  &  &  & $|H_{2} \rightarrow L + 13 \rangle$(0.3898) \tabularnewline
 &  &  &  & \tabularnewline
XI & 4.93 & 0.1943 & transverse & $|H_{1} \rightarrow L + 21\rangle$(0.3097) \tabularnewline
 &  &  &  & $|H - 2 \rightarrow L + 2 ; H_{1}  \rightarrow L + 2 \rangle$(0.2765) \tabularnewline
 & 4.91 & 0.1534 & in-plane & $| H_{2}  \rightarrow  L ; H_{1} \rightarrow L + 15 \rangle$(0.3197) \tabularnewline
 &  &  &  & $|H_{2} \rightarrow L ; H_{1} \rightarrow L + 6 \rangle$(0.2914) \tabularnewline
 &  &  &  & \tabularnewline
XII & 5.28 & 0.6223 & transverse & $|H_{1} \rightarrow L + 21\rangle$(0.5283) \tabularnewline
 &  &  &  & $|H_{2}  \rightarrow L + 7 \rangle$(0.5151) \tabularnewline
 & 5.35 & 0.7963 & in-plane & $| H_{2}  \rightarrow  L  + 13 \rangle$(0.3638) \tabularnewline
 &  &  &  & $|H - 2  \rightarrow L + 2 \rangle$(0.3360) \tabularnewline
 &  &  &  & \tabularnewline
XIII & 5.49 & 1.0220 & transverse & $|H_{2} \rightarrow L + 17\rangle$(0.3675) \tabularnewline
 &  &  &  & $|H_{2}  \rightarrow L + 7 \rangle$(0.3610) 
\end{longtable}

\LTcapwidth=\columnwidth
\begin{longtable}{ccccl}
\captionsetup{font=footnotesize}
 \caption{Excitation energies ($E$) and many-particle wavefunctions of excited
states corresponding to the peaks in the linear absorption spectrum
of Mg$_{4}$ pyramid isomer (\emph{cf}. Fig. \ref{fig:plot-mg4-pyra})
, along with the oscillator strength ($f_{12}$) of the transitions.
Absorption of x-,y- and z-polarized light is represented by x,y and z polarization respectively.
The rest of the information is the same as given in the caption for Table \ref{Tab:table_mg2_lin}.
} \label{Tab:table_mg4_pyra} \\

\hline \multicolumn{1}{c}{Peak} & \multicolumn{1}{c}{ $E$ (eV)} & \multicolumn{1}{c}{$f_{12}$ } 
 & \multicolumn{1}{c}{Polarization} &\multicolumn{1}{l}{Wave Function}     \\ \hline 
\endfirsthead

\multicolumn{5}{c}%
{{\scshape \tablename\ \thetable{} -- continued from previous page}} \\ 

\hline \multicolumn{1}{c}{Peak} & \multicolumn{1}{c}{ $E$ (eV)} & \multicolumn{1}{c}{$f_{12}$ } 
 & \multicolumn{1}{c}{Polarization} & \multicolumn{1}{l}{Wave Function}     \\ \hline 
\endhead

\hline \multicolumn{5}{r}{{{\scshape Continued on next page}}} \\ 
\endfoot

\hline \hline
\endlastfoot
GS
 &  &  &  & $|HF\rangle$ (0.8812)\tabularnewline
 &  &  &  & $|H-1\rightarrow L+8;H - 1 \rightarrow L + 17 \rangle$(0.0464)\tabularnewline
 &  &  &  & $|H-1\rightarrow L+17;H - 1 \rightarrow L + 17 \rangle$(0.0457)\tabularnewline
 &  &  &  & \tabularnewline
I  & 2.59 & 0.3403 & x,y,z & $|H - 1 \rightarrow L\rangle$(0.7570) \tabularnewline
   &  &  &  & $|H - 1 \rightarrow L + 9\rangle$(0.3405) \tabularnewline
   &  &  &  & $|H\rightarrow L\rangle$(0.7228) \tabularnewline
   &  &  &  & $|H\rightarrow L + 9\rangle$(0.3247) \tabularnewline
   &  &  &  & $|H - 2 \rightarrow L\rangle$(0.7228) \tabularnewline
   &  &  &  & $|H - 2 \rightarrow L + 9\rangle$(0.3247) \tabularnewline
   &  &  &  & \tabularnewline
II & 3.42 & 0.0024 & x,y,z & $|H - 2 \rightarrow L + 1 \rangle$(0.4094) \tabularnewline
    &  &  &  & $|H - 1 \rightarrow L + 2\rangle$(0.3429) \tabularnewline
    &  &  &  & $|H \rightarrow L  + 1 \rangle$(0.3327) \tabularnewline
    &  &  &  & \tabularnewline
III  & 4.54 & 5.6906 & x,y,z & $|H - 1 \rightarrow L + 8 \rangle$(0.3496) \tabularnewline
      &  &  &  & $|H - 2 \rightarrow L+3\rangle$(0.3390) \tabularnewline
      &  &  &  & $|H - 1 \rightarrow L+3 \rangle$(0.3619) \tabularnewline
      &  &  &  & $|H - 2 \rightarrow L + 2 \rangle$(0.3331) \tabularnewline
      &  &  &  & $|H \rightarrow L + 2 \rangle$(0.3875) \tabularnewline
      &  &  &  & $|H - 1 \rightarrow L + 1\rangle$(0.3777) \tabularnewline
      &  &  &  & \tabularnewline
IV & 4.94 & 0.2391 & x,y,z & $|H - 1 \rightarrow L + 6\rangle$(0.7194) \tabularnewline
     &  &  &  & $|H - 1 \rightarrow L + 20 \rangle$(0.2179) \tabularnewline
     &  &  &  & $|H \rightarrow L+ 5 \rangle$(0.1923) \tabularnewline
     &  &  &  & \tabularnewline
V  & 5.25 & 0.4525 & x,y,z & $|H - 1 \rightarrow L+9\rangle$(0.7656) \tabularnewline
    &  &  &  & $|H - 1 \rightarrow L \rangle$(0.3305) \tabularnewline
    &  &  &  & \tabularnewline
VI & 5.56 & 0.1642 & x,y,z & $|H - 2 \rightarrow L + 16 \rangle$(0.5953) \tabularnewline
     &  &  &  & $|H - 2 \rightarrow L + 7 \rangle$(0.4216) 
\end{longtable}

\begin{longtable}{ccccl}
\captionsetup{font=footnotesize}
 \caption{Excitation energies ($E$) and many-particle wavefunctions of excited
states corresponding to the peaks in the linear absorption spectrum
of Mg$_{4}$ rhombus isomer (\emph{cf}. Fig. \ref{fig:plot-mg4-rho}), 
along with the oscillator strength ($f_{12}$) of the transitions.
The rest of the information is the same as given in the caption for Table \ref{Tab:table_mg3_iso1}.
}\label{Tab:table_mg4_rho}  
\\

\hline \multicolumn{1}{c}{Peak} & \multicolumn{1}{c}{ $E$ (eV)} & \multicolumn{1}{c}{$f_{12}$ } 
 & \multicolumn{1}{c}{Polarization} &\multicolumn{1}{l}{Wave Function}     \\ \hline 
\endfirsthead

\multicolumn{5}{c}%
{{\scshape \tablename\ \thetable{} -- continued from previous page}} \\ 

\hline \multicolumn{1}{c}{Peak} & \multicolumn{1}{c}{ $E$ (eV)} & \multicolumn{1}{c}{$f_{12}$ } 
 & \multicolumn{1}{c}{Polarization} & \multicolumn{1}{l}{Wave Function}     \\ \hline 
\endhead

\hline \multicolumn{5}{r}{{{\scshape Continued on next page}}} \\ 
\endfoot

\hline \hline
\endlastfoot
GS
 &  &  &  & $|H_{1}^{1},H_{2}^{1}\rangle$ (0.8613)\tabularnewline
 &  &  &  & $|H-2\rightarrow L+8\rangle$(0.1004)\tabularnewline
 &  &  &  & $|H-2\rightarrow L+2\rangle$(0.0877)\tabularnewline
 &  &  &  & \tabularnewline
I & 0.98 & 0.0276 & transverse & $|H_{1}\rightarrow L+1\rangle$(0.7603) \tabularnewline
 &  &  &  & $|H_{1}\rightarrow L+10\rangle$(0.3529) \tabularnewline
 &  &  &  & $|H_{1}\rightarrow L+15\rangle$(0.2317) \tabularnewline
 &  &  &  & \tabularnewline
II & 1.71 & 0.0916 & in-plane & $|H_{1} \rightarrow L+16\rangle$(0.5460) \tabularnewline
 &  &  &  & $|H_{1}\rightarrow L + 4 \rangle$(0.5411) \tabularnewline
 &  &  &  & \tabularnewline
III  & 2.71 & 0.2111 & in-plane & $|H - 1 \rightarrow L \rangle$(0.6531) \tabularnewline
 &  &  &  & $|H-2\rightarrow H_{2} \rangle$(0.4453) \tabularnewline
 &  &  &  & \tabularnewline
IV  & 3.22 & 0.6283 & transverse & $|H_{1}\rightarrow L+15 \rangle$(0.7115) \tabularnewline
 &  &  &  & $|H_{1}\rightarrow L+27\rangle$(0.2382) \tabularnewline
 &  &  &  & \tabularnewline
V & 4.00 & 2.5521 & in-plane & $|H_{2} \rightarrow L + 1 ; H -1 \rightarrow L \rangle$(0.3614) \tabularnewline
 &  &  &  & $|H - 2 \rightarrow L + 1 \rangle$(0.3432) \tabularnewline
 &  &  &  & $|H_{1} \rightarrow  L + 2 \rangle$(0.3227) \tabularnewline
 &  &  &  & \tabularnewline
VI & 4.32 & 0.8095 & in-plane & $|H_{1}\rightarrow L + 16 \rangle$(0.4767) \tabularnewline
 &  &  &  & $|H_{1} \rightarrow L + 4 \rangle$(0.3712) \tabularnewline
 & 4.27 & 0.6899 & in-plane & $| H_{1} \rightarrow  L + 6 \rangle$(0.7104) \tabularnewline
 &  &  &  & $|H_{1} \rightarrow  L + 16 \rangle$(0.4365) \tabularnewline
 &  &  &  & \tabularnewline
VII & 4.58 & 2.8937 & in-plane & $|H_{2} \rightarrow L + 8 \rangle$(0.4339) \tabularnewline
 &  &  &  & $|H-1\rightarrow L  \rangle$(0.3119) \tabularnewline
 & 4.71 & 3.2991 & in-plane & $|H - 1 \rightarrow L + 1 \rangle$(0.4522) \tabularnewline
 &  &  &  & $|H-2\rightarrow L \rangle$(0.3518) \tabularnewline
 &  &  &  & \tabularnewline
VIII & 5.22 & 0.9862 & transverse & $|H_{2} \rightarrow  L + 13 \rangle$(0.4954) \tabularnewline
 &  &  &  & $|H_{1} \rightarrow L + 19 \rangle$(0.2824) \tabularnewline
 &  &  &  & $|H-2 \rightarrow  L + 4 \rangle$(0.2503) \tabularnewline
 &  &  &  & \tabularnewline
IX & 5.37 & 0.7381 & in-plane & $|H_{2} \rightarrow L + 1 ; H - 2 \rightarrow L \rangle$(0.4501) \tabularnewline
 &  &  &  & $|H - 1 \rightarrow L + 15 \rangle$(0.3246) \tabularnewline
 &  &  &  & $| H - 1 \rightarrow L + 5  \rangle$(0.2839) \tabularnewline
&  &  &  & \tabularnewline
X & 5.71 & 0.7749 & in-plane & $|H - 2 \rightarrow L + 1 ; H - 2 \rightarrow L \rangle$(0.3911) \tabularnewline
 &  &  &  & $|H - 1 \rightarrow L + 15 \rangle$(0.3156) \tabularnewline
&  &  &  & \tabularnewline
XI & 6.22 & 0.7608 & transverse & $|H_{2} \rightarrow L ; H - 1 \rightarrow L  + 4  \rangle$(0.3898) \tabularnewline
 &  &  &  & $|H_{2} \rightarrow L; H - 1 \rightarrow L + 16  \rangle$(0.3857) 

\end{longtable}

\begin{longtable}{ccccl}
\captionsetup{font=footnotesize}
 \caption{Excitation energies ($E$) and many-particle wavefunctions of excited
states corresponding to the peaks in the linear absorption spectrum
of Mg$_{4}$ square isomer (\emph{cf}. Fig. \ref{fig:plot-mg4-sqr}), 
along with the oscillator strength ($f_{12}$) of the transitions.
The rest of the information is the same as given in the caption for Table \ref{Tab:table_mg3_iso1}.
}\label{Tab:table_mg4_sqr}  
\\

\hline \multicolumn{1}{c}{Peak} & \multicolumn{1}{c}{ $E$ (eV)} & \multicolumn{1}{c}{$f_{12}$ } 
 & \multicolumn{1}{c}{Polarization} &\multicolumn{1}{l}{Wave Function}     \\ \hline 
\endfirsthead

\multicolumn{5}{c}%
{{\scshape \tablename\ \thetable{} -- continued from previous page}} \\ 

\hline \multicolumn{1}{c}{Peak} & \multicolumn{1}{c}{ $E$ (eV)} & \multicolumn{1}{c}{$f_{12}$ } 
 & \multicolumn{1}{c}{Polarization} & \multicolumn{1}{l}{Wave Function}     \\ \hline 
\endhead

\hline \multicolumn{5}{r}{{{\scshape Continued on next page}}} \\ 
\endfoot

\hline \hline
\endlastfoot
GS
 &  &  &  & $|H_{1}^{1},H_{2}^{1}\rangle$ (0.8049)\tabularnewline
 &  &  &  & $|H_{1}\rightarrow L+13\rangle$(0.2283)\tabularnewline
 &  &  &  & $|H_{1}\rightarrow L+7\rangle$(0.1969)\tabularnewline
 &  &  &  & \tabularnewline
I & 1.55 & 0.1823 & in-plane & $|H_{1}\rightarrow L+15\rangle$(0.4554) \tabularnewline
 &  &  &  & $|H_{1}\rightarrow L+10\rangle$(0.3800) \tabularnewline
 &  &  &  & $|H_{1}\rightarrow L+2\rangle$(0.3319) \tabularnewline
 &  &  &  & \tabularnewline
II & 2.56 & 0.1914 & in-plane & $|H - 1  \rightarrow L \rangle$(0.6101) \tabularnewline
 &  &  &  & $|H - 2 \rightarrow H_{2}  \rangle$(0.3295) \tabularnewline
 &  &  &  & $|H_{1}  \rightarrow L + 2   \rangle$(0.2585) \tabularnewline
 &  &  &  & \tabularnewline
III  & 3.30 & 0.1744 & transverse & $|H_{1}  \rightarrow L +  8 \rangle$(0.7723) \tabularnewline
 &  &  &  & $|H_{1} \rightarrow L + 1  \rangle$(0.2820) \tabularnewline
 & 3.34 & 0.6490 & in-plane & $|H_{2} \rightarrow L + 3 ; H_{1} \rightarrow L + 1 \rangle$(0.4620) \tabularnewline
 &  &  &  & $|H_{2} \rightarrow L + 21  \rangle$(0.3048) \tabularnewline
 &  &  &  & \tabularnewline
IV  & 3.62 & 0.5760 & in-plane & $|H_{1}\rightarrow L+2 \rangle$(0.3813) \tabularnewline
 &  &  &  & $|H- 1\rightarrow L\rangle$(0.3564) \tabularnewline
 &  &  &  & $|H_{2}\rightarrow L+21\rangle$(0.3177) \tabularnewline
 &  &  &  & \tabularnewline
V & 4.50 & 20.3052 & in-plane & $|H -1 \rightarrow L \rangle$(0.3918) \tabularnewline
 &  &  &  & $|H_{1} \rightarrow L + 15 \rangle$(0.3026) \tabularnewline
 &  &  &  & $|H_{1} \rightarrow  L + 2 \rangle$(0.3227) \tabularnewline
 &  &  &  & $|H_{2} \rightarrow  L + 1 ; H - 2 \rightarrow L + 1 \rangle$(0.2939) \tabularnewline
 &  &  &  & \tabularnewline
VI & 4.73 & 2.6321 & in-plane & $|H_{1}\rightarrow L + 24 \rangle$(0.5763) \tabularnewline
 &  &  &  & $|H_{1} \rightarrow L + 20 \rangle$(0.3061) \tabularnewline
 & 4.95 & 0.0082 & transverse & $| H - 2  \rightarrow  L ; H - 1 \rightarrow L + 1  \rangle$(0.3240) \tabularnewline
 &  &  &  & $|H_{1} \rightarrow  L + 1; H - 1 \rightarrow L + 3  \rangle$(0.2747) 

\end{longtable}

\LTcapwidth=\columnwidth
\begin{longtable}{ccccl}
\captionsetup{font=footnotesize}
 \caption{Excitation energies ($E$) and many-particle wavefunctions of excited
states corresponding to the peaks in the linear absorption spectrum
of Mg$_{5}$ bi-pyramid isomer (\emph{cf}. Fig. \ref{fig:plot-mg5-biprism})
, along with the oscillator strength ($f_{12}$) of the transitions.
Absorption of x-,y- and z-polarized light is represented by x,y and z polarization respectively.
The rest of the information is the same as given in the caption for Table \ref{Tab:table_mg2_lin}.
} \label{Tab:table_mg5_biprism} \\

\hline \multicolumn{1}{c}{Peak} & \multicolumn{1}{c}{ $E$ (eV)} & \multicolumn{1}{c}{$f_{12}$ } 
 & \multicolumn{1}{c}{Polarization} &\multicolumn{1}{l}{Wave Function}     \\ \hline 
\endfirsthead

\multicolumn{5}{c}%
{{\scshape \tablename\ \thetable{} -- continued from previous page}} \\ 

\hline \multicolumn{1}{c}{Peak} & \multicolumn{1}{c}{ $E$ (eV)} & \multicolumn{1}{c}{$f_{12}$ } 
 & \multicolumn{1}{c}{Polarization} & \multicolumn{1}{l}{Wave Function}     \\ \hline 
\endhead

\hline \multicolumn{5}{r}{{{\scshape Continued on next page}}} \\ 
\endfoot

\hline \hline
\endlastfoot
GS
 &  &  &  & $|HF\rangle$ (0.8806)\tabularnewline
 &  &  &  & $|H\rightarrow L+1;H \rightarrow L + 1 \rangle$(0.0981)\tabularnewline
 &  &  &  & $|H\rightarrow L+1;H  \rightarrow L + 9 \rangle$(0.0852)\tabularnewline
 &  &  &  & \tabularnewline
I  & 3.59 & 0.1291 & x & $|H - 1 \rightarrow L + 4 \rangle$(0.7779) \tabularnewline
   &  &  &  & $|H - 1 \rightarrow L + 18\rangle$(0.2218) \tabularnewline
   &  &  &  & $|H - 1 \rightarrow L \rangle$(0.2131) \tabularnewline
   &  &  &  & \tabularnewline
II & 4.06 & 0.2547 & z & $|H - 2 \rightarrow L \rangle$(0.6090) \tabularnewline
    &  &  &  & $|H - 3 \rightarrow L + 2 \rangle$(0.3980) \tabularnewline
    &  &  &  & $|H - 2 \rightarrow L  + 4 \rangle$(0.2984) \tabularnewline
    &  &  &  & \tabularnewline
III  & 4.30 & 0.2060 & x & $|H - 3 \rightarrow L + 3 \rangle$(0.6406) \tabularnewline
      &  &  &  & $|H - 2 \rightarrow L+5\rangle$(0.4396) \tabularnewline
      &  &  &  & $|H - 2 \rightarrow L+17 \rangle$(0.2348) \tabularnewline
      &  &  &  & \tabularnewline
IV & 4.50 & 0.5632 & z & $|H \rightarrow L + 24\rangle$(0.5166) \tabularnewline
     &  &  &  & $|H - 1 \rightarrow L + 5 \rangle$(0.4826) \tabularnewline
     &  &  &  & $|H - 3 \rightarrow L + 2 \rangle$(0.2873) \tabularnewline
     &  &  &  & \tabularnewline
V  & 5.37 & 13.1325 & y & $|H \rightarrow L+1\rangle$(0.6687) \tabularnewline
    &  &  &  & $|H - 1 \rightarrow L + 3\rangle$(0.3411) \tabularnewline
    &  &  &  & $|H - 2 \rightarrow L + 2\rangle$(0.3332) \tabularnewline
    &  &  &  & \tabularnewline
VI & 5.56 & 7.5501 & z & $|H - 1 \rightarrow L + 5 \rangle$(0.4801) \tabularnewline
     &  &  &  & $|H - 2 \rightarrow L  \rangle$(0.3236) \tabularnewline
     &  &  &  & $|H - 2 \rightarrow L + 4  \rangle$(0.3233) \tabularnewline
     &  &  &  & \tabularnewline
VII  & 6.03 & 2.9109 & x & $|H  - 1 \rightarrow L+6\rangle$(0.4562) \tabularnewline
    &  &  &  & $|H \rightarrow L + 27\rangle$(0.4149) \tabularnewline
    &  &  &  & $|H \rightarrow L + 12\rangle$(0.2982) 
%
\end{longtable}

\LTcapwidth=\columnwidth
\begin{longtable}{ccccl}
\captionsetup{font=footnotesize}
 \caption{Excitation energies ($E$) and many-particle wavefunctions of excited
states corresponding to the peaks in the linear absorption spectrum
of Mg$_{5}$ pyramid isomer (\emph{cf}. Fig. \ref{fig:plot-mg5-param})
, along with the oscillator strength ($f_{12}$) of the transitions.
In-plane and transverse polarization corresponds to the absorption
due to light polarized in and perpendicular to the base plane of the pyramidal isomer respectively. 
The rest of the information is the same as given in the caption for Table \ref{Tab:table_mg2_lin}.
} \label{Tab:table_mg5_param} \\

\hline \multicolumn{1}{c}{Peak} & \multicolumn{1}{c}{ $E$ (eV)} & \multicolumn{1}{c}{$f_{12}$ } 
 & \multicolumn{1}{c}{Polarization} &\multicolumn{1}{l}{Wave Function}     \\ \hline 
\endfirsthead

\multicolumn{5}{c}%
{{\scshape \tablename\ \thetable{} -- continued from previous page}} \\ 

\hline \multicolumn{1}{c}{Peak} & \multicolumn{1}{c}{ $E$ (eV)} & \multicolumn{1}{c}{$f_{12}$ } 
 & \multicolumn{1}{c}{Polarization} & \multicolumn{1}{l}{Wave Function}     \\ \hline 
\endhead

\hline \multicolumn{5}{r}{{{\scshape Continued on next page}}} \\ 
\endfoot

\hline \hline
\endlastfoot
GS
 &  &  &  & $|HF\rangle$ (0.8754)\tabularnewline
 &  &  &  & $|H\rightarrow L;H \rightarrow L \rangle$(0.0436)\tabularnewline
 &  &  &  & $|H\rightarrow L+1;H - 2   \rightarrow L + 17 \rangle$(0.0424)\tabularnewline
 &  &  &  & \tabularnewline
I  & 2.24 & 0.1278 & transverse & $|H - 1 \rightarrow L \rangle$(0.7746) \tabularnewline
   &  &  &  & $|H - 1 \rightarrow L + 7\rangle$(0.2334) \tabularnewline
   &  &  &  & $|H - 1 \rightarrow L + 3 \rangle$(0.1663) \tabularnewline
   & 2.40 & 0.1665 & in-plane & $|H \rightarrow L + 2 \rangle$(0.7422) \tabularnewline
   &  &  &  & $|H \rightarrow L + 12 \rangle$(0.2494) \tabularnewline
   &  &  &  & $|H \rightarrow L + 4 \rangle$(0.2412) \tabularnewline
   &  &  &  & \tabularnewline
II & 2.72 & 0.0954 & in-plane & $|H - 1 \rightarrow L + 2 \rangle$(0.6894) \tabularnewline
    &  &  &  & $|H - 1 \rightarrow L + 12 \rangle$(0.2402) \tabularnewline
    &  &  &  & $|H \rightarrow L  + 8 \rangle$(0.1888) \tabularnewline
    &  &  &  & \tabularnewline
III  & 3.47 & 1.0415 & transverse & $|H - 1 \rightarrow L + 3 \rangle$(0.5892) \tabularnewline
      &  &  &  & $|H - 1 \rightarrow L + 14\rangle$(0.3746) \tabularnewline
      &  &  &  & $|H - 1 \rightarrow L+5 \rangle$(0.2819) \tabularnewline
      & 3.50 & 2.1720 & in-plane & $|H - 2 \rightarrow L \rangle$(0.5507) \tabularnewline
      &  &  &  & $|H \rightarrow L + 4 \rangle$(0.3754) \tabularnewline
      &  &  &  & $|H \rightarrow L + 15 \rangle$(0.3084) \tabularnewline
      &  &  &  & \tabularnewline
IV & 4.22 & 3.6965 & in-plane & $|H - 2 \rightarrow L + 1\rangle$(0.4224) \tabularnewline
     &  &  &  & $|H - 2 \rightarrow L + 3 \rangle$(0.3548) \tabularnewline
     &  &  &  & $|H \rightarrow L + 8 \rangle$(0.2683) \tabularnewline
     &  &  &  & \tabularnewline
V  & 4.43 & 1.1481 & transverse & $|H  - 1 \rightarrow L+5\rangle$(0.5337) \tabularnewline
    &  &  &  & $|H - 1 \rightarrow L + 19\rangle$(0.2652) \tabularnewline
    &  &  &  & $|H - 2 \rightarrow L + 2\rangle$(0.2031) \tabularnewline
    &  &  &  & \tabularnewline
VI & 4.58 & 1.7660 & in-plane & $|H \rightarrow L ; H \rightarrow L + 2  \rangle$(0.3512) \tabularnewline
     &  &  &  & $|H - 1 \rightarrow L ; H \rightarrow L + 2 \rangle$(0.3056) \tabularnewline
     &  &  &  & $|H - 2 \rightarrow L + 3  \rangle$(0.2872) \tabularnewline
     &  &  &  & \tabularnewline
VII  & 4.80 & 1.3936 & transverse & $|H  - 1 \rightarrow L+5\rangle$(0.3849) \tabularnewline
       &  &  &  & $|H - 1 \rightarrow L + 3\rangle$(0.3234) \tabularnewline
       &  &  &  & $|H - 1 \rightarrow L + 7\rangle$(0.2948) \tabularnewline
       & 4.87 & 0.5048 & in-plane & $|H  - 1 \rightarrow L+10\rangle$(0.4101) \tabularnewline
       &  &  &  & $|H - 1 \rightarrow L + 21\rangle$(0.4013) \tabularnewline
       &  &  &  & $|H \rightarrow L + 12\rangle$(0.3200) \tabularnewline
      &  &  &  & \tabularnewline
VIII & 5.25 & 0.6565 & in-plane & $|H \rightarrow L + 10 \rangle$(0.4249) \tabularnewline
     &  &  &  & $|H \rightarrow L + 17 \rangle$(0.4106) \tabularnewline
     &  &  &  & $|H - 1 \rightarrow L ; H \rightarrow L + 2  \rangle$(0.3123) \tabularnewline
     &  &  &  & \tabularnewline
IX  & 5.46 & 1.1843 & transverse & $|H  \rightarrow L+20\rangle$(0.4637) \tabularnewline
       &  &  &  & $|H - 1 \rightarrow L + 14\rangle$(0.3195) \tabularnewline
       &  &  &  & $|H \rightarrow L + 9\rangle$(0.3166) \tabularnewline
      &  &  &  & \tabularnewline
 X   & 5.88 & 0.8339 & in-plane & $|H - 1 \rightarrow L+21\rangle$(0.4148) \tabularnewline
       &  &  &  & $|H - 3 \rightarrow L + 2\rangle$(0.3584) \tabularnewline
       &  &  &  & $|H - 1  \rightarrow L + 10\rangle$(0.3497) \tabularnewline
      &  &  &  & \tabularnewline
 XI   & 6.27 & 0.7081 & transverse & $|H \rightarrow L+1 ; H \rightarrow L + 1 \rangle$(0.3266) \tabularnewline
       &  &  &  & $|H - 2 \rightarrow L + 8\rangle$(0.2525) \tabularnewline
       &  &  &  & $|H - 2  \rightarrow L + 8\rangle$(0.2525) 
%
\end{longtable}

  \end{footnotesize}
 	\pagestyle{fancy}
 	\renewcommand{\sectionmark}[1]{\markright{\thesection\ #1}}
 	\fancyhf{}
 	\lhead{Bibliography}
  	\rhead{\href{home.iitb.ac.in/~ravindra.shinde}{Ravindra Shinde}}
 	\cfoot{Page \thepage\ of \pageref{LastPage}}

\begin{singlespace}
\bibliography{main}
\bibliographystyle{ravindra-thesis}
\end{singlespace}

\pagestyle{fancy}
\lhead{Acknowledgements}
\cfoot{Page \thepage\ of \pageref{LastPage}}
\cleardoublepage
\begin{center}
 {\bf {\Huge Acknowledgements}}
\end{center}
 I would like to express my deep gratitude to Professor Alok Shukla, my thesis advisor, for skillful guidance, encouragement and useful
critiques in the research which led to this thesis. He is a valuable source of insight and research advice while at the same 
time allowing me sufficient freedom to pursue new ideas. 
I thank my research progress committee members, Prof. P. P. Singh, Prof. S. Dhar and Prof. Kedar Damle, for their 
critical evaluation and constant support. \par
During this M.Sc - Ph.D dual degree tenure, I have learned a lot from a number of teachers. I feel very lucky to learn science from them.\par
I would also like to thank the members of the theoretical condensed group, my batchmates, and my
other friends at IIT. They made my graduate career enjoyable and they have contributed
greatly to my understanding of physics. \par
I also thank my research funding agencies, Council for Scientific and Industrial Research (09/087(0600)/2010-EMR-I) and Indian Institute
of Technology Bombay (IIT Bombay) for providing me research fellowship. I would like to thank CSIR, Department of Science
and Technology -- Government of India, IIT Bombay, CECAM, KU Leuven and  Centro de Ciencias de Benasque Pedro Pascual
 for providing travel grants for attending international conferences. I also acknowledge National Param Supercomputing 
Facility -- Param Yuva II, and IIT Bombays' High Performance Computing Facility -- SpaceTime. \par
I salute my mother and father, who worked very hard only to fulfil the dreams I had. I will always stay indebted to them.
I would also like to thank my wife and my colleague Meenakshi. She has been loving, supportive, patient, and someone who I 
could always depend upon for honest discussions. I also thank her for blessing me with the cutest daughter, Aquila.
\par
\vspace{1cm}
Ravindra Shinde 

\end{document}